\documentclass[seceq]{ptptex}

\usepackage{graphicx}
\usepackage{wrapft}
\usepackage{bm}

\title{
Physics Achievements from the Belle Experiment
}

\author{
Jolanta {Brodzicka}$^1$,
Thomas {Browder}$^2$,
Paoti {Chang}$^3$,
Simon {Eidelman}$^4$, 
Bostjan {Golob}$^{5, 6}$,
Kiyoshi {Hayasaka}$^7$, 
Hisaki {Hayashii}$^8$, 
Toru {Iijima}$^7$, 
Kenji {Inami}$^7$, 
Kay {Kinoshita}$^9$,
Youngjoon {Kwon}$^{10}$, 
Kenkichi {Miyabayashi}$^8$, 
Gagan {Mohanty}$^{11}$, 
Mikihiko {Nakao}$^{12}$, 
Hideyuki {Nakazawa}$^{13}$, 
Stephen {Olsen}$^{14}$,
Yoshihide {Sakai}$^{12}$, 
Christoph {Schwanda}$^{15}$, 
Alan {Schwartz}$^9$, 
Karim {Trabelsi}$^{12,}$\footnote{E-mail: karim.trabelsi@kek.jp},
Sadaharu {Uehara}$^{12}$, 
Shoji {Uno}$^{12}$, 
Yasushi {Watanabe}$^{16}$, 
Anze {Zupanc}$^{17}$ \\
(for the Belle Collaboration)
}

\inst{
$^1$H. Niewodniczanski Institute of Nuclear Physics, Krakow \\
$^2$University of Hawaii, Honolulu, Hawaii 96822\\
$^3$Department of Physics, National Taiwan University, Taipei\\
$^4$Budker Institute of Nuclear Physics SB RAS and Novosibirsk State 
University, Novosibirsk 630090\\
$^5$Faculty of Mathematics and Physics, University of Ljubljana, Ljubljana\\
$^6$Jozef Stefan Institute, Ljubljana\\
$^7$Kobayashi-Maskawa Institute, Nagoya University, Nagoya\\
$^8$Nara Women's University, Nara\\
$^9$University of Cincinnati, Cincinnati, Ohio 45221\\
$^{10}$Yonsei University, Seoul\\
$^{11}$Tata Institute of Fundamental Research, Mumbai\\
$^{12}$High Energy Accelerator Research Organization (KEK), Tsukuba\\
$^{13}$National Central University, Chung-li\\
$^{14}$Seoul National University, Seoul\\
$^{15}$Institute of High Energy Physics, Vienna\\
$^{16}$Kanagawa University, Yokohama\\
$^{17}$Institut f\"ur Experimentelle Kernphysik, Karlsruher Institut 
f\"ur Technologie, Karlsruhe
}
\abst{
The Belle experiment, running at the KEKB $e^+ e^-$ asymmetric energy 
collider during the first decade of the century, achieved its original 
objective of measuring precisely differences between particles and 
anti-particles in the $B$ system. After collecting 1000 fb$^{-1}$ of data 
at various $\Upsilon$ resonances, Belle also obtained the many other
physics results described in this article.
}

\begin{document}

\def\bz{{B^0}}
\def\bb{{\overline{B}{}^0}}
\def\pip{{\pi^+}}
\def\pim{{\pi^-}}
\def\rhop{{\rho^+}}
\def\rhom{{\rho^-}}
\def\calS{{\cal S}}
\def\calA{{\cal A}}
\def\dt{{\Delta t}}
\def\dmd{{\Delta m_d}}
\newcommand{\bbbar}{B\overline{B}{}}
\newcommand{\spipi}{{\calS_{\pi \pi}}}
\newcommand{\apipi}{{\calA_{\pi \pi}}}
\newcommand*{\dwl}{\ensuremath{{\Delta w_l}}}
\newcommand*{\fq}{\ensuremath{q}}

\def\slash#1{\not\!#1}
\def\slashb#1{\not\!\!#1}
\def\delsla{\not\!\partial}

\maketitle

\tableofcontents

\section{Introduction}

In the sections that follow, we describe the physics accomplishments
of the Belle experiment, which ran at the KEKB~\cite{detector_KEKB} 
$e^+ e^-$ asymmetric energy collider in Tsukuba, Japan
between 1999 and 2010. KEKB broke all records for integrated
and instantaneous luminosity for a high energy accelerator. 
As a result Belle was able to
integrate over 1000 fb$^{-1}$ or one inverse attobarn of data.

Belle was designed and optimized for the observation of $CP$ violation 
in the $B$ meson system. In 2001, Belle (along with BaBar, a competing
and similar experiment located in Stanford, California) was 
indeed able to observe
large $CP$ asymmetries in $B$ decays, which were expected and consistent with 
the theoretical proposal of Kobayashi and Maskawa. This experimental
result was explicitly recognized in the 2008 Physics Nobel Prize citation.

Nevertheless, the Belle spectrometer was a general
purpose device with reasonable solid coverage as well as high quality
vertexing with silicon strip detectors, charged particle tracking
with a central drift chamber, and excellent electromagnetic calorimetry
as well as muon and $K_L$ detection. These detector
capabilities allowed Belle to not only
cover most of the important topics in $B$ physics (in addition to
the $CP$ violation measurements) but also to make important
discoveries in charm physics, tau lepton physics, hadron 
spectroscopy, and two-photon physics.

Most of the Belle luminosity was recorded on or near the $\Upsilon(4S)$
resonance, which is the optimal center of mass (CM) energy for the
production of $B \bar{B}$ pairs used in $B$ physics analysis. However,
KEKB has some flexibility in energy and Belle also recorded a series
of {\it unique data sets} at the $\Upsilon(1S)$, $\Upsilon(2S)$, and
$\Upsilon(5S)$ resonances. The latter data set is of special interest
in hadron spectroscopy as a large number of new and some exotic
states were found in analyses of this sample.

\section{The Belle detector and its data samples}
\label{chap_detector}
\subsection{Overview} 
The Belle detector is located at the interaction region 
of an asymmetric energy $e^+e^-$ collider, called
KEKB~\cite{detector_KEKB}. 
Belle is optimized to measure time-dependent $CP$ violation
in $B$-meson decay. 
Therefore, the detector has good vertex resolution 
and good particle identification capabilities for leptons and
hadrons. The detector material is minimized to reduce multiple
scattering for charged particles 
and to maintain high efficiency and good resolution for low energy photons. 
The acceptance is asymmetric 
(covering the polar angle region from $17^{\circ}$ to $150^{\circ}$) 
to match the boost from the asymmetric 8 on 3.5 GeV energy collisions. 
Belle is a general purpose $4\pi$ detector, 
which can accommodate various physics programs, including studies of
$\tau$ pairs, two-photon physics, and $q\bar{q}$ continuum processes.
 
\begin{figure}[t]
\begin{center}
\includegraphics[width=0.95\textwidth]{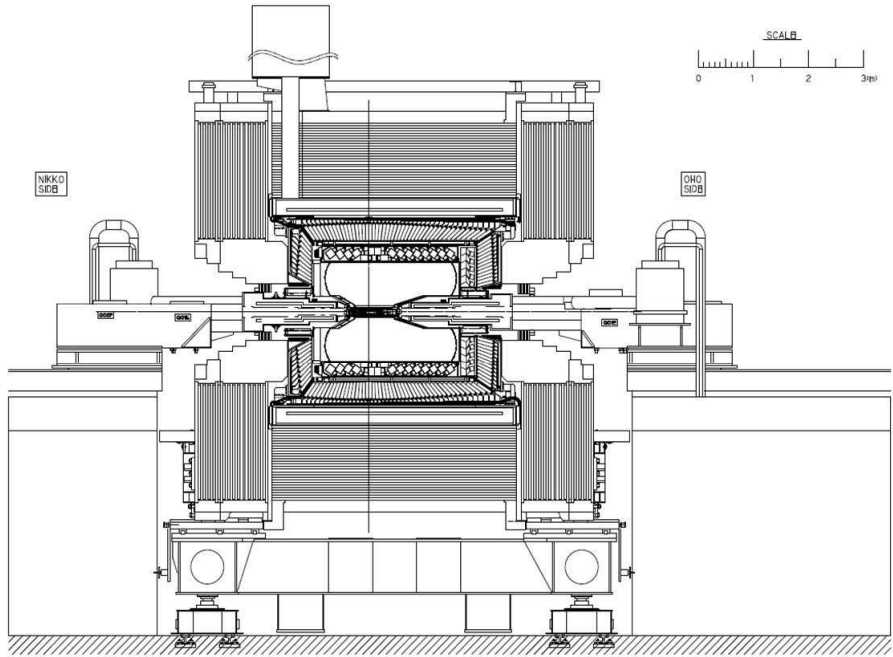}
\end{center}
 \caption{Side view of the Belle detector.}
\label{detector:Belle}
\end{figure}

Figure~\ref{detector:Belle} shows the Belle detector
configuration. The detector is built around a 1.5 Tesla 
superconducting solenoid and iron structure. 
The beam crossing angle is $\pm$11 mrad. 
$B$-meson decay vertices are measured by a double-sided silicon vertex detector
(SVD) situated around a cylindrical beryllium beam pipe. 
There are two inner detector configurations: 
SVD1 (three layers before the summer of 2003) and SVD2 (four layers). 
Charged particle tracking is provided by a central drift chamber
(CDC). Particle identification is provided 
by $dE/dx$ measurements in the CDC, 
aerogel Cerenkov counters (ACC), and time-of-flight counters (TOF) 
situated radially outside of the CDC. 
Electromagnetic showers are detected by an array of CsI(Tl) crystals
(ECL) located inside the solenoid coil. 
Muons and $K_L$ mesons are identified 
by arrays of resistive plate counters (KLM) interspersed in the iron
yoke. An array of bismuth germanate oxide (BGO) crystals called the
extreme forward calorimeter (EFC) is located on the surface 
of the cryostats of the compensation solenoid magnets 
in the forward and backward directions. 
The EFC is used as an active shield against beam background
and also measures the online luminosity.
Each subdetector is briefly described in the 
following subsections; more detailed information 
is available in Ref.~\cite{detector_NIM}.

\subsection{The beam pipe}
The Belle detector beam pipe~\cite{detector_bp} is connected 
to the KEKB accelerator beam pipe. 
The pipe is a double-wall beryllium structure with 
liquid paraffin cooling to remove the heat generated by the beams. 
The inner diameter is only 30 mm (40 mm) for the SVD2 (SVD1)
inner detector configuration to optimize vertex resolution. A 
10~$\mu$m-thick layer of gold is sputtered inside the beryllium wall 
to prevent synchrotron radiation photons from entering the detector.

\subsection{The SVD detector}
The SVD consists of four layers 
in a barrel-only design (SVD2~\cite{detector_SVD2}). 
Each layer is independently constructed and consists of ladders. 
Each ladder consists of double-sided silicon strip detectors (DSSDs) 
reinforced by support ribs. 
The design uses two types of DSSDs. For the second version, the DSSDs 
of the 4th layer are shorter and wider than those of the other layers.
The readout chain for the DSSDs is based on a VA1 (Viking architecture) 
integrated circuit. The back-end electronics is a system of flash 
analog-to-digital converters (FADCs) 
and field programmable gate arrays (FPGAs), 
which perform online common-mode noise subtraction, 
data sparsification, and data formatting. 

Before the summer of 2003, 
there were three DSSD layers with slightly less angular coverage
(SVD1). In addition, the beam pipe diameter was larger (40~mm versus
30 mm for the final data taking configuration). 
 
\subsection{The CDC detector}
The structure of the CDC is asymmetric 
in the $z$ direction in order to optimize the angular coverage. 
The longest wires are 2400~mm long. 
The inner radius extends to 80~mm without any walls 
in order to obtain good tracking efficiency for low momentum tracks with
minimal intervening material. The outer radius is 880~mm. 
The forward and backward smaller-$r$ regions have conical shapes 
in order to clear the accelerator components 
while maximizing the acceptance. 
A low-$Z$ gas, a $He$\---$C_2H_6(50/50)$ gas mixture, is used 
in order to minimize multiple scattering. 
The chamber has 50 cylindrical layers, 
each containing between three and six axial or small-angle stereo
layers, and three cathode strip layers. 
The CDC has a total of 8400 drift cells. 
We chose three layers for the two innermost stereo super-layers 
and four layers for the three outer stereo super-layers 
in order to provide a highly efficient and fast $z$-trigger, which
is combined with the information from the cathode strips. 

During the summer of 2003, the cathode part of the CDC was replaced 
by a compact small cell type drift chamber 
in order to make enough space for the SVD2 vertex detector. 
The cell sizes are only 5~mm in both the radial and azimuthal directions 
to accommodate two layers (128 cells per layer) in a limited space. 
The maximum drift time is rather small ($\sim$100~nsec); 
this feature can provide the first trigger signal for the SVD2 readout latch. 

\subsection{The ACC subsystem}
The ACC consists of 960 counter modules 
segmented into 60 cells in the $\phi$ direction 
for the barrel part and 228 modules arranged in 5 concentric layers 
for the forward end-cap part of the detector. 
All the counters are arranged in a semi-projective geometry, 
pointing to the interaction point (IP). 
In order to obtain good pion/kaon separation to cover the entire 
kinematical range of two-body $B$ decays, 
the refractive indices of the aerogel blocks vary between 1.01 and
1.03, depending on their polar angle region. 
Five aerogel tiles are stacked in a thin (0.2~mm thick) 
aluminum box of approximate dimensions
$12\times12\times\times12$~cm$^3$. 
In order to detect Cerenkov light effectively, 
one or two fine mesh-type photomultiplier tubes (FM-PMTs), 
which are operated in a 1.5 T magnetic field, 
are attached directly to the aerogel on the sides of the box. 
We use Hamamatsu Photonics PMTs 
of three different diameters: 3, 2.5, and 2 inches, 
depending on the refractive index of the aerogel block, 
in order to obtain a uniform response for relativistic particles. 

\subsection{The TOF subsystem}
The TOF system consists of 
128 TOF counters and 64 thin trigger scintillation counters (TSC). 
Two trapezoidal shaped TOF counters 
and one TSC counter, with a 1.5~cm intervening radial gap, form a
single module. In total, 64 TOF/TSC modules 
located at a radius of 1.2 m from the IP cover 
a polar angle range from $34^{\circ}$ to $120^{\circ}$. 
The thicknesses of the scintillators (BC408, Bicron) 
are 4~cm and 0.5~cm for the TOF and TSC counters, respectively. 
The fine mesh PMTs operating inside the
1.5 T magnetic field, with a 2-inch diameter and 24 stages, 
were attached to both ends of the 
TOF counter with an air gap of 0.1~mm. 
For the TSCs, the tubes were glued 
to the light guides at the backward ends of the counters.

\subsection{The ECL detector subsystem}
	A highly segmented array of CsI(Tl) crystals 
with silicon photodiode readout were selected for the 
ECL~\cite{detector_ECL}.
Each crystal has a tower-like shape and 
is arranged so that it nearly points to the IP. 
The calorimeter covers the full Belle angular region. 
A small gap between the barrel and end-cap crystals 
provides a pathway for the cables and room for 
supporting members of the inner detectors. 
The entire system contains 8736 counters. 
The size of each crystal is typically 55~mm $\times$ 55~mm (front
face) and 65~mm $\times$ 65~mm (rear face). 
The 30~cm length (16 radiation lengths) is chosen to 
avoid deterioration of the energy resolution for high energy gammas 
due to fluctuations in the shower leakage out the rear of the counter. 
Each counter is read out by an independent pair of silicon PIN
photodiodes 
and charge sensitive preamplifiers attached at the end of the crystal. 

\subsection{The KLM detector}
The KLM consists of alternating layers of charged particle 
detectors and 4.7~cm-thick iron plates, 
which are the magnetic flux return in the barrel and 
endcap regions~\cite{detector_KLM}. There are 15 detector layers 
and 14 iron layers in the octagonal barrel region 
and 14 detector layers in each of the forward and backward
end-caps. The iron plates provide a total of 3.9 interaction lengths 
of material for a particle traveling normal to the detector planes. 
The detection of charged particles is provided by 
glass-electrode resistive plate counters (RPCs). 
The resistive plate counters have two parallel plate electrodes 
of 2.4~mm-thick commercially available float glass. 
The bulk resistivity of the glass is $10^{12}-10^{13} \Omega$ cm at room
temperature. To distribute the high voltage on the glass, 
the outer surface was coated with carbon ink, which achieves a
surface resistivity of $10^6 -10^7 \Omega/$square. The discharge
signal can then be obtained from external pickup strips. 
The readout of 38K pickup strips is accomplished 
with the use of custom-made VME-based discriminator/time multiplexing boards. 

\subsection{Trigger and data acquisition}
The Belle trigger system consists of a Level-1 hardware trigger and
a Level-3 software trigger~\cite{trigger_DAQ}. 
The latter is implemented in the online computer farm. 
The Level-1 trigger system consists of a subdetector trigger
system and a central trigger system called the global decision logic
(GDL). The subdetector trigger systems are based on two categories: 
track triggers and energy triggers. 
The CDC and TOF are used to yield trigger signals for charged
particles. The ECL trigger system~\cite{trigger_ECL} provides 
triggers based on total energy deposit and cluster counting 
of crystal hits. 
These two categories have sufficient redundancy.
The KLM trigger gives additional information on muons. 
The EFC triggers are used for tagging two-photon events 
as well as Bhabha events.   

The Belle data acquisition system used one type of 
multi-hit TDC modules for all subsystems except for the SVD. 
The signal pulse height is recorded as timing information using 
a charge to time conversion chip (Q-to-T chip). 
Precise timing information in the TOF is recorded 
by commercial TDC modules with special time expansion modules. 
The TDC modules did not have a pipe-line readout scheme. 
Therefore, the readout deadtime is large (around 30~$\mu$sec). 
There were several electronics upgrades in order to reduce deadtime
carried out during the latter parts of the experiment. The TDC modules 
were gradually replaced with pipe-lined TDCs (2.8~$\mu$sec) for
most of the subdetectors in the 2007\---2009 running
period~\cite{detector_COPPER}.
It was carefully checked that these electronics upgrades did not
affect the data quality.

Belle turned off the detector high voltage during beam injection, 
as do other experiments. 
The KEKB injection time was slightly longer than at PEP-II (the 
collider hosting the BaBar experiment) and 
the average efficiency was lower. In order to reduce such losses, 
a continuous injection scheme~\cite{detector_KEKB} was implemented 
in January 2004. The detector high voltage was kept on 
and the trigger signals were vetoed for a 3.5~msec interval
just after each beam injection. 
This scheme leads to 3.5\% deadtime only in the case of 
a 10 Hz injection rate. After adopting continuous injection, 
the KEKB machine became more stable and the peak luminosity 
improved due to the leveling of the beam currents. 

\subsection{Detector performance}
The charged track reconstruction mainly uses the CDC.  
Good momentum resolution is obtained by combining CDC tracks
together with SVD hit information, especially for low momentum 
tracks, thanks to the limited amount of intervening material.
The following expression gives the momentum resolution 
for a charged track as a function of its
transverse momentum:
\begin{equation}
\sigma_{p_t}/p_t = 0.0019 \; p_t[{\rm GeV}/c] \oplus 0.0030/\beta 
\end{equation}
The typical mass resolution of $D^0$ mesons is 5~MeV in hadronic
events. The $z$-vertex resolution is 61~$\mu$m in the 
$J/\psi \to \mu^+\mu^-$ mode. 
A similar resolution is also obtained in the $r-\phi$ plane. 
The energy resolution of the ECL is 1.7\% for Bhabha events. 
A $\pi^0$ mass resolution of 4.8~MeV is obtained 
for low momentum photons in hadronic events. 
\begin{figure}[htb]
\parbox{\halftext}{
\centerline{\includegraphics[width=6.5cm]{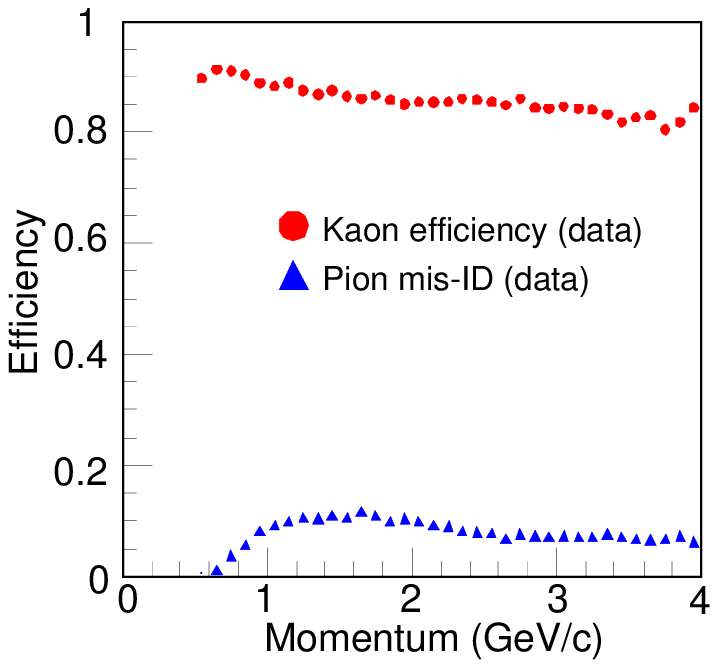}}
\caption{Kaon identification efficiency and fake rate as a function 
of momentum.}
\label{detector:kpi}
}
\hfill
\parbox{\halftext}{
\centerline{\includegraphics[width=6.5cm]{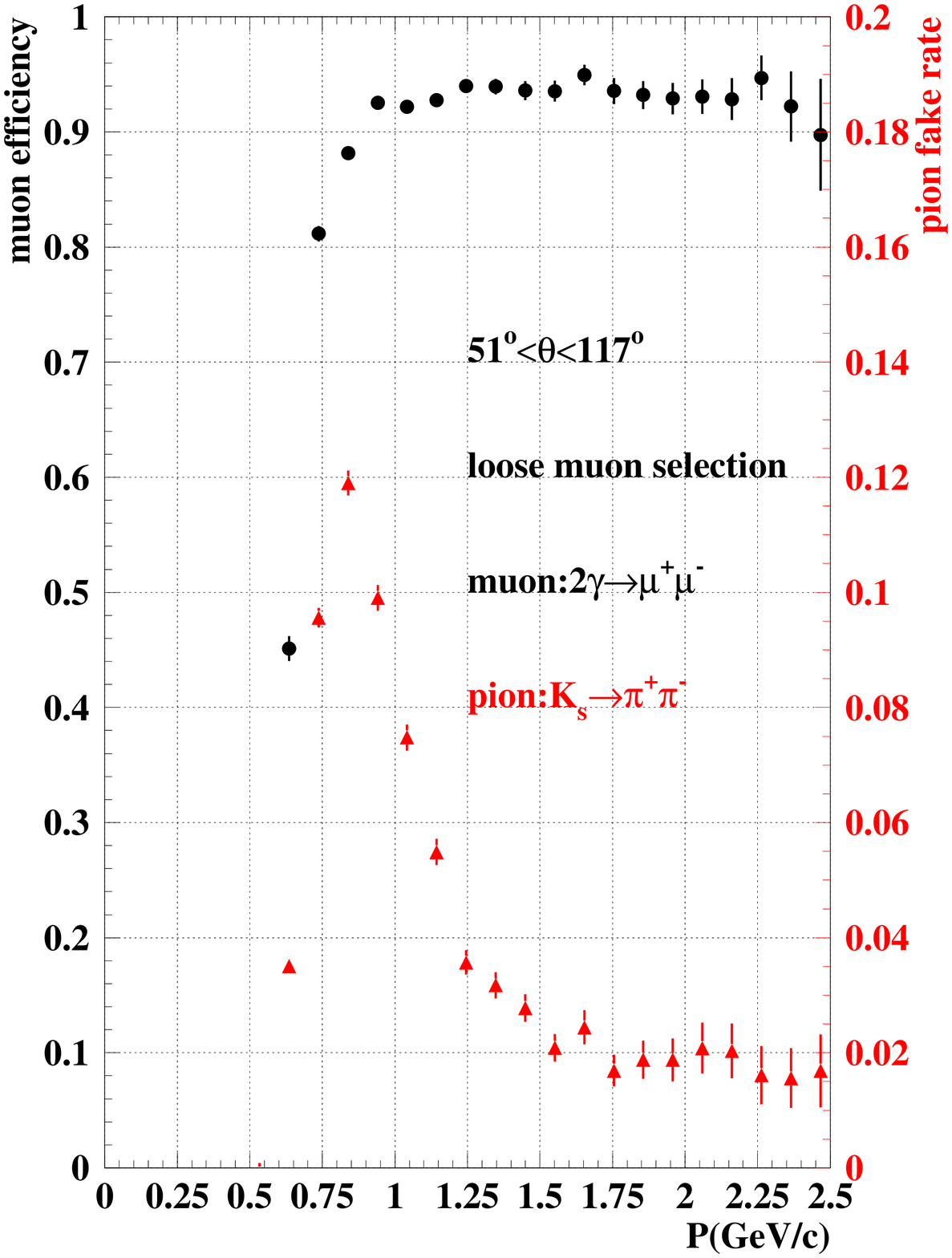}}
\caption{Muon identification efficiency and fake rate 
as a function of momentum.}
\label{detector:muid}}
\end{figure}

Pion/kaon/proton separation is obtained by
combining ACC, TOF, and CDC $dE/dx$ information. 
The kaon efficiency and the fake rate are shown in
Fig.~\ref{detector:kpi}. 
The typical electron identification efficiency is 90\% 
with a small fake rate (0.3\%). Muons are also identified 
with 90\% efficiency (2\% fake rate) for 
charged tracks with momenta larger than 0.8~GeV (Fig.~\ref{detector:muid}). 
More detailed information is available in 
Refs.~\cite{detector_identification}. 

\subsection{Luminosity}
          
\begin{wrapfigure}{l}{6.1cm}
\centerline{\includegraphics[width=0.5\textwidth]
{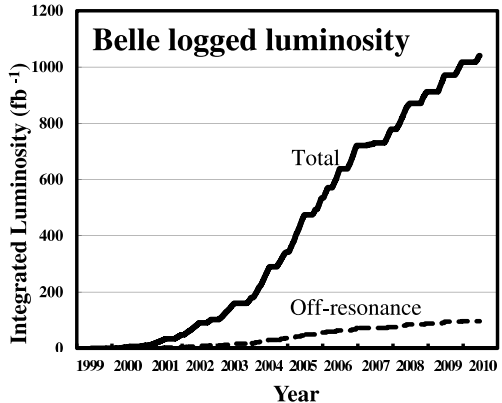}}
\caption{Integrated luminosity taken by Belle.}
\label{detector:luminosity}
\end{wrapfigure}

Belle started data taking on 1 June 1999. After that, 
data runs were taken for 6\---9 months every year 
until the final shutdown on 30 June 2010.
The total integrated luminosity reached 1040 fb$^{-1}$ , 
as shown in Fig.~\ref{detector:luminosity}. 
Belle took most of its data at the energy of the $\Upsilon(4S)$
resonance in order to study $B$-meson decay. 
Off-resonance data were collected 60 MeV below the resonance peak 
energy for 10\% of the running time about every two months in order 
to determine the non-$B\bar{B}$ background. 
%
%
The first non-$\Upsilon(4S)$ data were taken at the energy of 
the $\Upsilon(5S) $ resonance for just three days in 2005. 
In the same year, $\Upsilon(3S)$ resonance data were taken to search 
for invisible decay modes of the $\Upsilon(1S)$ resonance.  
The last $\Upsilon(4S)$ resonance data were taken in June 2008.  
During the last two years of operation,
$\Upsilon(1S)$, $\Upsilon(2S)$, and $\Upsilon(5S)$ resonance data
samples were taken as well as energy scans between the 
$\Upsilon(4S)$ and $\Upsilon(6S)$ resonances.
The integrated luminosity collected by Belle for each CM 
energy is listed in Table~\ref{detector:luminosity_Belle} 
and is calculated using barrel Bhabha events after removing bad
runs, which could not be used in physics analysis 
due to serious detector problems. 
The systematic error in the luminosity measurement 
is about 1.4\%; the statistical error is usually small 
compared with the systematic error.  
Integrated luminosities for $\Upsilon(4S)$ data  are shown separately
for the SVD1 and SVD2 data sets, 
which were taken with different inner detector hardware configurations
as described in the previous subsection. 
Other resonance and scan data were taken in the SVD2 configuration. 
\begin{table} 
\caption{Summary of the luminosity integrated by Belle, broken down by 
CM energy.}
\medskip
\label{detector:luminosity_Belle}
\begin{center}
\begin{tabular}{cccc}
\hline\hline
Resonance & On-peak &       Off-peak  & Number of resonances     \\
& luminosity (fb$^{-1}$)  &  luminosity (fb$^{-1})$  \\
\hline
$\Upsilon(1S)$   & 5.7        &     1.8     & 102 $\times 10^6$\\
$\Upsilon(2S)$   & 24.9       &     1.7   & 158 $\times 10^6$   \\
$\Upsilon(3S)$   &  2.9       &     0.25    & 11 $\times 10^6$ \\
$\Upsilon(4S)$ SVD1  & 140.0  &  15.6      & 152 $\times 10^6$ $B\bar{B}$\\
$\Upsilon(4S)$ SVD2  & 571.0  &  73.8      & 620 $\times 10^6$ $B\bar{B}$ \\
$\Upsilon(5S)$   & 121.4   &      1.7      & 7.1 $\times 10^6$ $B_s \bar{B}_s$   \\
Scan           &             &    27.6         \\
\hline\hline
\end{tabular}
\end{center}
\end{table}

\section{CKM angle measurements}
\label{chap_angles}
\subsection{The Kobayashi\---Maskawa model and unitarity triangle }
\label{section_km}
\begin{wrapfigure}{r}{6.1cm}
  \centerline{\includegraphics[width=0.45\textwidth]
                                {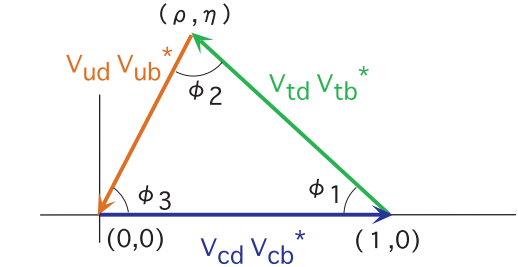}}
                              \caption{The unitarity triangle relevant to $B$ decays. 
The $CP$ violation parameters are defined 
as the angles $\phi_1$, $\phi_2$ and $\phi_3$.}
\label{fig_unitaritytriangle}
\end{wrapfigure}

The phenomenon of $CP$ violation was one of the major unresolved 
issues in elementary particle physics
after its discovery in 1964 in neutral kaon decay~\cite{bib_K_CP}.
In 1973, M.~Kobayashi and T.~Maskawa proposed a model in which
a quark-mixing matrix among six quark flavors includes
a single irreducible complex phase that causes $CP$ violation~\cite{bib_KM}.
Conventionally the quark-mixing matrix is written as~\cite{bib_Wolfenstein}:
\begin{eqnarray}
 \left( \begin{array}{ccc}
   V_{ud} & V_{us} & V_{ub} \\
   V_{cd} & V_{cs} & V_{cb} \\
   V_{td} & V_{ts} & V_{tb} \\
   \end{array} \right) 
 = 
\left( \begin{array}{ccc}
   1-\lambda^2/2 & \lambda & A\lambda^3(\rho - i \eta ) \\
   -\lambda & 1-\lambda^2/2 & A\lambda^2 \\
   A\lambda^3(1 - \rho -i\eta ) & -A\lambda^2 & 1 \\
   \end{array} \right) + o(\lambda^4)
\end{eqnarray}
where the nontrivial complex phases are assigned to $V_{ub}$ and $V_{td}$.
Due to the unitarity of this matrix, 
the following relation is expected to hold, in particular for the terms 
involving the $b$-quark:
\begin{eqnarray}
 V_{td}V_{tb}^* + V_{cd}V_{cb}^* + V_{ud}V_{ub}^* = 0.
\label{unitarityofB}
\end{eqnarray}
This expression can be visualized as a closed triangle in the complex plane 
as shown in Fig.~\ref{fig_unitaritytriangle}.
Here the phase of $V_{td}$ plays a fundamental role and induces
time-dependent $CP$ asymmetries via interference with amplitudes containing
$V_{cb}$ and $V_{ub}$. Measurements of the relevant time-dependent 
$CP$ violation parameters are used to determine 
the $CP$-violating angles, 
$\phi_1$ and $\phi_2$~\cite{footnote:phi1phi2phi3}, that are described in 
Sects.~\ref{section_phi1} and~\ref{section_phi2}.
In contrast, the angle $\phi_3$ is determined by the direct $CP$ asymmetries in
$B \to D K^{(*)}$ decays and is discussed in Sect.~\ref{section_phi3}.

\subsection{$CP$ violation and $\bz$\---$\bb$ mixing }

\begin{figure}
\begin{center}
\resizebox{1.0\textwidth}{!}{\rotatebox{0}{\includegraphics{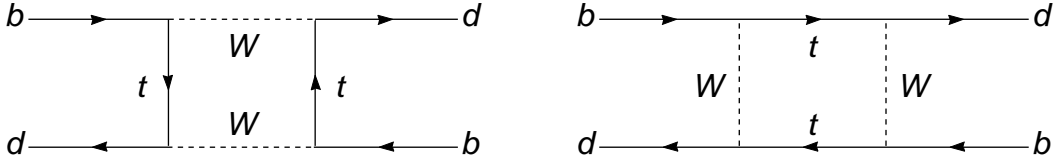}}}
\caption{ Box diagrams that contribute to $\bz$\---$\bb$ mixing. }
\label{fig:box_diagram}
\end{center}
\end{figure}

Neutral $B$ mesons, $\bz$ and $\bb$, can transform or mix 
into their antiparticles
through box diagrams as shown in Fig.~\ref{fig:box_diagram}.
The frequency of the mixing transition (oscillation) 
is $\Delta m_d = (0.507 \pm 0.004)$ ps$^{-1}$~\cite{pdg2012}, while 
the lifetime ($\tau_{\bz}$) is $1.519 \pm 0.007$ ps~\cite{pdg2012}. 

A.I.~Sanda, A.R.~Carter, and I.I.~Bigi showed that a sizable $CP$ violation
can appear in $B$-meson decays if $\bz$\---$\bb$ mixing is large~\cite{Sanda}.
In neutral $B$ decays to $CP$ eigenstate ($f_{CP}$), both $\bz$ and $\bb$ can 
decay to the same final state.  
Because of $\bz$\---$\bb$ mixing, the decay proceeds
through two paths; one from direct decay, $\bz \to f_{CP}$, and the
other through $\bz$\---$\bb$ mixing, $\bz \to \bb \to f_{CP}$.
These two amplitudes have a phase difference of $\phi_{\rm mix} - 2\phi_D$
where $\phi_{\rm mix}$ is the weak phase of $\bz$\---$\bb$ mixing,
arg($V_{td}V_{tb}^*/V_{td}^*V_{tb}$), and $\phi_D$ is the weak phase of
the $\bz \to f_{CP}$ decay.  
In the Wolfenstein representation, $\phi_{\rm mix} = 2\phi_1$ and the phase
difference is given as $2(\phi_1 - \phi_D)$.
The interference term for the two amplitudes has
opposite signs for $\bz$ and $\bb$ decays and leads to $CP$ violation
effects proportional to $\sin2(\phi_1 - \phi_D)$.

\subsection{ Experimental approach at a $B$-factory }

At a $B$-factory, pairs of neutral $B$ mesons 
in a coherent state with $C = -1$ are produced by
$\Upsilon(4S) \to \bz\bb$ decays 
In a decay in which one $B$ meson decays to $f_{CP}$ and the other
$B$ meson decays to a flavor specific final state, $f_{\rm tag}$,
the decay rate is given as
\begin{equation}
 {\cal P}(\Delta t, q; \calS, \calA) = 
 \frac{ e^{-|\Delta t|/\tau_\bz}}{4\tau_\bz}
 \biggl\{ 1 + q \cdot \Bigl[\calS \sin(\dmd\Delta t) + 
  \calA \cos(\dmd\dt)\Bigr]
 \biggr\}. \label{eq:physics}
\end{equation}
Here $\Delta t = t_{CP} - t_{\rm tag}$  is the difference between the proper 
decay times of $f_{CP}$ and $f_{\rm tag}$, $q=\pm1$ is the flavor of
$f_{\rm tag}$ ($+1$ for $\bz\to f_{\rm tag}$).  The quantities 
$\calS$ and $\calA$ are $CP$ violation parameters that
are dependent on the decay mode.
The parameter $\calS$ describes mixing-induced $CP$ violation and 
is given by $\calS = -\eta_{CP} \sin2(\phi_1 - \phi_D)$, where 
$\eta_{CP}$ is the $CP$ eigenvalue of $f_{CP}$. The other parameter,
$\calA$, corresponds to direct $CP$ violation (i.e. no $CP$ violation
in the $\bz \leftrightarrow \bb$ transition rates).
It should be noted that, depending on the weak phase of the decay,
$CP$ violation measurements give information on the
various angles of the unitarity triangle.
The asymmetry in the rate of $\bz$ and $\bb$ decays is given by
\begin{equation}
 A(\dt) \equiv 
 \frac{{\cal P}(\dt,+1;\calS,\calA)-{\cal P}(\dt,-1;\calS,\calA)}
      {{\cal P}(\dt,+1;\calS,\calA)+{\cal P}(\dt,-1;\calS,\calA)}
 =\calS \sin\dmd \dt + \calA \cos\dmd \dt
\end{equation}
%

An experimental measurement of time-dependent $CP$ violation 
at a $B$-factory includes the following steps: 
\begin{enumerate}
\item Reconstruct one $B$ decaying to $f_{CP}$.
\item Determine $q$ using all available information on the $B\to f_{\rm tag}$
      decay.
\item Reconstruct vertices for $f_{CP}$ and $f_{\rm tag}$ and determine
      $\dt$ from the distance between the two $B$ vertices.
\item Obtain $\calS$ and $\calA$ by fitting the $\dt$ distribution of
      reconstructed signal candidates.
\end{enumerate}
Each step is described in more detail below.

\subsection{Measurement of $\phi_1$}
\label{section_phi1}
At the quark level neutral $B$ meson decays into $(c\bar{c}) K^0$ are 
induced by a $b \to c\bar{c}s$ transition.
Since both leading and sub-leading order diagrams of this process contain 
neither $V_{ub}$ nor $V_{td}$, 
there is no complex phase in the decay amplitude. 
Thus $\phi_D$ is zero and the mixing-induced $CP$ violation 
parameter ${\cal S}$ is directly related to one of the $CP$-violating 
angles, $\phi_1$. In the SM,
\begin{eqnarray}
{\cal S} = -\eta_{CP} \cdot \sin 2 \phi_1 \quad \mbox{and} \quad
{\cal A} \approx 0
\label{eq_stosin2phi1anda}
\end{eqnarray}
are expected.

\subsubsection{$B^0 \to (c\bar{c}) K^0$ reconstruction}
We reconstruct $J/\psi K^0_S$, $J/\psi K^0_L$, $\psi(2S) K^0_S$, 
and $\chi_{c1} K^0_S$ as the $f_{CP}$ in neutral $B$ meson decays 
to $(c\bar{c}) K^0$. 
$J/\psi$ mesons are reconstructed via their decay into oppositely 
charged lepton pairs ($e^+e^-$ or $\mu^+\mu^-$) while 
$\psi(2S)$ mesons are reconstructed by lepton pairs as well as 
$J/\psi \pi^+\pi^-$ final states. 
We  reconstruct $\chi_{c1}$ mesons in the $J/\psi \gamma$ mode and
$K^0_S$ mesons in the  $\pi^+\pi^-$ final state. 

For $B^0 \to J/\psi K^0_S$, $\psi(2S) K^0_S$, and $\chi_{c1} K^0_S$
candidates, the $B$ signal is identified using 
two kinematic variables calculated in the $\Upsilon(4S)$ CM:
the energy difference $\Delta E \equiv E_B^* - E_{\rm beam}^*$ 
and the beam-energy constrained mass 
$M_{\rm bc} \equiv \sqrt{(E_{\rm beam}^*)^2 - (p_B^*)^2}$,
where $E_{\rm beam}^*$ is the beam energy in the CM of 
the $\Upsilon(4S)$ resonance, and $E_B^*$ and $p_B^*$
are the CM energy and momentum of the reconstructed $B$ candidate, 
respectively. In the $B^0 \to J/\psi K^0_L$ case, 
candidate $K^0_L$ mesons are selected using information 
recorded in the ECL and/or the KLM.
Since the $K^0_L$ energy cannot be measured, we determine only its 
direction. Thus $B^0 \to J/\psi K^0_L$ candidates are identified by the value 
of $p_B^*$ calculated using a two-body decay kinematic assumption.

\begin{figure}[htb]
\begin{center}
\includegraphics[width=0.44\textwidth,clip]{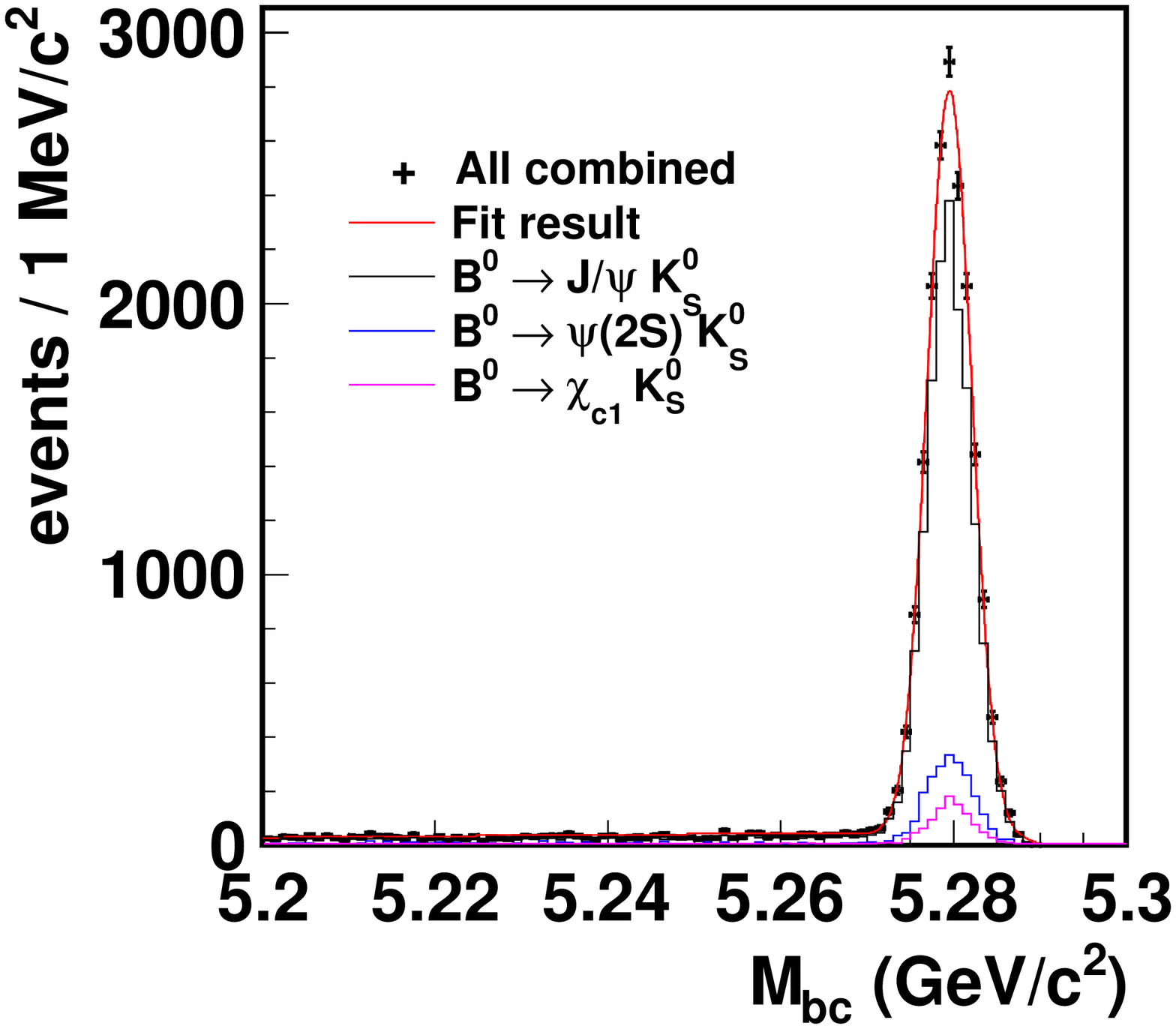}
\includegraphics[width=0.40\textwidth,clip]{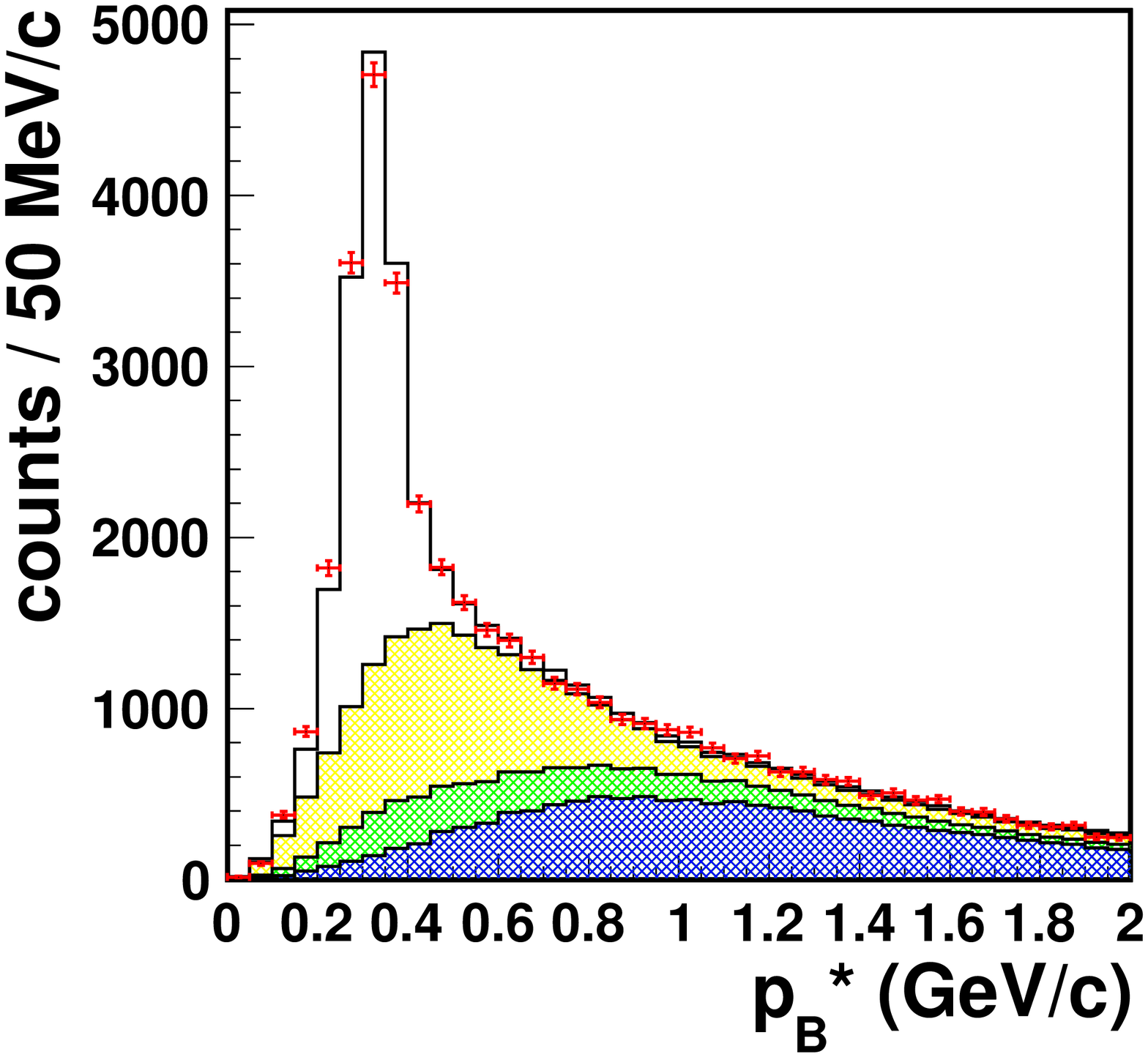}
\end{center}
\caption{$M_{\rm bc}$ distribution within the $\Delta E$ 
signal region for $B^0 \rightarrow J/\psi K^0_S$ (black), 
$\psi(2S) K^0_S$ (blue), and $\chi_{c1} K^0_S$ (magenta);
the superimposed curve (red) shows the fit result for all modes 
combined (left) and the $p_B^*$ distribution for $B^0 \rightarrow J/\psi K^0_L$
candidates with the results of the fit separately shown as 
signal (open histogram), background with a real $J/\psi$ and a 
real $K^0_L$ (yellow), background
with a real $J/\psi$ but without a real $K^0_L$ (green), 
and background without a real $J/\psi$ (blue) (right).}
\label{fig_mbc_pb}
\end{figure}

The $M_{\rm bc}$ distribution for signal candidates with 
a stringent $\Delta E$  requirement
($| \Delta E |<$ 40~MeV for $J/\psi K^0_S$, 
$| \Delta E |<$ 30~MeV for $\psi(2S) K^0_S$, and 
$| \Delta E |<$ 25~MeV for $\chi_{c1} K^0_S$)
as well as the $p_B^*$ distribution for $J/\psi K^0_L$ 
candidates are shown in Fig.~\ref{fig_mbc_pb}.
The signal yields and purities are estimated for each $f_{CP}$ mode and 
given in Table~\ref{tab_sig_and_purity}. 
\begin{table}[htb]
\caption{Signal yield ($N_{\rm sig}$), $CP$ eigenvalue ($\eta_{CP}$),  
and purity for each $B^0 \to f_{CP}$ mode.}
\label{tab_sig_and_purity}
\begin{center}
\begin{tabular}{lccc}
\hline \hline
$B$ decay mode & $\eta_{CP}$ & $N_{\rm sig}$ & Purity (\%) \\
\hline
$J/\psi K^0_S$ & $-1$ & $12 649 \pm 114$ & 97 \\
$\psi(2S)(\ell^+\ell^-) K^0_S$ & $-1$ & $904 \pm 31$ & 92 \\
$\psi(2S)(J/\psi \pi^+\pi^-) K^0_S$ & $-1$ & $1067 \pm 33$ & 90 \\
$\chi_{c1} K^0_S$ & $-1$ & $940 \pm 33$ & 86 \\
$J/\psi K^0_L$ & $+1$ & $10 040 \pm 154$ & 63 \\
\hline \hline
\end{tabular}
\end{center}
\end{table}

\subsubsection{Flavor tagging}

For the events in which we reconstructed
 $B^0 \to f_{CP}$ candidates, the neutral $B$  flavor is identified
 from the decay products of the accompanying $B$ meson.
The available information is obtained from leptons, kaons, $\Lambda$ 
baryons, and pions.
Leptons directly coming from $B$ decay and secondary leptons and 
strange particles in the cascade decays carry the mother 
$b$-flavor information. Low momentum tagging
pions may come from $D^{*\pm}$ decays. In addition, there are
high momentum pions originating from 
$\bb \to D^{(*)+} \pi^-$ or $D^{(*)^+} \rho^-$ decays.
Both types of tagging pions give some information
about $b$-flavor. The information from all the decay products is handled by 
a multi-dimensional likelihood approach with corresponding 
look-up tables~\cite{bib:fbtg_nim}. 

To calibrate $w$,
we select a flavor specific final state of neutral $B$ meson decays
such as semileptonic $\overline{B^0} \to D^{*+} \ell^- \bar{\nu}$ decays 
and hadronic $\overline{B^0} \to D^{(*+)} \pi^-$ and $D^{+*} \rho^-$ decays. 
We then determine the wrong tag fraction $w$ by measuring the
 time evolution of the opposite-sign flavor asymmetry, 
as it exhibits a $\Delta t$ dependence proportional to
$(1 -2w) \cos(\Delta m_d \Delta t)$. 
We also determine $\Delta w$, which is the difference in $w$ between 
$\fq = +1$ and $-1$ events.
For $B^0 \to J/\psi K^0_S$ decay, 
we obtain the effective tagging efficiency, 
$\varepsilon_{\rm eff} = \varepsilon(1-2w)^2 = (30.1\pm0.4)\mbox{\%}$,
where $\varepsilon$ is the tagging efficiency.

\subsubsection{$\Delta t$ determination and its resolution}
\label{sec_dtresol}

In energy-asymmetric $e^+e^-$ collisions at KEKB, the
$\Upsilon(4S)$ is produced with a Lorentz boost of 
$\beta \gamma = 0.425$ nearly along  the $z$-axis, which is
 defined as the direction anti-parallel to the $e^+$ beam at Belle.
Since $B$ mesons are approximately at rest with respect to the $\Upsilon(4S)$,
we can measure $\Delta t$ by measuring the displacement between 
the two $B$ meson decay vertices in the $z$ direction, $\Delta z$,
\begin{eqnarray}
\Delta t \simeq \frac{\Delta z}{\beta \gamma c}.
\label{eq_dztodt}
\end{eqnarray}

The $B$ meson decay vertex is reconstructed by a Lagrange multiplier 
approach, which minimizes the $\chi^2$ calculated 
from the decay vertex position
and the daughter particle tracks~\cite{bib:resol_nim}. 
We call this procedure a ``vertex fit''.
The vertex fit is carried out using daughter tracks with a
sufficient (minimal)
number of SVD hits and a constraint on the interaction-region 
profile in the plane perpendicular to the beam axis.

Because of the negligible flight length of $J/\psi$ or $\psi(2S)$ mesons,
the vertex reconstructed from their daughter lepton tracks can represent 
the $B^0 \to f_{CP}$ decay vertex; its resolution is 
found to be approximately 75 $\mu$m.
On the other hand, the $B^0 \to f_{\rm tag}$ vertex is obtained with 
well-reconstructed tracks that are not assigned to $f_{CP}$.
Here, high momentum leptons are always retained because they usually
come directly from semileptonic $B$ meson decays.
Since $f_{\rm tag}$ may contain long-lived particles 
such as $D^+$, $D^0$, $K^0_S$, and so on, the vertex reconstructed using 
the daughter tracks coming from these intermediate particles can deviate 
from the true $B^0 \to f_{\rm tag}$ vertex. 
This effect is minimized by removing tracks that are identified by a large 
contribution to the vertex fit $\chi^2$.
The $f_{\rm tag}$ vertex position resolution 
is found to be approximately 165 $\mu$m.

In the Belle experiment, the contributions to $\Delta t$ measurement error are
divided into three categories: detector measurement error, the  
effect of secondary particles in $f_{\rm tag}$ vertex reconstruction, 
and the kinematical approximation, $\Delta t \simeq \Delta z/(\beta \gamma c)$.
These three effects are convoluted on an event-by-event basis to obtain the 
$\Delta t$ resolution function, which is used in a maximum likelihood 
fit to extract ${\cal S}$ and ${\cal A}$ as discussed in the next section.

\subsubsection{Extracting $CP$ violation parameters}

We determine $\sin 2 \phi_1$ and ${\cal A}$ from a maximum likelihood fit 
using $\Delta t$ and $\fq$ information obtained on an 
event-by-event basis from signal candidates.
By taking the effect of incorrect flavor assignment into account, 
the probability density function (PDF) 
expected for the signal distribution is given by 
\begin{eqnarray}
\lefteqn{{\cal P}_{\rm sig}(\Delta t)} \nonumber \\
&=& 
 \frac{e^{-|\Delta t|/{\tau_{B^0}}}}{4{\tau_{B^0}}}
\bigg\{1 -q\Delta w_l + q(1-2w_l)\nonumber \\
& & \times \Big[ (-\eta_{CP})\sin 2 \phi_1 \sin(\Delta m_d \Delta t)
   + {\cal A} \cos(\Delta m_d \Delta t)
\Big]
\bigg\}.
\end{eqnarray}
The distribution is convoluted with 
the $\Delta t$ resolution function $R_{\rm sig}(\Delta t)$,
which takes into account the finite vertex resolution as described in
Sect.~\ref{sec_dtresol}.
The background PDF ${\cal P}_{\rm bkg}(\Delta t)$ is determined by the 
events found in a sideband region well away from the signal region in
$M_{\rm bc}$\---$\Delta E$ space as well as Monte Carlo (MC) events.
A small component of broad outliers in the $\Delta z$ distribution, caused
by misreconstruction, is represented by a Gaussian function
${\cal P}_{\rm ol}(\Delta t)$ with $\sigma \approx 30$ ps.
We determine the following likelihood value for each
event indexed by $i$:
\begin{eqnarray}
\lefteqn{{\cal P}_i(\Delta t_i, \fq_i; \sin 2 \phi_1, {\cal A})}
\nonumber \\
&=& (1-f_{\rm ol})f_{\rm sig}\int_{-\infty}^{\infty}
{\cal P}_{\rm sig}(\Delta t')R_{\rm sig}(\Delta t_i-\Delta t')d(\Delta t') \\ \nonumber
& & + (1-f_{\rm ol})f_{\rm bkg}{\cal P}_{\rm bkg}(\Delta t_i) 
+ f_{\rm ol} P_{\rm ol}(\Delta t_i),
\end{eqnarray}
where $f_{\rm ol}$ is the outlier fraction,
$f_{\rm sig}$ and $f_{\rm bkg}$ are 
the signal and background probabilities 
calculated as functions of $\Delta E$ and $M_{\rm bc}$.
The $CP$ violation parameters, $\sin 2 \phi_1$ and ${\cal A}$,
are determined by maximizing the likelihood function
\begin{eqnarray}
{L}(\sin 2 \phi_1, {\cal A}) = \prod_i
{\cal P}_i(\Delta t_i, \fq_i; \sin 2 \phi_1, {\cal A}),
\end{eqnarray}
where the product runs over all events.
A fit to the candidate events results in 
the $CP$ violation parameters~\cite{bib_belle_sin2phi1_final},
\begin{eqnarray}
\sin 2 \phi_1 &=& 0.667 \pm 0.023 \mbox{(stat)} \pm 0.012 \mbox{(syst)}, \\ \nonumber
{\cal A} &=& 0.006 \pm 0.016 \mbox{(stat)} \pm 0.012 \mbox{(syst)}.
\end{eqnarray}
\begin{figure}[htb]
\begin{center}
\includegraphics[width=0.40\textwidth,clip]{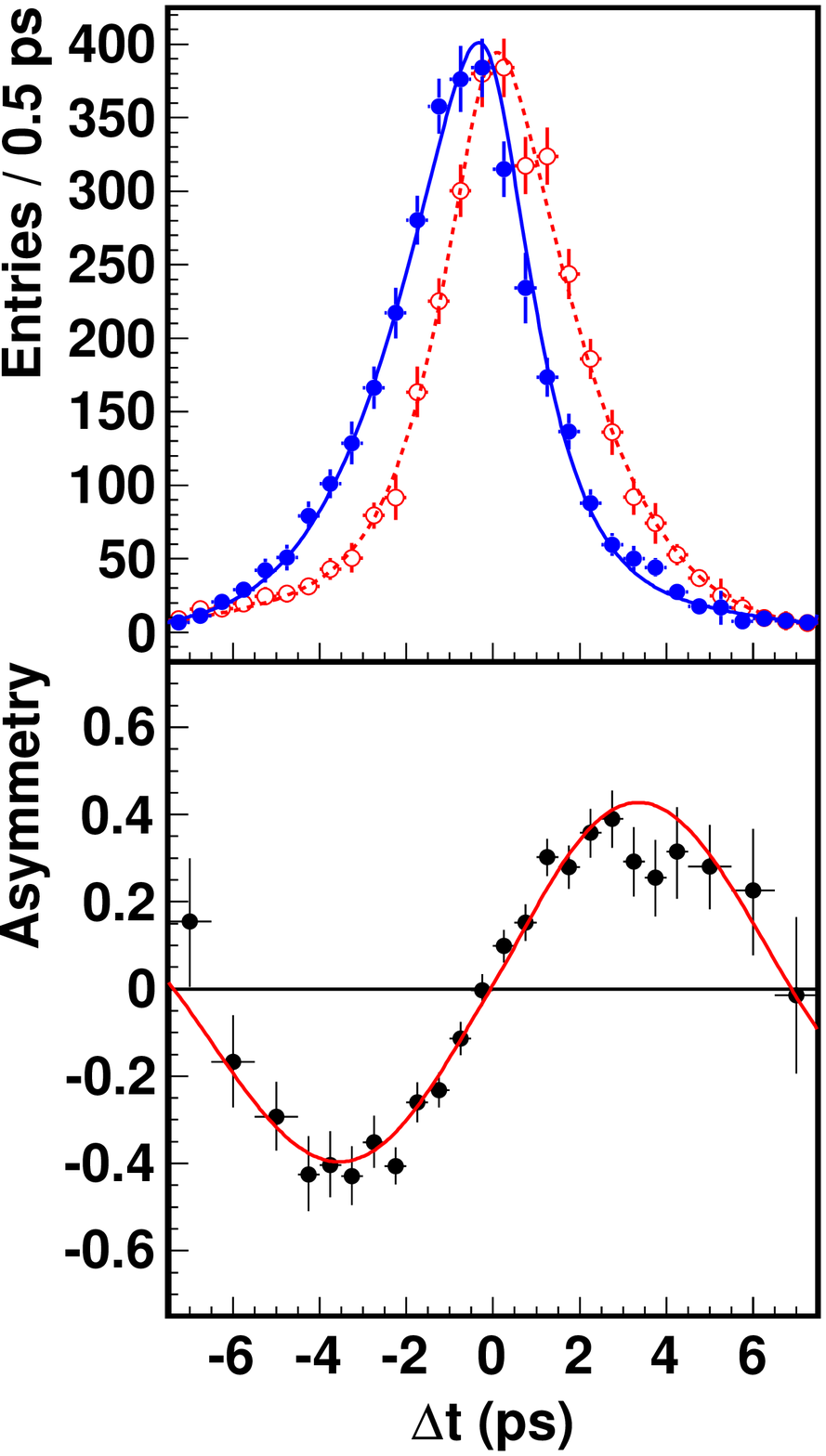}
\includegraphics[width=0.40\textwidth,clip]{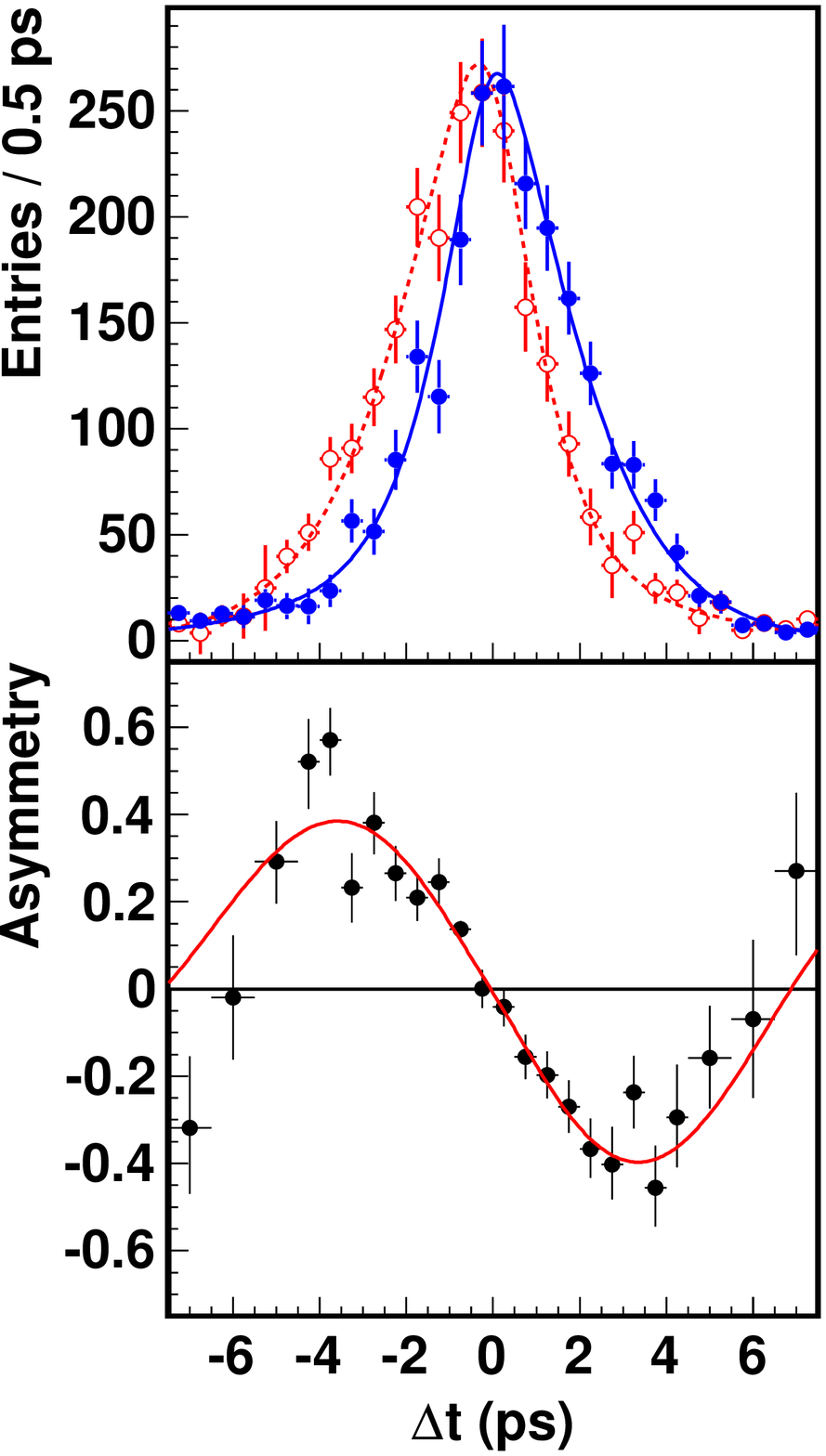}
\end{center}
\caption{The background-subtracted $\Delta t$ distribution for 
$q = +1$ (red) and $q = -1$ (blue) events and asymmetry 
for events with good quality tags in $(c\bar{c}) K^0_S$ (left) and 
$J/\psi K^0_L$ (right) decays.}
\label{fig_dt_asym_sin2phi1}
\end{figure}
The background-subtracted $\Delta t$ distribution for 
$q = +1$ and $q = -1$ events and the asymmetry 
for events with good quality tags are shown in Fig.~\ref{fig_dt_asym_sin2phi1}.
The world average of $\sin 2 \phi_1$ is now $0.68 \pm 0.02$, which is a 
firm SM reference.

\subsubsection{Search for new physics using $CP$ violation measurements 
in $b \to s$ penguin modes.}

$B$ meson decays involving penguin diagrams are thought to be a 
sensitive probe for new physics (NP) beyond the SM
because of the one-loop nature of penguins. NP could appear as 
deviations of $CP$ violation parameters from the SM expectation.
In this section, some highlight results for penguin modes are reviewed.

In SM $b \to s \bar{q} q$ hadronic $B$ decays, the
relevant coupling is $V_{tb}^* V_{ts}$ and the
weak phase is the same as in the $b \to c\bar{c}s$ transition, 
e.g. $B^0 \to (c\bar{c}) K^0$ decay. Therefore, the main point 
is to check whether the penguin $CP$ violation 
results deviate from the SM expectation,
${\cal S}=-\eta_{CP} \sin 2 \phi_1$ and ${\cal A}=0$.
In this context, the time-dependent $CP$-violating parameters are 
denoted as $\sin 2 \phi_1^{\rm eff}$ and ${\cal A}$.
The modes $B^0 \to \phi K^0$, $\eta' K^0$, and $K^0 K^0 K^0$ that involve
only $b \to s \bar{s} s$ processes are of special interest, 
since the SM theoretical uncertainty for $CP$ violation
is small for these decay processes.

In the Belle experiment, attempts to perform measurements of time-dependent 
$CP$ violation in $b \to s \bar{q} q$ induced decays with 
$B^0 \to \eta' K^0_S$ and $\phi K^0_S$ modes were made from 
the earliest stage of data taking, starting in 2002. 
In 2003, using a $152 \times 10^6$ $B\overline{B}$ data sample, 
the value of $\cal{S}$
in $B^0 \to \phi K^0_S$ flipped sign and exhibited a 3.5$\sigma$
deviation  from the S parameter 
measured in $B^0 \to (c\bar{c}) K^0$ modes~\cite{bib_phik02003}.
This was very striking and suggestive of an NP effect.
In 2006, with a larger statistics data sample 
corresponding to $535 \times 10^6$
$B\overline{B}$,  updated measurements were reported.
These measurements added $B^0 \to \eta' K^0_L$
and $\phi K^0_L$ decays to the $B^0 \to \eta' K^0$ and $\phi K^0$ 
sample~\cite{bib_etapk02006}. 
The results are summarized in Table~\ref{tab_three_penguins}. 
\begin{table}[htb]
\caption{Measurements of $CP$ violation parameters, 
$\sin 2 \phi_1^{\rm eff}$ and ${\cal A}$, 
in $B^0 \to \eta' K^0$, $\phi K^0$, and 
$K^0_S K^0_S K^0_S$ modes with a $535 \times 10^6$ $B\overline{B}$ data sample.
The first and second errors are statistical and systematic errors,
respectively.}
\label{tab_three_penguins}
\begin{center}
\begin{tabular}{lcc}
\hline \hline
$B$ decay mode & $\sin 2 \phi_1^{\rm eff}$ & ${\cal A}$ \\
\hline
$\eta' K^0$ & $+0.64 \pm 0.10 \pm 0.04$ & $-0.01 \pm 0.07 \pm 0.05$ \\
$\phi K^0$  & $+0.50 \pm 0.21 \pm 0.06$ & $+0.07 \pm 0.15 \pm 0.05$ \\
$K^0_S K^0_S K^0_S$  & $+0.30 \pm 0.32 \pm 0.08$ & $-0.31 \pm 0.20 \pm 0.07$ \\
\hline \hline
\end{tabular}
\end{center}
\end{table}
In $B^0 \to \eta' K^0$ decay, $CP$ violation is observed with a 
statistical significance of 5.6$\sigma$.
In all these three $B$ decay modes, 
the large deviation from $B^0 \to (c\bar{c}) K^0$ has disappeared.

In spite of the small theoretical uncertainty, 
experimentally, several contributions overlap in $B^0 \to \phi K^0$ 
because of the relatively wide natural widths of the resonances 
that contribute in the $K^+K^-$ final state.
In order to resolve these interfering contributions, Belle fits 
the time-dependent Dalitz distribution by expressing each contribution 
at the amplitude level for the $B^0 \to K^+ K^- K^0_S$ candidate events.
With this technique, the extracted parameter is not $\sin 2 \phi_1^{\rm eff}$
but rather the angle $\phi_1^{\rm eff}$ itself and ${\cal A}$.
Therefore the result does not have a two-fold ambiguity between
 $\phi_1^{\rm eff}$ and $\pi/2 - \phi_1^{\rm eff}$.
In $B^0 \to K^+ K^- K^0_S$ decays, we find four solutions related
to resonant amplitude interference. The preferred one is identified using 
external information related to $f_0(980)$ 
and $f_X$ (assumed to be $f_0(1500)$) branching fractions.
The obtained $CP$ violation parameters are summarized in 
Table~\ref{tab_kkksdalitz}~\cite{bib_kkksdalitz}.
\begin{table}[htb]
\caption{$CP$ violation parameters in $B^0 \to K^+ K^- K^0_S$
time-dependent Dalitz analysis, 
$\phi_1^{\rm eff}$ and ${\cal A}$. 
The first, second, and third errors are statistical, 
experimental systematic, and Dalitz model uncertainties,
respectively.}
\label{tab_kkksdalitz}
\begin{center}
\begin{tabular}{lcc}
\hline \hline
$B$ decay mode & $\phi_1^{\rm eff}$ & ${\cal A}$ \\
\hline
$\phi K^0_S$  & $(32.2 \pm 9.0 \pm 2.6 \pm 1.4)^{\circ}$ & $+0.04 \pm 0.20 \pm 0.10 \pm 0.02$ \\
$f_0 K^0_S$  & $(31.3 \pm 9.0 \pm 3.4 \pm 4.0)^{\circ}$ & $-0.30 \pm 0.29 \pm 0.11 \pm 0.09$ \\
\hline \hline
\end{tabular}
\end{center}
\end{table}
These are consistent with the $CP$ violation in $B^0 \to c\bar{c} K^0$
decays at the 1$\sigma$ level.

Including other $b \rightarrow s$ mediated $B$ decays,
the precision of $\sin 2 \phi_1^{\rm eff}$ is still statistically 
limited, typically $0.1 \sim 0.2$. 
Obtaining ${\cal O}(10^{-2})$ sensitivity 
requires an integrated luminosity of ${\cal O}$(10 ab$^{-1}$), 
and a Super $B$-factory experiment.

\subsection{ Measurement of $\phi_2$ }
\label{section_phi2}

After the first observation of $CP$ violation in $B$ meson decays,
which gave 
a measurement of $\phi_1$, a precise measurement of $\phi_2$ became
the next target of $CP$ violation measurements for the validation of 
the Kobayashi\---Maskawa model.
The first Belle measurement of $CP$ asymmetry parameters
in $B^0 \to \pi^+\pi^-$ decay~\cite{pipi2002} was reported in March 2002,
representing the second decay mode (after $B \to c\bar{c}K^0$) with a  
time-dependent $CP$ violation measurement.

\begin{figure}[hpt]
\begin{center}
\includegraphics[width=0.35\textwidth,clip]{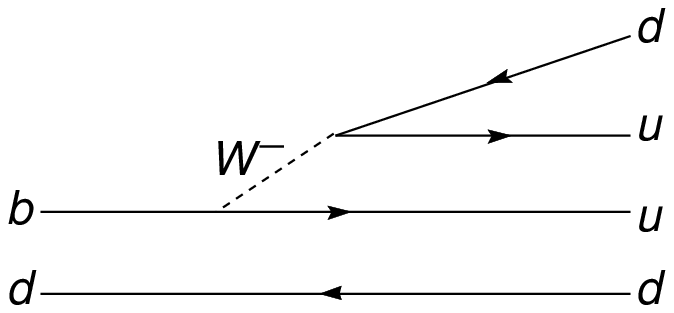} 
\includegraphics[width=0.35\textwidth,clip]{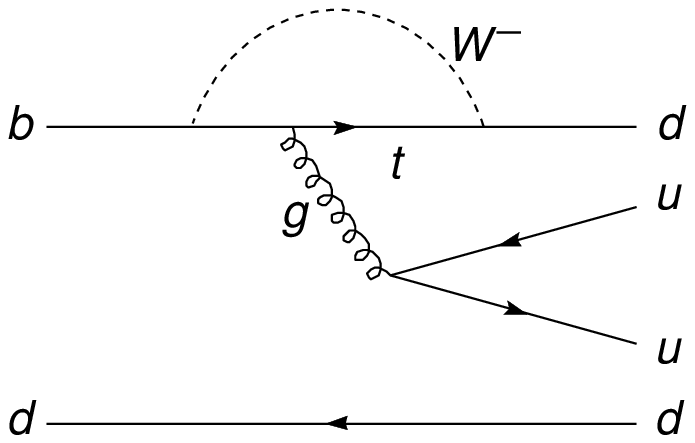} 
\end{center}
\caption{
 Tree (left) and penguin (right) diagrams for $B^0 \to \pi^+\pi^-$ decay.}
\label{fig_pipi_diagram}
\end{figure}

The decay modes used for $\phi_2$ measurements are those proceeding 
via $b \to u$ transition, such as $\bz \to \pip \pim$,
$\bz \to \rhop \pim$, $B^0 \to \rhop \rhom$.
The $b \to u$ transition is shown in Fig.~\ref{fig_pipi_diagram}(left)
and includes the Cabibbo\---Kobayashi\---Maskawa (CKM) element, $V_{ub}$; 
it can be shown that the time dependent $CP$ asymmetry is then given
as $\calS=\sin2\phi_2$ and $\calA \simeq 0$.
However, an additional amplitude, a ``penguin 
diagram'' (Fig.~\ref{fig_pipi_diagram}(right)), contributes and has a 
phase that is different from the tree 
diagram ($V_{td}$ instead of $V_{ub}$). 
This causes a deviation of $\calS$ from $\sin2\phi_2$
and a non-zero $\calA$.

The first $\phi_2$ measurement was  attempted using the 
$\bz \to \pip \pim$ decay mode.  This decay has the simplest two-body topology 
and was one of the first well established charmless $B$ decays. 
The reconstruction of the decay is straightforward: a pair of 
oppositely charged pions with an invariant mass consistent with the $B$-meson 
mass ($M_{\rm bc} = m_B$) is selected; the $B$ meson energy in CM  
is required to be consistent with the beam energy ($\Delta E = 0$).
However, the selected sample 
suffers from a very large background from the $e^+e^- \to q\bar q$
($q = u, d, s, c$) continuum process since the same kinematic properties can 
easily be faked by two oppositely charged pions
fragmented from primary quarks and carrying about half of their momentum.
Another significant background is from $B^0 \to K^+\pi^-$ decay, 
where the kaon is misidentified as a pion. The  
branching fraction for the former 
decay mode is about four times higher than that of
$\bz \to \pip \pim$.  In this case, the reconstructed $\Delta E$ is 
shifted by $-40$ MeV and good $K/\pi$ separation and good momentum resolution
are important to reduce this background.

The continuum background is suppressed utilizing a difference in the global
event topology for the two classes of events; 
continuum events have a two-jet like shape 
while $B\bar B$ events have an isotropic shape 
as the two $B$ mesons are produced almost at rest in the CM.
To quantify the event shape, we use a 
Fisher discriminant~\cite{Fisher} combining
modified Fox\---Wolfram moments~\cite{FW}.
We form a likelihood ${\cal L}_s$ (${\cal L}_b$) for signal (continuum 
background) using the Fisher 
discriminant and the angle between the 
flight direction of the $B$ candidate and the 
beam direction in the CM,  $\cos\theta_B$.
The likelihood ratio ${\cal R} = {\cal L}_s / ( {\cal L}_s + {\cal L}_b) $
is used as the final continuum suppression parameter.
In the early analyses~\cite{pipi2002}, we imposed 
a tight requirement on ${\cal R}$ by 
optimizing $S/\sqrt{S + B}$, where $S$ and $B$ are the expected number of
signal and background events, respectively.  
In a later analysis~\cite{pipi535}, 
we optimized the ${\cal R}$ requirement 
depending on the flavor tagging quality.  

\begin{figure}[htb]
\parbox{\halftext}{
\centerline{\includegraphics[width=6.5cm]{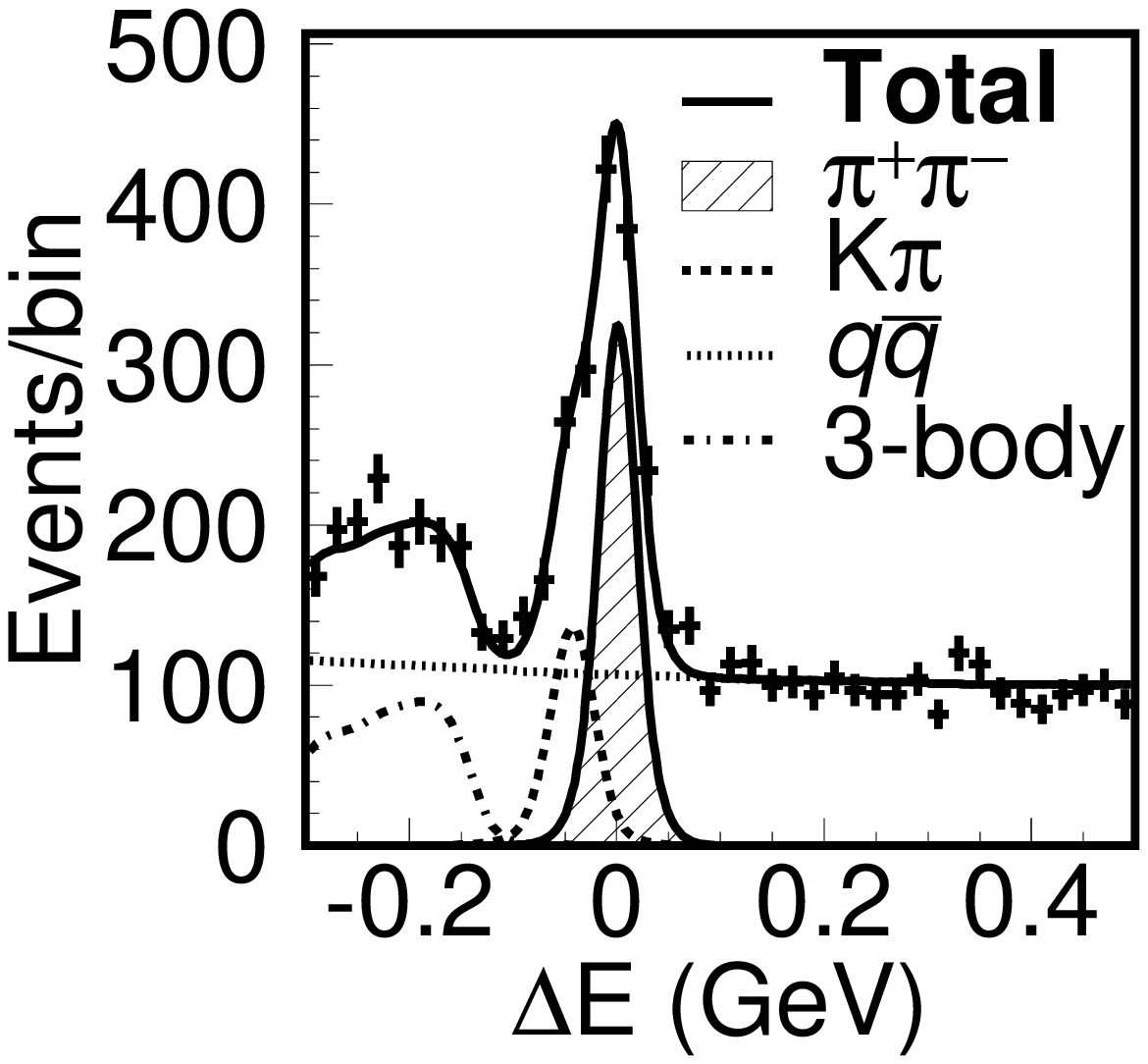}}
\caption{$\Delta E$ distribution of $\bz \to \pip \pim$ candidates.
In order to enhance the signal, requirements are imposed on
the two other variables, $M_{\rm bc}$ and ${\cal R}$.
}
\label{fig:pipi_de}
}
\hfill
\parbox{\halftext}{
\centerline{\includegraphics[width=6.5cm]{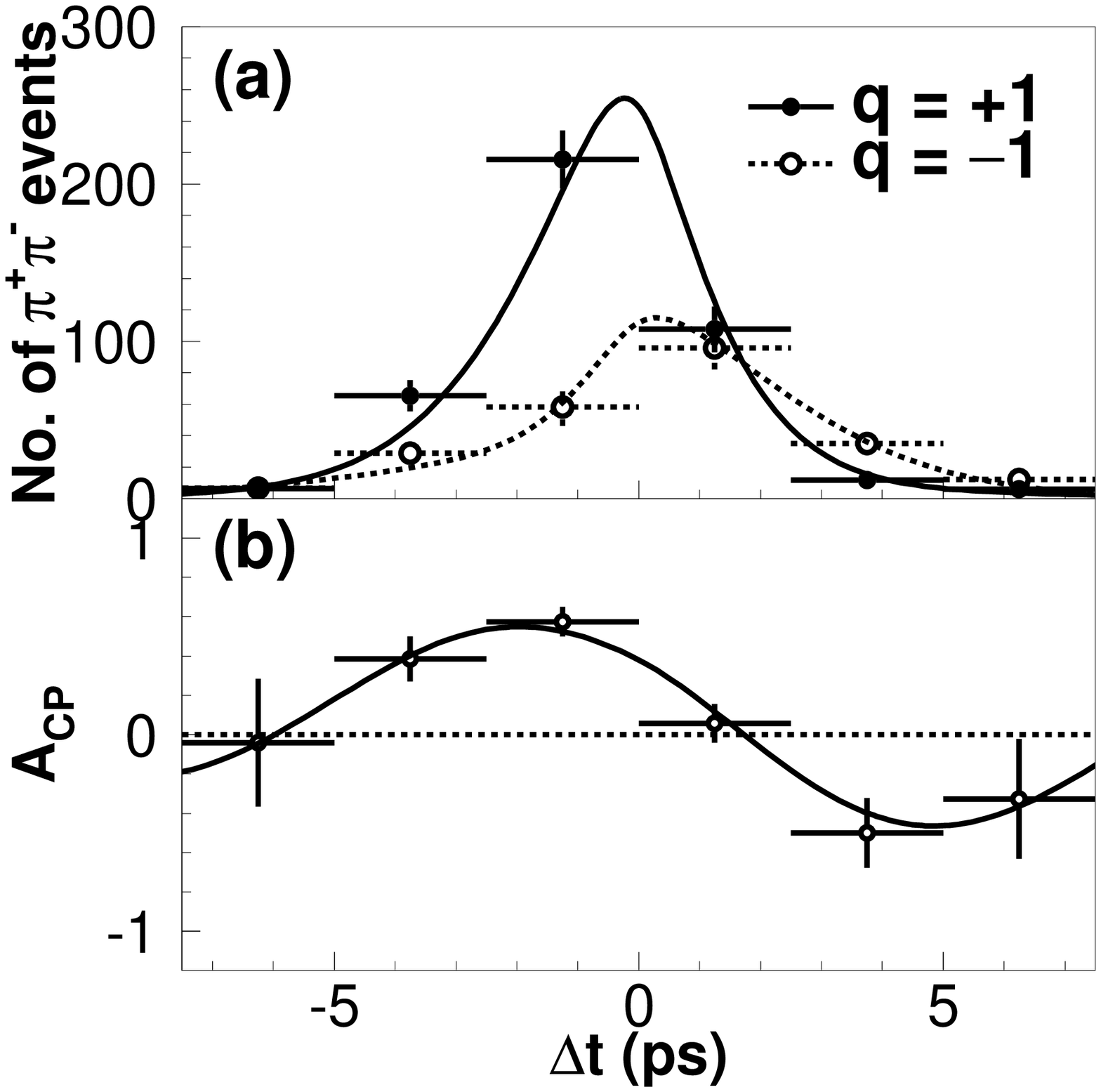}}
\caption{$\Delta t$ distribution for $B^0$ and $\bar{B}^0$ tagged 
$B\to\pi^+\pi^-$ events (top) 
and the $CP$ asymmetry together with 
the fit result (bottom). }
\label{fig:pipi_dt}}
\end{figure}

The $\Delta E$ distribution of $\bz \to \pip \pim$ candidates is
shown in Fig.~\ref{fig:pipi_de}.  Background events due to three-body decays
populate the negative $\Delta E$ region but they do not contribute in
the $\bz \to \pip \pim$ signal region ($|\Delta E| < 0.064$ GeV).  

The vertex reconstruction and the flavor tagging are performed in the same way
as for the $\sin2\phi_1$ measurements.
The $CP$ violation parameters are extracted from a fit to the $\Delta t$
distribution for the events in the signal region in $\Delta E$ and
$M_{\rm bc}$ ([5.271, 5.287] GeV/c$^2$).  The PDFs include the signal,
continuum background, and $\bz \to K^+\pi^-$ background.
The first result was reported using $48 \times 10^6$ $\bbbar$
pairs:~\cite{pipi2002}
\begin{eqnarray}
 \spipi = -1.21^{+0.38}_{-0.27}(\rm{stat})^{+0.16}_{-0.13}(\rm{syst}) 
 \nonumber \\
 \apipi = +0.94^{+0.25}_{-0.31}(\rm{stat})\pm 0.09(\rm{syst})
\end{eqnarray}
In the latest analysis 
using $535 \times 10^6$ $\bbbar$ pairs~\cite{pipi535}, a
stringent selection on $K/\pi$ particle identification 
is not imposed and instead the $B^0 \to K^+\pi^-$ decays are 
included as a component in the fit
to extract the $CP$ violation parameters.  This increases the signal
detection efficiency by 23\% and improves the measurement errors by 10\%.
The results are~\cite{pipi535}
\begin{eqnarray}
 \spipi = -0.61\pm 0.10(\rm{stat})\pm 0.04(\rm{syst})
 \nonumber \\
 \apipi = +0.55\pm 0.08(\rm{stat})\pm 0.05(\rm{syst}).
\end{eqnarray}

The $\Delta t$ distribution and the asymmetry together with fit results
are shown in Fig.~\ref{fig:pipi_dt}.
\begin{figure}[htb]
\parbox{\halftext}{
\centerline{\includegraphics[width=6.5cm]{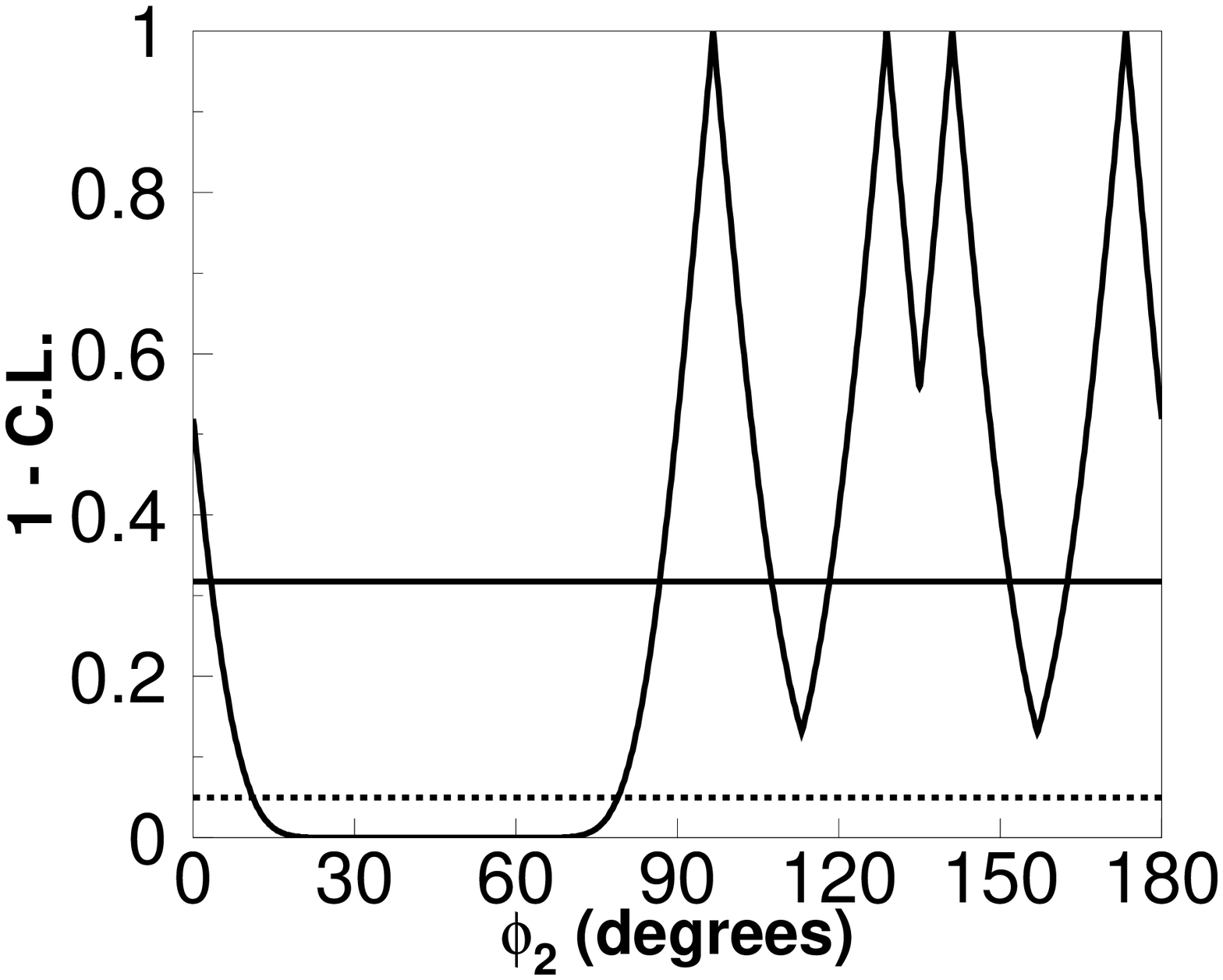}}
\caption{1$-$C.L. for a range of $\phi_2$ values
as obtained with an isospin analysis of $B\to\pi\pi$ decays.
The solid and dashed lines indicate C.L.\ = 68.3\% and 95\%, respectively.}
\label{fig:pipi_phi2_1}
}
\hfill
\parbox{\halftext}{
\centerline{\includegraphics[width=6.5cm]{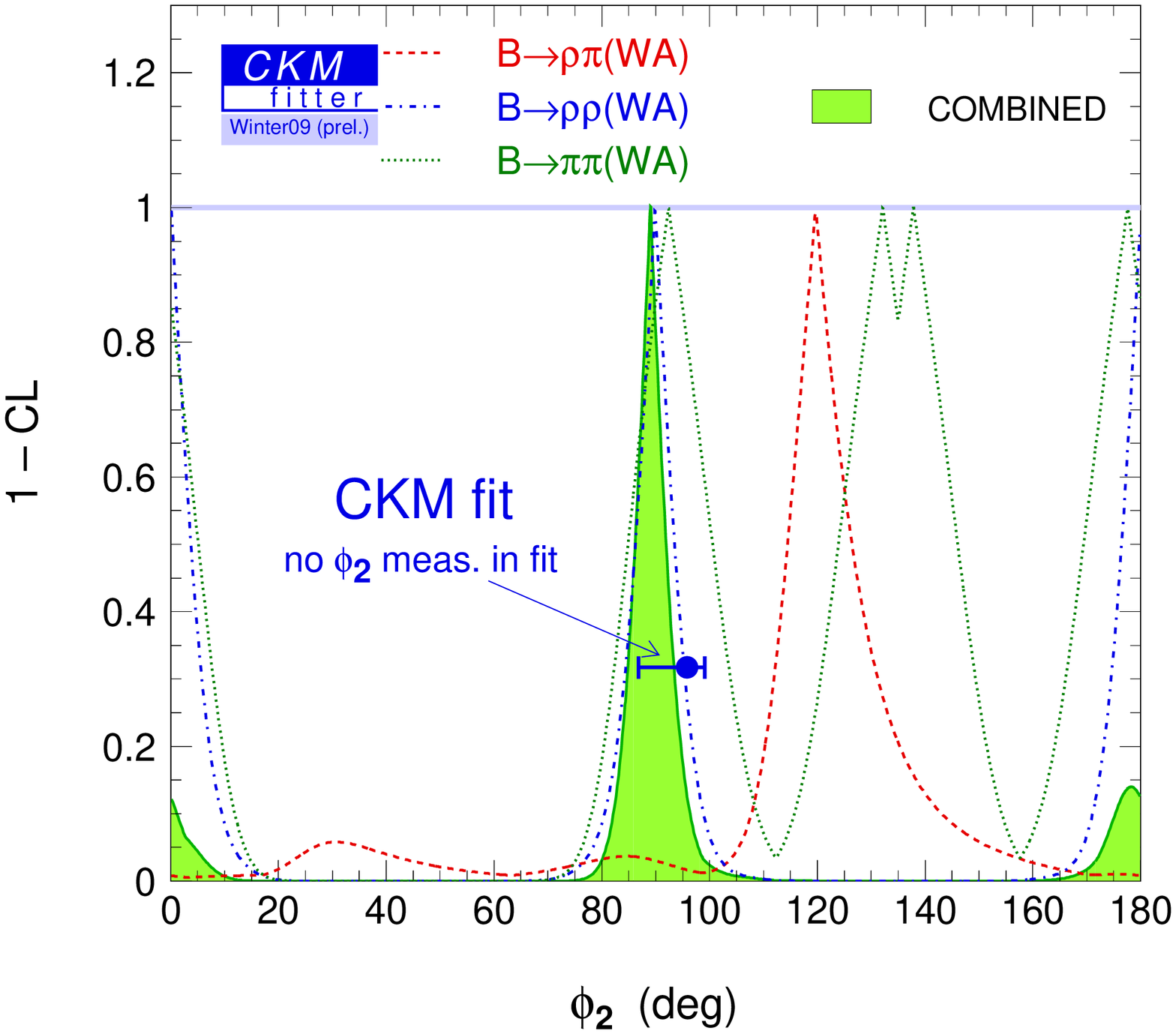}}
\caption{1$-$C.L. as a function of $\phi_2$ from the average of 
the Belle and BaBar results of 
$\bz \to \pip \pim$, $\bz \to \rho \pi$, $B^0 \to \rho \rho$.}
\label{fig:pipi_phi2_2}}
\end{figure}
A clear non-zero $\apipi$, i.e. a clear direct $CP$ violation, is seen 
(the asymmetry exhibits a significant cosine term).
As shown above, the first measurement already indicated $CP$
violation in decays with a significance of $2.9\sigma$. The first evidence
of direct $CP$ violation in a  $B$ decay mode was reported with $3.2\sigma$ 
significance in January 2004 using a sample of 
$152 \times 10^6$ $\bbbar$ pairs ~\cite{pipi152}.
Although this claim was not widely accepted at that time 
because the result of the BaBar collaboration showed a rather small
$\apipi$ value, the latest world average, 
$\apipi = 0.38 \pm 0.06$~\cite{hfag}, establishes $CP$ violation in 
 $B^0 \to \pi^+\pi^-$ decays with a significance above $5\sigma$.

The large direct $CP$ violation indicates that the contribution of
the penguin diagram is sizable and the deviation of 
$\spipi$ from $\sin2\phi_2$ may be significant.
The angle $\phi_2$ can be extracted using the isospin relation among 
branching fractions and $CP$ asymmetries of $B^0 \to \pi^+\pi^-$,
$B^0 \to \pi^0\pi^0$, and $B^+ \to \pi^+\pi^0$ decays; this was
first proposed by M.~Gronau and D.~London~\cite{GL}.
The result, shown in Fig.~\ref{fig:pipi_phi2_1}, is obtained using 
the results for  
$\spipi$ and $\apipi$ given above and the world average values of 
branching fractions of the three $B \to \pi \pi$ modes and direct
 $CP$ asymmetry  in $\bz \to \pi^0\pi^0$. 
Using this method, there are multiple discrete ambiguities for the
angle $\phi_2$.  
The solution that is closest to the global fit result~\cite{CKMFitter}
gives $\phi_2 = (97\pm 11)^\circ$.
%
%

The final state in $B^0 \to \rho^+\pi^-$ decay is not a $CP$ eigenstate,
but the decay proceeds through the same quark diagrams as $B^0 \to \pi^+\pi^-$.
Since $B^0$ and $\bb$ can decay to $\rho^+\pi^-$, time-dependent
$CP$ violation can occur and provide information on $\phi_2$.
Here the final state is $\pi^+\pi^-\pi^0$ and the decay 
$B^0 \to \pi^+\pi^-\pi^0$ contains three intermediate states;
$\bz \to \rho^+\pi^-$, $\rho^-\pi^+$, and $\rho^0\pi^0$. 
These three amplitudes interfere and their magnitudes and relative strong 
phases can be extracted from a Dalitz plot amplitude analysis. 
Knowing the hadronic phases of these amplitudes in the Dalitz plane, a 
time-dependent Dalitz plot analysis allows 
the determination of $\phi_2$~\cite{QS}.
This method provides $\phi_2$ without ambiguities 
(assuming large signal statistics) except for $\phi_2 \to \phi_2 + \pi$.

The reconstruction and continuum suppression are similar to the 
$B^0 \to \pip\pim$
analysis with an additional $\pi^0$ 
reconstructed in the $\pi^0 \to \gamma \gamma$ decay mode. 
$CP$ violation parameters are obtained from a three-dimensional fit
to the distribution of $\Delta t$ and two Dalitz distribution parameters, 
$M^2_{\pi^+\pi^0}$ and $M^2_{\pi^-\pi^0}$. 
Belle performed the analysis using $449 \times 10^6$ $\bbbar$ 
pairs~\cite{rhopi}.
The amplitudes include $\rho(770)$ and higher mass resonances, $\rho(1450)$
and $\rho(1700)$.
The time-dependent Dalitz plot distribution is parameterized with 27 
real parameters describing the 
components that have different time- and Dalitz plot behaviors. 
$CP$ violation parameters for $B^0 \to \rho^\pm \pi^\mp$, $B^0 \to \pi^0\pi^0$
decays and $\phi_2$ are extracted from these parameters.
We obtain a $68^\circ<\phi_2<95^\circ$ at a 68.3\% confidence level (C.L.) 
interval for the solution consistent with the global fit result.
A large region ($0^\circ<\phi_2<5^\circ$, $23^\circ<\phi_2<34^\circ$, 
and $109^\circ<\phi_2<180^\circ$) also remains.
With a larger data sample, a more restrictive constraint without
ambiguities is expected from this measurement.

In the $B^0 \to \rho^+\rho^-$ mode a pseudoscalar decays into two
vector particles and the final state is a mixture of $CP$-even and
$CP$-odd amplitudes.
In order to extract the fraction of each $CP$ component, an angular
analysis is required.  Fortunately, the 
fraction of the longitudinal polarization
turns out to be close to 100\%~\cite{Zhang:2003up,Somov:2006sg,rhorho_pol}, 
simplifying the measurement. 
The signal candidates are reconstructed in $\rho^\pm \to \pi^\pm \pi^0$ 
decays.  Because of two $\pi^0$s in the final state, the combinatorial 
background due to fake $\pi^0$ candidates is very large.
The results using $535 \times 10^6$ $\bbbar$ pairs are~\cite{rhorho}:
\begin{eqnarray}
 \calA_{\rho^+\rho^-} = +0.16\pm 0.21(\rm{stat})\pm 0.07(\rm{syst}) 
 \nonumber \\
 \calS_{\rho^+\rho^-} = +0.19\pm 0.30(\rm{stat})\pm 0.07(\rm{syst}) 
\end{eqnarray}
In this mode, $\phi_2$ can be obtained using an isospin relation similar
to that in the $B^0 \to \pi^+\pi^-$ case.
Because the branching fraction for $B^0 \to \rho^0 \rho^0$ is much smaller
than those of $B^0 \to \rho^+\rho^-$ and $B^+ \to \rho^+\rho^0$,
the deviation of $\phi_2$ from the measured value is small and some 
ambiguities are degenerate. 
So far only an upper limit on  
${\cal{B}}(B^0 \to \rho^0 \rho^0)$ has been obtained; this is used in
the isospin analysis. The isospin analysis gives 
$62^\circ\le\phi_2\le 106^\circ$ at the 68.3\% C.L.

All of the above results and results from the BaBar collaboration 
can be combined to obtain
the $\phi_2$ constraint shown in Fig.~\ref{fig:pipi_phi2_2}~\cite{CKMFitter}.
$\phi_2 = (89.0^{+4.4}_{-4.2})^\circ$ is obtained at a 68.3\% C.L.

\subsection{ Measurement of $\phi_3$ }
\label{section_phi3}
The angle $\phi_1$ has been now measured with high precision 
(Sect.~\ref{section_phi1}).
Measurement of the angle $\phi_2$ is more difficult due to theoretical
uncertainties from the contributions 
of penguin diagrams ~(Sect.~\ref{section_phi2}).
Precise determination of the third angle of the unitarity triangle, $\phi_3$,
is possible using $B^\pm \to DK^\pm$ decays.
However, it requires much more data than determinations of the other angles. 
The determination of $\phi_3$  
is theoretically clean due to the absence of loop contributions; 
$\phi_3$ can be determined using tree-level processes only, exploiting the
interference between $b \to c \overline{u} d$ and $b \to u \overline{c} d$
transitions that occurs when a process involves a neutral $D$ meson
reconstructed in a final state accessible to both $D^0$ and ${\overline{D}}^0$
decays. Therefore, $\phi_3$ provides an SM benchmark, 
and its precise measurement
is crucial in order to disentangle non-SM contributions to other 
processes, via global CKM fits.

\begin{figure}[hpt]
\begin{center}
\includegraphics[width=0.65\textwidth,clip]{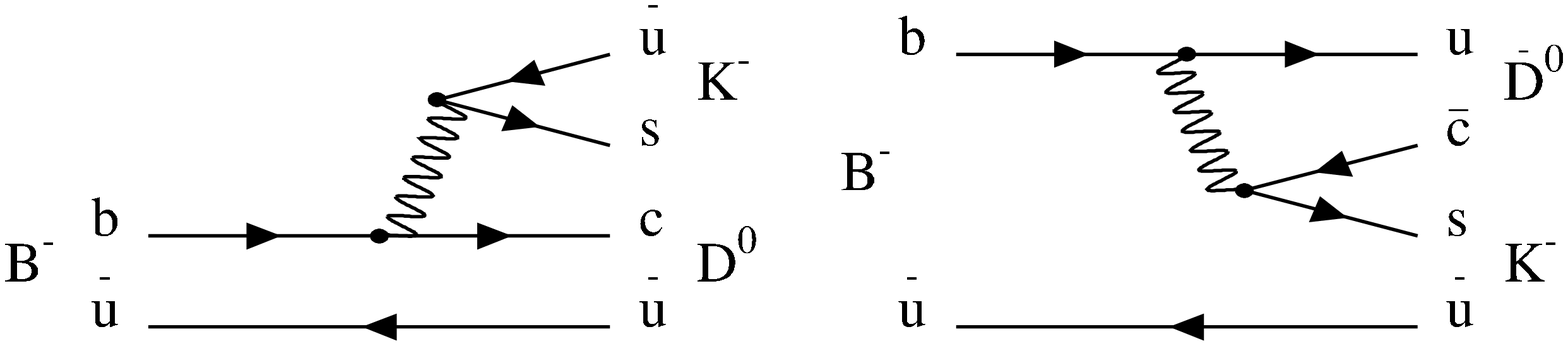} 
\end{center}
\caption{Feynman diagrams for $B^- \to D^0 K^-$ and  $B^- \to 
{\overline{D}}^0 K^-$. }
\label{fig_phi3_diagram}
\end{figure}

\begin{figure}[hpt]
\begin{center}
\includegraphics[width=0.35\textwidth,clip,angle=90]{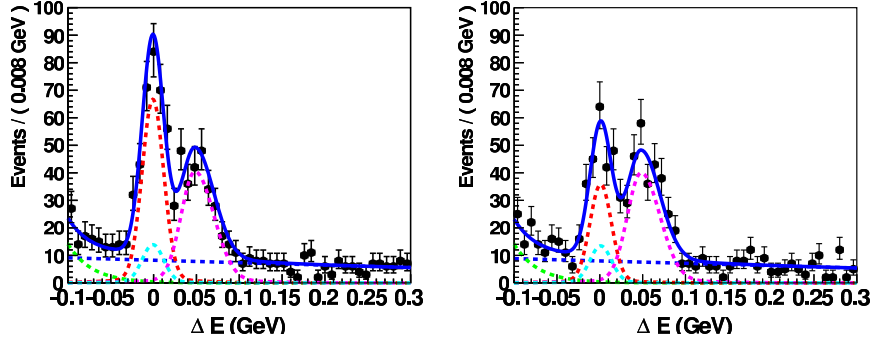} 
\end{center}
\caption{Signals for $B^\pm \to D_1 K^\pm$ decays. The left 
(right) figure is for $B^-$ ($B^+$) decays. The plotted variable, $\Delta E$,
peaks at zero for signal decays, while background from $B^\pm \to D\pi^\pm$
appears as a satellite peak at positive values.}
\label{fig_glw}
\end{figure}

Several different $D$ decays have been studied in order to maximize the
sensitivity to $\phi_3$. The archetype is the use of $D$ decays to $CP$
eigenstates, a method proposed by M.~Gronau, D.~London, and D.~Wyler 
(and called the GLW method)~\cite{GLW}. 
Belle makes use of $CP$-even modes ($D_1$), 
such as $K^+ K^-$, and $CP$-odd modes ($D_2$), such as $K_S^0 \pi^0$.
To extract $\phi_3$ using the GLW method, the following observables
sensitive to $CP$ violation are used:
the asymmetries
\begin{equation}
\label{eq:a12}
{\cal{A}}_{1, 2} \equiv \frac{{\cal{B}}(B^- \to D_{1, 2} K^-) - {\cal{B}}(B^+ \to D_{1, 2} K^+)}{{\cal{B}}(B^- \to D_{1, 2} K^-) + {\cal{B}}(B^+ \to D_{1, 2} K^+)} \\
= \pm\frac{2 r_B \sin \delta_B \sin \phi_3 }{1 +r_B^2 \pm 2 r_B \cos \delta_B \cos \phi_3}
\end{equation}
and the ratios
\begin{equation}
\label{eq:r12}
{\cal{R}}_{1, 2} \equiv \frac{{\cal{B}}(B^- \to D_{1, 2} K^-) + {\cal{B}}(B^+ \to D_{1, 2} K^+)}{{\cal{B}}(B^- \to D^0 K^-) + {\cal{B}}(B^+ \to D^0 K^+)} \\
= 1 +r_B^2 \pm 2 r_B \cos \delta_B \cos \phi_3
\end{equation}
where 
$r_B$ is the ratio of the magnitudes of the two tree diagrams
shown in Fig.~\ref{fig_phi3_diagram} and $\delta_B$ is their strong-phase 
difference. The value of $r_B$ is given by the 
product of the ratio of the CKM matrix elements
$|V_{ub}^* V_{cs}|/|V_{cb}^* V_{us}| \sim 0.38$ and the color suppression
factor, which altogether results in a value of around 0.1. 
In the expressions above, mixing and $CP$ violation in
the neutral $D$ meson system are neglected.
Among these four observables, ${\cal{R}}_{1, 2}$ and ${\cal{A}}_{1, 2}$, 
only three are independent (since ${\cal{A}}_{1} {\cal{R}}_{1} = -
{\cal{A}}_{2} {\cal{R}}_{2}$).
\begin{wraptable}{r}{\halftext}
\caption{Results of the GLW analysis for $B^\pm \to  D K^\pm$ mode.}
\label{tab_r12_a12}
\begin{center}
\begin{tabular}{lccc}
\hline \hline
${\cal{R}}_1$ & $1.03 \pm 0.07 \pm 0.03$ \\
${\cal{R}}_2$ & $1.13 \pm 0.09 \pm 0.05$ \\
${\cal{A}}_1$ & $+0.29 \pm 0.06 \pm 0.02$ \\
${\cal{A}}_2$ & $-0.12 \pm 0.06 \pm 0.01$ \\
\hline \hline
\end{tabular}
\end{center}
\end{wraptable}
Recently, Belle updated their GLW analysis using their final data sample of 
$772 \times 10^6$ $B\overline{B}$ pairs~\cite{Belle_GLW}. The analysis 
uses $D^0$ decays to $K^+ K^-$ and $\pi^+ \pi^-$ as $CP$-even modes 
(Fig.~\ref{fig_glw}), $K_S^0 \pi^0$ and $K_S^0 \eta$ as $CP$-odd modes.
From Eqs.~\ref{eq:a12}\---\ref{eq:r12}, the signs of the ${\cal{A}}_1$
and ${\cal{A}}_2$ asymmetries should be opposite, which is confirmed by
experiment (Table~\ref{tab_r12_a12}).

\begin{figure}[hpt]
\begin{center}
\includegraphics[width=0.85\textwidth,clip]{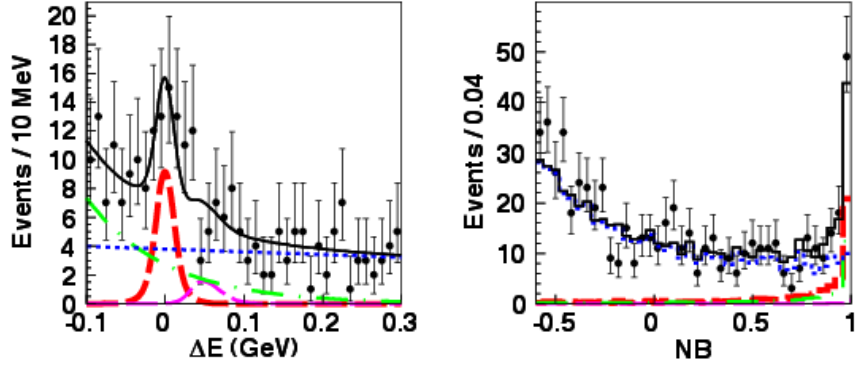}  
\end{center}
\caption{Signal for $B^\pm \to D K^\pm$ decays from Belle ADS analysis.
In these $\Delta E$ and $NB$ (continuum suppression variable) distributions, 
$[K^+ \pi^-]_D K^-$ components are shown by thicker dashed curves (red).}
\label{fig_ads}
\end{figure}

The difficulties in the application of the GLW methods arise primarily
due to the small magnitude of the $CP$ asymmetry of the $B^\pm \to D_{CP}K^\pm$
decay, which may lead to significant systematic uncertainties
in the observation of the $CP$ violation. An alternative approach was
proposed by D.~Atwood, I.~Dunietz, and A.~Soni~\cite{ADS}. 
Instead of using $D^0$ decays to $CP$ eigenstates, 
the ADS method uses Cabibbo-favored and doubly
Cabibbo-suppressed decays: $\overline{D}^0 \to K^- \pi^+$ and 
$D^0 \to K^- \pi^+$. In the decays $B^+ \to [K^-\pi^+]_D K^+$ and 
$B^- \to [K^+\pi^-]_D K^-$, the suppressed $B$ decay is followed by a
Cabibbo-allowed $D^0$ decay, and vice versa. Therefore, the interfering
amplitudes are of similar magnitude, and one can expect a large 
$CP$ asymmetry. 
Unfortunately, the branching ratios of the decays mentioned above are
small. The observable that is measured in the ADS method is the ratio
of the suppressed and allowed branching fractions:
\begin{equation}
{\cal{R}}_{\rm ADS} = \frac{{\cal{B}}(B^\pm \to [K^\mp \pi^\pm]_D K^\pm)}{{\cal{B}}(B^\pm \to [K^\pm \pi^\mp]_D K^\pm)}\\
= r_B^2 + r_D^2 + 2 r_B r_D \cos \phi_3 \cos \delta,
\end{equation}
and
\begin{equation}
{\cal{A}}_{\rm ADS} = \frac{{\cal{B}}(B^- \to [K^+ \pi^-]_D K^-) - {\cal{B}}(B^+ \to [K^- \pi^+]_D K^+)}{{\cal{B}}(B^- \to [K^+ \pi^-]_D K^-) + {\cal{B}}(B^+ \to [K^- \pi^+]_D K^+)}\\
= 2 r_B r_D \sin \phi_3 \sin \delta/{\cal{R}}_{\rm ADS},
\end{equation}
where $r_D$ is the ratio of the doubly Cabibbo-suppressed and Cabibbo-allowed
$D^0$ decay amplitudes and $\delta$ is the sum of strong phase differences
in $B$ and $D$ decays: $\delta = \delta_B + \delta_D$.
The ADS analysis~\cite{Belle_ADS} using the full $\Upsilon(4S)$ data 
sample was reported by the Belle collaboration (Fig.~\ref{fig_ads}).
The analysis uses $B^\pm \to D K^\pm$ decays with $D^0$ decaying to $K^+\pi^-$
and $K^-\pi^+$ (and their charge-conjugated partners).
The signal yield obtained is $56^{+15}_{-14}$ events, which
corresponds to the first evidence of an ADS signal
(with a significance of 4.1$\sigma$); the ratio of
the suppressed and allowed modes is summarized in Table~\ref{tab_ads}.
Although the analyses
with $B^\pm \to D K^\pm$ decays give the most precise results, different $B$
decays have also been studied. The use of two additional decay modes,
$D^* \to D \pi^0$ and $D^* \to D \gamma$, provides an extra handle on the 
extraction of $\phi_3$ from $B^\pm \to D^* K^\pm$, which is becoming visible
in the most recent results~\cite{Belle_GLW}. 
\begin{table}[htb]
\caption{Results of the Belle ADS analyses.}
\label{tab_ads}
\begin{center}
\begin{tabular}{lcc}
\hline \hline
Mode & ${\cal{R}}_{\rm ADS}$ & ${\cal{A}}_{\rm ADS}$\\
\hline
$B \to DK$ & $0.0163 {}^{+0.0044}_{-0.0041} {}^{+0.0007}_{-0.0013}$ & 
$-0.39 {}^{+0.26}_{-0.28} {}^{+0.04}_{-0.03}$ \\
$B \to D^\star K$, $D^\star \to D \pi^0$ & 
$0.010 {}^{+0.008}_{-0.007} {}^{+0.001}_{-0.002}$ & 
$+0.4 {}^{+1.1}_{-0.7} {}^{+0.2}_{-0.1}$ \\
$B \to D^\star K$, $D^\star \to D \gamma$ & 
$0.036 {}^{+0.014}_{-0.012} \pm 0.002$ & 
$-0.51{}^{+0.33}_{-0.29} \pm 0.08$ \\
\hline
\hline \hline
\end{tabular}
\end{center}
\end{table}

A Dalitz plot analysis of a three-body $D$ meson final state allows
one to obtain all the information required for determination of $\phi_3$
in a single decay mode. Three-body final states such as $K_S^0 \pi^+ \pi^-$
have been suggested as promising modes~\cite{GGSZ} for the extraction of 
$\phi_3$.
As in the GLW and ADS methods, the two amplitudes interfere if the $D^0$ and
$\overline{D}^0$ mesons decay into the same final state $K_S^0 \pi^+ \pi^-$. 
Assuming no $CP$ asymmetry in neutral
$D$ decays, the amplitude for $B^+ \to D[K_S\pi^+\pi^-]K^+$ decay 
as a function 
of Dalitz plot variables 
$m_+^2 = m^2_{K_S^0 \pi^+}$ and $m_-^2 = m^2_{K_S^0 \pi^-}$ is 
\begin{equation}
f_{B^+} = f_D (m^2_+, m^2_-) + r_B e^{i\phi_3 + i\delta_B} f_D (m^2_-, m^2_+)
\end{equation}
where $f_D (m^2_+, m^2_-)$ is the amplitude of the $\overline{D}^0 \to
K_S^0 \pi^+ \pi^-$ decay.
Similarly, the amplitude for $B^- \to D[K_S\pi^+\pi^-]K^-$ decay 
is 
\begin{equation}
f_{B^-} = f_D (m^2_-, m^2_+) + r_B e^{-i\phi_3 + i\delta_B} f_D (m^2_+, m^2_-).
\end{equation}
The $\overline{D}^0 \to K_S^0 \pi^+ \pi^-$ decay amplitude $f_D$ can be
determined from a large sample of flavor-tagged 
$\overline{D}^0 \to K_S^0 \pi^+ \pi^-$ decays produced 
in continuum $e^+ e^-$ annihilation. Once $f_D$ is
known, a simultaneous fit to $B^+$ and $B^-$ data allows the contributions
of $r_B$, $\phi_3$ and $\delta_B$ to be separated. The method has only 
two-fold ambiguity: $(\phi_3, \delta_B)$ and $(\phi_3 + 180^\circ, \delta_B
+ 180^\circ)$ solutions cannot be distinguished. 
To test the consistency of the fit, the same procedure was applied 
to the $B^\pm \to D^{(*)} \pi^\pm$ control samples and the 
$B^\pm \to D^{(*)} K^\pm$ signal. 
A combined unbinned maximum likelihood fit to the $B^+$ and $B^-$ samples
with free parameters $r_B$, $\phi_3$, and $\delta_B$ yields the 
values given in Table~\ref{tab_ggsz}.
Combining $B^\pm \to D K^\pm$ and $B^\pm \to D^* K^\pm$, we 
obtain~\cite{Belle_GGSZ}
the value $\phi_3 = (78 {}^{+11}_{-12} \pm 4 \pm 9)^\circ$,
where the sources of uncertainties are statistical, systematic, and due to
imperfect knowledge of the amplitude model 
that describes $D \to K_S^0 \pi^+ \pi^-$ decays. 
\begin{table}[htb]
\caption{Results of Belle Dalitz plot analyses.}
\label{tab_ggsz}
\begin{center}
\begin{tabular}{lccc}
\hline \hline
Mode &  $\phi_3$ $({}^\circ)$ & $\delta_B$  $({}^\circ)$ &  $r_B$ \\
\hline
$B \to DK$ & $81 {}^{+13}_{-15} \pm 5$ & 
$137 {}^{+13}_{-16} \pm 4$ & $0.16 \pm 0.04 \pm 0.01$ \\
$B \to D^\star K$ & $74 {}^{+19}_{-20} \pm 4$ & $342 {}^{+19}_{-21} \pm 3$
& $0.20 \pm 0.07 \pm 0.01$ \\
\hline \hline
\end{tabular}
\end{center}
\end{table}
The last source of uncertainty can be eliminated by binning the Dalitz 
plot (Refs.~\cite{GGSZ, GGSZ_2}),
using information on the average strong phase difference between $D^0$ and
${\overline{D}}^0$ decays in each bin that can be determined using quantum
correlated $\psi(3770)$ data.  Results have been published recently
by CLEO-c~\cite{CLEOc}. The measured strong phase difference is used 
to obtain a model-independent result~\cite{Belle_GGSZ_2}:
\begin{equation}
\phi_3 = (77 \pm 15 \pm 4 \pm 4)^\circ,
\end{equation}
where the last uncertainty is due to the statistical precision of the CLEO-c
results.

\section{Measurement of $|V_{cb}|$ and $|V_{ub}|$, semileptonic, and
  leptonic $B$ decays}
\label{section_slb}

\subsection{Introduction}

\begin{wrapfigure}{r}{6.1cm}
  \centerline{\includegraphics[width=0.45\textwidth]{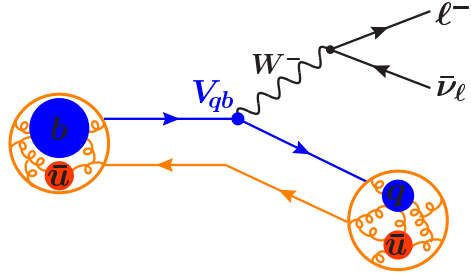}}
  \caption{Illustration of the semileptonic $B$~meson decay $B\to
    X\ell\nu$.} \label{fig:feyn_sl}
\end{wrapfigure}

The Cabibbo\---Kobayashi\---Maskawa (CKM) matrix elements $|V_{cb}|$ and
$|V_{ub}|$ are determined from semileptonic $B\to X\ell\nu$ ($\ell=e,\mu$)
decays to charmed and charmless final states,
respectively (Fig.~\ref{fig:feyn_sl}).
These decays are chosen because semileptonic decays proceed via
leading-order weak interactions and thus are free of possible
non-Standard Model contributions. Their branching fractions are
sizable compared to purely leptonic $B\to\ell\nu$ decays, and have
hadronic uncertainties that are well controlled by various theoretical
techniques.

In this section, we also discuss purely leptonic and semileptonic
$B$~decays involving
a heavy $\tau$ lepton. At present these decays are not relevant for
the determination of $|V_{cb}|$ and $|V_{ub}|$ but are studied because
of their sensitivity to the charged Higgs boson and other manifestations
of new physics.

There are two orthogonal approaches to measuring semileptonic decays and
determining $|V_{cb}|$ and $|V_{ub}|$: Analyses can either be
\emph{exclusive}, {\it i.e.},
these reconstruct only a specific semileptonic final state, such
as $D^*\ell\nu$, $\pi\ell\nu$, \dots.
Alternatively, the analysis can be \emph{inclusive}, which means
that it is sensitive to all semileptonic final states, $X_c\ell\nu$ or
$X_u\ell\nu$, in a given region of phase space, where $X_c$ and $X_u$
refer to a hadronic system with charm or without charm,
respectively.

Exclusive and inclusive analyses are affected by different
experimental uncertainties. In addition, different and largely independent
theoretical approaches are used to describe the QCD contributions in
exclusive and inclusive decays. Since both approaches rely on different
experimental techniques and involve different theoretical
approximations, they complement each other and provide largely
independent determinations of comparable
accuracy for $|V_{cb}|$ and $|V_{ub}|$. This in turn provides a crucial
cross check of the methods and our understanding of semileptonic
$B$~decays in general.

\subsection{$|V_{cb}|$}

\subsubsection{$|V_{cb}|$ from exclusive semileptonic decays}

The determination of $|V_{cb}|$ from exclusive decays is based on 
the $B\to D^*\ell\nu$ or $B\to D\ell\nu$ decay modes.
Experimentally, one has to measure the
differential decay rate as a function of the velocity transfer $w$,
defined as
\begin{equation}
  w=\frac{P_B\cdot P_{D^{(*)}}}{m_B
    m_{D^{(*)}}}=\frac{m_B^2+m_{D^{(*)}}^2-q^2}{2m_Bm_{D^{(*)}}}~,
\end{equation}
where $m_B$ and $m_{D^{(*)}}$ are the masses of the $B$ and the
charmed mesons, $P_B$ and $P_{D^{(*)}}$ are their four-momenta, and
$q^2=(P_\ell+P_\nu)^2$. The point~$w=1$ is referred to as zero recoil,
because there the charmed meson is at rest in the $B$~meson frame. To
determine $|V_{cb}|$, the experimental analyses extrapolate the decay
rate to $w=1$, as theory can determine the decay form factors with
greater accuracy at this kinematic point.
When neglecting the lepton mass, {\it i.e.}, considering only
electrons and muons, the differential decay rate of $B\to D^*\ell\nu$
as a function of $w$ is given by~\cite{Neubert:1993mb} 
\begin{equation}
  \frac{d\Gamma}{dw}=\frac{G^2_Fm^3_{D^*}}{48\pi^3}\big(m_B-m_{D^*}\big)^2\sqrt{w^2-1}\,\chi(w)\mathcal{F}^2(w)|V_{cb}|^2~.
\end{equation}
Here, $G_F$ is Fermi's constant equal to $(1.16637\pm 0.00001)\times
10^{-5}$~GeV$^{-2}$ and $\chi(w)$ is a known phase space factor,
\begin{equation}
  \chi(w)=(w+1)^2\left[1+4\frac{w}{w+1}\frac{1-2wr+r^2}{(1-r)^2}\right]~,
\end{equation}
where $r=m_{D^*}/m_B$. The dynamics of the decay are contained in the
form factor $\mathcal{F}(w)$, which can be parameterized by the
normalization $\mathcal{F}(1)$, the slope $\rho^2_{D^*}$, and the
amplitude ratios $R_1(1)$ and $R_2(1)$ in the framework of the heavy
quark effective theory (HQET)~\cite{Caprini:1997mu}.

A similar expression can be derived for the differential rate of the
decay $B\to D\ell\nu$,
\begin{equation}
  \frac{d\Gamma}{dw} = \frac{G_F^2
    m^3_{D}}{48\pi^3}(m_B+m_{D})^2(w^2-1)^{3/2}\mathcal{G}^2(w)|V_{cb}|^2~.
\end{equation}
As the $D$~meson is a pseudoscalar, the form factor $\mathcal{G}(w)$
of this decay is simpler than $\mathcal{F}(w)$ and can be
parameterized by the normalization $\mathcal{G}(1)$ and the slope
$\rho^2_D$ only~\cite{Caprini:1997mu}.

In the limit of infinite quark masses, the form factors $\mathcal{F}(w)$
and $\mathcal{G}(w)$ coincide with the Isgur\---Wise
function~\cite{Isgur:1989ed}, which is normalized to unity at zero
recoil, $w=1$. Corrections to the heavy quark limit have been
calculated in the framework of lattice QCD (LQCD). In LQCD, the QCD
action is discretized on a Euclidean spacetime lattice and
calculations are performed numerically on computers using Monte Carlo
methods. Physical results are then recovered in the limit of zero lattice
spacing. Because lattice results are obtained from QCD first
principles, they can be improved to arbitrary precision, given
sufficient computing resources.

The form factor values at $w=1$ are the main theoretical input needed
for the determination of $|V_{cb}|$ from exclusive decays and also the
main source of theoretical uncertainty. The current LQCD value of
$\mathcal{F}(1)$, describing the decays $B\to D^*\ell\nu$,
is~\cite{Bailey:2010gb}
\begin{equation}
  \mathcal{F}(1) = 0.908\pm0.017~. \label{eq:1_1}
\end{equation}
The LQCD $B\to D\ell\nu$ form factor at zero recoil is calculated to
be~\cite{Okamoto:2004xg}
\begin{equation}
  \mathcal{G}(1) = 1.074\pm 0.024~. \label{eq:1_2}
\end{equation}

The Belle measurement of $B^0\to D^{*-}\ell^+\nu$~\cite{Dungel:2010uk}
is based on $772\times 10^6$ $B\bar B$~events, resulting in about 120000
reconstructed decays. In this analysis the
decay chain $D^{*-}\to\bar D^0\pi^-$ followed by $\bar D^0\to K^+\pi^-$ is
reconstructed and $D^*$ candidates are combined with a charged
lepton~$\ell$ ($\ell=e,\mu$) with momentum between 0.8~GeV and
2.4~GeV. As the analysis makes no requirement on the second $B$~meson
in the event, the direction of the neutrino is not
precisely known. However, using the $\cos\theta_{BY}$~variable,
\begin{equation}
  \cos\theta_{BY}= \frac{2 E_B E_Y - m^2_B - m^2_Y}{2 P_B P_Y}~, \label{eq:1_3}
\end{equation}
with $Y=D^*\ell$, the $B$~momentum vector is constrained to a cone 
centered on the $D^*\ell$~direction. By averaging over the possible 
$B$~directions one can approximate the neutrino momentum and calculate
the kinematic variables of the decay ($w$ and three decay angles). The
parameters of the form factor $\mathcal{F}(w)$ are obtained by fitting
these four kinematic distributions. The very large data sample led
to much reduced statistical and systematic uncertainties.

The result of the Belle analysis (after rescaling input parameters to
their most recent values~\cite{pdg2012}) is
\begin{equation}
  \mathcal{F}(1)|V_{cb}| = (34.7\pm 0.2 (\rm stat)\pm 1.0(\rm syst))\times
  10^{-3}~,
\end{equation}
where the dominant systematic uncertainties stem from charged track
reconstruction. Assuming the form factor normalization of
Eq.~\ref{eq:1_1}, we obtain
\begin{equation}
  |V_{cb}|=(38.2\pm 1.1(\rm exp)\pm 0.7(\rm th))\times 10^{-3}~.
\end{equation}
 The experimental uncertainty is at the level of 3.0\% while the
 theoretical uncertainty from lattice QCD amounts to 1.9\%.

In addition, the decay $B\to D\ell\nu$ has been studied at
Belle using $10.8\times 10^6$ $B\bar B$~events~\cite{Abe:2001yf}. For the
determination of $|V_{cb}|$, the
decay $B\to D^*\ell\nu$ is preferred over $B\to D\ell\nu$ for both
theoretical and experimental reasons: On the theory side, the rate at
zero recoil is lower for $B\to D\ell\nu$ than for $B\to D^*\ell\nu$
due to the factor $(w^2-1)^{3/2}$ in the expression for the width
(instead of $\sqrt{w^2-1}$ in the $D^*$~case). Experimentally, due to
the presence of the slow pion in the decay $D^*\to D\pi$, the
$D^*$~signal is cleaner than the $D$~signal and backgrounds in the
analysis of $B\to D\ell\nu$ are typically the limiting factor.

The result of the Belle analysis (after rescaling input parameters to
their most recent values~\cite{pdg2012}) is
\begin{equation}
  \mathcal{G}(1)|V_{cb}| = (40.8\pm 4.4(\rm stat)\pm 5.2(\rm syst))\times
  10^{-3}~,
\end{equation}
with the dominant systematic uncertainties from background
estimation. Assuming the $\mathcal{G}(1)$~value from Eq.~\ref{eq:1_2},
we obtain
\begin{equation}
  |V_{cb}|=(38.0\pm 6.3(\rm exp)\pm 0.8(\rm th))\times 10^{-3}~.
\end{equation}
This determination of $|V_{cb}|$ is consistent with the $B\to
D^*\ell\nu$ value but has a significantly larger uncertainty.

\subsubsection{$|V_{cb}|$ from inclusive semileptonic decays}

The theoretical tool for calculating the inclusive semileptonic decay
width $\Gamma(B\to X_c\ell\nu)$ of the $B$~meson is the operator product
expansion (OPE). In this framework, a simplified form
reads~\cite{Benson:2003kp,Bauer:2004ve}
\begin{equation}
  \Gamma(B\to
  X_c\ell\nu)=\frac{G^2_Fm^5_b}{192\pi^3}|V_{cb}|^2\big(1+\frac{c_5(\mu)\,\langle
    O_5\rangle(\mu)}{m^2_b}+\frac{c_6(\mu)\,\langle
    O_6\rangle(\mu)}{m^3_b}+\mathcal{O}(\frac{1}{m^4_b})\big)~, \label{eq:2_1}
\end{equation}
where the expansion parameter is the $b$-quark mass $m_b$.
At leading order in $1/m_b$, the OPE result coincides with the parton
model, {\it i.e.}, with the decay width of a (hypothetical) free
$b$-quark. Corrections to the free $b$-quark decay arise at order
$1/m^2_b$: the term $\langle O_5\rangle(\mu)$ denotes the expectation values of
local dimension 5 operators, which depend on the renormalization scale
$\mu$. A detailed analysis shows that only two operators appear at
$\mathcal{O}(1/m^2_b)$: the kinetic operator, related to the kinetic
energy of the $b$-quark inside the $B$~hadron, and the chromomagnetic
operator, related to the $B^*$-$B$ hyperfine mass splitting. At
$\mathcal{O}(1/m^3_b)$, new operators appear. These
expectation values of local operators describe basic hadronic
properties of the $B$~meson and do not depend on the observable (here
$\Gamma(B\to X_c\ell\nu)$) calculated using the OPE. As they contain
soft hadronic physics, they cannot be calculated by perturbative QCD.

These matrix elements are multiplied by the Wilson coefficients $c_5$,
$c_6$, \dots, which encode the short-distance QCD contributions to the
process and thus can be calculated in perturbation theory as a series
in powers of $\alpha_s$. Hence, the OPE factorizes the calculable and
the non-calculable contributions to the semileptonic width. Even more
interestingly, the hadronic matrix elements in the non-calculable part
also appear in similar OPE expressions for other inclusive observables
in semileptonic $B$~decays. By measuring these additional observables, one
can determine the non-perturbative OPE parameters, substitute them into
the expression of the semileptonic width, and measure $|V_{cb}|$ with a
total precision of about $1$\---$2\%$. This is the basic idea underlying
the global fit analysis of $|V_{cb}|$ discussed in this section.

These other observables are the (truncated)
moments of the lepton energy $E_\ell$ (in the $B$~rest frame) and the
hadronic mass squared $m^2_X$ spectra in $B\to X\ell\nu$. The
quantity $m^2_X$ is the invariant mass squared 
of the hadronic system $X_c$ accompanying the
lepton\---neutrino pair. The lepton energy moments are defined as
\begin{equation}
  \langle
  E^n_\ell\rangle_{E_\mathrm{cut}}=\frac{R_n(E_\mathrm{cut})}{R_0(E_\mathrm{cut})}~,
\end{equation}
where $E_\mathrm{cut}$ is the lower lepton energy threshold and
\begin{equation}
  R_n(E_\mathrm{cut})=\int_{E_\ell>E_\mathrm{cut}}E^n_\ell\frac{d\Gamma}{dE_\ell}dE_\ell~.
\end{equation}
Here, $d\Gamma/dE_\ell$ is the partial semileptonic width as a
function of the lepton energy. The hadronic mass moments are
\begin{equation}
  \langle m^{2n}_X\rangle_{E_\mathrm{cut}}=\frac{S_n(E_\mathrm{cut})}{S_0(E_\mathrm{cut})}~,
\end{equation}
with
\begin{equation}
  S_n(E_\mathrm{cut})=\int_{E_\ell>E_\mathrm{cut}}m^{2n}_X\frac{d\Gamma}{dm^2_X}dm^2_X~.
\end{equation}
Here the integration over the $B\to X_c\ell\nu$ phase space
is restricted by the requirement $E_\ell>E_\mathrm{cut}$. These
observables can be expanded in OPEs similar to Eq.~\ref{eq:2_1},
containing the \emph{same} non-perturbative parameters.

In practice, the semileptonic width and moments in $B\to X_c\ell\nu$
have been calculated in two theoretical frameworks, referred to by the
name of the renormalization scheme used for the quark masses (though
this is not the only difference in the calculations): The calculations
in the \emph{kinetic scheme} are now available at
next-to-next-to-leading (NNLO) order in
$\alpha_s$~\cite{Benson:2003kp,Gambino:2004qm}. At leading order in
the OPE, the non-perturbative parameters are the quark masses $m_b$
and $m_c$. At $\mathcal{O}(1/m^2_b)$ the parameters are $\mu^2_\pi$
and $\mu^2_G$, and at $\mathcal{O}(1/m^3_b)$ the parameters $\rho^3_D$
and $\rho^3_{LS}$ appear. Independent expressions have been obtained
in the \emph{1S scheme}~\cite{Bauer:2004ve}. Here, the long-distance
parameters are $m_b$ at leading order, $\lambda_1$ and $\lambda_2$
at $\mathcal{O}(1/m^2_b)$ and $\rho_1$, $\tau_{1-3}$ at
$\mathcal{O}(1/m^3_b)$. Note that the numerical values of the quark
masses in the two schemes cannot be compared directly due to their
different definitions.

Belle has measured moments of inclusive observables in $B\to
X_c\ell\nu$~decays~\cite{Urquijo:2006wd,Schwanda:2006nf}. The 
lepton energy~$E_\ell$ and hadronic mass squared~$m^2_X$ spectra in
$B\to X_c\ell\nu$ are based on $152\times 10^6$ $\Upsilon(4S)\to B\bar
B$~events. These analyses first fully reconstruct the decay of one
$B$~meson ($B_\mathrm{tag}$) in the event
in a hadronic mode (or a hadronic tag).  
The tracks and clusters associated with $B_\mathrm{tag}$
 are removed from the event.
The semileptonic decay of the second $B$~meson in the
event ($B_\mathrm{sig}$) is then identified by searching for a charged
lepton among the remaining particles in the event. In the lepton
energy analysis~\cite{Urquijo:2006wd}, the electron momentum
spectrum~$p^*_e$ in
the $B$~meson rest frame is measured down to 0.4~GeV/$c$. In the
hadronic mass study~\cite{Schwanda:2006nf}, all remaining particles in
the event,
after excluding the charged lepton (either an electron or muon), are
combined to reconstruct the hadronic $X$~system. The $m^2_X$ spectrum
is measured for lepton energies above 0.7~GeV in the $B$~meson rest frame.

The observed spectra are distorted by resolution and
acceptance effects and cannot be used directly to obtain the
moments. In the Belle analyses, acceptance and finite resolution
effects are corrected by unfolding the observed spectra using the
singular value decomposition (SVD)
algorithm~\cite{Hocker:1995kb}. Belle measures the lepton energy
moments $\langle E^k_\ell\rangle$ for $k=0,1,2,3,4$ and minimum lepton
energies ranging from 0.4 to 2.0~GeV. Moments of the hadronic mass~$\langle
m^k_X\rangle$ are measured for $k=2,4$ and minimum lepton energies
between 0.7 and 1.9~GeV.

\begin{figure}
  \begin{center}
    \includegraphics[width=0.32\columnwidth]{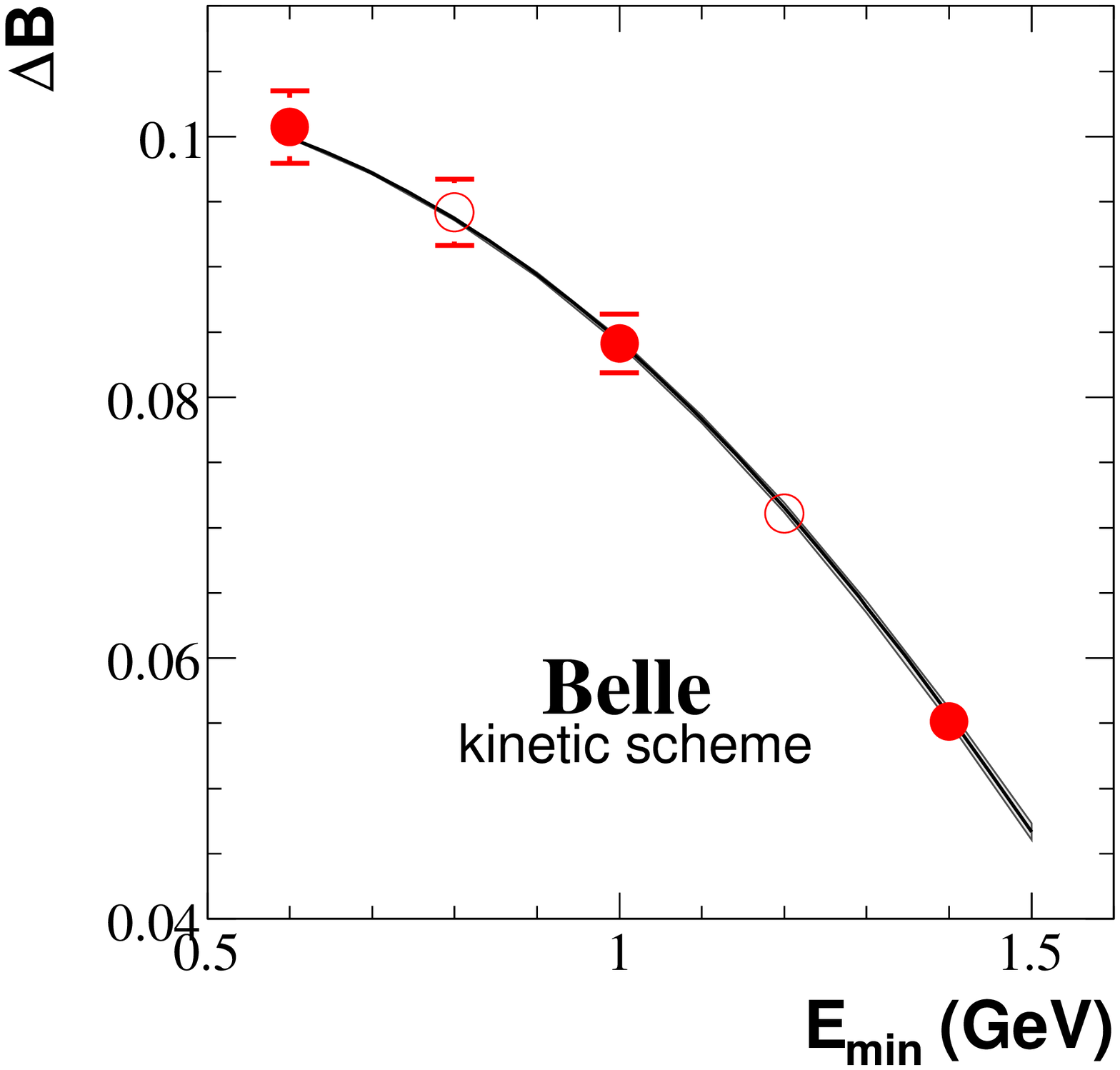}
    \includegraphics[width=0.32\columnwidth]{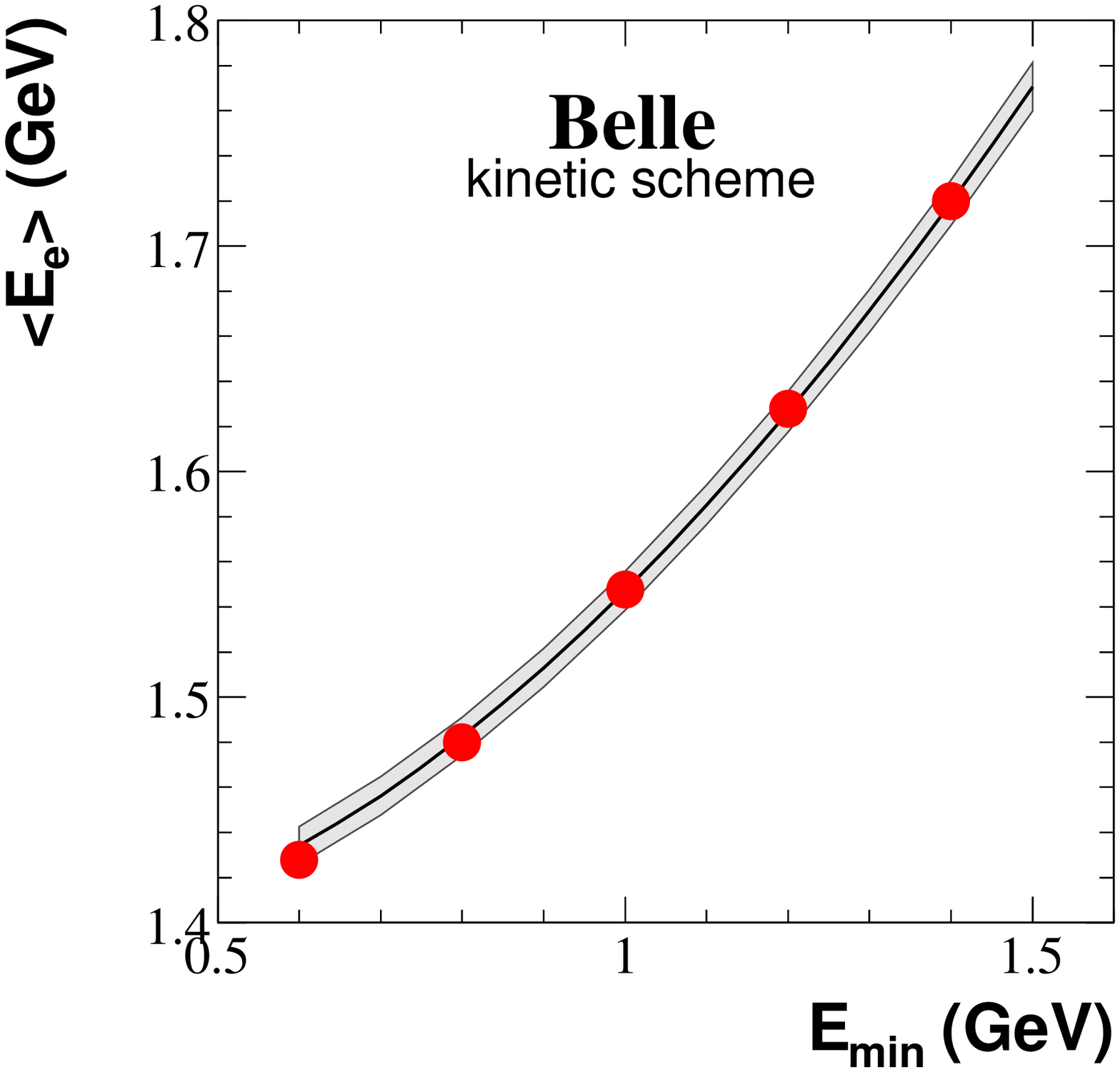}
    \includegraphics[width=0.32\columnwidth]{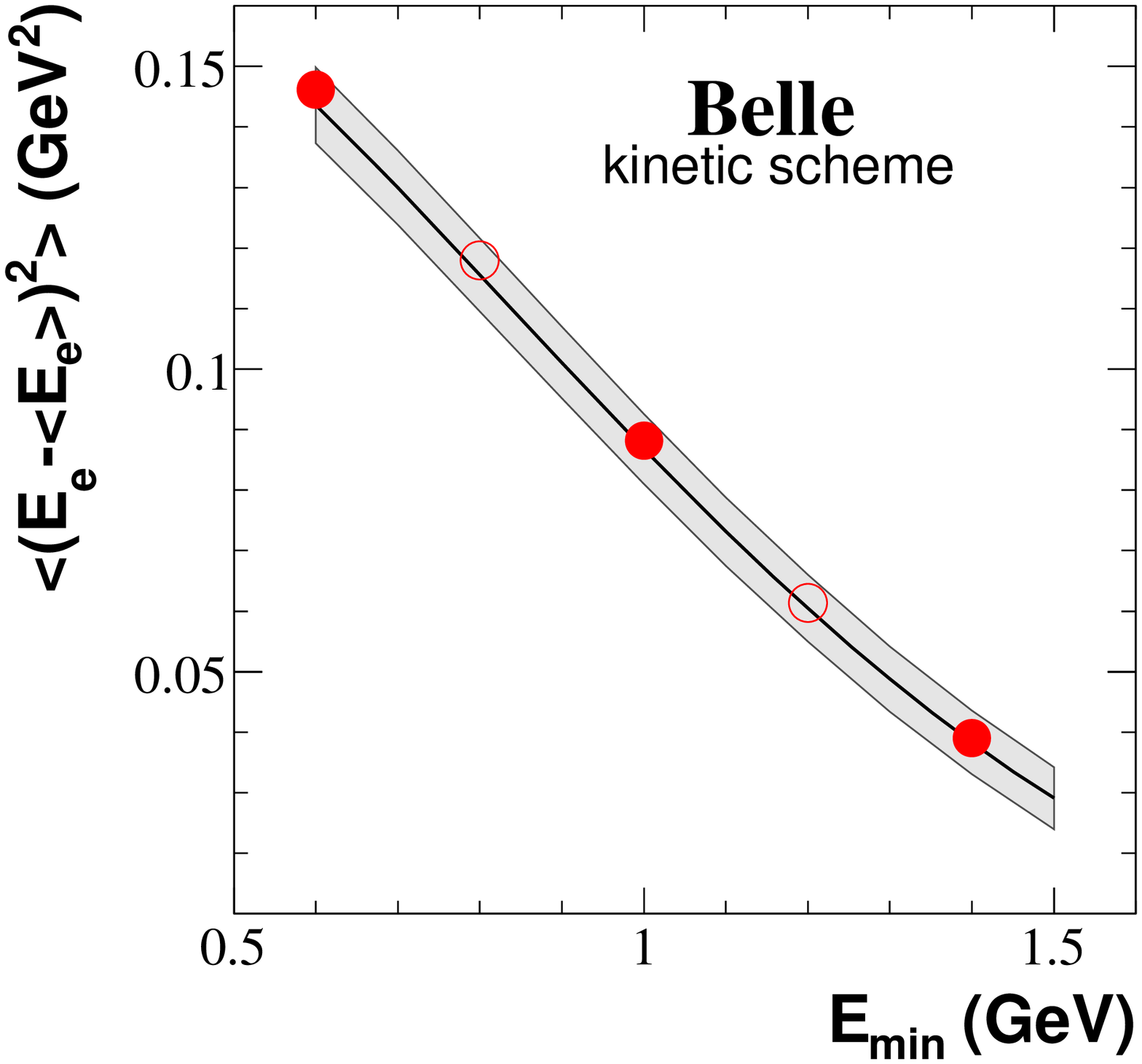}\\
    \includegraphics[width=0.32\columnwidth]{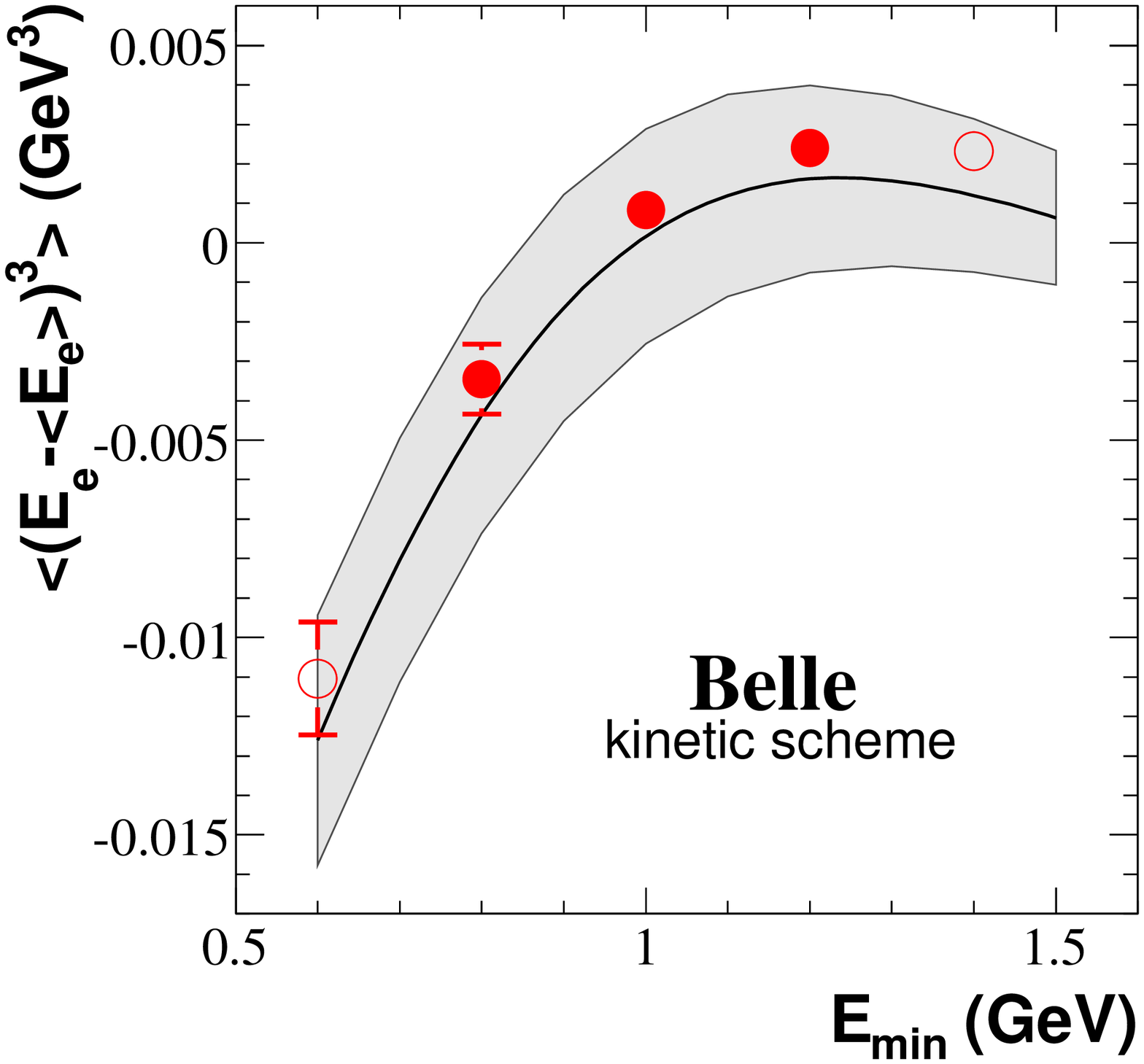}
    \includegraphics[width=0.32\columnwidth]{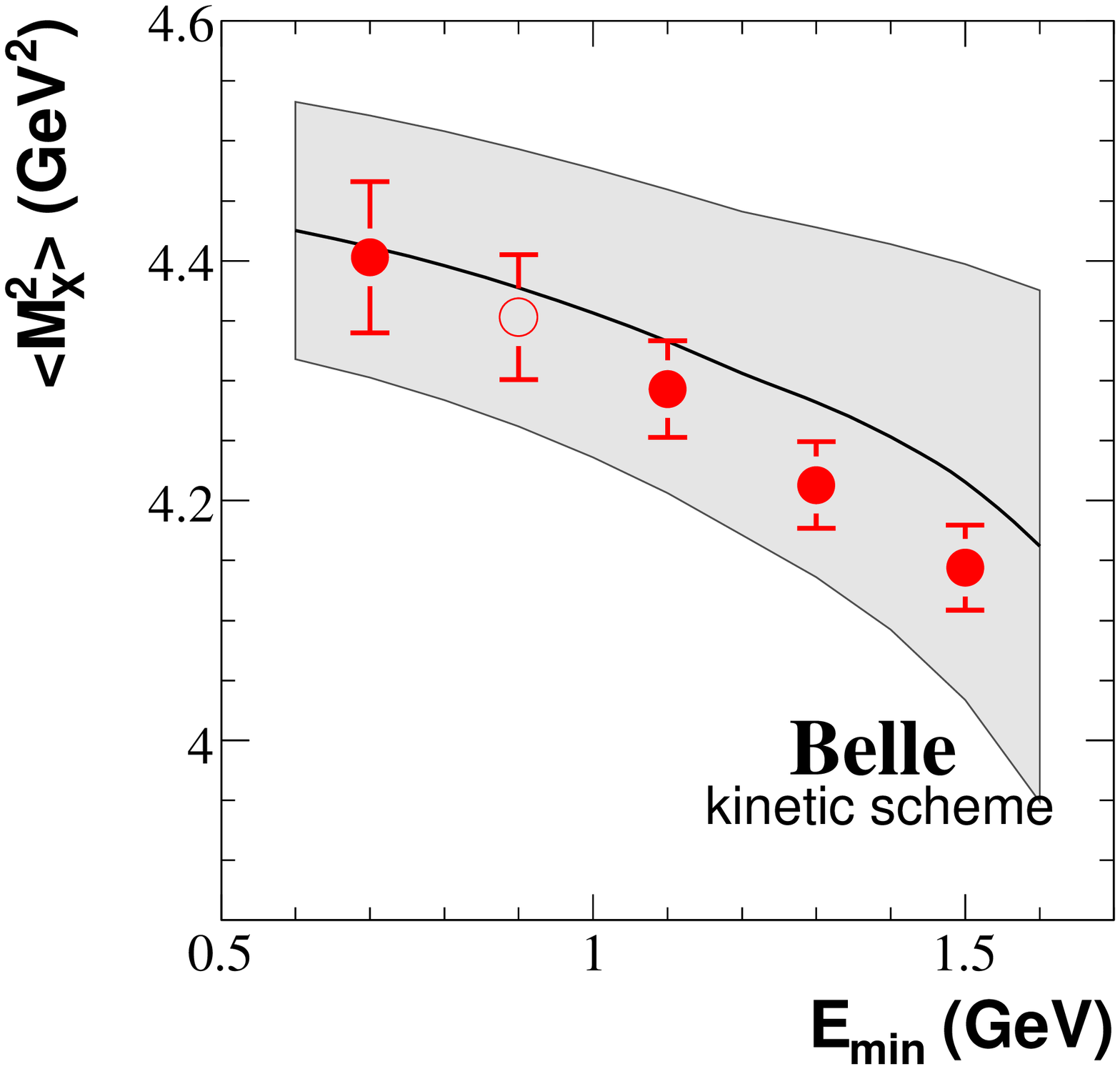}
    \includegraphics[width=0.32\columnwidth]{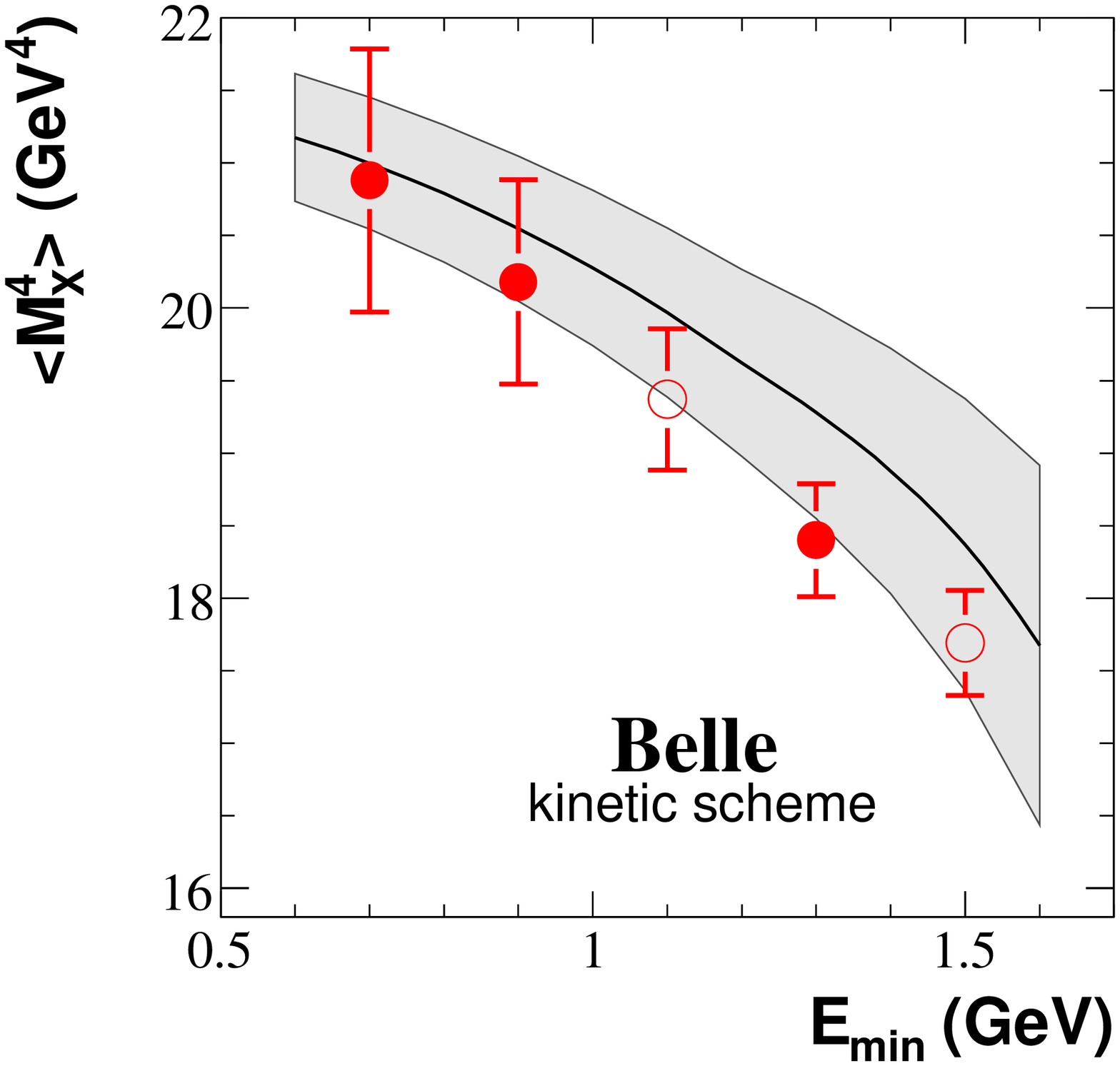}
  \end{center}
  \caption{Global fit of moments in $B\to X_c\ell\nu$ decays
    measured at Belle to theoretical expressions obtained in the
    kinetic scheme. The error bars show the experimental
    uncertainties. The
    error bands represent the theory error. Filled circles are data
    points used in the fit, and open circles are unused
    measurements.} \label{fig:xclnu_fit}
\end{figure}

To determine $|V_{cb}|$, Belle performs fits~\cite{Schwanda:2008kw} to
14 lepton energy moments, 7 hadronic mass moments and 4 moments of the
photon energy spectrum in $B\to X_s\gamma$ based on OPE
expressions derived in the kinetic
\cite{Benson:2003kp,Gambino:2004qm,Benson:2004sg} and
1S~schemes~\cite{Bauer:2004ve} (Fig.~\ref{fig:xclnu_fit}). Both
theoretical frameworks are considered independently and yield very
consistent results: The fit to the Belle data in the kinetic scheme yields
\begin{equation}
  |V_{cb}| = (41.58\pm 0.90)\times 10^{-3}~,
\end{equation}
while in the 1S scheme we obtain 
\begin{equation}
  |V_{cb}| = (41.56\pm 0.68)\times 10^{-3}~.
\end{equation}
While the result in the 1S scheme is more precise (1.6\% uncertainty
compared to 2.2\% in the kinetic scheme), it should be noted that the
assumptions on the dominant theory error are significantly different.

\subsection{$|V_{ub}|$}

\subsubsection{$|V_{ub}|$ from exclusive $B \to X_u \ell \nu$ decays}
The absolute value of $V_{ub}$, one of the least known CKM elements, can be
determined from rate measurements of exclusive charmless semileptonic decays,
such as 
$B \to \pi \ell \nu$, $B \to \rho \ell \nu$ and $B \to \omega \ell \nu$.
Of these, $B^0 \to \pi^- \ell^+ \nu$ decay has been the most extensively 
studied both theoretically and experimentally.
The decay rates and $|V_{ub}|$ are related as\,
\begin{equation}
\frac{d\Gamma(B^0 \to \pi^- \ell^+ \nu)}{dq^2} = 
\frac{G_F^2}{24 \pi^3} |V_{ub}|^2 p_{\pi}^3 |f_{+}(q^2)|^2~,
\label{eqn:pilnu-vub}
\end{equation} 
where $G_F$ is the Fermi coupling constant, and $f_+(q^2)$ is the $B \to \pi$
transition form factor, which is calculated in lattice QCD and 
by QCD sum rules.
Compared to the inclusive measurements, described below, the exclusive 
measurements are relatively straightforward 
experimentally, but suffer from large
theoretical uncertainties in the form factors, 
which must be determined from non-perturbative QCD calculations.

\begin{figure}[htb]
\parbox{\halftext}{
\centerline{\includegraphics[width=6.5cm]{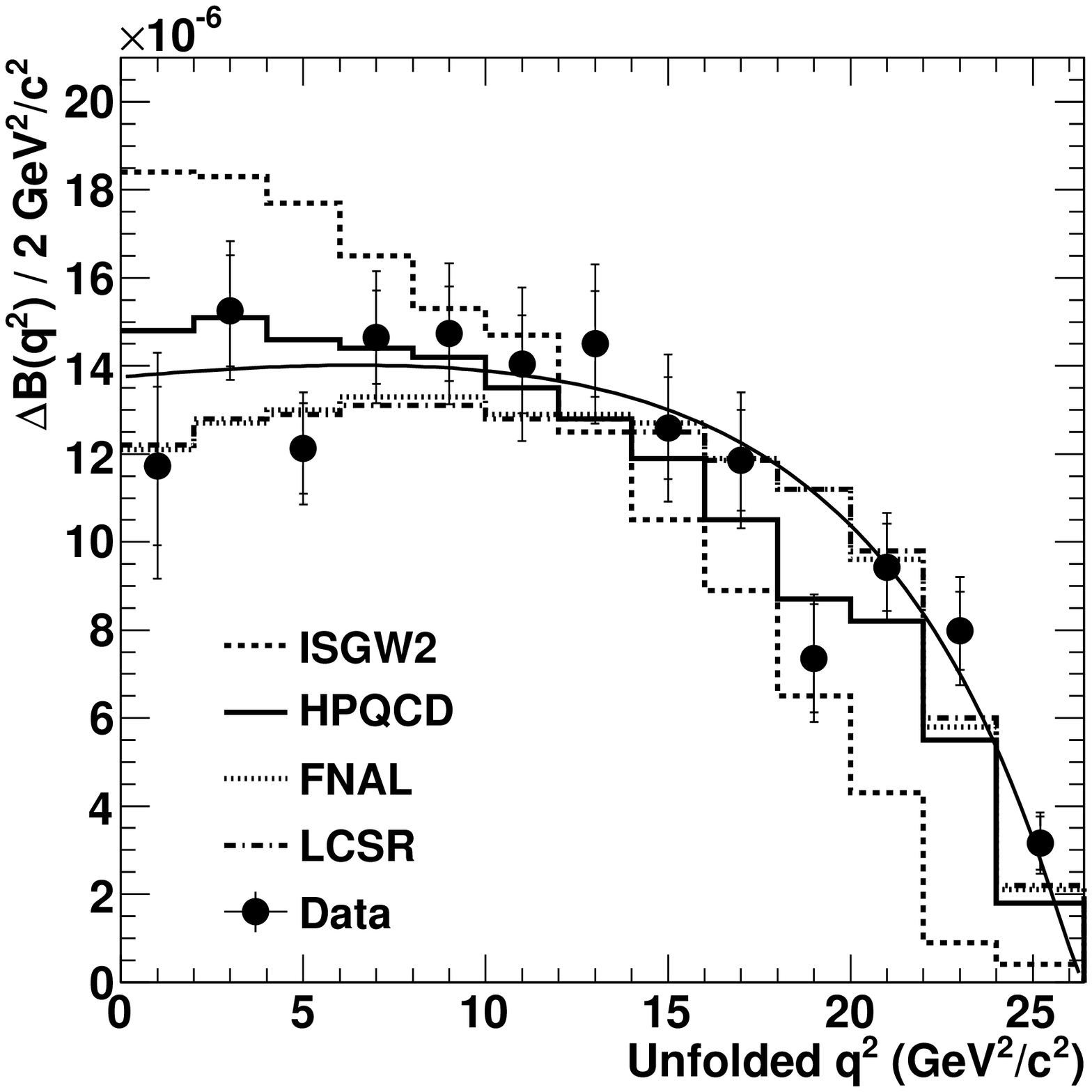}}
\caption{
Measured $q^2$ distribution for the $B^0 \to \pi^- \ell \nu$ decay. 
The curve represents a fit to an empirical form factor parameterization.
The four histograms show various form factor predictions (dashed: ISGW2;
plain: HPQCD; dotted: FNAL; dot-dashed: LCSR).
}
\label{fig:pilnu_1}
}
\hfill
\parbox{\halftext}{
\centerline{\includegraphics[width=6.5cm]{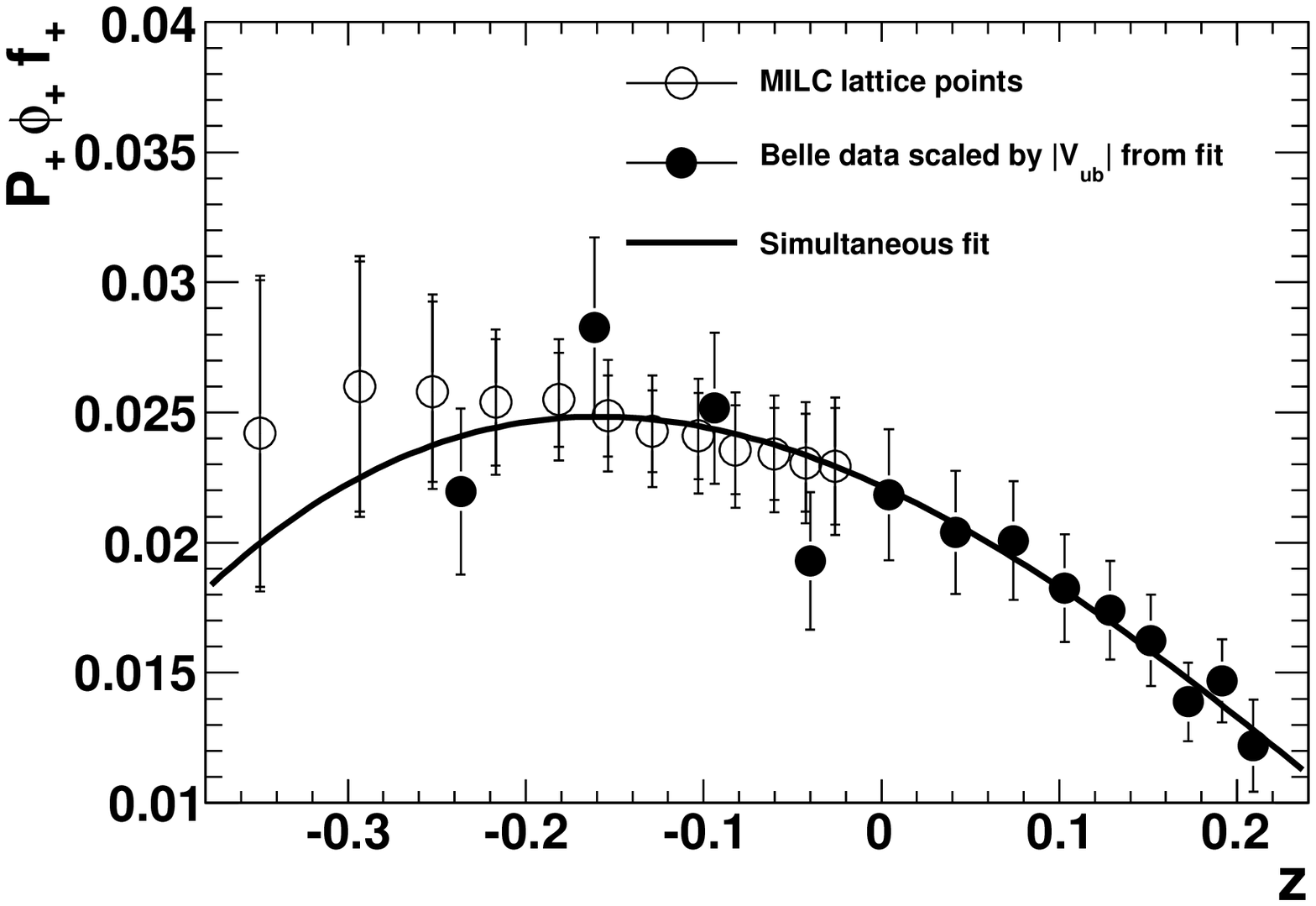}}
\caption{
$|V_{ub}|$ extraction from a simultaneous fit to experimental (closed
circles) and FNAL/MILC lattice QCD results (open circles). 
}
\label{fig:pilnu_2}}
\end{figure}

The Belle collaboration has measured $B \to \pi/\rho/\omega \ell\nu$ 
decays~\cite{Schwanda:2004, Hokuue:2007, Ha:2011}.
The most recent measurement of the $B^0 \to \pi^- \ell^+ \nu$ decay
\cite{Ha:2011} uses a data sample containing $657 \times 10^6$ $B\bar{B}$ 
pairs, and has the best precision for the $|V_{ub}|$ determination.
In this analysis, signals are reconstructed by combining an oppositely charged 
pion and lepton (either electron or muon), originating from a common vertex.
For the reconstruction of the undetected neutrino, the missing energy and 
momentum in the c.m. frame are defined as 
$E_{\rm miss} \equiv 2 E_{\rm beam} - \Sigma_i E_i$ and
$\vec{p}_{\rm miss} \equiv - \Sigma_i \vec{p_i}$, respectively, where 
$E_{\rm beam}$ is the beam energy in the c.m. frame, and the sums include all 
charged and neutral particle candidates in the event. 
We require $E_{\rm miss} > 0$ GeV, 
and the neutrino 4-momentum is taken to be 
$p_{\nu} = (|\vec{p}_{\rm miss}|,\vec{p}_{\rm miss})$, 
since the determination of 
$\vec{p}_{\rm miss}$ is more accurate than that of the missing energy.
As in the analysis of $B^0 \to D^{*-} \ell^+ \nu$, using the variable
cos$\theta_{BY}$ (Eq.~\ref{eq:1_3}), 
the $B$ momentum vector is constrained to lie on
a cone centered on the $\pi^- \ell^+$
direction; signals can then be selected by requiring $|\cos\theta_{BY}| < 1$.
Background from continuum 
$e^+e^- \to q \bar{q} (q = u, d, s, c)$ jets are reduced using 
an event topology requirement based on the second Fox\---Wolfram moment.
Signals are extracted, for each of 13 $q^2$ bins ranging from 0 to 26 
GeV$^2/c^2$, by fitting the two-dimensional distribution of the beam energy
constrained mass
$M_{\rm bc} = \sqrt{E_{\rm beam}^2-|\vec{p_{\pi}}+\vec{p_{\ell}}+\vec{p_{\nu}}|^2}$ 
and the energy difference 
$\Delta E = E_{\rm beam} - (E_{\pi} + E_{\ell} + E_{\nu})$.
Figure~\ref{fig:pilnu_1} shows the obtained $q^2$ distribution.
The total branching fraction, 
integrated over the entire $q^2$ region, is 
\begin{equation}
{\cal B}(B^0 \to \pi^- \ell^+ \nu) 
= [1.49 \pm 0.04 ({\rm stat}) \pm 0.07 ({\rm syst})] \times 10^{-4}~.
\end{equation}

The value of $|V_{ub}|$ can be determined from the measured 
differential $q^2$ distribution
using Eq.(\ref{eqn:pilnu-vub}).
Following the procedure proposed by the FNAL/MILC collaboration
~\cite{FNAL/MILC:2009}, $|V_{ub}|$ can be 
extracted from a simultaneous fit to experimental and lattice QCD results from
the FNAL/MILC collaboration, as shown in Fig.~\ref{fig:pilnu_2}.
In this approach, $q^2$ is transformed to a dimensionless quantity $z$, and
both the 
experimental and lattice QCD distributions are fit to a third-order polynomial
with $|V_{ub}|$ determined as a relative normalization between the lattice 
QCD and experimental results.
We find $|V_{ub}| = (3.43 \pm 0.33) \times 10^{-3}$, as shown in 
Table~\ref{tbl:vub_pilnu}.
The table also lists $|V_{ub}|$ values determined using only a fraction 
of the overall phase space, 
leading to less precise but statistically compatible results. 
The form factor $f_+(q^2)$ predictions
are based on the light cone sum rule (LCSR) 
and lattice QCD (LQCD), which can be applied in the regions
$q^2 < 16$~GeV$^2/c^4$ and $q^2 > 16$~GeV$^2/c^4$, respectively.

\begin{table}[htbp]
\begin{center}
\caption{Summary of $|V_{ub}|$ results
from a recent $B^0 \to \pi^- \ell \nu$ measurement 
by Belle.}
\label{tbl:vub_pilnu}
\begin{tabular}{ccc}
\hline
\hline
Theory & $q^2$ (GeV$^2/c^4$) & $|V_{ub}|$ ($\times 10^{-3}$) \\
\hline
LCSR~\cite{LCSR:2005}   & $ < 16 $ & $3.64 \pm 0.11 ^{+0.60}_{-0.40}$ \\
HPQCD~\cite{HPQCD:2006} & $ > 16 $ & $3.55 \pm 0.13 ^{+0.62}_{-0.41}$ \\
FNAL~\cite{Okamoto:2004xg}   & $ > 16 $ & $3.78 \pm 0.14 ^{+0.65}_{-0.43}$ \\
\hline
FNAL/MILC~\cite{FNAL/MILC:2009} & all regions & $3.43 \pm 0.33$ \\
\hline
\hline
\end{tabular}
\end{center}
\end{table}

\subsubsection{$|V_{ub}|$ from inclusive $B \to X_u \ell \nu$ decays}
For inclusive $B \to X_u \ell \nu$ decays, the theoretical description relies
on the OPE, as in the case of inclusive 
$B \to X_c \ell \nu$ decays.
However, $B \to X_u \ell \nu$ decays are about 50 times less abundant than
$B \to X_c \ell \nu$ decays, and thus the experimental sensitivity to
$B \to X_u \ell \nu$ and $|V_{ub}|$ is highest in the region of phase space
that is less impacted by the dominant background
from $B \to X_c \ell \nu$ decays.
In this phase space region, however, non-perturbative corrections are
kinematically enhanced, and as a result non-perturbative dynamics become
an O(1) effect. Extracting $|V_{ub}|$
 requires the use of theoretical parameterizations called 
{\it shape functions (SF)} to describe the unmeasured regions of phase space.

A classical method is to measure the lepton momentum spectrum at the end-point 
of the spectrum 
($p_{\ell}^{cm} > 2.3$GeV/$c$), where the $b \to c$ decay is forbidden.
This method allows the measurement of $|V_{ub}|$ with small data samples, but 
suffers from a large extrapolation error, because only 
a limited portion of the phase space 
($\sim$10 \% of the total) is measured.
Belle reported a result using this method in 2005~\cite{Limosani:2005}.
The high luminosity data at Belle enable us to also measure kinematic
variables such as the invariant mass of the $X_u$ hadronic system, $m_X$,
and the four-momentum transfer of the $B$ meson to the $X_u$ system, $q$.
This enables us to control the experimental and theoretical errors by 
optimizing the region of phase space for the measurement.
Belle reported the first measurement using $m_x - q^2$ for the inclusive 
$B \to X_u \ell \nu$ decay~\cite{Kakuno:2004}. 

\begin{figure}[htbp]
\centerline{\includegraphics[width=14.0cm]{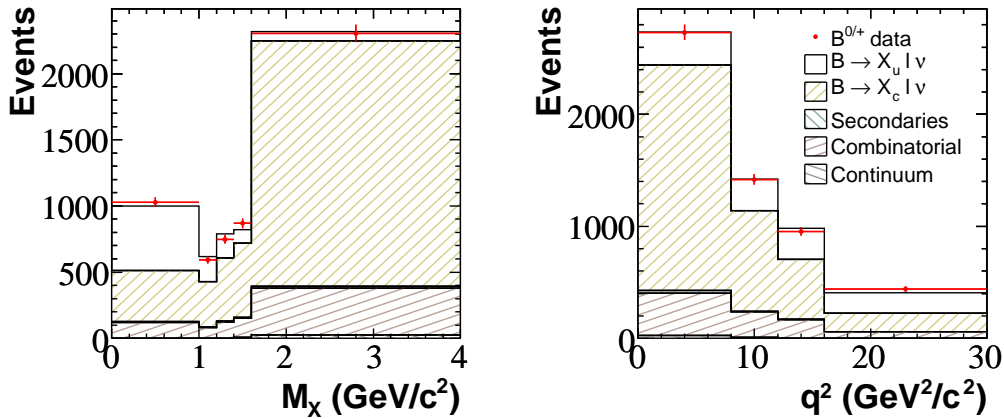}}
\caption{
Projections of the $m_X - q^2$ fit in bins of $m_X$ (left) and $q^2$ (right).
}
\label{fig:xulnu}
\end{figure}

More recently, Belle reported a measurement of the partial branching fraction 
of $B \to X_u \ell \nu$ decays with a lepton momentum threshold of 1 GeV/$c$ 
using a multivariate data mining technique, with a data sample containing 
$657 \times 10^6 B\bar{B}$ pairs~\cite{Urquijo:2010}.
This method allows us to access $\sim 90 \%$ of the $B \to X_u \ell \nu$ phase 
space and minimizes the dependence on an SF.
The measurement is made 
by fully reconstructing one $B$ meson ($B_{\rm tag}$) in
hadronic decays, and measuring the semileptonic decay 
of the other $B$ meson 
($B_{\rm sig}$) with a high momentum electron or muon.
The $B \to X_u \ell \nu$ decays are selected based on a nonlinear 
multivariate boosted decision tree (BDT), which incorporates a total of 17 
discriminating variables, such as the kinematical quantities of candidate 
semileptonic decays, number of kaons in the event, $M_{\rm bc}$ of 
$B_{\rm tag}$, etc.
The candidates passing the selection of the BDT classifier are analyzed in a 
two-dimensional fit in the ($m_X$, $q^2$) plane.
The hadronic invariant mass $m_X$ is calculated 
from the measured momenta of 
all charged tracks and neutral clusters 
that are not associated to $B_{\rm tag}$ 
reconstruction or used as a lepton candidate. The momentum transfer is 
calculated as $q = p_{\Upsilon(4S)} - p_{B_{\rm tag}} - p_X$.
Figure~\ref{fig:xulnu} shows the one-dimensional 
projections of the $(m_X, q^2)$ 
distribution with a lepton momentum requirement
of $p_{\ell}^{*B} > 1.0$GeV/$c$, fitted 
with distributions for the $B \to X_u \ell \nu$ signal, $B \to X_c \ell \nu$ 
and other backgrounds mainly from secondary and misidentified leptons.
The partial branching fraction for $p_{\ell}^{*B} > 1.0$GeV/$c$ is 
\begin{equation}
\Delta {\cal B}(B \to X_u \ell \nu; p_{\ell}^{*B} > 1.0 {\rm GeV}/c)
= 1.963 \times (1 \pm 0.088 ({\rm stat}) \pm 0.081 ({\rm syst})) 
\times 10^{-3}.
\end{equation}     
A $|V_{ub}|$ value is obtained from the partial branching fraction using 
$|V_{ub}|^2 = \Delta {\cal B}_{u\ell\nu}/(\tau_B \Delta R)$, where $\Delta R$ 
is the predicted $B \to X_u \ell \nu$ partial rate in the given phase space 
region, and $\tau_B$ is the average $B$ lifetime.
Table~\ref{tbl:vub_xulnu} presents $|V_{ub}|$ results
based on different theoretical
prescriptions that predict $\Delta R$.
Here the results were obtained by the Heavy Flavor Averaging Group (HFAG) 
using the most recent calculations and input parameters~\cite{hfag}.
The results are consistent within their stated theoretical uncertainties, 
and have an overall uncertainty of $\sim 7\%$.
   
As described above, there is a tension between the $|V_{ub}|$ values extracted
from the exclusive and inclusive methods, which are subject to further 
clarification with improved experimental and theoretical errors in
the future.

\begin{table}[htbp]
\begin{center}
\caption{$|V_{ub}|$ values obtained using the inclusive $B \to X_u \ell nu$ 
measurement by Belle and input parameters ($m_b$ and $\mu^2_{\pi}$). 
The errors quoted on $|V_{ub}|$ correspond to experimental and theoretical 
uncertainties, respectively.}
\label{tbl:vub_xulnu}
\begin{tabular}{cccc}
\hline
\hline
Theory & $m_b$ (GeV) & $\mu^2_{\pi}$ (GeV$^2$) & $|V_{ub}|$ ($\times 10^{-3}$) \\
\hline
BLNP~\cite{BLNP:2005} & $4.588 \pm 0.025$ & $0.189^{+0.046}_{-0.057}$ & 
$4.47 \pm 0.27 ^{+0.19}_{-0.21}$ \\ 
DGE~\cite{DGE:2006}   & $4.194 \pm 0.043$ & --- & 
$4.60 \pm 0.27 ^{+0.11}_{-0.13}$ \\
GGOU~\cite{GGOU:2007} & $4.560 \pm 0.023$ & $0.453 \pm 0.036$ & 
$4.54 \pm 0.27 ^{+0.10}_{-0.11}$ \\
ADFR~\cite{ADFR:2009} & $4.194 \pm 0.043$ & --- & 
$4.48 \pm 0.30 ^{+0.19}_{-0.19}$ \\
\hline
\hline
\end{tabular}
\end{center}
\end{table}

\subsection{Purely leptonic $B^- \to \ell^- \bar{\nu}_\ell
  ~(\ell = e$, $\mu$, or $\tau)$ decays}

\def\BF{{\cal B}}
\def\EECL{E_{\rm ECL}}
\def\MM2{M_{\rm miss}^2}
\def\BBbar{B\overline{B}}
\def\lnu{\ell^- \bar{\nu}_\ell}
\def\enu{e^- \bar{\nu}_e}
\def\munu{\mu^- \bar{\nu}_\mu}
\def\taunu{\tau^- \bar{\nu}_\tau}
\def\Btolnu{B^- \to \lnu}
\def\Btoenu{B^- \to \enu}
\def\Btomunu{B^- \to \munu}
\def\Btotaunu{B^- \to \taunu}
\def\Btag{B_{\rm tag}}
\def\Bsig{B_{\rm sig}}
\def\CosBDl{\cos \theta_{B, D^{(*)}\ell}}
\def\plB{p_\ell^B}
\def\BztoDstTaunu{\overline{B}^0 \to D^{*+} \taunu}
\def\BmtoDzTaunu{B^- \to D^{0} \taunu}
\def\BmtoDstzTaunu{B^- \to D^{*0} \taunu}

\begin{wrapfigure}{r}{6.1cm}
  \centerline{\includegraphics[width=0.45\textwidth]
                                {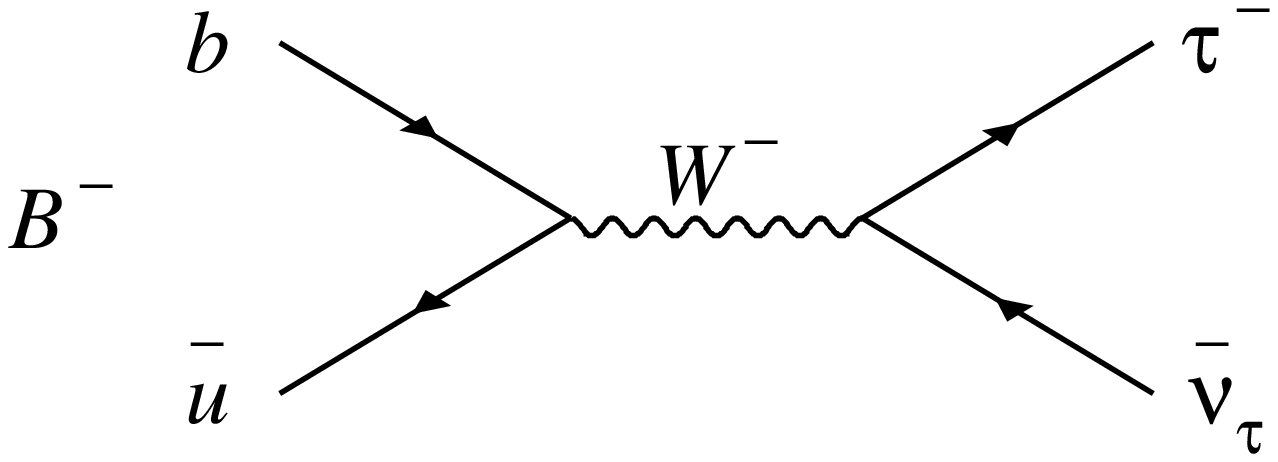}}
                              \caption{A Feynman diagram for the SM
                                $\Btotaunu$ process.}
\label{fig:B2taunuDiagram}
\end{wrapfigure}

In the SM, $\Btolnu$ decays to purely leptonic final states 
($\ell = e,$ $\mu$ or $\tau$) occur via annihilation of the two quarks
in the initial state, $b$ and $\bar{u}$, to a $W^-$ boson
(Fig.~\ref{fig:B2taunuDiagram}). The branching fraction for a
$\Btolnu$ decay is given by
\begin{equation}
\BF (\Btolnu) = \frac{G_F^2 m_B m_\ell^2}{8\pi} \left( 1 -
  \frac{m_\ell^2}{m_B^2} \right)^2
f_B^2 |V_{ub}|^2 \tau_B \ ,
\label{eq:B2lnu}
\end{equation}
where $G_F$ is the weak interaction coupling constant, 
$m_\ell$ and $m_B$
are the lepton and $B^+$ meson masses, respectively, $\tau_B$ is the
$B^-$ lifetime, $|V_{ub}|$ is the magnitude of a CKM matrix element,
and $f_B$ is the $B^-$ meson decay constant. All these input
parameters have been directly measured with good precision except for
$f_B$. The value of $f_B$ can be obtained using LQCD calculations.
Since LQCD calculations are based on first
principles of QCD, it is possible to calculate 
the SM expectation for $\BF (\Btolnu)$ with high precision.  
Therefore, measurement
of $f_B$ via $\Btolnu$ decays can provide a stringent test of the
LQCD, within the framework of the SM.

On the other hand, particles from physics beyond the SM, for
example, a
charged Higgs boson in supersymmetry or a generic two-Higgs doublet model,
may take the place of the $W^-$ in Fig.~\ref{fig:B2taunuDiagram} and
modify the branching fraction.  Moreover, in the minimum flavor
violation NP scheme, it is expected that the relative branching fractions
of charged lepton modes will remain the same as those predicted by the SM.
Accordingly, measuring the branching fractions 
of $\Btolnu$ ($\ell = e$, $\mu$ or $\tau$) modes
and their relative ratios can provide a very
sensitive probe for NP beyond the SM.

Due to helicity suppression, the branching fraction
(Eq.~\ref{eq:B2lnu}) is proportional to the square of the charged
lepton mass, $m_\ell^2$.  As a result, the SM branching
fractions for $\enu$ and $\munu$ modes are suppressed in comparison to
the $\taunu$ mode by factors of $\sim 10^7$ and $\sim 200$,
respectively.  At the time of this report, there exists evidence for
$\Btotaunu$ from
Belle~\cite{Belle:taunuFirst,Belle:taunuSemilep} and
BaBar\cite{BaBar:taunuResults}, 
but no evidence has yet been found for the $\Btoenu$ and $\Btomunu$ modes.

\subsubsection{$\Btotaunu$}

While the large mass of the $\tau$ lepton significantly 
enhances the branching
fraction of $\Btotaunu$ compared to other modes, the presence of one or more
neutrinos from the $\tau$ decay make it difficult to cleanly detect
$\Btotaunu$ decays.  In the process $e^+ e^- \to \Upsilon(4S) \to
B\overline{B}$, signal sensitivity is greatly improved by 
completely reconstructing or ``tagging'' one
$B$ meson ($\Btag$); the signature of the signal is then 
searched for in the other $B$ meson ($\Bsig$).  Experimentally, two
different tagging methods have been applied to measure $\BF
(\Btotaunu)$: reconstructing a full decay chain of a hadronic final state
(``hadronic tagging'') or reconstructing all particles except for a
neutrino in semileptonic $\Btag \to D^{(*)} \ell\nu$ decays 
(``semileptonic tagging''). 

\subsubsection{Hadronic tagging analysis}
\label{sec:FRB2taunu}

The first evidence for $\Btotaunu$ decays was obtained in a hadronic
tagging analysis by Belle~\cite{Belle:taunuFirst} using $449 \times
10^6$ $\BBbar$ events, which obtained $\BF(\Btotaunu) =
(1.79_{~-0.49-0.51}^{~+0.56+0.46})\times 10^{-4}$.  Recently, Belle
has updated the hadronic tagging analysis of $\Btotaunu$, analyzing
the full Belle data sample containing $772 \times 10^6$ $\BBbar$
events.~\cite{Belle:newtaunu}

\begin{figure}[htb]
  \parbox{\halftext}{
    \centerline{\includegraphics[width=0.45\textwidth]
      {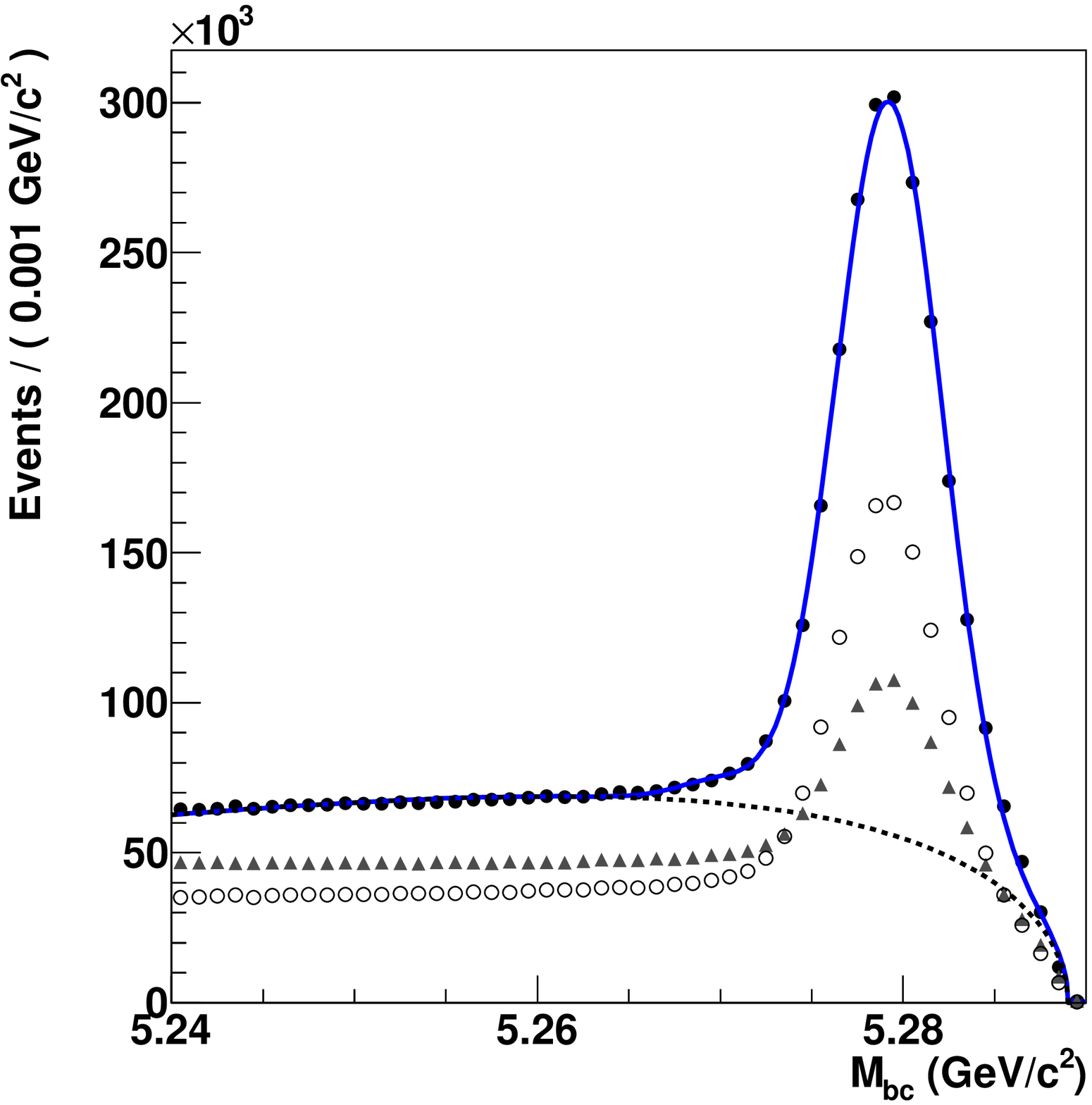}} \caption{$M_{\rm bc}$
      distributions for the $\Btag$ candidate events. The triangles, open
      circles, and solid circles represent the distributions obtained
      by applying the original tagging
      algorithm~\cite{Belle:taunuFirst} to the previous data set,
      applying improved hadronic 
      tagging to the previous data set, and applying improved
      tagging to the latest fully reprocessed data set, respectively.  The
      solid and dotted curves show the sum and the background
      component, respectively, of the fit to the full data sample.} 
  \label{fig:BtagNewOld}}
\hfill
  \parbox{\halftext}{
  \centerline{\includegraphics[width=0.4\textwidth]
                                {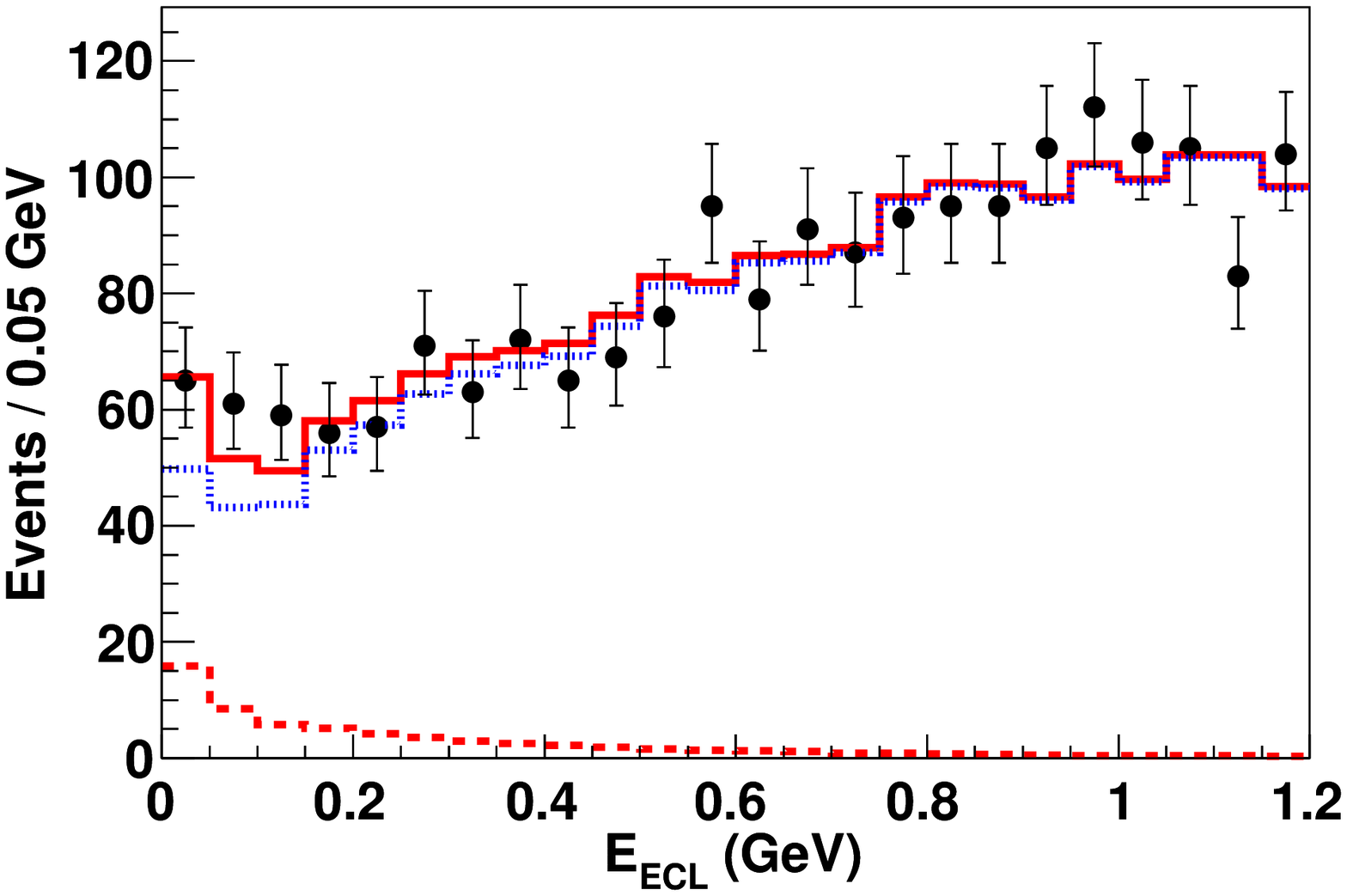}}

  \centerline{\includegraphics[width=0.4\textwidth]
                                {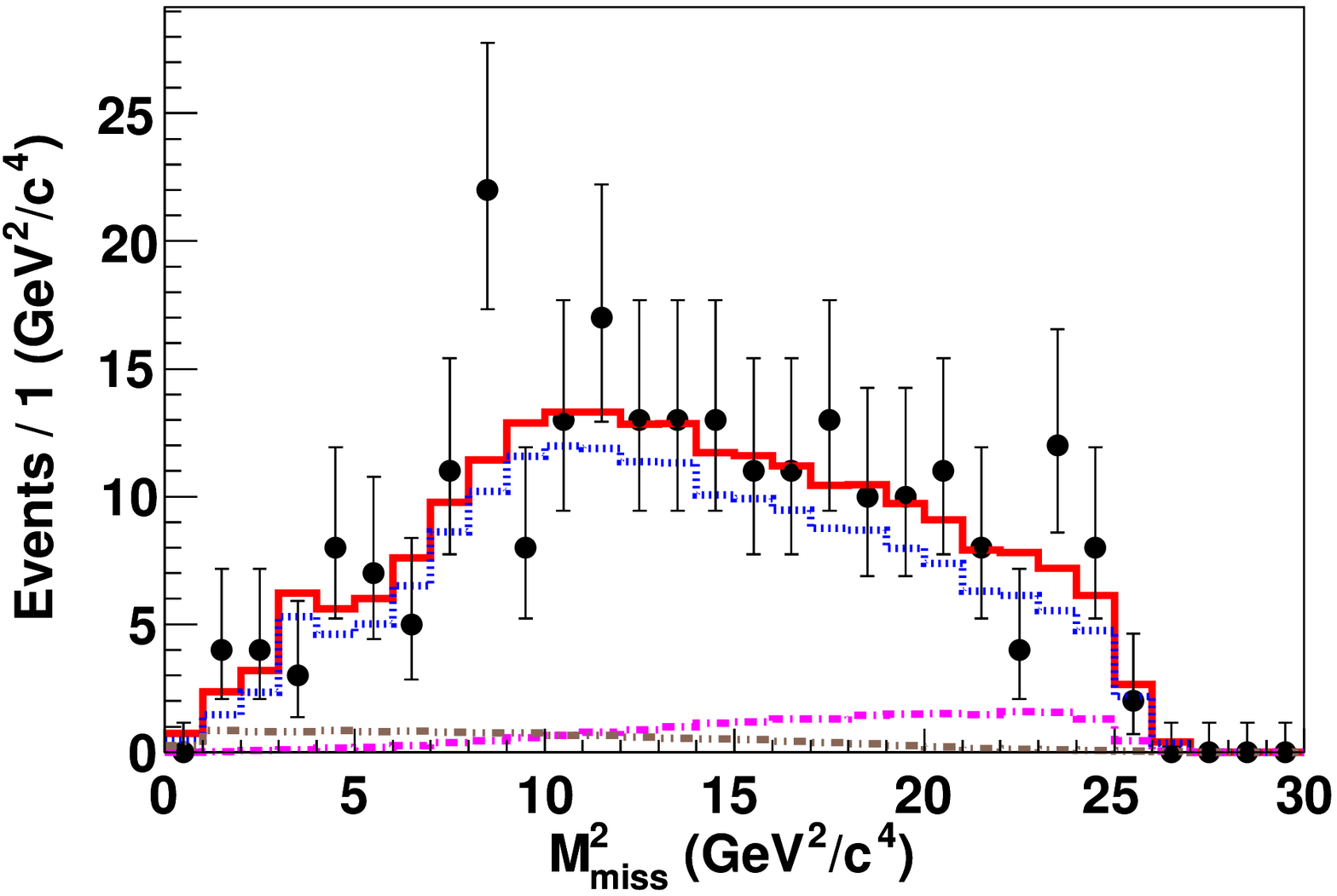}}
  \caption{Distributions of $\EECL$ (top) and $\MM2$ (bottom) combined
  for all the $\tau^-$ decays. The $\MM2$ distribution is shown for the
  signal region $\EECL < 0.2~{\rm GeV}$.  The solid circles with
  error bars are data.  The solid histograms show the projections of
  the fits.  The dashed and dotted histograms show the signal and
  background components, respectively.}
  \label{fig:EECLMM2}}
\end{figure}

In the most recent analysis, the data sample is fully reprocessed with 
much improved tracking and slightly
improved neutral cluster detection.  A new
hadronic tagging algorithm using a Bayesian artificial neural network
has been developed and applied to the
analysis.~\cite{Ref:NeuroBayesFullRecon} As a result of all these
improvements, the
statistics of the $\Btag$ sample has increased by nearly a factor of
three.  Figure~\ref{fig:BtagNewOld} shows the $M_{\rm bc}$
distribution of $\Btag$ candidate events, in comparison with that 
from the previous analysis.

Once the $\Btag$ candidates are selected, we search for $\Btotaunu$
decays using the particles not belonging to $\Btag$ in these events.
The $\tau^-$ lepton is identified in four decay modes: $\tau^- \to
e^- \bar{\nu}_e \nu_\tau,\ \mu^- \bar{\nu}_\mu \nu_\tau,\ \pi^-
\nu_\tau$, and $\pi^- \pi^0 \nu_\tau$.  Signal candidate events are
required to have only one track with charge opposite to 
$\Btag$. The charged tracks are required to be consistent with being
either an electron, muon, or pion.  For the $\tau^- \to \pi^- \pi^0 \nu_\tau$
mode, with $\pi^0 \to \gamma \gamma$, the invariant mass of the $\pi^-
\pi^0$ system must be within $0.15~{\rm GeV}\!/c^2$ of the nominal
$\rho^-$ mass.  There should be no other detected particles after
removing the particles from the $\Btag$ and the charged tracks and
$\pi^0$'s from the $\Bsig$. In particular, events containing extra $\pi^0$
and $K_L^0$ candidates are rejected.

The signal yield is evaluated by fitting the two-dimensional
distribution of $\EECL$ and $\MM2$, where $\EECL$ is the sum of the
energies of neutral clusters that are not associated with either the
$\Btag$ or the $\pi^0$ candidate in the $\tau^- \to \pi^- \pi^0
\nu_\tau$ decay and $\MM2$ is the missing mass squared defined by
$\MM2 = (E_{\rm CM} - E_{\Btag} - E_{\Bsig})^2 - |\vec{p}_{\Btag}
+\vec{p}_{\Bsig}|^2 $ with the energies and momenta measured in the CM
frame. To reduce background, we require $\MM2 > 0.7~({\rm
  GeV}\!/c^2)^2$. Figure~\ref{fig:EECLMM2} shows the projections of the
result of the fit on
$\EECL$ and $\MM2$ where the four $\tau$ decay modes are combined. The
preliminary fitted signal yield is 
$62_{~-22}^{~+23} \pm 6$ events and
the branching fraction $\BF_{\rm had}$ (in the hadronic
tagging analysis) is
\begin{equation}
\BF_{\rm had} (\Btotaunu) = (0.72_{~-0.25}^{~+0.27} \pm 0.11) \times
10^{-4} ~. 
\end{equation}
The signal significance is $3.0\sigma$ including systematic
uncertainty.  This result is consistent with the previous
measurement considering the overlap of the event samples.

\subsubsection{Semileptonic tagging analysis}

In the semileptonic tagging analysis of $\Btotaunu$, $\Btag$ is
reconstructed in $B^+ \to \overline{D}^{*0} \ell^+ \nu$ and $B^+ \to
\overline{D}^{0} \ell^+ \nu$ decays, where $\ell$ is an electron or muon.
Since semileptonic tagging imposes fewer constraints on the $\Bsig$
kinematics, only $\tau^-$ decays to $\ell^- \bar{\nu}_{\ell} \nu_\tau$
$(\ell=e, \mu)$ and $\pi^- \nu_\tau$ are used for 
$\Bsig$ reconstruction. Except for
$\overline{D}^0$ or $\overline{D}^{*0}$, one $\ell^+$ for $\Btag$, 
and one $\ell^-$ or $\pi^-$ for $\Bsig$, we allow no other charged track
or neutral particle in the event. 

One of the main variables to suppress background events is the cosine
of the angle, $\CosBDl$, between the momentum of $\Btag$ and that of
$\overline{D}^{(*)0}$ and $\ell^+$ system.  This variable is defined
in the same way as the variable
$\cos \theta_{BY}$ discussed in the previous section,
but with $Y = \overline{D}^{(*)0} \ell^+$. Correctly reconstructed
$\Btag$ candidates populate the physical range $-1 \leq
\CosBDl \leq 1$.  Signal candidates are selected based on
$P_\ell^{\rm cm}$ (the lepton momentum of $\Btag$ in the CM frame),
$\CosBDl$, and $P_{\rm sig}^{\rm cm}$ (the CM-frame momentum of the
charged track from $\Bsig$). 
The selection criteria depend on the $\tau$ decay mode of $\Bsig$. 
After all selections, the signal yield ($n_{\rm s}$) is obtained 
by fitting the
$\EECL$ distribution.  From a combined fit to the three $\tau^-$ decay
modes, $n_{\rm s} = 143_{~-35}^{~+36}$ events is obtained. The signal
significance is found to be $3.6\sigma$ including the systematic uncertainty.
The branching fraction $\BF_{\rm SL}$ (in the semileptonic tagging
analysis) is
\begin{equation}
\BF_{\rm SL} (\Btotaunu) = (1.54_{~-0.37-0.31}^{~+0.38+0.29}) \times
10^{-4} ~. 
\end{equation}

\subsubsection{The combined result}

The two results, $\BF_{\rm had}$ and $\BF_{\rm SL}$, are combined after
taking the correlation in the systematic uncertainties between the two
results into account\cite{Ref:taunuavg}. The signal significance for
the combined result is $4.0\sigma$ and the average branching fraction
is
\begin{equation}
\BF (\Btotaunu) = (0.96 \pm 0.22 \pm 0.13) \times
10^{-4} ~. 
\end{equation}
The result is consistent with the SM expectation obtained from other
experimental constraints.  Using this result along with the input
values found from the most recent world averages\cite{pdg2012}, 
we obtain $f_B
|V_{ub}| = (7.4\pm 0.8\pm 0.5) \times 10^{-4}~{\rm GeV}$.
This result sets stringent constraints on the parameters of various
models involving  charged Higgs bosons.

\subsubsection{$\Btolnu~(\ell = e, \mu)$}

As discussed above, the $\Btoenu$ and $\Btomunu$ decays are suppressed
compared to $\Btotaunu$ due to helicity suppression.  On the other
hand, these decays have a clear experimental signature: the monochromatic
energy of the charged lepton in the rest frame of the signal $B$.  Two
methods have been applied to measure these decays: a loose
reconstruction analysis and a hadronic tagging analysis.  

In the loose reconstruction analysis, where a data sample containing
$277 \times 10^6$ $\BBbar$ pairs is used\cite{Belle:B2lnuloose},
the signal candidates are selected mainly via a tight requirement on
$\plB$, which is the charged lepton momentum (magnitude) in the signal
$B$ rest frame.  The signal yield is then obtained by fitting the
$M_{\rm bc}$ distribution, where $M_{\rm bc}$ is calculated by
including all detected particles in the event except for the signal
charged lepton.  No significant excess of signal in any mode is found.
We set the following upper limits on the corresponding branching
fractions at the 90\% C.L.:
\begin{eqnarray}
 & ~& \BF (\Btoenu) < 0.98 \times 10^{-6}, \\
 & ~& \BF (\Btomunu) < 1.7 \times 10^{-6}. 
\end{eqnarray}

The hadronic tagging analysis is based on a method similar to that
described in Sect.~\ref{sec:FRB2taunu} and uses the full 
data set of Belle containing $772 \times 10^6$ $\BBbar$ pairs.  After
selecting signal candidates primarily using 
the $M_{\rm bc}$ and $\Delta E$ variables of the $\Btag$ 
and requiring that the
$\Bsig$ be consistent with $\Btolnu$, including a requirement on
$\EECL$, the expected background in the
signal region, $2.6 < \plB < 2.7~{\rm GeV}\!/c$, is much less than one
event. The background estimate
 is determined by examining data and MC events in the
sideband of $\plB$ below the signal region.

The signal yield is obtained by counting the events in the $\plB$
signal region.  No events are found in any mode and we set 90\% C.L.
upper limits on the branching fractions using the POLE~\cite{Ref:POLE}
program taking the uncertainty in signal efficiency and the
expected background with its uncertainty into account. The preliminary
upper limits (at 90\% C.L.) for the branching fractions $\BF_{\rm had}$
(by hadronic tagging analysis) are\cite{Belle:YookICHEP}:
\begin{eqnarray}
 & ~& \BF_{\rm had} (\Btoenu) < 3.5 \times 10^{-6}, \\
 & ~& \BF_{\rm had} (\Btomunu) < 2.5 \times 10^{-6}. 
\end{eqnarray}
Although the constraints are not as stringent as those obtained in the
loose reconstruction analysis, the amount of background is much
smaller, nearly zero; hence it is anticipated that the sensitivity may
improve almost linearly with the increase of statistics.
Therefore, the hadronic tagging analysis will be very interesting in
the next-generation super $B$-factory experiments such as Belle II.

\subsection{$B \to D^{(*)} \tau \nu$ decays}

Compared to ordinary semileptonic decays $\overline{B}^0 \to D^{*+}
\lnu$ with $\ell = e$ or $\mu$, $B\to D^{(*)}\taunu$ decays,
occurring through a quark-level $b \to c\taunu$ process, are
suppressed because of the large $\tau$ mass.  The predicted branching
fractions, based on the SM, are approximately $1.4\%$ and $0.7\%$ for
$B\to D^* \taunu$ and $B \to D \taunu$ decays,
respectively.\cite{ref:ChenGeng} On the other hand, the large $\tau$
lepton mass makes them sensitive to interactions with a charged Higgs, where
the $H^+$ may replace the virtual $W$, thereby modifying the branching
fraction.  Therefore, these $B\to D^{(*)}\taunu$ 
 modes can be a very effective probe to
search for indirect evidence of charged Higgs or other NP 
hypotheses beyond the SM.  Moreover, compared with $\Btotaunu$, these
decay modes provide more observables to search for NP, e.g. the
polarization of the $\tau$ lepton.  On the experimental side,
however, it is very difficult to measure these modes because of the multiple
neutrinos in the final state, the low
lepton momenta, and the large associated background
contamination.

The first observation of $B\to D^{(*)}\taunu$ decays was reported by
Belle in the $\BztoDstTaunu$ mode using an event sample of $535 \times
10^6$ $B\bar{B}$ pairs.~\cite{Belle:DstTaunuFirst} In contrast to the
hadronic tagging analysis (see Sect.~\ref{sec:FRB2taunu}), a loose
reconstruction of the accompanying $B$ ($\Btag$), where all particles
not belonging to the signal decay chain are included without taking
subdecay information into account, was used and tighter kinematic
constraints were applied for improved background suppression. The
signal yield was obtained by fitting the distribution of the
beam-constrained mass $M_{\rm bc}$ of the $\Btag$.  A clear signal
excess of $60_{~-11}^{~+12}$ events was observed with a significance of
$5.2\sigma$ including systematic uncertainties.  The measured
branching fraction was $\BF(\BztoDstTaunu) = (2.02_{~-0.37}^{~+0.40}
\pm 0.37)\%$.

Belle has also published measurements of other $B\to D^{(*)}\taunu$
decay modes.  Analyzing a data sample of $657 \times 10^6$
$B\overline{B}$ pairs, using a similar analysis to that described above,
$446_{~-56}^{~+58}$ events of the $\BmtoDstzTaunu$ decay mode are observed
with a significance of $8.1\sigma$ and $146_{~-41}^{~+42}$ events of the 
$\BmtoDzTaunu$ decay mode are obtained, providing the first evidence of
this mode with a significance of
$3.5\sigma$.~\cite{Belle:DstTaunu2010} The branching fractions are
$\BF (\BmtoDstzTaunu) = (2.12_{~-0.27}^{~+0.28} \pm 0.29)\%$ and $\BF
(\BmtoDzTaunu) = (0.77 \pm 0.22 \pm 0.12)\%$.

A preliminary branching fraction of the $\overline{B}^0 \to D^+
\taunu$ mode is measured by an analysis that uses a
 hadronic tagging method similar to the one
described in Sect.~\ref{sec:FRB2taunu}: $\BF (\overline{B}^0 \to D^+
\taunu) = (1.01_{~-0.41-0.11}^{~+0.46+0.13} \pm 0.10)
\%$,~\cite{Belle:DstTaunu2009} where the third error comes from the
branching fraction uncertainty of the normalization mode,
$\overline{B}^0 \to D^+ \ell^- \bar{\nu}_\ell$.  The branching
fractions of the other $B \to D^{(*)} \taunu $ decay modes are also
obtained in this analysis; the results are consistent with
published results.\cite{Belle:DstTaunuFirst,Belle:DstTaunu2010}

Recently, BaBar has claimed that the branching fractions of $B\to
D^*\taunu$ and $B\to D\taunu$ are larger than SM expectations at a
combined significance of $3.4\sigma$.\cite{Babar:DstTaunu2012} We note
that all the branching fractions of $B\to D^{(*)}\taunu$ modes
measured by Belle are also larger than the SM-predicted
values.\cite{ref:ChenGeng} It will be interesting to see the final 
Belle results on these modes using improved hadronic tagging and
the full data sample of $772\times 10^6$ $B\overline{B}$ pairs.

\section{Rare $B$ decays}
\def\CP{\ensuremath{C\!P}}
\def\mbc{\mbox{$M_{\rm bc}$}}
\def\DeltaE{\mbox{$\Delta E$}}
\def\fz{\ensuremath{f_0(980)}}
\def\Bbar{\kern 0.18em\overline{\kern -0.18em B}{}}
\def\BB{\ensuremath{B\Bbar}} 
\def\BR{{\ensuremath{\cal B}}}
\def\Kstar{\ensuremath{K^*}}
\def\Kstarp{\ensuremath{K^{*+}}}
\def\Kstarz{\ensuremath{K^{*0}}}
\def\Kbar{\kern 0.2em\overline{\kern -0.2em K}{}}
\def\KS{\ensuremath{K^0_{\scriptscriptstyle S}}} 
\def\Sigbar{\kern 0.2em\overline{\kern -0.2em \Sigma}{}}
\def\jpsi{\ensuremath{{J\mskip -3mu/\mskip -2mu\psi\mskip 2mu}}}
\def\Lbar{\kern 0.2em\overline{\kern -0.2em\Lambda\kern 0.05em}\kern-0.05em{}}
\newcommand{\etapr}{\ensuremath{\eta^{\prime}}}
\newcommand{\err}[3]{\ensuremath{#1 \pm #2 \pm #3}}
\newcommand{\aerr}[4]{\ensuremath{#1 ^{+#2}_{-#3} \pm {#4}}}
\newcommand{\berr}[4]{\ensuremath{#1 \pm {#2} ^{+#3}_{-#4}}}
\newcommand{\aerrsy}[5]{\ensuremath{#1 {^{+#2+#4}_{-#3-#5}}}}
\label{chap_rare}
\subsection{Charmless hadronic decays}

\begin{figure}[htb]
\begin{center}
\includegraphics[width=.32\textwidth]{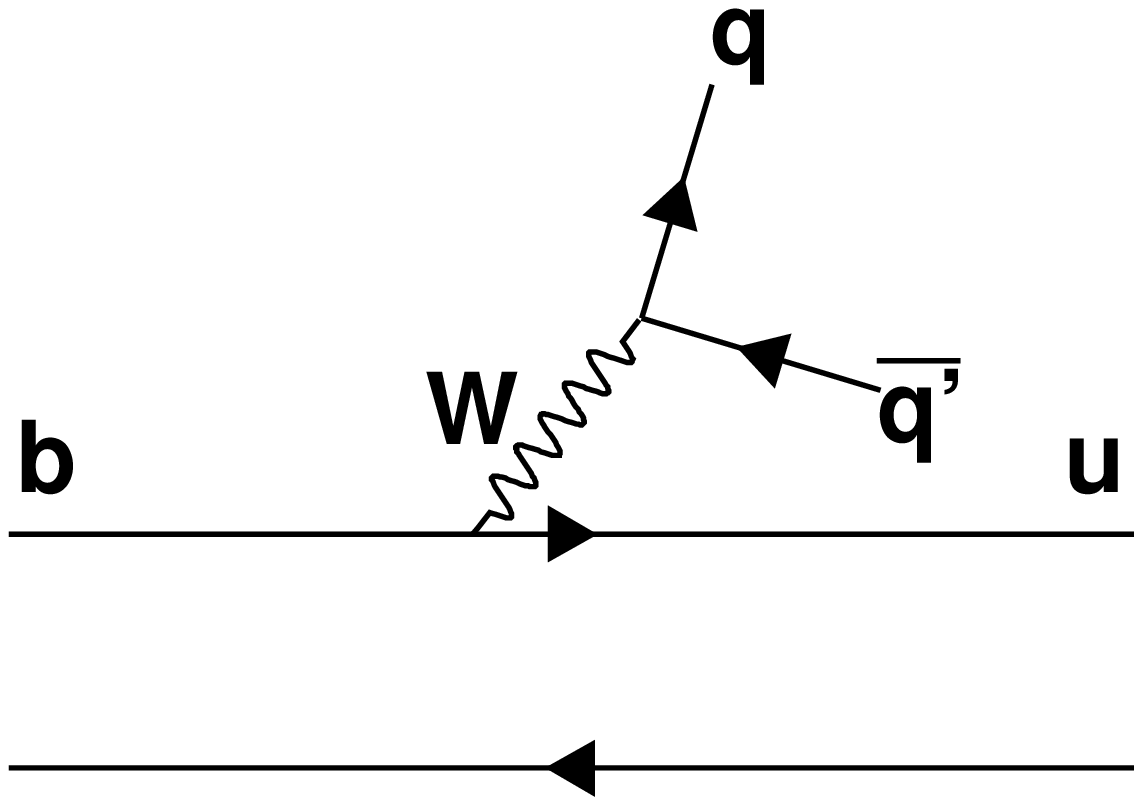}
\includegraphics[width=.32\textwidth]{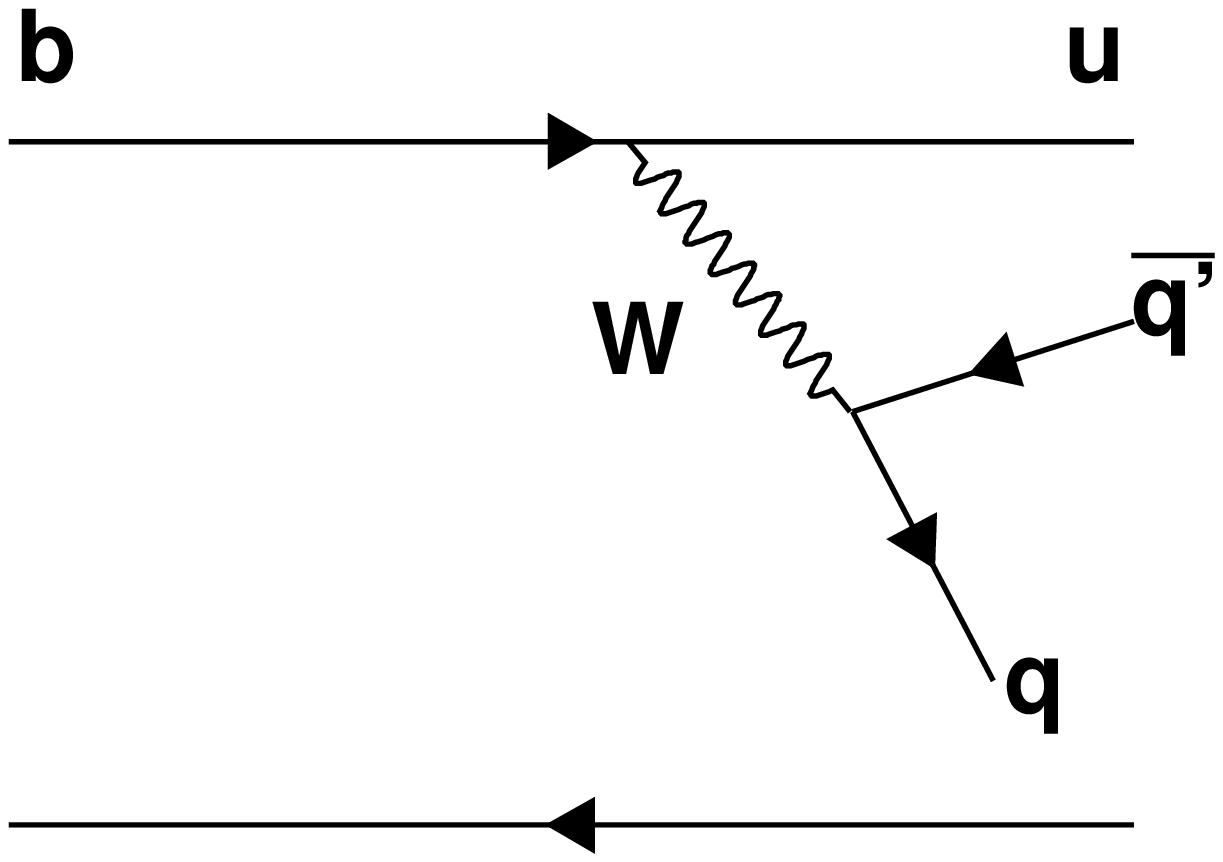}
\includegraphics[width=.32\textwidth]{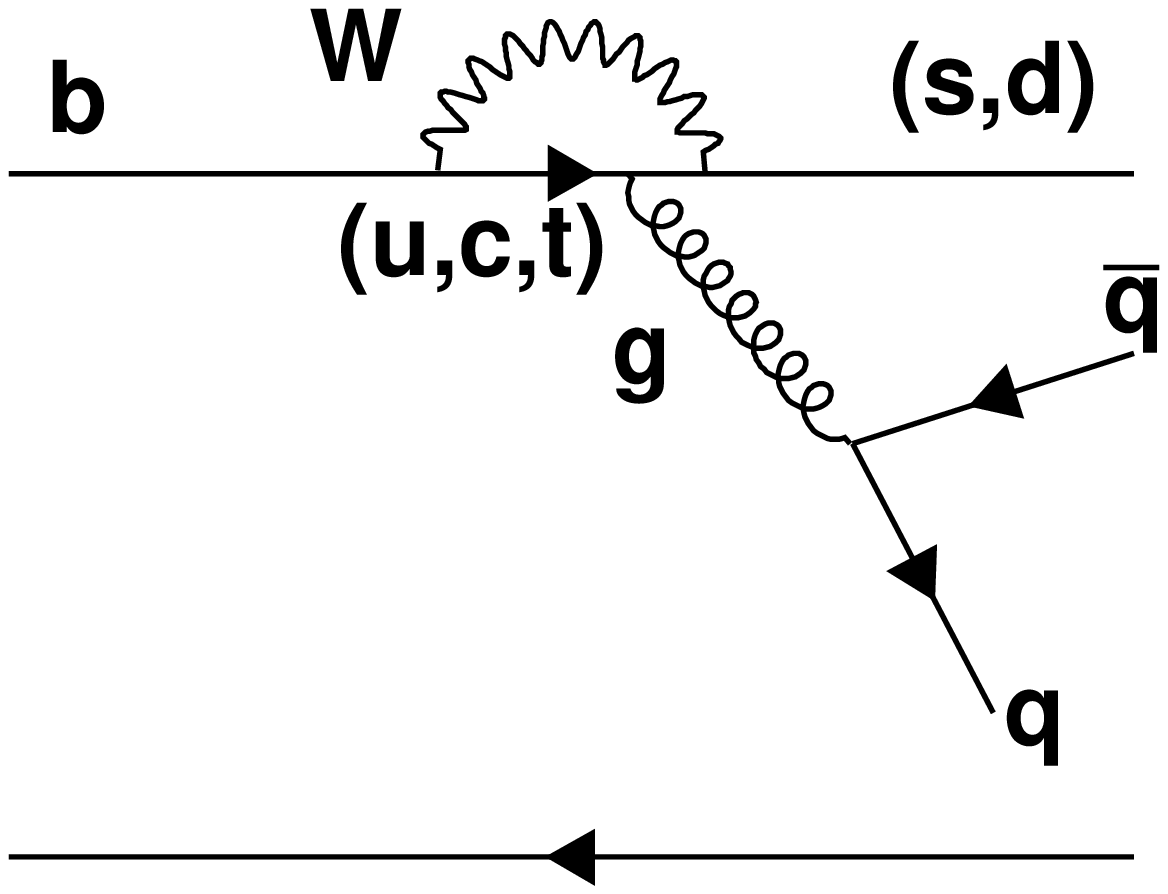}
\end{center}
\caption{Feynman diagrams for the quark-level transitions that
 dominantly contribute to charmless hadronic $B$ decays: (left)
 color allowed and (middle) color suppressed $b\to u$ tree diagrams,
 and (right) $b\to (s,d)g$ penguin diagrams.}
\label{fig:feyn}
\end{figure}
Charmless hadronic $B$ decays give rise to final states with
two or more hadrons that do not contain any charm quark. These decays
are suppressed in the SM, mostly proceeding via the CKM-suppressed
$b\to u$ tree level transition and $b\to (s,d)g$ penguin diagrams, 
as shown in Fig.~\ref{fig:feyn}. Compared to the CKM-favored $b\to
c$ transition such as $B^0\to\jpsi K^0$, the golden channel for
determining the angle $\phi_1$ of the unitarity triangle (Sect. 3),
their branching fractions are about two to four orders of magnitude
lower. By virtue of this suppression, charmless decays provide a
good window to probe new physics beyond the SM. For instance,
these rare decays have the potential to reveal the contribution
of heavy, non-SM virtual particles in penguin loops. Branching
fraction calculations within the SM ---whether they are based on
QCD factorization~\cite{qcdf}, SU(3) flavor symmetry~\cite{su3fit},
or perturbative QCD~\cite{pqcd} ---suffer from large theoretical
uncertainties. However, one can examine physics observables in
which theory errors as well as common experimental systematic
uncertainties largely cancel out. Such observables include direct
$\CP$ asymmetries, ratios of branching fractions, and longitudinal
polarization fractions (in the case in which the decay final
states consist of two vector particles). In this section, we
summarize Belle's results on charmless $B$ decays, which have
resulted in close to $100$ journal publications.

\subsubsection{Experimental methodology}
\label{chmls-exp-meth}

Before examining various categories of charmless decays, we wish to
describe the important experimental considerations. As these decays
are suppressed in the SM one needs to be extremely careful in
devising selection algorithms to select candidate events and fitting
methods used for extracting the final signal yields. We identify $B$
mesons using two kinematical variables: the beam-energy constrained
mass, $\mbc\equiv\sqrt{E_{\rm beam}^2-|\vec{p}_B|^2}$, and
the energy difference $\DeltaE\equiv E_B-E_{\rm beam}$, where $E_{\rm
beam}$ is the beam energy, and $E_B$ and $\vec{p}_B$ are the energy
and momentum of $B$ candidates in the CM frame,
respectively. The dominant background contribution is from $e^+e^-
\to q\bar{q}$ $(q=u,d,s,c)$ continuum processes. To suppress this
background, we use variables based on the event topology; these
selections rely on the fact that $B$ decays are nearly isotropic, in
contrast to jet-like continuum events. In some analyses, additional
discrimination is provided by variables pertaining to the nature of
$B$ decay, e.g., the flight-length difference along the beam direction
between the signal and recoil $B$ decay. All this information is
combined into either a likelihood ratio or a neural network to optimize
the sensitivity. $B$ decays proceeding via a CKM-favored $b\to c$
transition can have a final state that is either the same as our signal
or mis-reconstructed. Since branching fractions for $b\to c$ decays
are much larger and the charm mesons involved are quite narrow, we
suppress their contributions by applying a veto on the reconstructed
invariant mass of daughter particles of the charm meson. Backgrounds
from other $B$ decays, especially those due to particle
misidentification, pose a special challenge. The $\DeltaE$ and
charged-hadron identification variables help in discriminating such
backgrounds. The final signal yield is extracted by means of an
unbinned maximum likelihood fit to the discriminating variables, $\mbc$,
$\DeltaE$, and continuum suppression variable (in some analyses we have
used an optimized and tight requirement on the latter using the expected
signal significance from Monte Carlo simulations as a figure of merit). 
For decays involving narrow or non-zero spin resonances in the final
state, we employ the invariant mass and helicity angle distributions
to further enhance the sensitivity of our results.

\subsubsection{Results on charmless decays}
\label{chmls-chmls}

Charmless hadronic $B$ decays can be roughly divided into
``two-body'', ``quasi-two-body (Q2B)'', and ``three-body'' categories.
Although two-body decays are easily identifiable by the presence of
two long-lived final state particles, such as $K\pi$, the latter two
classes of decays are somewhat intertwined. For instance, when one
performs the Dalitz plot analysis of a three-body final state, one can
access information on related Q2B decays along with the three-body
nonresonant decay. The Dalitz plot approach is the most appropriate
method when dealing with broad intermediate resonances, e.g.,
$\rho(770)$. However, if the intermediate resonances are narrow or
the resonances decaying to the same final state do not interfere,
we can use a Q2B approach where the interference is accounted for
as an additional source of systematic error. Since the aforementioned
distinction between the two categories is not ``black and white'',
we begin the discussion with two-body decays, then Q2B decays are
described together with intermediate resonance final states from
three-body Dalitz analyses, and finally we move on to results on
three-body nonresonant and inclusive final states.

\medskip
\noindent
\underline{Two-body decays}
\label{chmls-chmls2b}

$B$ meson decays to two stable hadrons are kinematically
easy to identify, having average momenta larger than a
typical $B$ decay. The kinematics also provides a good handle
on the continuum background. A formidable challenge is posed by
the feed-across background arising due to particle misidentification
(mostly, kaons misidentified as pions). We tackle this issue by
performing a simultaneous fit to the event samples that can cross
feed into each other.

Table~\ref{chmls2b} summarizes the branching fraction and $\CP$
asymmetry results for various charmless two-body $B$ decays from
Belle. The most notable result here has been the observation of
a non-zero difference of $\CP$ violation asymmetry in the $B^0\to
K^+\pi^-$ and $B^+\to K^+\pi^0$ decays: $\Delta A_{K\pi}=A_{\CP}
(K^+\pi^0)-A_{\CP}(K^+\pi^-)=\err{+0.112}{0.027}{0.007}$~\cite{Duh:2012ie}.
This discrepancy, also called the {\em $\Delta A_{K\pi}$ puzzle}, 
may be explained either by a large contribution from the
color-suppressed tree diagram~\cite{supress} or a new physics
contribution in the electroweak penguin~\cite{ewp}. Before concluding
on this issue, we must improve the uncertainties on $\CP$ violation
results for the decay $B^0\to K^0\pi^0$. This would allow us to precisely
test the prediction of an isospin sum rule~\cite{sumrule} given by
$A_{\CP}(K^+\pi^-)+A_{\CP}(K^0\pi^+)\frac{\Gamma(K^0\pi^+)}{\Gamma(K^+\pi^-)}
-A_{\CP}(K^+\pi^0)\frac{2\Gamma(K^+\pi^0)}{\Gamma(K^+\pi^-)}-A_{\CP}(K^0\pi^0)
\frac{2\Gamma(K^0\pi^0)}{\Gamma(K^+\pi^-)}=0$, where $\Gamma$ is the
partial width. Belle's latest update~\cite{Duh:2012ie} reports a sum
of $\err{-0.270}{0.132}{0.060}$ with $1.9\sigma$ significance.

\begin{table}[!htb]
\caption{Data samples used ($N_{\BB}$), branching fractions ($\BR$), $90\%$
 confidence-level upper limits on $\BR$ (UL), and direct $\CP$ asymmetries
 ($A_{\CP}$) obtained for various charmless two-body $B$ decays. The two
 uncertainties quoted here and elsewhere are statistical and systematic,
 respectively.}
\label{chmls2b}
\begin{center}
\renewcommand{\arraystretch}{1.15}
\begin{tabular}{l|clclc}
\hline\hline
Final state&$N_{\BB}~(10^6)$&\multicolumn{1}{c}{$\BR~(10^{-6})$}&
UL $(10^{-6})$&\multicolumn{1}{c}{$A_{\CP}$}&Ref.\\
\hline
$K^+K^-$&$772$&$\err{0.10}{0.08}{0.04}$&$0.20$&&\cite{Duh:2012ie}\\
$K^+\Kbar^0$&$772$&$\err{1.11}{0.19}{0.05}$&&$\err{+0.014}{0.168}{0.002}$&\cite{Duh:2012ie}\\
$K^0\Kbar^0$&$772$&$\err{1.26}{0.19}{0.05}$&&&\cite{Duh:2012ie}\\
$K^+\pi^-$&$772$&$\err{20.00}{0.34}{0.60}$&&$\err{-0.069}{0.014}{0.007}$&\cite{Duh:2012ie}\\
$K^+\pi^0$&$772$&$\err{12.62}{0.31}{0.56}$&&$\err{+0.043}{0.024}{0.002}$&\cite{Duh:2012ie}\\
$K^0\pi^+$&$772$&$\err{23.97}{0.53}{0.71}$&&$\err{-0.011}{0.021}{0.006}$&\cite{Duh:2012ie}\\
$K^0\pi^0$&$772$&$\err{9.68}{0.46}{0.50}$&&&\cite{Duh:2012ie}\\
$\pi^+\pi^-$&$772$&$\err{5.04}{0.21}{0.18}$&&&\cite{Duh:2012ie}\\
$\pi^+\pi^0$&$772$&$\err{5.86}{0.26}{0.38}$&&$\err{+0.025}{0.043}{0.007}$&\cite{Duh:2012ie}\\
$\pi^0\pi^0$&$275$&$\aerrsy{2.3}{0.4}{0.5}{0.2}{0.3}$&&$\aerr{+0.44}{0.53}{0.52}{0.17}$&\cite{Abe:2004mp}\\
$p\overline{p}$&$449$&&$0.41$&&\cite{Tsai:2007pp}\\
$p\Lbar$&$449$&&$0.49$&&\cite{Tsai:2007pp}\\
$\Lambda\Lbar$&$449$&&$0.69$&&\cite{Tsai:2007pp}\\
\hline\hline
\end{tabular}
\end{center}
\end{table}

\medskip
\noindent
\underline{Quasi-two-body (Q2B) decays}

Q2B analyses assume that the intermediate resonances decaying to
the same final state (such as $\rho(770)$ and $\fz$ decaying to
$\pi^+\pi^-$) do not interfere. This treatment allows us to compare
branching fraction results with measurements from earlier experiments,
in which the effects of interference were treated as a part of the
systematic error. The extent and nature of the background (as most
charmless $B$ decays suffer from a low signal-to-background ratio)
and of the nonresonant signal component strongly influence our
analysis strategy. The helicity angle plays an important role in
such Q2B analyses when the intermediate resonances have a non-zero
spin; we can use it either as a simple selection criterion or to
extract physics observables, such as the fraction of longitudinal
polarization $f_L$, directly from the fit. The helicity angle $\theta_H$
for a resonance is defined as the angle between the momentum vector
of one of its daughter particles and the direction opposite to the
$B$-meson momentum in the resonance rest frame~\cite{Kramer:1991xw}.

In the discussions that follow, the decays have been grouped according to
their spin. For each spin grouping, the results for branching fractions,
direct $\CP$ asymmetries, and longitudinal polarization fractions (where
applicable) are listed in the accompanying tables. In Table~\ref{chmlsq2beta}
we start with final states comprising at least an $\eta$, or $\etapr$ meson
that is reconstructed in the two channels $\eta\to\gamma\gamma$ and $\eta
\to\pi^+\pi^-\pi^0$, or $\etapr\to\eta\pi^+\pi^-$ and $\etapr\to\rho^0\gamma$,
respectively. Among the highlighted results are the first observation of
$B^0\to\eta K^0$, and evidence for direct $\CP$ violation in the decays
$B^+\to\eta K^+$ and $B^+\to\eta\pi^+$ with significances of $3.8\sigma$
and $3.0\sigma$, respectively~\cite{Hoi:2012pk}. The latter results call
for a large interference between the $b\to s$ penguin process and the
CKM-suppressed, color-favored $b\to u$ tree transition, both of which
contribute to $B^+\to\eta h^+$ $(h=K,\pi)$.

\begin{table}[!htb]
\caption{Data samples used ($N_{\BB}$), branching fractions ($\BR$), $90\%$
 confidence-level upper limits on $\BR$ (UL), and direct $\CP$ asymmetries
 ($A_{\CP}$) obtained for various charmless Q2B decays with an $\eta$ or
 $\etapr$ meson in the final state.}
\label{chmlsq2beta}
\begin{center}
\renewcommand{\arraystretch}{1.15}
\begin{tabular}{l|clclc}
\hline\hline
Final state&$N_{\BB}~(10^6)$&\multicolumn{1}{c}{$\BR~(10^{-6})$}&
UL $(10^{-6})$&\multicolumn{1}{c}{$A_{\CP}$}&Ref.\\
\hline
$\eta K^+$&$772$&$\err{2.12}{0.23}{0.11}$&&$\err{-0.38}{0.11}{0.01}$&\cite{Hoi:2012pk}\\
$\eta K^0$&$772$&$\aerr{1.27}{0.33}{0.29}{0.08}$&&&\cite{Hoi:2012pk}\\
$\eta\pi^+$&$772$&$\err{4.07}{0.26}{0.21}$&&$\err{-0.19}{0.06}{0.01}$&\cite{Hoi:2012pk}\\
$\eta\pi^0$&$152$&$\err{1.2}{0.7}{0.1}$&$2.5$&&\cite{Chang:2004fz}\\
$\eta\eta$&$152$&$\aerr{0.7}{0.7}{0.4}{0.1}$&$2.0$&&\cite{Chang:2004fz}\\
$\etapr K^+$&$386$&$\err{69.2}{2.2}{3.7}$&&$\err{+0.028}{0.028}{0.021}$&\cite{Schumann:2006bg}\\
$\etapr K^0$&$386$&$\aerr{58.9}{3.6}{3.5}{4.3}$&&&\cite{Schumann:2006bg}\\
$\etapr\pi^+$&$386$&$\berr{1.76}{0.67}{0.62}{0.15}{0.14}$&&$\aerr{+0.20}{0.37}{0.36}{0.04}$&\cite{Schumann:2006bg}\\
$\etapr\pi^0$&$386$&$\berr{2.79}{1.02}{0.96}{0.25}{0.34}$&&&\cite{Schumann:2006bg}\\
$\etapr\eta$&$535$&&$4.5$&&\cite{Schumann:2007qv}\\
$\etapr\etapr$&$535$&&$6.5$&&\cite{Schumann:2007qv}\\
$\eta K^{*+}$&$449$&$\aerr{19.3}{2.0}{1.9}{1.5}$&&$\err{+0.03}{0.10}{0.01}$&\cite{Wang:2007rzb}\\
$\eta K^{*0}$&$449$&$\err{15.2}{1.2}{1.0}$&&$\err{+0.17}{0.08}{0.01}$&\cite{Wang:2007rzb}\\
$\eta\rho^+$&$449$&$\aerr{4.1}{1.4}{1.3}{0.4}$&$6.5$&$\aerr{-0.04}{0.34}{0.32}{0.01}$&\cite{Wang:2007rzb}\\
$\eta\rho^0$&$449$&$\aerr{0.84}{0.56}{0.51}{0.19}$&$1.9$&&\cite{Wang:2007rzb}\\
$\etapr K^{*+}$&$535$&&$2.9$&&\cite{Schumann:2007qv}\\
$\etapr K^{*0}$&$535$&&$2.6$&&\cite{Schumann:2007qv}\\
$\etapr\rho^+$&$535$&&$5.8$&&\cite{Schumann:2007qv}\\
$\etapr\rho^0$&$535$&&$1.3$&&\cite{Schumann:2007qv}\\
$\etapr\omega$&$535$&&$2.2$&&\cite{Schumann:2007qv}\\
$\etapr\phi$&$535$&&$0.5$&&\cite{Schumann:2007qv}\\
\hline\hline
\end{tabular}
\end{center}
\end{table}

The branching fractions and $\CP$ asymmetries for other Q2B decays
without an $\eta$ or $\etapr$ meson in the final state are summarized
in Table~\ref{chmlspv}. Most of the results are obtained as a by-product
of a three-body Dalitz plot analysis. The systematic uncertainties in
the table include the experimental systematic as well as Dalitz-plot
model dependence, where applicable. We report the first evidence of
$\CP$ violation in the Q2B decay $B^+\to\rho^0K^+$ exceeding the
$3\sigma$ level. Note that this was the first evidence for direct
$\CP$ violation in a charged meson decay, a phenomenon that was
already observed in decays of neutral $K$~\cite{dcpv-k0} and
$B$~\cite{Duh:2012ie,dcpv-b0} mesons, and very recently in $D^0$
decays~\cite{dcpv-d0}.

In Table~\ref{chmlsvv} we present results obtained from the
vector\---vector final states. One naively expects $B\to VV$ decays to
be dominated by longitudinal polarization amplitudes since $f_L=
1-4m_V/m_B\sim 0.9$~\cite{polar-sm}, where $m_V$ $(m_B)$ is the mass
of the vector ($B$) meson. Contrary to this expectation, it is found
out that the decays dominated by the $b\to s$ penguin transition such
as $B\to\phi K^*$ have $f_L$ values closer to $0.5$. However, decays
proceeding via the $b\to u$ tree diagram, notably $B\to\rho\rho$, follow
the expected trend. This so-called {\em polarization puzzle} could
be explained by the presence of new particles in the penguin
loop~\cite{polar-np}. However, large SM corrections appear to be a
more plausible explanation.

\begin{table}[!htb]
\caption{Data samples used ($N_{\BB}$), branching fractions ($\BR$), $90\%$
 confidence-level upper limits on $\BR$ (UL), and direct $\CP$ asymmetries
 ($A_{\CP}$) obtained for various charmless Q2B decays without an $\eta$
 or $\etapr$ meson in the final state.}
\label{chmlspv}
\begin{center}
\renewcommand{\arraystretch}{1.15}
\begin{tabular}{l|clclc}
\hline\hline
Final state&$N_{\BB}~(10^6)$&\multicolumn{1}{c}{$\BR~(10^{-6})$}&
UL $(10^{-6})$&\multicolumn{1}{c}{$A_{\CP}$}&Ref.\\
\hline
$f_0(980)K^0$&$388$&$\berr{7.6}{1.7}{0.9}{1.3}$&&&\cite{Garmash:2006fh}\\
$f_0(980)K^+$&$386$&$\berr{8.78}{0.82}{0.85}{1.76}$&&$\berr{-0.077}{0.065}{0.046}{0.026}$&\cite{Garmash:2005rv}\\
$f_2(1270)K^+$&$386$&$\berr{1.33}{0.30}{0.23}{0.34}$&&$\err{-0.59}{0.22}{0.04}$&\cite{Garmash:2005rv}\\
$f_2(1270)K^0$&$388$&&$2.5$&&\cite{Garmash:2006fh}\\
$K_0^*(1430)^+\pi^-$&$388$&$\berr{49.7}{3.8}{6.8}{8.2}$&&&\cite{Garmash:2006fh}\\
$K_0^*(1430)^0\pi^+$&$386$&$\berr{51.6}{1.7}{7.0}{7.5}$&&$\berr{+0.076}{0.038}{0.028}{0.022}$&\cite{Garmash:2005rv}\\
$K^{*+}\pi^-$&$388$&$\berr{8.4}{1.1}{1.0}{0.9}$&&$\err{-0.21}{0.11}{0.07}$&\cite{Garmash:2006fh}\\
$K^{*0}\pi^+$&$386$&$\berr{9.67}{0.64}{0.81}{0.89}$&&$\err{-0.149}{0.064}{0.022}$&\cite{Garmash:2005rv}\\
$K^{*0}\pi^0$&$85$&$\aerr{0.4}{1.9}{1.7}{0.1}$&$3.5$&&\cite{Chang:2004um}\\
$\omega K^+$&$388$&$\err{8.1}{0.6}{0.6}$&&$\aerr{+0.05}{0.08}{0.07}{0.01}$&\cite{Jen:2006in}\\
$\omega K^0$&$388$&$\aerr{4.4}{0.8}{0.7}{0.4}$&&&\cite{Jen:2006in}\\
$\omega\pi^+$&$388$&$\err{6.9}{0.6}{0.5}$&&$\err{-0.02}{0.09}{0.01}$&\cite{Jen:2006in}\\
$\omega\pi^0$&$388$&$\aerr{0.5}{0.4}{0.3}{0.1}$&$2.0$&&\cite{Jen:2006in}\\
$\phi K^+$&$152$&$\berr{9.60}{0.92}{1.05}{0.84}$&&$\err{+0.01}{0.12}{0.05}$&\cite{Garmash:2004wa}\\
$\phi K^0$&$85$&$\aerr{9.0}{2.2}{1.8}{0.7}$&&&\cite{Chen:2003jfa}\\
$\phi\pi^+$&$657$&$\aerrsy{0.08}{0.09}{0.08}{0.06}{0.03}$&$0.33$&&\cite{Kim:2012gt}\\
$\phi\pi^0$&$657$&$\aerrsy{-0.07}{0.06}{0.04}{0.04}{0.08}$&$0.15$&&\cite{Kim:2012gt}\\
$\phi(1680)K^+$&$152$&&$0.8$&&\cite{Garmash:2004wa}\\
$\rho^-K^+$&$85$&$\aerrsy{15.1}{3.4}{3.3}{2.4}{2.6}$&&$\aerrsy{+0.22}{0.22}{0.23}{0.06}{0.02}$&\cite{Chang:2004um}\\
$\rho^0K^+$&$386$&$\berr{3.89}{0.47}{0.43}{0.41}$&&$\berr{+0.30}{0.11}{0.11}{0.05}$&\cite{Garmash:2005rv}\\
$\rho^0K^0$&$388$&$\berr{6.1}{1.0}{1.1}{1.2}$&&&\cite{Garmash:2006fh}\\
$\rho^+\pi^0$&$152$&$\berr{13.2}{2.3}{1.4}{1.9}$&&$\berr{+0.06}{0.17}{0.04}{0.05}$&\cite{Zhang:2004wza}\\
$\rho^0\pi^+$&$32$&$\aerr{8.0}{2.3}{2.0}{0.7}$&&&\cite{Gordon:2002yt}\\
$\rho^0\pi^0$&$449$&$\err{3.0}{0.5}{0.7}$&&&\cite{rhopi}\\
$\rho^{\mp}\pi^{\pm}$&$449$&$\err{22.6}{1.1}{4.4}$&&&\cite{rhopi}\\
\hline\hline
\end{tabular}
\end{center}
\end{table}

\begin{table}[!htb]
\caption{Data samples used ($N_{\BB}$), branching fractions ($\BR$), $90\%$
 confidence-level upper limits on $\BR$ (UL), longitudinal polarization fraction
 ($f_L$), and direct $\CP$ asymmetries ($A_{\CP}$) obtained for various charmless
 Q2B decays with vector\---vector final states.}
\label{chmlsvv}
\begin{center}
\renewcommand{\arraystretch}{1.15}
\begin{tabular}{l|clcllc}
\hline\hline
Final state&$N_{\BB}~(10^6)$&\multicolumn{1}{c}{$\BR~(10^{-6})$}&UL $(10^{-6})$&
\multicolumn{1}{c}{$f_L$}&\multicolumn{1}{c}{$A_{\CP}$}&Ref.\\
\hline
$K^{*0}K^{*0}$&$657$&&$0.2$&&&\cite{Chiang:2010ga}\\
$K^{*0}\Kbar^{*0}$&$657$&$\aerrsy{0.26}{0.33}{0.29}{0.10}{0.08}$&$0.8$&&&\cite{Chiang:2010ga}\\
$K^{*0}\rho^+$&$275$&$\err{8.9}{1.7}{1.2}$&&$\berr{0.43}{0.11}{0.05}{0.02}$&&\cite{Abe:2004mq}\\
$K^{*0}\rho^0$&$657$&$\aerrsy{2.1}{0.8}{0.7}{0.9}{0.5}$&$3.4$&&&\cite{Kyeong:2009qx}\\
$\omega K^{*0}$&$657$&$\berr{1.8}{0.7}{0.3}{0.2}$&&$\berr{0.56}{0.29}{0.18}{0.08}$&&\cite{Goldenzweig:2008sz}\\
$\phi K^{*+}$&$85,257$&$\aerrsy{6.7}{2.1}{1.9}{0.7}{1.0}$&&$\err{0.52}{0.08}{0.03}$&$\err{-0.02}{0.14}{0.03}$&
\cite{Chen:2003jfa,Chen:2005zv}\\
$\phi K^{*0}$&$85,257$&$\aerrsy{10.0}{1.6}{1.5}{0.7}{0.8}$&&$\err{0.45}{0.05}{0.02}$&$\err{+0.02}{0.09}{0.02}$&
\cite{Chen:2003jfa,Chen:2005zv}\\
$\rho^+\rho^-$&$275$&$\berr{22.8}{3.8}{2.3}{2.6}$&&$\aerr{0.94}{0.03}{0.04}{0.03}$&&\cite{Somov:2006sg}\\
$\rho^+\rho^0$&$85$&$\berr{31.7}{7.1}{3.8}{6.7}$&&$\err{0.95}{0.11}{0.02}$&$\err{+0.00}{0.22}{0.03}$&
\cite{Zhang:2003up}\\
$\rho^0\rho^0$&$657$&$\berr{0.4}{0.4}{0.2}{0.3}$&$1.0$&&&\cite{Chiang:2008et}\\
\hline\hline
\end{tabular}
\end{center}
\end{table}

\medskip
\noindent
\underline{Three-body decays}

As was discussed above, the Dalitz-plot method is the most
robust analysis technique for a three-body decay, especially for
$B\to 3$ pseudoscalars. This method has greater complexity
but at the same time provides a better understanding of the
underlying physics. A subtle point here is that one needs a good
deal of statistics before carrying out a full-fledged Dalitz plot
analysis. As the integrated luminosity continued to increase at
Belle, starting with measurements of inclusive branching fractions
and charge asymmetries, various rare decay analyses slowly evolved
into a detailed study of the three-body phase space, e.g., $B^+\to
K^+\pi^+\pi^-$. At times, study of Q2B final states served as an
intermediate step. The choice of analysis technique is mostly
dictated by the luminosity, expected signal and background, and
the understanding of the intermediate resonances involved.
Table~\ref{chmls3body} summarizes results on the branching
fraction and $\CP$ asymmetry for various decays with three-body
mesonic final states.

\begin{table}[!htb]
\caption{Data samples used ($N_{\BB}$), branching fractions ($\BR$), $90\%$
 confidence-level upper limits on $\BR$ (UL), and direct $\CP$ asymmetries
 ($A_{\CP}$) obtained for various charmless decays with three-body mesonic
 final states.}
\label{chmls3body}
\begin{center}
\renewcommand{\arraystretch}{1.15}
\begin{tabular}{l|clclc}
\hline\hline
Final state&$N_{\BB}~(10^6)$&\multicolumn{1}{c}{$\BR~(10^{-6})$}&
UL $(10^{-6})$&\multicolumn{1}{c}{$A_{\CP}$}&Ref.\\
\hline
$K^+K^+\pi^-$&$85$&&$2.4$&&\cite{Garmash:2003er}\\
$K^+K^-K^+$&$152$&$\err{30.6}{1.2}{2.3}$&&&\cite{Garmash:2004wa}\\
$K^+K^-K^0$&$85$&$\err{28.3}{3.3}{4.0}$&&&\cite{Garmash:2003er}\\
$K^+K^-\pi^+$&$85$&$\err{9.3}{2.3}{1.1}$&$13$&&\cite{Garmash:2003er}\\
$K^+\KS\KS$&$85$&$\err{13.4}{1.9}{1.5}$&&&\cite{Garmash:2003er}\\
$K^+\pi^+\pi^-$ nonres.&$386$&$\berr{16.9}{1.3}{1.7}{1.6}$&&&\cite{Garmash:2005rv}\\
$K^+\pi^+\pi^-$&$386$&$\err{48.8}{1.1}{3.6}$&&$\err{+0.049}{0.026}{0.020}$&\cite{Garmash:2005rv}\\
$K^+\pi^-\pi^0$ nonres.&$85$&$\aerrsy{5.7}{2.7}{2.5}{0.5}{0.4}$&$9.4$&&\cite{Chang:2004um}\\
$K^+\pi^-\pi^0$&$85$&$\aerr{36.6}{4.2}{4.3}{3.0}$&&$\err{+0.07}{0.11}{0.01}$&\cite{Chang:2004um}\\
$K^-\pi^+\pi^+$&$85$&&$4.5$&&\cite{Garmash:2003er}\\
$K^0K^-\pi^+$&$85$&&$18$&&\cite{Garmash:2003er}\\
$K^0\pi^+\pi^-$ nonres.&$388$&$\berr{19.9}{2.5}{1.7}{2.0}$&&&\cite{Garmash:2006fh}\\
$K^0\pi^+\pi^-$&$388$&$\err{47.5}{2.4}{3.7}$&&&\cite{Garmash:2006fh}\\
$\KS\KS\KS$&$85$&$\aerr{4.2}{1.6}{1.3}{0.8}$&&&\cite{Garmash:2003er}\\
$\KS\KS\pi^+$&$85$&&$3.2$&&\cite{Garmash:2003er}\\
$\omega K^+\pi^-$ nonres.&$657$&$\err{5.1}{0.7}{0.7}$&&&\cite{Goldenzweig:2008sz}\\
$\phi\phi K^+$&$449$&$\aerr{3.2}{0.6}{0.5}{0.3}$&&$\aerr{+0.01}{0.19}{0.16}{0.02}$&\cite{Abe:2099ab}\\
$\phi\phi K^0$&$449$&$\aerr{2.3}{1.0}{0.7}{0.2}$&&&\cite{Abe:2099ab}\\
$\rho^0K^+\pi^-$&$657$&$\err{2.8}{0.5}{0.5}$&&&\cite{Kyeong:2009qx}\\
$\rho^0\pi^+\pi^-$&$657$&$\aerr{5.9}{3.5}{3.4}{2.7}$&$12$&&\cite{Chiang:2008et}\\
$f_0(980)K^+\pi^-$&$657$&$\berr{1.4}{0.4}{0.3}{0.4}$&$2.1$&&\cite{Kyeong:2009qx}\\
$f_0(980)\pi^+\pi^-$&$657$&$\aerr{0.3}{1.9}{1.8}{0.9}$&$3.8$&&\cite{Chiang:2008et}\\
\hline\hline
\end{tabular}
\end{center}
\end{table}

A great deal of effort has also been applied to studying charmless
three-body baryonic decays. Quite often Belle has reported results
before its sister experiment, BaBar, in these kind of studies. In
Table~\ref{chmls3body-baryon} we attempt to summarize the results
obtained in these baryonic decays. An intriguing feature of the
results is the peaking of baryon\---antibaryon pair mass distributions
toward threshold. These enhancements have generated much theoretical
interest~\cite{theo-baryon}.

\begin{table}[!htb]
\caption{Data samples used ($N_{\BB}$), branching fractions ($\BR$), $90\%$
 confidence-level upper limits on $\BR$ (UL), and direct $\CP$ asymmetries
 ($A_{\CP}$) obtained for various charmless decays with three-body baryonic
 final states.}
\label{chmls3body-baryon}
\begin{center}
\renewcommand{\arraystretch}{1.15}
\begin{tabular}{l|clclc}
\hline\hline
Final state&$N_{\BB}~(10^6)$&\multicolumn{1}{c}{$\BR~(10^{-6})$}&
UL $(10^{-6})$&\multicolumn{1}{c}{$A_{\CP}$}&Ref.\\
\hline
$p\overline{p}\pi^+$&$449$&$\aerr{1.57}{0.17}{0.15}{0.12}$&&$\err{-0.17}{0.10}{0.02}$&\cite{Wei:2007fg}\\
$p\overline{p}K^+$&$449$&$\aerr{5.00}{0.24}{0.22}{0.32}$&&$\err{-0.02}{0.05}{0.02}$&\cite{Wei:2007fg}\\
$p\overline{p}K^0$&$535$&$\aerr{2.51}{0.35}{0.29}{0.21}$&&&\cite{Chen:2008jy}\\
$p\overline{p}\Kstarp$&$535$&$\aerr{3.38}{0.73}{0.60}{0.39}$&&$\err{-0.01}{0.19}{0.02}$&\cite{Chen:2008jy}\\
$p\overline{p}\Kstarz$&$535$&$\aerr{1.18}{0.29}{0.25}{0.11}$&&$\err{-0.08}{0.20}{0.02}$&\cite{Chen:2008jy}\\
$p\Lbar\pi^0$&$449$&$\aerr{3.00}{0.61}{0.53}{0.33}$&&$\err{+0.01}{0.17}{0.04}$&\cite{Wang:2007as}\\
$p\Lbar\pi^-$&$449$&$\aerr{3.23}{0.33}{0.29}{0.29}$&&$\err{-0.02}{0.10}{0.03}$&\cite{Wang:2007as}\\
$p\Lbar K^-$&$85$&&$0.82$&&\cite{Wang:2003yi}\\
$p\Sigbar^0\pi^-$&$85$&$\aerr{3.97}{1.00}{0.80}{0.56}$&&&\cite{Wang:2003yi}\\
$\Lambda\Lbar\pi^+$&$152$&&$2.80$&&\cite{Lee:2004mg}\\
$\Lambda\Lbar K^+$&$152$&$\aerr{2.91}{0.90}{0.70}{0.38}$&&&\cite{Lee:2004mg}\\
\hline\hline
\end{tabular}
\end{center}
\end{table}

\def\qqbar{q\overline{q}}
\def\BBbar{B\overline{B}}
\def\Bbar{\overline{B}}
\def\pbar{\overline{p}}
\def\bbar{\overline{b}}
\def\KS{K_S^0}
\def\GeV{\;{\rm GeV}}
\def\Br{{\cal B}}
\def\Mbc{M_{\rm bc}}
\def\DeltaE{\Delta E}
\def\Vtd{V_{td}}
\def\Vts{V_{ts}}

\clearpage
\subsection{Radiative penguin decays}

\begin{wrapfigure}{r}{5.1cm}
  \centerline{\includegraphics[width=0.35\textwidth]
     {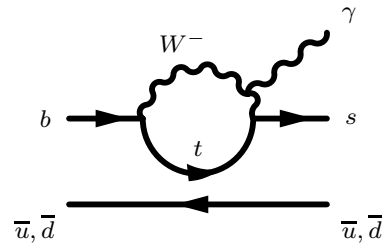}}
     \caption{Feynman diagram for the $b\to s\gamma$ process.}
\label{fig:feynmp_xsgam}
\end{wrapfigure}
Decay processes of a $b$ quark that emit a photon
are not allowed at the tree level in the SM, and require a so-called
radiative ``penguin'' loop (Fig.~\ref{fig:feynmp_xsgam}).  
The dominant contribution in the
SM is from a loop with a top quark and a weak boson. However, these heavy
SM particles may be replaced by hypothetical particles such as a charged
Higgs boson or supersymmetric particles.  In such a scenario, the decay
rate or other observables could be drastically modified. Hence, radiative
decays have been extensively studied to search for and to constrain
physics beyond the SM.

\subsubsection{Inclusive $B\to X_s\gamma$ measurement}

At the hadron level, the quark-level 
$b\to s\gamma$ transition is represented by a radiative $B$ meson
decay into a high energy photon and an inclusive hadronic final state
with a unit strangeness denoted by the symbol $X_s$.  The clean signature
of the high energy photon makes it possible to measure the decay rate
without reconstructing the $X_s$.  The SM transition rate is calculated
including  next-to-next-to-leading logarithmic corrections to 7\%
precision~\cite{Misiak:2006zs}.

\begin{wrapfigure}{r}{6.1cm}
  \centerline{\includegraphics[width=0.45\textwidth]
     {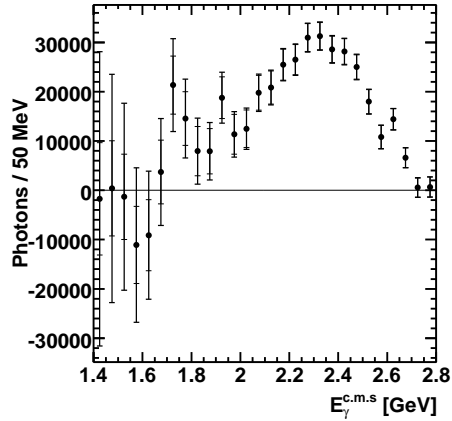}}
     \caption{Photon energy spectrum from $B\to X_s\gamma$.}
\label{fig:xsgam}
\end{wrapfigure}

A large  and dominant background to $B\to X_s\gamma$ 
is from the $\pi^0$s (and to a lesser extent
$\eta$) in the $e^+e^-\to\qqbar$ continuum, which subsequently
decay into a pair of photons. Although this background
is several orders of magnitude larger than the inclusive photon signal, 
backgrounds that are not from a $B$ meson decay can be
statistically subtracted by using the off-resonance data sample taken at
60 MeV below the $\Upsilon(4S)$ resonance.  However, since only 10\%
of integrated luminosity is taken off-resonance, this 
continuum background remains
the main source of the statistical and systematic error.  The remaining
backgrounds are from $B$ meson decays, where the photon backgrounds are
dominantly (in order of their importance)
from the $\pi^0$s, $\eta$s, radiative decays of other
hadrons, final state radiation and electron bremsstrahlung, and
mis-reconstructed $K^0_L$s and (anti-)neutrons.
The inclusive $\pi^0$ and $\eta$
production rate from a $B$ meson is directly measured in data, and used
to subtract the corresponding background
contribution, while other sub-dominant contributions
are subtracted using MC expectations after correcting for the measured
data-MC differences.  The photon energy spectrum, which is monochromatic
if $b\to s\gamma$ is strictly a two-body process, is broadened by QCD
corrections and the Fermi motion of the $b$ quark in the $B$
meson~\cite{Kagan:1998ym,Buchmuller:2005zv}. The measured spectrum in
the $\Upsilon(4S)$ rest frame is further broadened by the small momentum
of the $B$ meson and the detector resolution. 
The branching fraction
has to be integrated over the entire photon energy range. It
becomes more difficult to do so for lower energies as the signal
contribution becomes smaller and the background becomes insurmountably
large.  It is now customary to compare the 
extrapolated branching fraction in the range
$E_\gamma>1.6\GeV$ to theoretical predictions.
Experimental efforts to lower this bound have been the focus of most 
past $B\to X_s\gamma$ measurements. Using
$657 \times 10^6$ $\BBbar$ events, Belle measured $B\to X_s\gamma$
with $E_\gamma>1.7\GeV$~\cite{Limosani:2009qg}. This should cover
$(98.5\pm0.4)\%$ of the spectrum above
$1.6\GeV$~\cite{Buchmuller:2005zv}.  The spectrum is shown in
Fig.~\ref{fig:xsgam} and the branching fraction was measured to be
\begin{equation}
\Br(B\to X_s\gamma;\; E_\gamma>1.7\GeV)
=(3.45\pm0.15{\rm(stat)}\pm0.40{\rm(syst)})\times10^{-4},
\end{equation}
where the errors are statistical and systematic.  
Together with 
BaBar's measurement, the world average~\cite{hfag} extrapolated for
$E_\gamma>1.6\GeV$ is $\Br(B\to
X_s\gamma)=(3.55\pm0.24{\rm(exp)}\pm0.09{\rm(model)}\times10^{-4}$,
where the first error is a combined experimental (statistical and
systematic) uncertainty and the second is the model error in the
extrapolation.  This can be compared with the theory
prediction~\cite{Misiak:2006zs} of
 $\Br(B\to X_s\gamma)=(3.15\pm0.23)\times10^{-4}$.
The results are consistent, and have been used to constrain new physics
scenarios.  For example, the charged Higgs mass is bounded to be above
$295\GeV$.

\subsubsection{Exclusive radiative $B$ decays with $b \to s \gamma$}

Exclusive radiative $B$ meson decay modes, such as $B\to
K^*(892)\gamma$~\cite{Nakao:2004th}, have been more precisely measured,
since one can fully constrain and effectively suppress the background
of the decay kinematics using the beam-energy
constrained mass ($\Mbc$) and the energy difference ($\DeltaE$). 
However, theoretical predictions
suffer from large uncertainties in the exclusive form factors, 
which cannot be
reliably determined~\cite{Keum:2004is,Bosch:2001gv,Ali:2001ez}.

The $B\to K^*(892)\gamma$ constitutes about 15\% of the total 
$B\to X_s\gamma$ branching fraction.  Since the $B$ meson has spin zero and the
photon has spin one and is longitudinally polarized, the $X_s$ system cannot
be a single kaon (with spin zero), a resonance, or an 
S-wave $K\pi$ system.  Of the
higher kaonic resonances, only $B\to K_2^*(1430)\gamma$~\cite{Nishida:2002me}
and $B\to K_1(1270)\gamma$~\cite{Yang:2004as} have been measured.  In
particular, higher kaonic resonances around 1.4 GeV have a complicated
structure, and among these the $K_1(1270)$ contribution was found to be
 dominant~\cite{Li:2008qma}.  In the multi-body final states, many
modes have been measured: $B\to K\pi\pi\gamma$~\cite{Nishida:2002me},
$B\to K\eta\gamma$~\cite{Nishida:2004fk}, $B\to
K\eta'\gamma$~\cite{Wedd:2008ru}, $B\to
K\rho\gamma$~\cite{Nishida:2002me}, $B\to
K\phi\gamma$~\cite{Sahoo:2011zd}, and $B\to \Lambda \pbar
\gamma$~\cite{Lee:2005fba}.

One way to reduce the theoretical uncertainty is to take ratios or
asymmetries.  In particular, the time-dependent $CP$ asymmetry for a
radiative decay
into a self-conjugate final state has a unique feature.  In the SM,
the final state, e.g. $\KS\pi^0\gamma$, is not a $CP$ eigenstate since
the photon is dominantly left-handed from $\Bbar^0$ (with a $b$ quark)
decay and thus does not mix with the decay from $B^0$ (with a $\bbar$ quark)
with a right-handed photon.  The spin flip is suppressed by the
quark mass ratio $2m_s/m_b$ and hence the time-dependent $CP$ asymmetry
is also suppressed in the SM to a few per cent~\cite{Atwood:1997zr}.
Therefore, this asymmetry in the $b\to s\gamma$ process 
is sensitive to non-SM right-handed currents. 

In $B\to K^*(892)^0\gamma$, the rate to the
$\KS(\to\pi^+\pi^-)\pi^0\gamma$ final state is only 1/9 of that for
$K^+\pi^-\gamma$.  The time-dependent asymmetry is measured by extrapolating
the $\KS$ momentum from the $\KS$ decay vertex to the interaction
region.  Therefore, the detection efficiency and
statistics of the final signal sample are not large.  The coefficient to
the sine term is measured with $535 \times 10^6$ $\BBbar$ to be~\cite{Ushiroda:2006fi}

\begin{equation}
{\cal S}_{K^{*0}\gamma} = -0.32^{+0.36}_{-0.33}{\rm(stat)}\pm0.05{\rm(syst)}.
\end{equation}

This study can be extended to $B^0\to P^0 Q^0\gamma$, where $P^0$ and
$Q^0$ are any pseudoscalars~\cite{Atwood:2004jj}, or to the 
$P^0 V^0\gamma$ state
if the spin parity of the $P^0 V^0$ system is determined.  Time-dependent
asymmetries have been measured for $\KS\pi^0\gamma$, $\KS\rho^0\gamma$, 
and $\KS\phi\gamma$ states, although none of them is yet able to constrain
the right-handed current.  This study is one of the promising modes in the 
search for physics beyond the SM with the high statistics data 
samples expected at Belle II.

\subsubsection{Radiative $B$ decays with $b\to d\gamma$}

The $b\to d\gamma$ penguin loop is suppressed
with respect to $b\to s \gamma$ by $|\Vtd/\Vts|^2$, and therefore is
sensitive to this ratio.  It is particularly interesting because a
more precise determination of $|\Vtd/\Vts|$ was not available until the
$B_s$ mixing rate was measured~\cite{Abulencia:2006ze} and even
after that it provided an independent test of this ratio of CKM parameters.

Since the dominant diagram is suppressed, there are more contributions
from subleading diagrams.  These could lead to a large direct
$CP$ violation or large isospin asymmetry, although they also modify the
determination of $|\Vtd/\Vts|$.  On the other hand, time-dependent
asymmetry is expected to be even smaller, 
since the phase from $\Vtd$ in mixing and $b\to d\gamma$ transition
cancel~\cite{Atwood:1997zr,Ushiroda:2007jf}.
Contributions from non-SM physics can therefore be relatively enhanced
and may be more clearly visible than in the $b\to s\gamma$ case.

\begin{figure}
  \begin{center}
  \resizebox{0.32\textwidth}{!}{\includegraphics{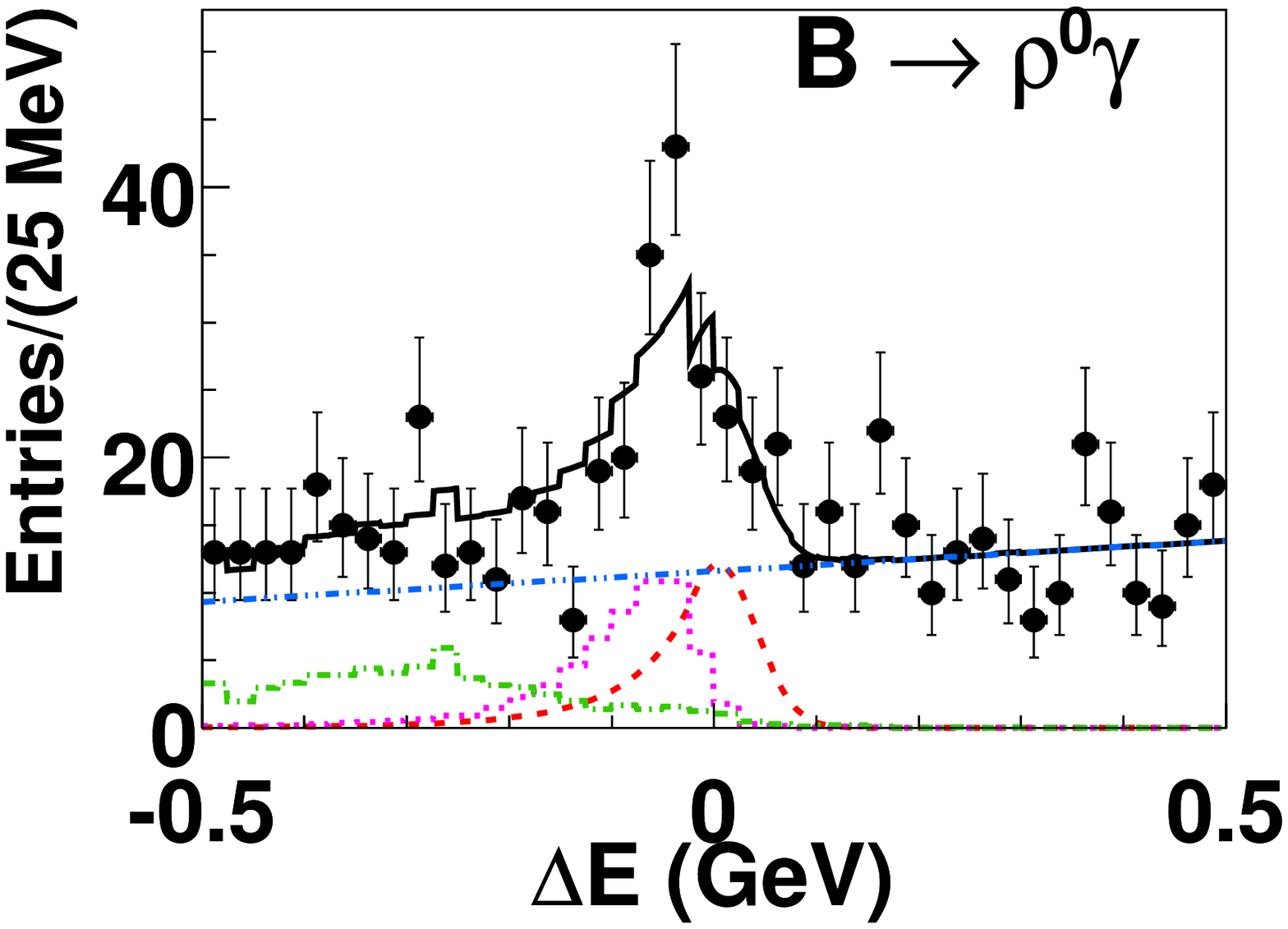}}
  \resizebox{0.32\textwidth}{!}{\includegraphics{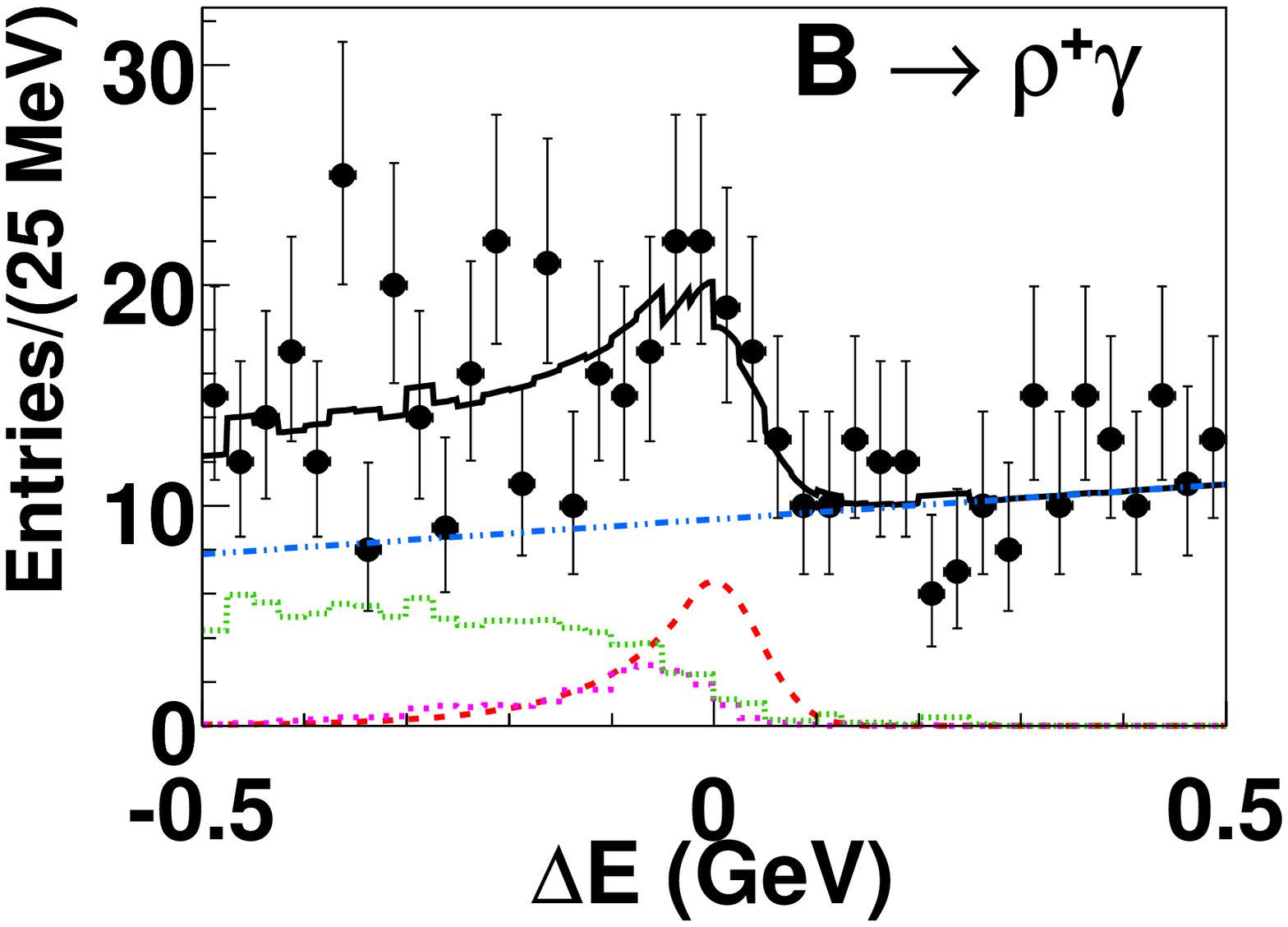}}
  \resizebox{0.32\textwidth}{!}{\includegraphics{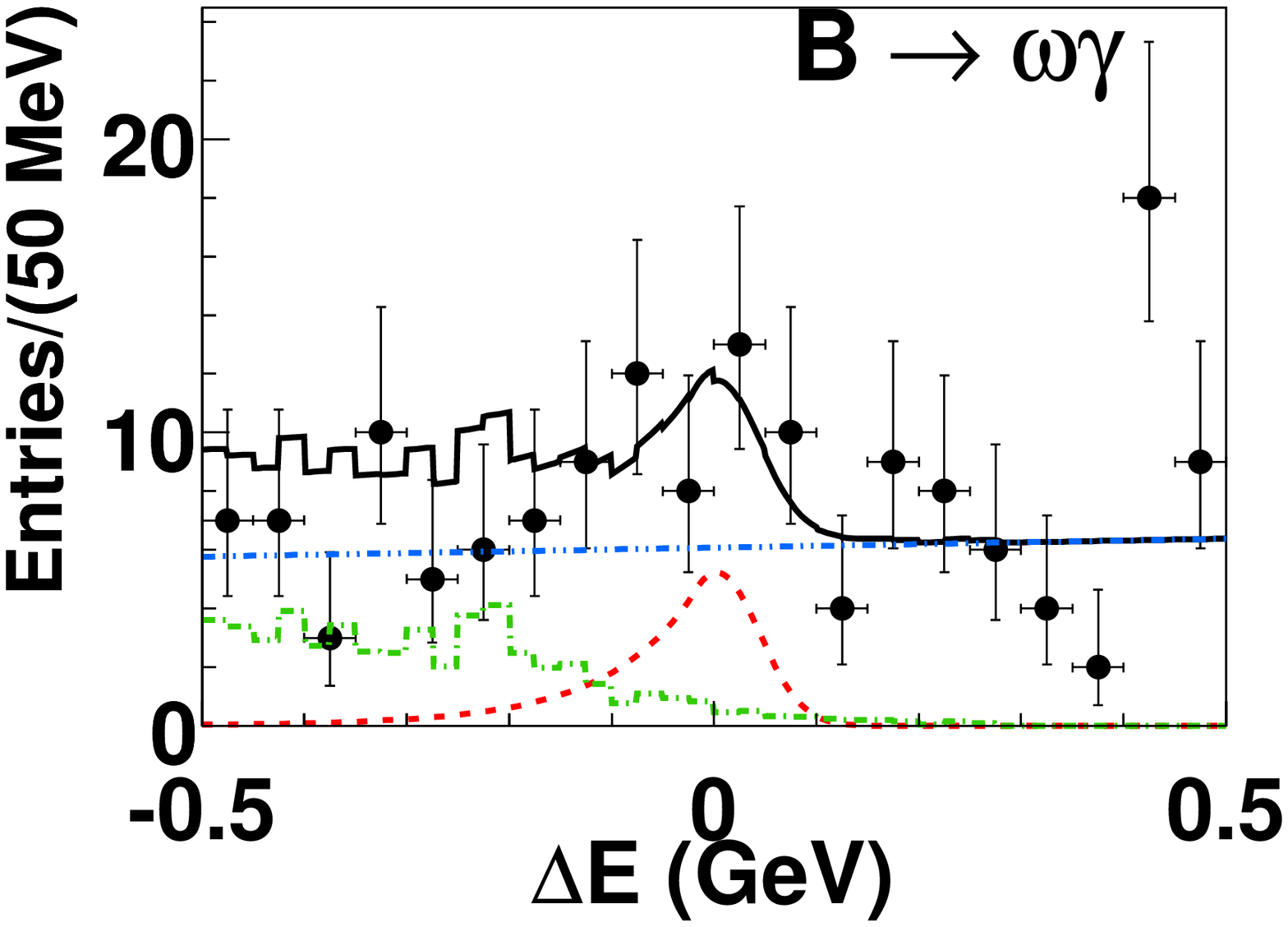}}
  \caption{$\Delta E$ distributions of $B^0\to\rho^0\gamma$ (left),
    $B^+\to\rho^+\gamma$ (middle), and $B^0\to\omega\gamma$ (right).}
  \label{fig:rhogam}
  \end{center}
\end{figure}

Because of the similarity of the kinematics, the large $b\to s\gamma$
process is a severe background to the suppressed $b\to d\gamma$
process.  In the reconstruction of an exclusive decay mode, particle
identification devices are crucial to separate the kaon in $b\to
s\gamma$ from the pion in $b\to d\gamma$.  Exclusive $b\to d\gamma$
decay modes such as $B\to\rho\gamma$ and $B\to\omega\gamma$ have been 
searched for since the start of Belle, and 
were finally observed with $386 \times 10^6$ $\BBbar$ pairs in a
combined measurement~\cite{Mohapatra:2005rj}.  Charged and neutral modes
are combined assuming isospin symmetry
$\Br(B^+\to\rho^+\gamma)=2\Br(B^0\to\rho^0\gamma)$ and
$\Br(B^0\to\rho^0\gamma)=\Br(B^0\to\omega\gamma)$.  The latest result
with $657 \times 10^6$ $\BBbar$ pairs 
is shown in Fig.~\ref{fig:rhogam} and the combined
branching fraction is measured to be~\cite{Taniguchi:2008ty}

\begin{equation}
\Br(B\to(\rho,\omega)\gamma)=(1.14\pm0.20{\rm(stat)}^{+0.10}_{-0.12}{\rm(syst)})\times10^{-6}.
\end{equation}

As $B\to\rho\gamma$ is suppressed compared to $B\to K^*\gamma$ by
$|\Vtd/\Vts|^2$, known kinematic corrections, and less-known form factor
ratios and corrections for subleading diagrams, the result is combined 
with a corresponding analysis on $B\to K^*\gamma$ to constrain
$|\Vtd/\Vts|$.  The result is

\begin{equation}
|\Vtd/\Vts|=0.195^{+0.020}_{-0.019}{\rm(exp)}\pm0.015{\rm(theo)}
\end{equation}
where the first error is a combined statistical and systematic 
uncertainty and the second error is the theory uncertainty on the ratio.

The $B^0\to\rho^0\gamma$ signal is found to be stronger than
$B^+\to\rho^+\gamma$.  This corresponds to a large isospin asymmetry,
which is defined as
$\Delta(\rho\gamma)=
{\tau_{B^0}\over2\tau_{B^+}}\Br(B^+\to\rho^+\gamma)/\Br(B^0\to\rho^0\gamma)-1$.
The isospin asymmetry is calculated as

\begin{equation}
\Delta(\rho\gamma)=-0.48^{+0.21}_{-0.19}{\rm(stat)}^{+0.08}_{-0.09}{\rm(syst)}.
\end{equation}
BaBar also measures this ratio and finds the same tendency; the combined
isospin asymmetry is $\sim3\sigma$ away from the SM expectation, which could
be at most $\sim$10\%.  As the statistical error is still large,
the high statistics expected at Belle II will be necessary to clarify this
tension.

\subsection{Electroweak penguin decays} 

\begin{figure}
  \centerline{\includegraphics[width=0.3\textwidth]
     {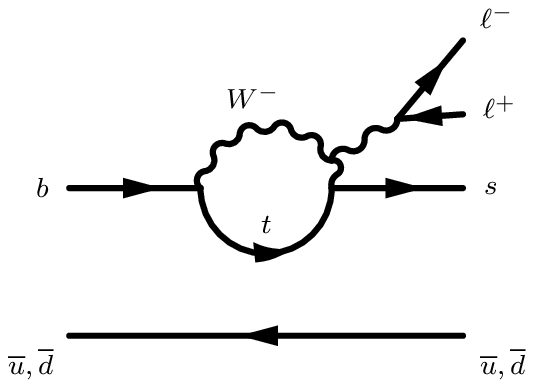}
     \includegraphics[width=0.3\textwidth]
     {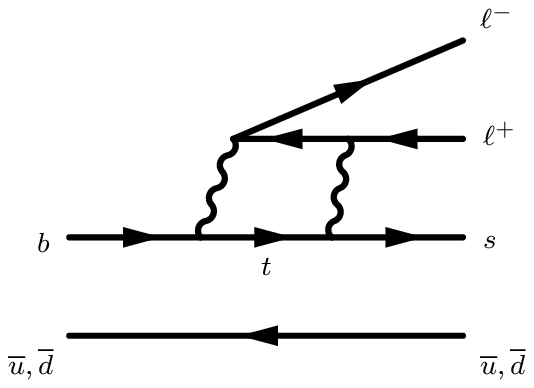}}
     \caption{Feynman diagrams for the $b\to s l^+ l^-$ process.}
\label{fig:feynmp_xsll}
\end{figure}
%
The $b\to s(d) l^+l^-$ transitions proceed at 
lowest order in the SM via  $Z/\gamma$ penguin diagrams or a $W$
box diagram (Fig.~\ref{fig:feynmp_xsll}).
The $b\to s(d)\nu_l\overline{\nu}_l$ transitions also proceed
through similar diagrams except for the $\gamma$ penguin diagram.
NP mediated by SUSY particles or a possible fourth generation may 
contribute to the penguin loop or box diagram and as a result branching 
fractions and other properties could be modified \cite{npzpen}. 
Such NP contributions may change the Wilson coefficients 
that parameterize the strength of the short distance interactions. 
This is similar to $b\to s(d)\gamma$, but has a richer structure.
The decay $B\to K^* l^+l^-$ is of particular interest
since its large branching fraction facilitates
the examination of various observables that are sensitive to NP.
For instance, the lepton forward\---backward asymmetry ($A_{FB}$), 
the $K^*$ polarization ($F_L$),  
and the $K^* l^+ l^-$ isospin asymmetry ($A_I$) as functions of dilepton
mass squared ($q^2$) differ from the SM expectations in various NP
models~\cite{var}.  

 The neutral pure leptonic decays $B^0 \to l^+ l^-$ and 
 $B^0\to \nu_l \overline{\nu}_l$ proceed mainly through the box and 
 $Z$ boson mediated annihilation diagrams, which are equivalent
  to the diagrams for $b\to d l^+l^-$ and $b\to d\nu_l\overline{\nu}_l$. 
In the SM these decays are also helicity suppressed and, compared to the 
charged purely leptonic decays (see Sect.~4.3), the branching fractions are 
about three orders of magnitude smaller for the 
corresponding generation~\cite{ckmf}. The SM branching fraction of 
$B^0\to \nu_l \overline{\nu}_l$ is at the level of
$10^{-20}$~\cite{bburas}. 
The lepton-flavor-violating decay $B^0\to e^\pm\mu^\mp$ is 
not an electroweak penguin decay and is
forbidden in the SM, but can occur in the Pati\---Salam model~\cite{pati} or 
supersymmetric models~\cite{susy}, and can be searched for
simultaneously. A positive signal for any of these decay modes  
with the current Belle data sample would demonstrate NP in the loop. 
 
\subsubsection{Exclusive $b\to s(d) \; l^+ l^-$ decays}
The study of the decay $B\to K^{(*)} l^+ l^-$ started at the 
beginning of Belle and was updated  several 
times.  We reported the first observations of 
$B\to K l^+ l^-$ \cite{ishi1} and $B\to K^* l^+ l^-$ \cite{ishi2}
 with $31.3\times 10^6$ 
and  $152\times 10^6$ $B\overline{B}$ pairs, respectively. 
In 2006, Belle 
published the first measurements of the forward\---backward asymmetry and 
the ratios of Wilson coefficients $A_9/A_7$ and $A_{10}/A_7$ using $386\times 
10^6 B\overline{B}$ pairs\cite{ishi3}.  An unbinned maximum likelihood fit to 
$q^2$ and $\cos\theta_l$ was used to extract the ratios of the Wilson 
coefficients, where $\theta_l$ is the angle between 
the momenta of a negative (positive) lepton 
and the $B$ ($\overline{B}$) meson in the dilepton rest frame.

The latest analysis in 2008~\cite{weisll} used $657\times10^6$ $B\overline{B}$ 
pairs; more observables were measured.
Candidate $B\to K^{(*)} l^+ l^-$ decays were reconstructed in 10
channels: $K^+\pi^-$, $K^0_S\pi^+$, $K^+\pi^0$ for $K^*$, $K^+$, 
 and $K^0_S$ for $K$, with $e^+e^-$ and $\mu^+\mu^-$ lepton pairs.
The dilepton mass of each candidate was required to be outside 
of the $J/\psi$ and
$\psi(2S)$ mass regions to avoid the large charmonium background, and
 above the $\pi^0$ mass for $e^+e^-$ pairs to avoid the $\pi^0$ Dalitz decay,
 photon conversion, and the pole at $q^2=0$.
 Two major backgrounds were considered: the continuum and 
$B\overline B$ events in which both $B$ mesons decay semileptonically. 
These backgrounds were suppressed by imposing requirements on the 
signal\---continuum and signal\---$B\overline{B}$  likelihood ratios. 

After requiring the candidate $\Delta E$ to lie in the signal region,  
the signal yields in each $q^2$ bin were extracted from an unbinned 
likelihood fit to $M_{\rm bc}$ and the $K\pi$ mass ($M_{K\pi}$) for the 
$K^* l^+l^-$ mode and $M_{\rm bc}$ only for the $K l^+l^-$ mode. The 
corresponding branching fractions were thus obtained.
 The $F_L$ and $A_{FB}$ parameters 
were extracted from fits to $\cos\theta_{K^*}$
and $\cos\theta_l$ in the signal region, where $\theta_{K^*}$ is the angle 
between the kaon direction and the direction opposite to the $B$ meson in the
$K^*$ rest frame. The signal PDFs for the 
$\cos\theta_{K^*}$ and $\cos\theta_l$ variables
are the product of the following two functions,
\begin{equation*}
 [\frac{3}{2}F_L\cos^2\theta_{K^*} + \frac{4}{3}(1-F_L)(1-\cos^2\theta_{K^*})]
  \times \epsilon(\cos\theta_{K^*})
\end{equation*}
and 
\begin{equation*}
  [\frac{3}{4}F_L(1-\cos^2\theta_l) + \frac{3}{8}(1-F_L)(1+\cos^2\theta_l)
   + A_{FB}\cos\theta_l] \times \epsilon(\cos\theta_l), 
\end{equation*}
where $\epsilon(\cos\theta_{K^*})$ and $\epsilon(\cos\theta_l)$ are the 
reconstruction efficiencies. For the $B\to K l^+ l^-$ modes, $F_L$ is set to 1.
Furthermore, this analysis also reported the isospin asymmetry defined as 
\begin{equation*}
    A_I = \frac{(\tau_{B^+}/\tau_{B^0})\times{\cal B}(K^{(*)0} l^+l^-) -
                 {\cal B}(K^{(*)\pm} l^+ l^-) } 
           {(\tau_{B^+}/\tau_{B^0})\times{\cal B}(K^{(*)0} l^+l^-)  +
                 {\cal B}(K^{(*)\pm} l^+ l^-) },
\end{equation*}
\begin{wrapfigure}{r}{6.0cm} 
 \begin{center} 
  \includegraphics[width=6.0cm]{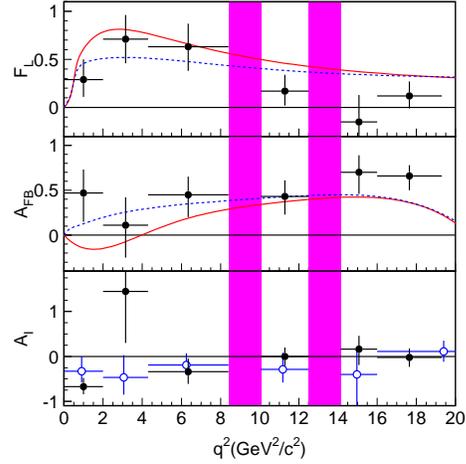}
  \end{center}
\caption{The $q^2$ dependence of $F_L$ (top), $A_{FB}$ (middle). and
   $A_I$ (bottom) in six bins. The results for
  $B\to K^*l^+l^-$ are the filled circles 
   and those for $B\to Kl^+l^-$ are open circles ($A_I$ only).}
   \label{fig:kstarll}
\end{wrapfigure}
where $\tau_{B^+}/\tau_{B^0}$ is the ratio of $B^+$ to $B^0$ lifetimes.
These observables were measured for the first time in six $q^2$ bins as
shown in Fig.~\ref{fig:kstarll}.
Although the uncertainties in the $A_{FB}$ values are still large, 
the positive central values in all $q^2$ bins suggested a non-zero 
$A_{FB}(q^2)$. This phenomenon would have been 
an undeniable signature of NP, but unfortunately did not
persist with larger data samples at the LHC hadron collider~\cite{lhcb}.
Two more observables, the direct $CP$-violating asymmetry and the 
lepton flavor ratio of the muon to electron modes, were also measured. 
The latter is sensitive to Higgs emission and could be larger than 
the SM expectation in the two Higgs doublet model at large 
$\tan\beta$~\cite{susysll}. 

The observed values are $A_{CP}(K^*l^+l^-) = -0.10\pm 0.10 \pm 0.01$ and 
$A_{CP}(Kl^+l^-) = 0.04\pm 0.10\pm 0.02$, consistent with no asymmetry, and 
$R_{K^*} =0.83\pm 0.18 \pm 0.08$ and $R_K = 1.03\pm 0.19\pm 0.06$, similar to
the SM values. The measurements of so many observables demonstrate the 
richness and potential of the $B\to K^{(*)} l^+l^-$ decay.

A search for the exclusive $b\to dl^+l^-$ process, 
$B\to \pi l^+ l^-$ $(\pi = \pi^+ \; {\rm or}\; \pi^0)$,  was performed
using  $657\times 10^6 B\overline{B}$ pairs \cite{weidll}.  
No obvious signal
was observed and upper limits on the branching fractions 
at the 90\% C.L. were obtained:
${\cal B}(B^+\to \pi^+ l^+ l^-)< 4.9\times 10^{-8}$ and 
${\cal B}(B^0\to \pi^0 l^+ l^-)< 15.4\times 10^{-8}$. These limits are 
approaching the SM expectations, which are $O(10^{-8}$).

\subsubsection{Inclusive $B\to X_s l^+ l^-$ decay} 

  The inclusive measurement of the $b\to s l^+ l^-$ process is
experimentally challenging, but can be compared with
theoretically clean predictions. The standard technique is to analyze 
 $B\to X_s l^+ l^-$ events with a semi-inclusive approach, 
where the $X_s$ is reconstructed in 18 different
combinations of either a $K^+$ or $K^0_S$ combined with 0 to 4 pions, of which
up to one $\pi^0$ is allowed.  
This set of final states covers around 62\% of $X_s$ decay states. 
The missing states were taken into account 
in the signal efficiency obtained from MC simulations.

The first observation of $B\to X_s l^+ l^-$  was reported by Belle using
$65.4\times 10^6$ $B\overline{B}$ pairs in 2003 \cite{kaneko}. The
latest Belle results in 2009 used $657\times 10^6$ 
$B\overline{B}$ pairs~\cite{sll}. As in the exclusive analysis, signal 
candidates were selected with $\Delta E$
and then the $M_{\rm bc}$ distribution is
used to extract the signal yield. The dilepton mass was required to
be outside of the $J/\psi$ and $\psi(2S)$ regions and the low mass
region below 0.2 GeV/$c^2$.
\begin{figure}[ht] 
 \begin{center} 
  \includegraphics[width=5.5 cm,height=4.4cm]
   {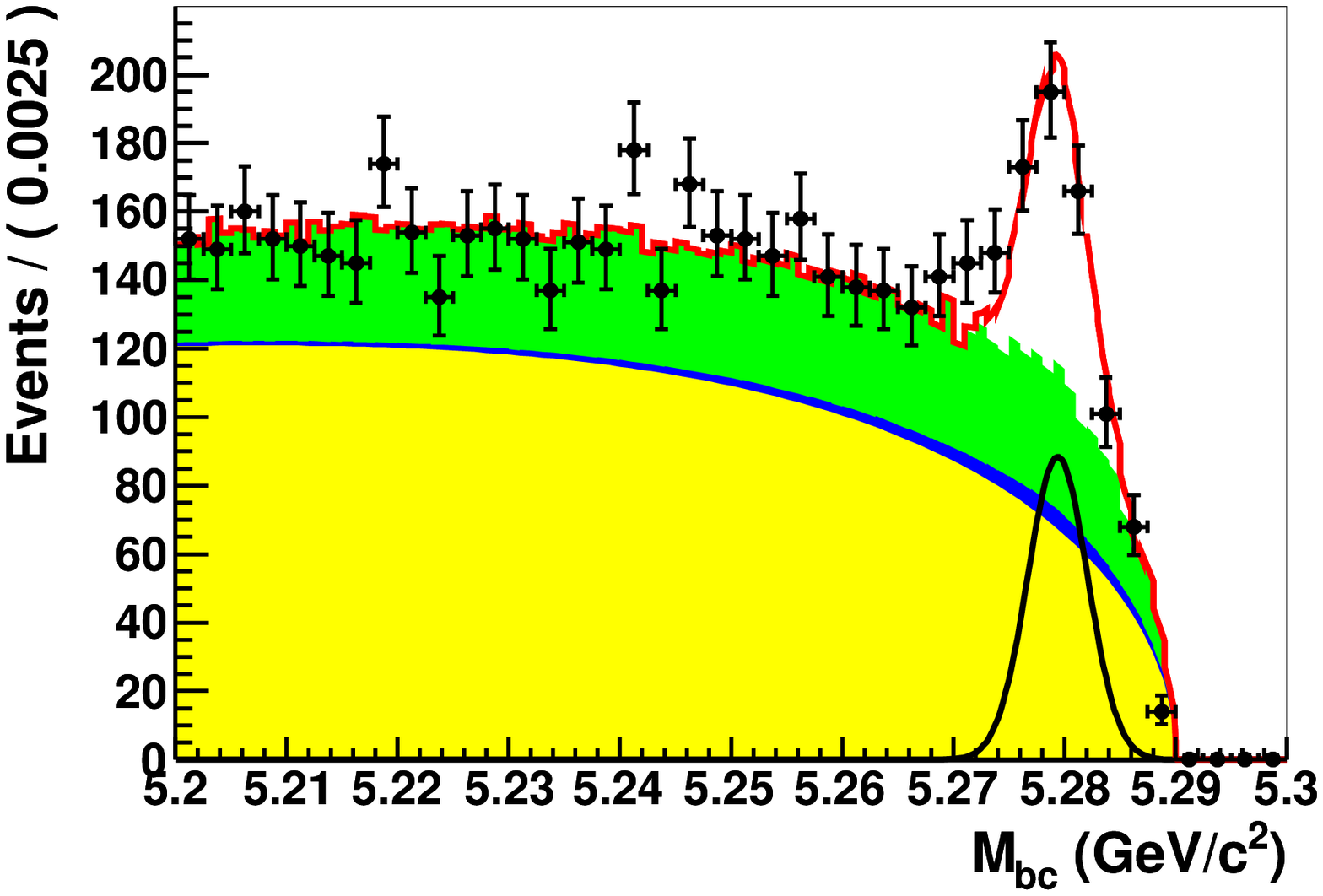}
   \includegraphics[width=5.5 cm,height=4.4cm]
   {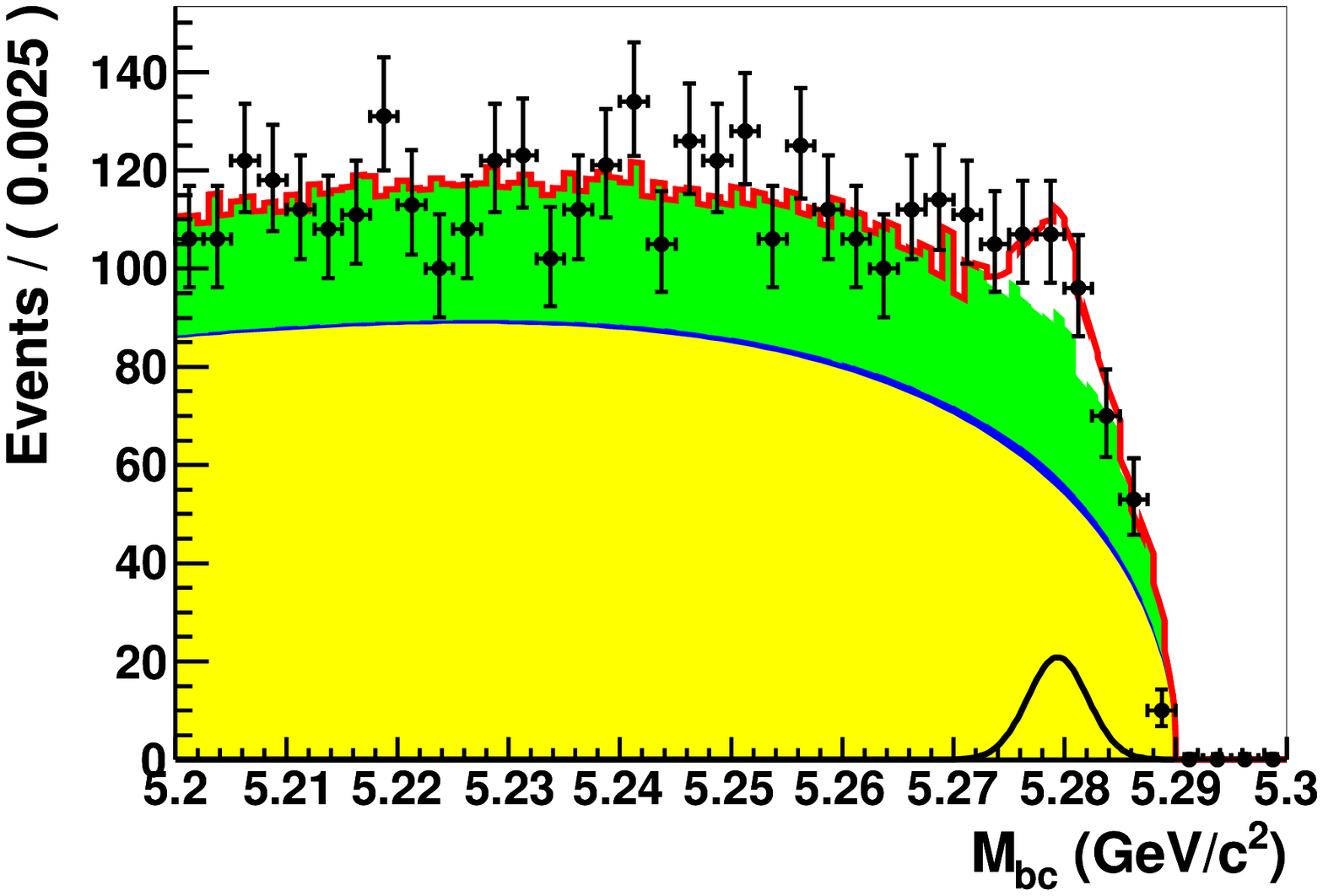}
   
   \includegraphics[width=5.5 cm,height=4.4cm]
   {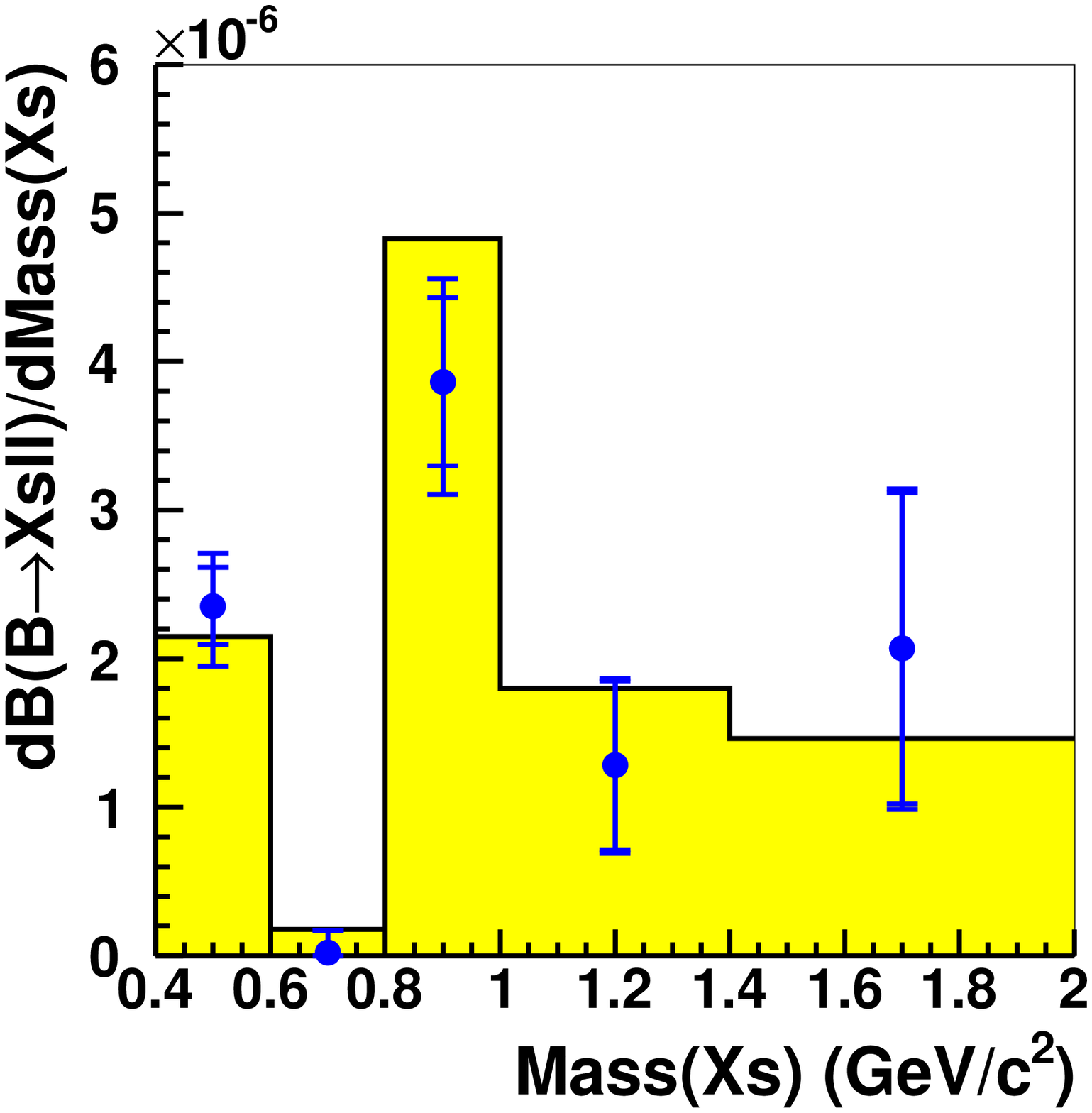}
   \includegraphics[width=5.5 cm,height=4.4cm]
   {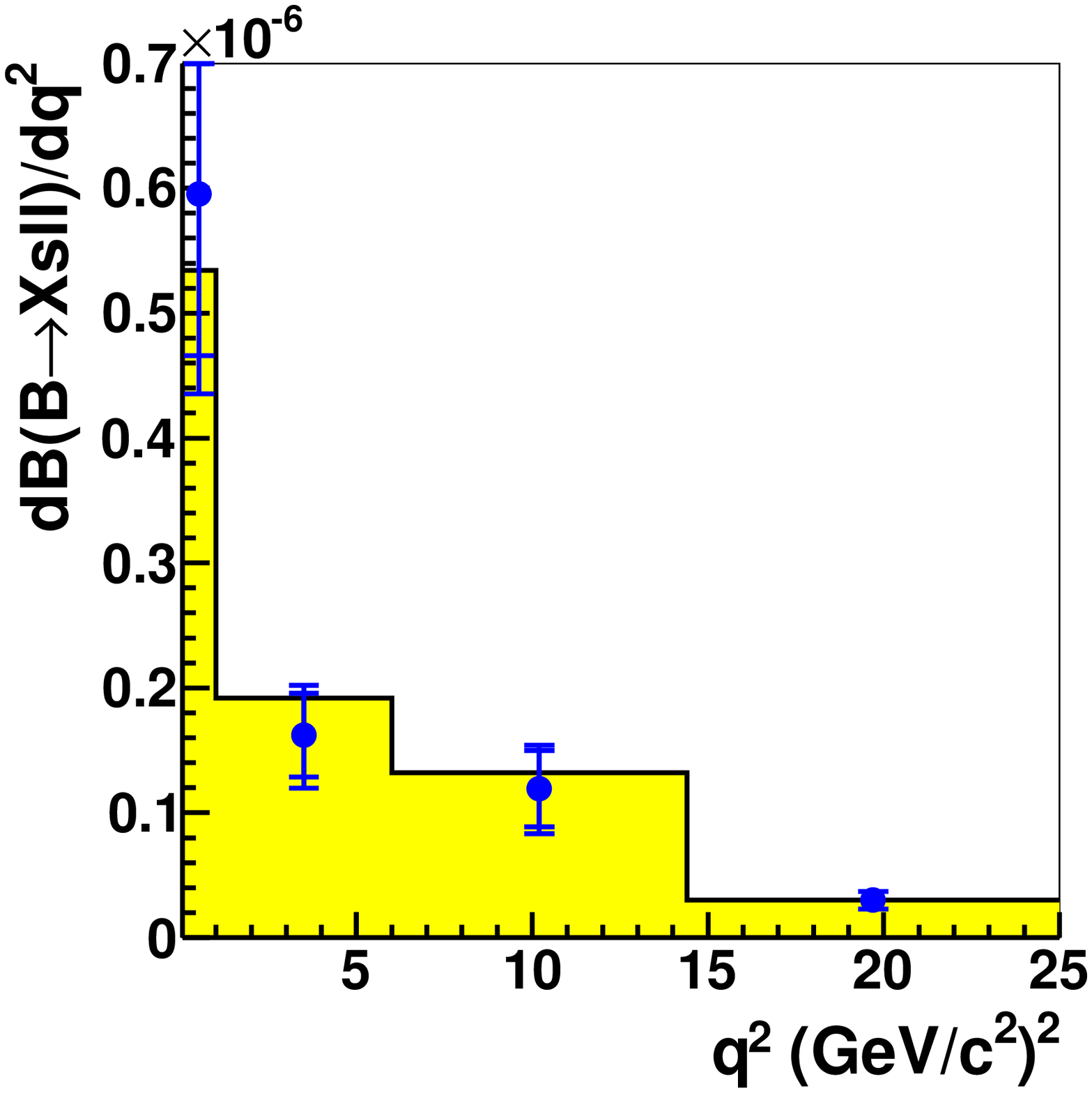}
  \end{center}
\caption{The top two plots show the $M_{\rm bc}$ distributions with the fit
          curves superimposed for the entire sample (left)
         and for $1.0 \;{\rm GeV}/c^2 < M_{X_s}< 2.0$ GeV/$c^2$ (right).
         Points with error bars are data; the dominant background, peaking 
         background, and self-cross-feed components 
         are the yellow, green, and blue solid shaded
         areas, respectively. The bottom plots show 
         the $d{\cal B}$/$dM_{X_s}$ (left) and $d{\cal B}$/$dq^2$ (right) 
         distributions for data (points) and the SM expectation (histograms).  
          }
   \label{fig:mbc_xsll}

\end{figure}
%
The signal yields were extracted from 
an unbinned maximum likelihood fit to the $M_{\rm bc}$ distribution.  
In addition to the dominant backgrounds from $B\overline{B}$ pairs and 
continuum, an effort was made to investigate the peaking background 
and include it in the fit. Two kinds of peaking background were 
considered: charmonium peaking background and hadronic peaking background. 
The former includes the residual of $B\to J/\psi X_s, B\to \psi(2S) X_s$ events
after the $J/\psi$ and $\psi(2S)$ vetoes, and a possible  
contribution from higher $\psi$ resonances, such as the 
$\psi(3770), \psi(4140)$, and $\psi(4160)$. 
The hadronic peaking background contains $B\to X_s h h$ and
$B\to X_s h l\nu$ events in which
 one or two hadrons are misidentified as leptons.
The peaking backgrounds were estimated directly from data or simulations and
the corresponding yields were fixed in the fit. 
Finally, the last component considered is the self-cross-feed  
background, in which the $B$ daughter particles are not correctly 
selected. Its probability density function was modeled as a histogram with
the ratio of the normalization of the self-cross-feed background to signal 
fixed to the MC simulation value in the fit. The probability density function 
for the dominant background was modeled by an ARGUS function with the 
parameters floated in the fit. A simultaneous fit to the $X_s l^+ l^-$ 
and $X_s e^\pm \mu^\mp$ samples was performed 
with the same ARGUS parameters for the dominant background.        

The branching fractions of $B\to X_s l^+ l^-$ were reported as a
function of $M_{X_s}$ and $q^2$
separately. Fit results for the total sample and a subset
with $M_{X_s}>1.0$~GeV$/c^2$ are shown in the top two plots of
Fig.~\ref{fig:mbc_xsll}, and the differential branching
fractions as functions of $M_{X_s}$ and $q^2$ are shown in the bottom
two plots.  The differential distributions are compared with the SM
expectation~\cite{sm_expect}, and found to be in good agreement.
The branching fraction
of $B\to X_s l^+l^-$ in the entire $M_{X_s}$ range was obtained by summing
the branching fraction in each $M_{X_s}$ region and correcting 
for the $X_s l^+l^-$ fraction in the $J/\psi$, $\psi(2S)$, and 
$M_{X_s}>2.0$ GeV/$c^2$ regions. The
branching fraction with the dilepton mass above 0.2 GeV/$c^2$ is thus measured
to be 
${\cal B}(B\to X_s l^+l^-) =(3.33\pm 0.80^{+0.19}_{-0.24})\times
10^{-6}$. 
We also reported the branching fractions separately for the electron and muon
 modes using the same analysis procedure, ${\cal B}(B\to X_s e^+e^-) 
=(4.56\pm 1.15^{+0.33}_{-0.40})\times 10^{-6}$  and ${\cal B}(B\to X_s 
\mu^+\mu^-) =(1.91\pm 1.02^{+0.16}_{-0.18})\times 10^{-6}$.

\subsubsection{Searches for $B^0$ decays to invisible final states}
      
 Searches for $B^0$ decays to invisible final states are rather challenging.  
The same strategy used in the $B^+\to \tau^+ \nu_\tau$ analysis 
was applied to identify the signal (Sect.~\ref{sec:FRB2taunu}). 
Candidate events were selected by fully reconstructing a $B^0$ meson 
and requiring no additional charged, $\pi^0$, or $K^0_L$ particles
in the rest of the event. 
The signal can be identified by requiring no or very little extra 
calorimeter energy ($E_{\rm ECL}$) in the event.
Furthermore, two  variables
were used to distinguish 
the signal and the continuum background: $\cos\theta_B$  
and $\cos\theta_T$, where the latter is the cosine of 
the angle of the $B_{\rm tag}$ thrust axis with respect 
to the beam axis in the CM frame. The continuum was suppressed by making   
requirements on $\cos\theta_T$ and $\cos\theta_B$.

The signal yield was extracted from 
an unbinned extended likelihood fit
to $E_{\rm ECL}$ and $\cos\theta_B$. Candidate events in the fit were 
categorized as signal, $B\overline{B}$, and non-$B$ backgrounds,
where the latter includes the continuum and a small $e^+e^- \to \tau^+\tau^-$
background. Using a sample of 
$657 \times 10^6$ $B\overline{B}$ pairs, the signal
yield obtained in the fit was $8.9^{+6.3}_{-5.5}$ events. 
Since no significant signal was observed, 
we provide the branching fraction upper limit including
systematics at the 90\% C.L. of 
${\cal B}(B^0\to {\rm invisible}) < 1.3 \times 10^{-4}$~\cite{capio}. 
The expected upper limit from the MC study is $1.1\times 10^{-4}$.  

\subsubsection {Search for $B^0\to l^+ l^-$}

The results of searches for the decays $B^0\to e^+e^-$, $\mu^+\mu^-$ 
and $e^\pm \mu^\mp$ (collectively denoted by $B^0\to l^+ l^-$) 
were reported at the beginning of Belle using only 
$85\times 10^6$ $B \bar{B}$ pairs \cite{jerry}. 
Since the background for the two 
energetic leptons is relatively small, the Belle analysis was able to 
suppress the background effectively while maintaining 
a high reconstruction efficiency.  
After all the selection criteria, 
no events were found in any of the three
modes~\cite{jerry}.
The upper limits are: ${\cal B}(B^0\to e^+e^-)<1.9\times 10^{-7}$, 
${\cal B}(B^0\to \mu^+\mu^-)<1.6\times 10^{-7},$ and 
${\cal B}(B^0\to e^\pm \mu^\mp)<1.7\times 10^{-7}$. Furthermore, a lower bound
on the mass of the Pati\---Salam leptoquark model of 46 TeV/$c^2$ was obtained 
using the upper limit for the $e\mu$ mode.

\section{Tau physics}
\label{chap_tau}
The tau lepton is an extremely convenient probe to search for NP beyond 
the SM because of the well-understood 
mechanisms that govern its production and decay in electroweak interactions. 
With its large mass, it is the only lepton that can decay into hadrons,
thus providing a clean environment to study QCD effects in the 1 GeV
energy region.
Tau physics at Belle is categorized by two themes; NP searches and
SM precision measurements. To probe NP, we  search for lepton-flavor 
violating (LFV) decays, $CPV$ in the charged lepton sector, and the
electric dipole moment (EDM) of the tau lepton.
For SM precision measurements, we measure the $\tau$ lepton mass,
the branching fractions of various hadronic decay modes, and their 
invariant mass distributions.
In this section, we summarize the results obtained from
the world's largest data sample (about $10^9~\tau^+\tau^-$ pairs) accumulated 
at the Belle experiment.

\subsection{New physics searches}
\subsubsection{Tau lepton flavor violation}

An observation of LFV would be a clear signature of NP 
since LFV in charged leptons has a negligibly 
small probability in the SM, $O(10^{-54} ) - O(10^{-52})$, 
even if neutrino oscillations are taken into account~\cite{tau:LFVneutrino}. 
Since the $\tau$ is the most massive charged lepton,
it has many possible LFV decay modes.
Belle has examined as many decay modes as possible in the LFV searches, 
since the specific mechanisms of NP are unknown.

Models including supersymmetry (SUSY), 
which is the most popular scenario beyond the SM,
can naturally induce LFV at one loop.
In many SUSY models, including  see-saw extensions and  
grand unified theories, $\tau\rightarrow\mu\gamma$
is expected to have the largest 
branching fraction of all the possible $\tau$ LFV decays.
In some cases, however, 
such as the Higgs-mediated scenario, $\tau$ decay into $\mu\eta$ or 
$\mu\mu\mu$ can become more probable. 
By measuring the branching fractions for various 
$\tau$ LFV decays, one may be able
to determine the NP model favored by nature. 
Among the various modes studied in Belle, we focus here on three possibilities,
$\tau\rightarrow\ell \gamma$, $\ell \ell' \ell''$, and $\ell P^0$, 
where $\ell$ stands for $e$ or $\mu$ and $P^0$ is $\pi^0$, $\eta$, or $\eta'$.

In an LFV analysis, in order to evaluate the signal yield,
two independent variables are used:
the reconstructed mass of the signal
and the difference  between the sum of  energies  of the signal 
$\tau$ daughters and the beam energy  ($\Delta E$) 
in the CM frame. In the $\tau\rightarrow\mu\gamma$ case,
these variables are defined as
\begin{eqnarray}
 M_{\mu\gamma}=\sqrt{E_{\mu\gamma}^2-P_{\mu\gamma}^2},\\
 \Delta E=E_{\mu\gamma}^{\rm CM}-E_{\rm beam}^{\rm CM},
\end{eqnarray}
where $E_{\mu\gamma}$ and $P_{\mu\gamma}$ are the sum of the energies and 
the magnitude of the vector sum of the momenta for
 the $\mu$ and the $\gamma$, respectively.
The superscript $\rm CM$ indicates that the
variable is defined in the CM frame, e.g.
$E_{\rm beam}^{\rm CM}$ is the beam energy in the CM frame.
For signal,  $M_{\mu\gamma}$ and $\Delta {E}$ should be 
in the vicinity of  $M_{\mu\gamma}\sim m_{\tau}$ and $\Delta E\sim0$~(GeV),
while for the background,  $M_{\mu\gamma}$ and $\Delta {E}$ will
smoothly vary without any special peaking structure.
Taking into account the resolution of the detector and the correlation
between $M_{\mu\gamma}$ and $\Delta {E}$, we use an elliptical
signal region.
To avoid bias, we perform a blind analysis:
the data in the signal region are blinded when determining
the selection criteria and the systematic uncertainties.
After fixing these quantities, 
we open the blind and evaluate the number of signal events 
in the signal region.

\medskip
\noindent
\underline{$\tau\rightarrow\ell\gamma$}

\begin{figure}[!ht]
\begin{center}
\includegraphics[height=5cm]{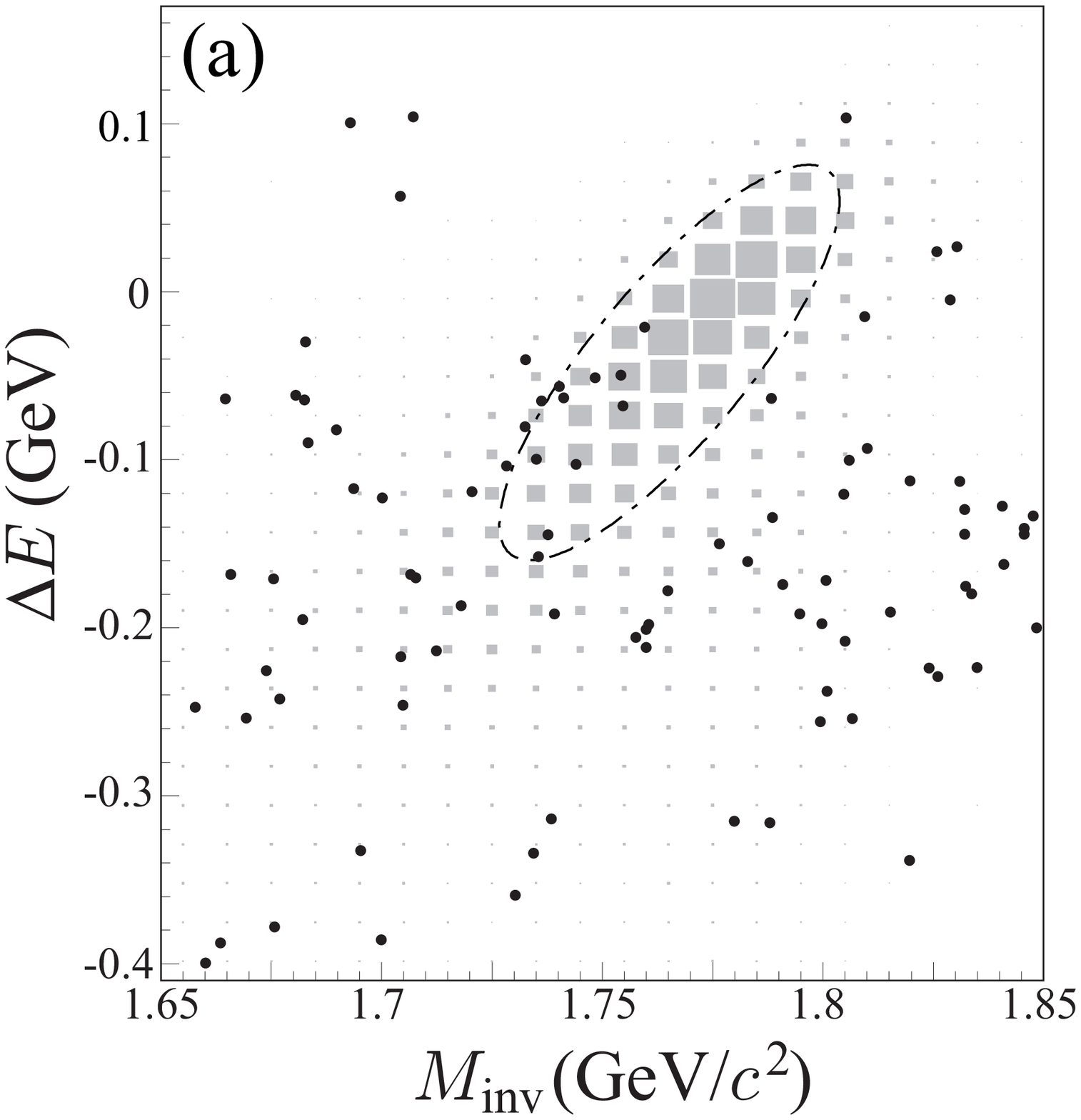}
\includegraphics[height=5cm]{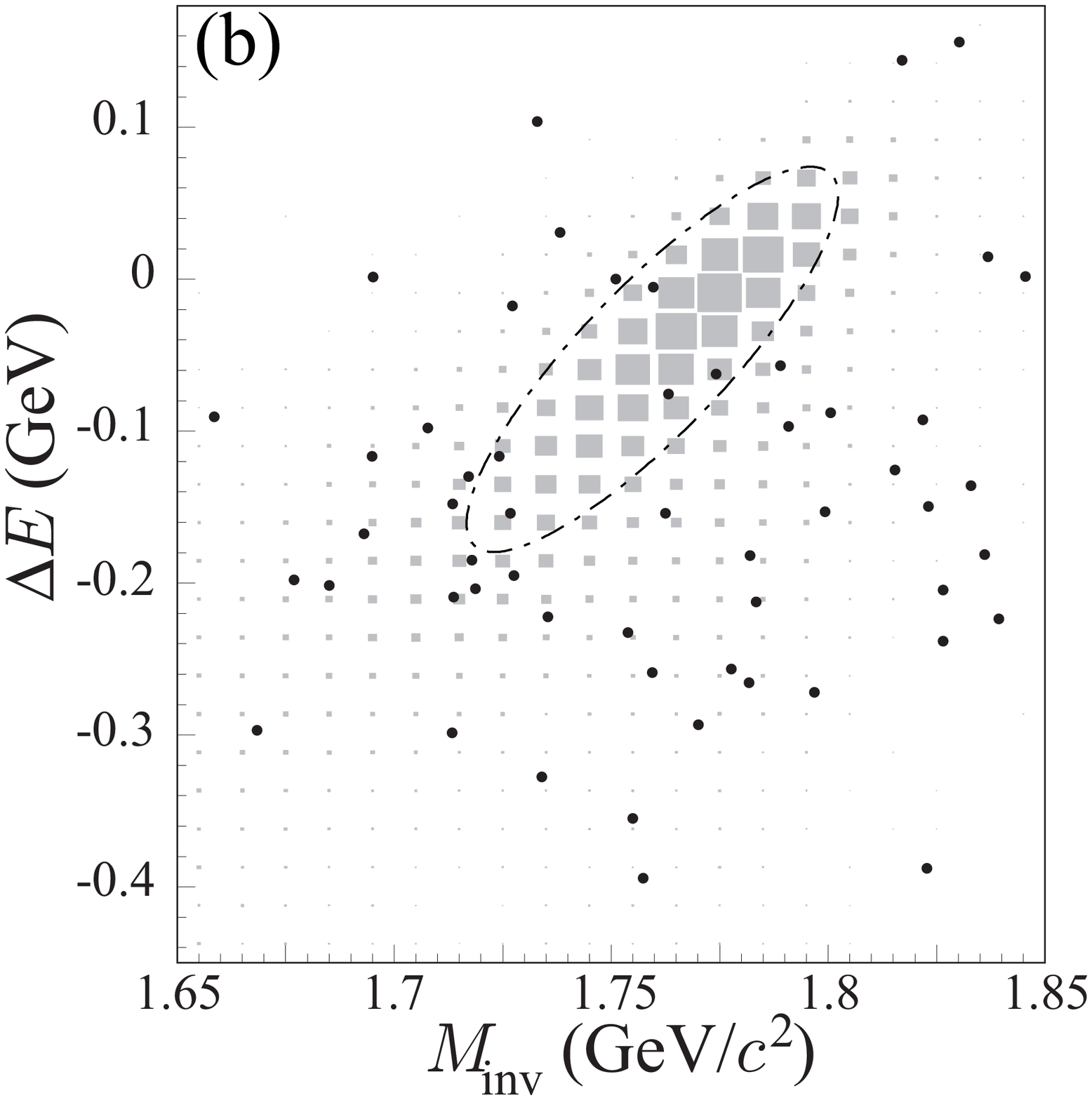}
\end{center}
\caption{ $M_{\mu\gamma}$--$\Delta E$  distributions in the search 
for (a) $\tau \to \mu \gamma$ and
(b) $\tau \to e \gamma$~\cite{tau:mg}.
The black dots and shaded boxes show 
the data and signal MC, respectively,
and the ellipse is the $2\sigma$ signal region.}
\label{fig:TauLG}
\end{figure}

We have searched for $\tau\rightarrow\ell\gamma$
with a data set corresponding to produced
$4.9\times10^8$ $\tau^+\tau^-$ pairs~\cite{tau:mg}.
The main background (BG) is from
$\tau\rightarrow\ell\nu_{\ell}\nu_{\tau}$ + extra $\gamma$ events
and radiative di-muon (for $\mu\gamma$) or Bhabha (for $e\gamma$) events.
The observed $M_{\mu\gamma}$--$\Delta E$  distributions are shown 
in Figs.~\ref{fig:TauLG}(a) and (b) for $\tau\to \mu \gamma$ and
$\tau\to e\gamma$, respectively.
The signal yield is evaluated   
from an extended unbinned maximum-likelihood fit to the
  $M_{\mu\gamma}$--$\Delta E$ distribution.
We found no excess in the signal region. We thus obtain
an upper limit on the branching
fraction for $\tau\rightarrow\mu\gamma$ $(e\gamma)$
of $4.5\times10^{-8}$ ($1.2\times10^{-7}$) at 90\% C.L.

\medskip

\noindent
\underline{$\tau\rightarrow\ell\ell'\ell''$}

\begin{figure}[!ht]
\begin{center}
\includegraphics[height=5cm]{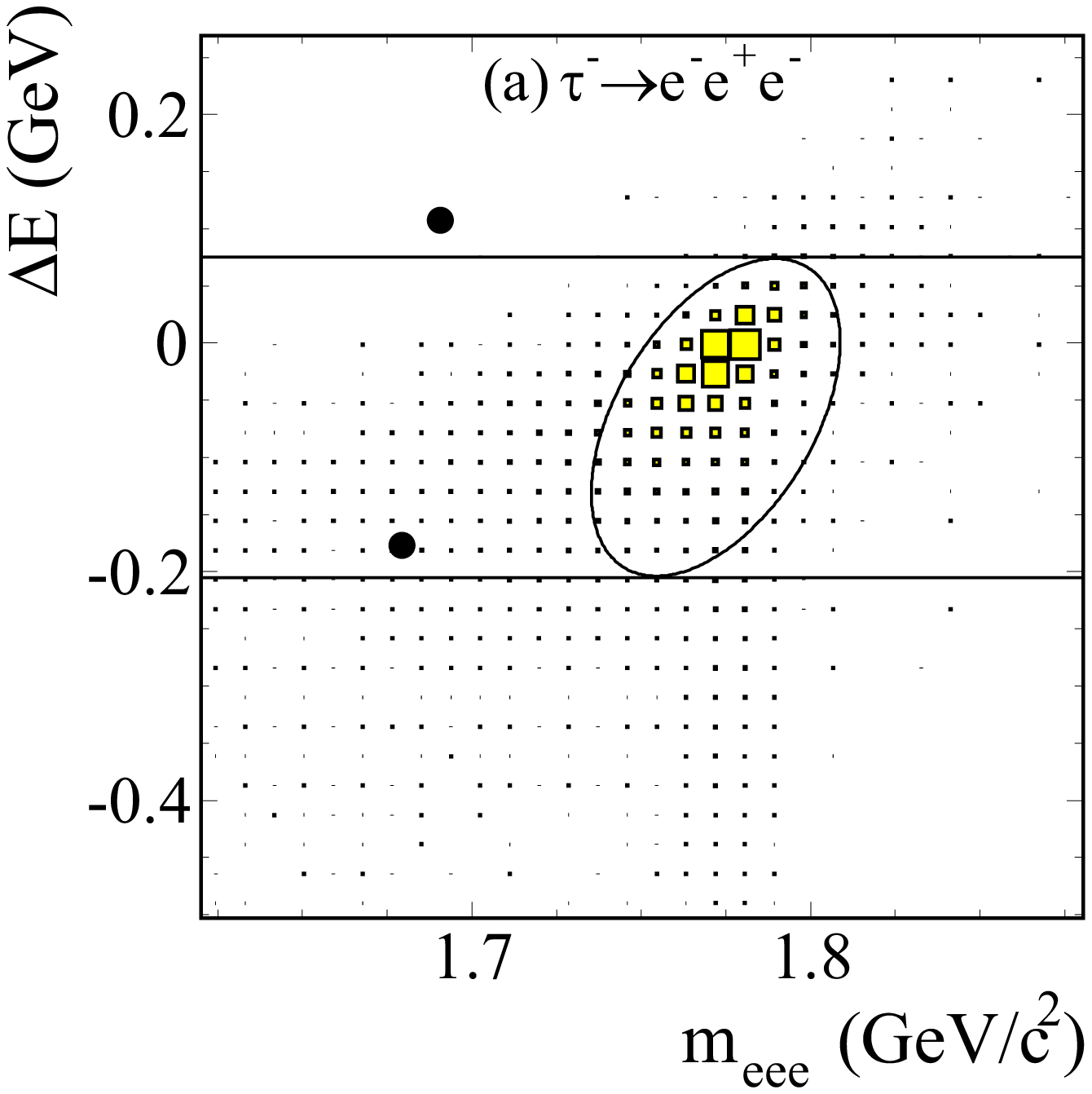}
\includegraphics[height=5cm]{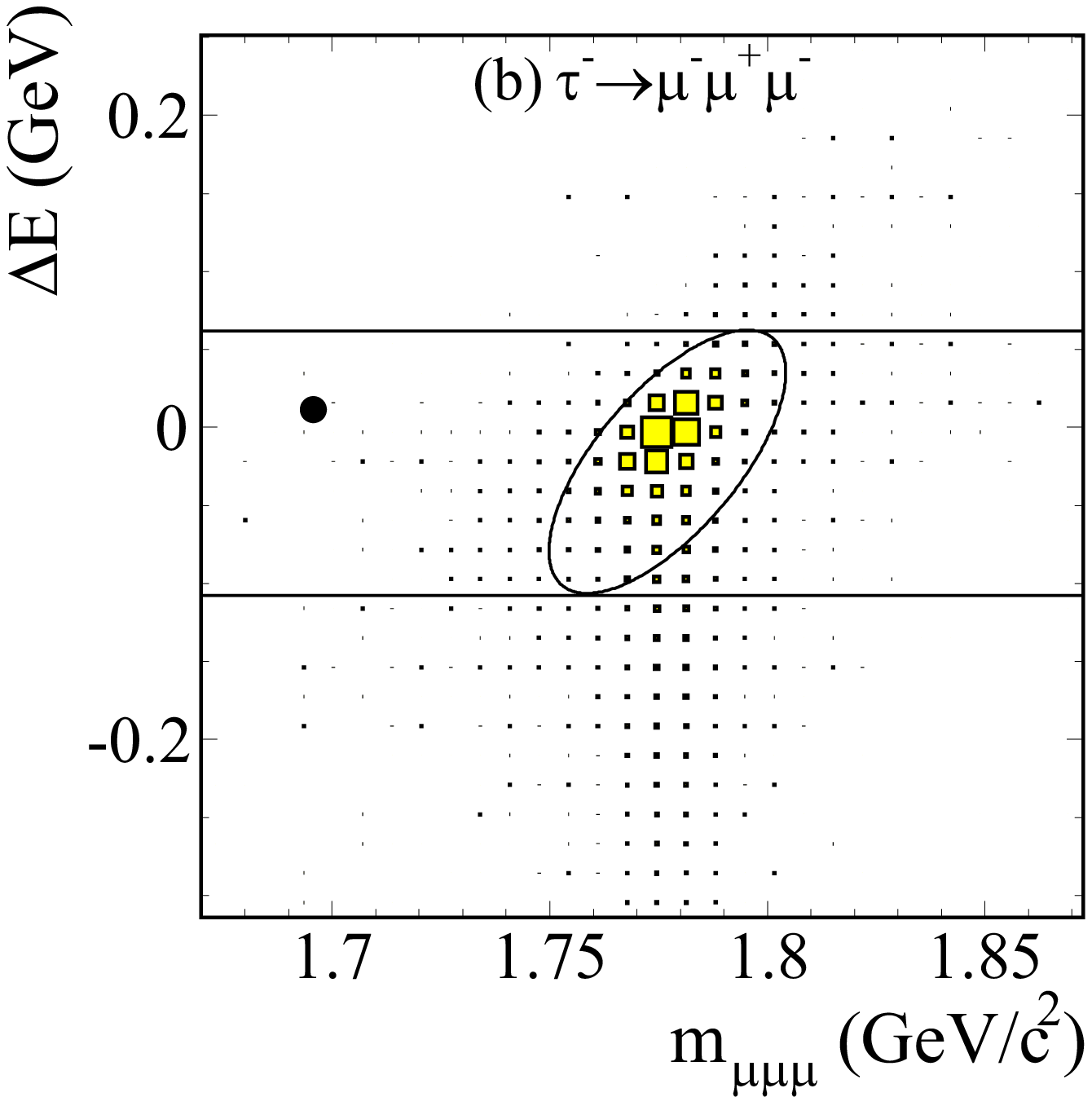}
\end{center}
\caption{  $M_{\ell\ell\ell}$--$\Delta E$  distributions for 
(a) $\tau^- \to e^-e^+e^-$ and 
(b) $\tau^- \to \mu^- \mu^+ \mu^-$ modes~\cite{tau:3l}.
The black dots and shaded boxes show 
the data and signal MC, respectively.
The ellipse is the signal region.
The region formed by the two parallel lines, excluding the signal 
ellipse region, is the side-band region used to evaluate 
the expected number of backgrounds in the signal region.}
\label{fig:TauLLL}
\end{figure}

The decays $\tau\rightarrow\ell\ell'\ell''$ have been searched for
with nearly the entire data sample of
$7.2\times10^8$ $\tau^+\tau^-$ pairs obtained by Belle~\cite{tau:3l}. 
Figures~\ref{fig:TauLLL}(a) and (b) show the three-lepton 
invariant mass versus $\Delta E$ ($M_{\ell\ell\ell}$--$\Delta E$) 
distributions for the $\tau^- \to e^-e^+e^-$ and
$\tau^- \to \mu^-\mu^+\mu^-$ candidates after selection, respectively.
 No events in the signal region have been found in any of the six modes;
the 90\% C.L. upper limits on the branching
fractions in units of $10^{-8}$ are given in Table~\ref{table:3l}.
The obtained upper limits are two or three times more restrictive
than those obtained previously~\cite{tau:3l_others}. 

\begin{table}[!ht]
\caption{Summary of the efficiency (Eff.),
the expected number of BG events  ($N_{BG}^{\rm exp}$),
and the upper limit on the branching fraction (UL) at 90\% C.L. for 
$\tau^-\rightarrow\ell^-\ell'^+\ell''^-$.}
\label{table:3l}
\footnotesize
\begin{tabular}[t]{lccc|lccc}
\hline
\hline
 Mode & Eff.(\%) &$N_{BG}^{\rm exp}$ & UL ($10^{-8}$)& 
 Mode & Eff.(\%) &$N_{BG}^{\rm exp}$ & UL ($10^{-8}$)\\
\hline
$e^-e^+e^-$ &6.0&$0.21\pm0.15$&2.7&   $e^-\mu^+\mu^-$ &6.1&$0.10\pm0.04$&2.7\\
$e^-e^+\mu^-$ &9.3&$0.04\pm0.04$&1.8& $\mu^-e^+\mu^-$ &10.1&$0.02\pm0.02$&1.7\\
$e^-\mu^+e^-$ &11.5&$0.01\pm0.01$&1.5& $\mu^-\mu^+\mu^-$&7.6&$0.13\pm0.06$&2.1\\
\hline
\hline
\end{tabular}

\end{table}

\medskip 

\noindent
\underline{$\tau\rightarrow\ell P^0$ $(P^0=\pi^0, \eta, \eta')$}

Early Belle results on the search for the $\tau$ decays
into a lepton and a neutral pseudoscalar $(\pi^0, \eta, \eta')$, were
based on a data sample of $3.6\times10^8$ $\tau^+\tau^-$ pairs; the 
resulting upper limits were in the range $(0.8-2.4)\times10^{-7}$
at 90\% C.L.~\cite{tau:leta}. Recently, we have updated the results with 
a data set two times larger. By studying the backgrounds in detail, 
we obtain on average about 1.5 times higher efficiency
than in our previous study while maintaining a background level 
in the signal region of less than one event in all modes.
The results are summarized in Table~\ref{table:lp0}.
A single event is found in 
$\tau\rightarrow e\eta(\rightarrow\gamma\gamma)$
while no events are observed in other modes.
The obtained 90\% C.L. ULs on the 
branching fraction are in the range $(2.2-4.4)\times10^{-8}$.

\begin{table}[!ht]
\caption{Summary of the efficiency (Eff.),
the expected number of BG events ($N_{BG}^{\rm exp}$),
and the upper limit on the branching fraction (UL) for 
$\tau\rightarrow\ell P^0$,
where (comb.) means the combined result from subdecay modes.}
\label{table:lp0}

\footnotesize
\begin{tabular}[t]{lccc|lccc}
\hline
\hline
 Mode & Eff.(\%) & $N_{BG}^{\rm exp}$ & UL ($10^{-8}$)& 
 Mode & Eff.(\%) & $N_{BG}^{\rm exp}$ & UL ($10^{-8}$)\\
\hline
$\mu\eta(\rightarrow\gamma\gamma)$ &8.2&$0.63\pm0.37$&3.6&
$e\eta(\rightarrow\gamma\gamma)$ &7.0&$0.66\pm0.38$&8.2\\
$\mu\eta(\rightarrow\pi\pi\pi^0)$ &6.9&$0.23\pm0.23$&8.6&
$e\eta(\rightarrow\pi\pi\pi^0)$ &6.3&$0.69\pm0.40$&8.1\\
\hline
$\mu\eta$(comb.) &&&2.3&
$e\eta$(comb.) &&&4.4\\
\hline
$\mu\eta'(\rightarrow\pi\pi\eta)$ &8.1&$0.00^{+0.16}_{-0.00}$
&10.0&
$e\eta'(\rightarrow\pi\pi\eta)$ &7.3&$0.63\pm0.45$&9.4\\
$\mu\eta'(\rightarrow\gamma\rho^0)$ &6.2&$0.59\pm0.41$&6.6&
$e\eta'(\rightarrow\gamma\rho^0)$ &7.5&$0.29\pm0.29$&6.8\\
\hline
$\mu\eta'$(comb.) &&&3.8&
$e\eta'$(comb.) &&&3.6\\
\hline
$\mu\pi^0$ &4.2&$0.64\pm0.32$&2.7&
$e\pi^0$ &4.7&$0.89\pm0.40$&2.2\\
\hline
\hline
\end{tabular}

\end{table}

\medskip

\begin{figure}[!ht]
\begin{center}
\centerline{\includegraphics[height=6.5cm]{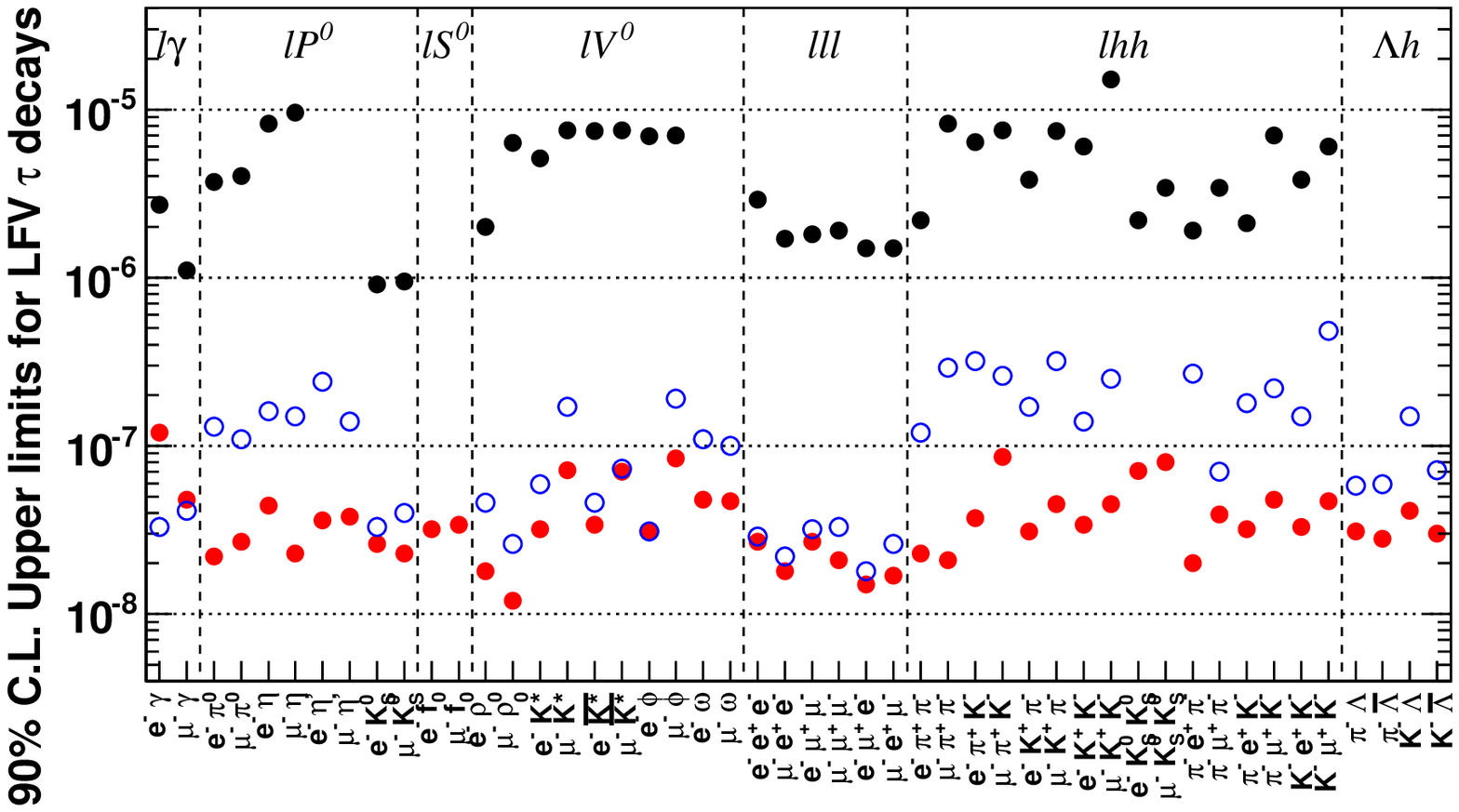}}
\caption{\label{uls}\small
Current 90\% C.L. upper limits for the 
branching fraction of $\tau$ LFV decays obtained in the CLEO, BaBar, and 
Belle experiments. Red, blue, and black
circles show Belle, BaBar, and CLEO results, respectively.}
\end{center}
\end{figure}

In total, Belle has completed searches for 46 lepton-flavor-violating
$\tau$ decay modes  
using nearly the entire data sample of 1000 fb${}^{-1}$ 
except for the ongoing analysis of the $\ell\gamma$ modes.
No evidence for LFV decays has been observed in any of the modes studied and
we set 90\% C.L. ULs on the branching fractions at the $O(10^{-8})$ level.
The current status of $\tau$ LFV searches in $B$-factory experiments 
and in the CLEO experiment
is summarized in Fig.~\ref{uls}.
The sensitivity for LFV searches has been improved by two orders of 
magnitude 
in comparison with the CLEO results. This is  due to the effective background
rejection as well as the increase in the size of the data sample.
In the near future, SuperKEKB/Belle II at KEK will reach a sensitivity at 
the $O(10^{-9}) \mbox{--} O(10^{-10})$ level and explore
a wider region of parameters in various NP scenarios.

\subsubsection{$CP$-violating $\tau$ decays}

To date $CPV$ has been observed only in the $K$ and $B$ meson systems. 
In the SM, all observed $CPV$ effects can be explained by a single irreducible 
complex phase in the CKM quark mixing matrix. It is important to look for
other $CP$-violating effects where SM $CPV$ is not expected in order to find NP.
One such system is the $\tau$ lepton.
In hadronic $\tau$ decays, no $CP$-violating effects from the SM are
expected except for cases in which the decay products contain $K_{S}^{0}$ mesons.
In other words, the $\tau$ decay is an ideal place to look for other
$CP$-violating effects that could originate from new physics scenarios,
such as  the minimal supersymmetric model~\cite{tau:CPVMSSM} or from 
multi-Higgs-doublet models~\cite{tau:CPVMHDM}.

If there is a $CP$-violating NP amplitude in a $\tau$ decay,  
interference between the SM and NP amplitudes should occur.
Even in this case, as was emphasized by 
J.H.~K\"uhn and E.~Mirkes~\cite{tau:Kuhn},
one  cannot observe the $CPV$ effects as a difference of the total 
decay rates between $\tau^-$ and $\tau^+$,  but instead one needs to measure 
the  difference between the
decay-angular distributions of the hadronic system for $\tau^-$ and $\tau^+$.
The analysis of the decay-angular distribution is therefore crucial.

We  searched for  $CP$ violation in $\tau^\pm \to K^0_S \pi^\pm \nu_\tau$ 
using a 699 fb$^{-1}$ data sample~\cite{tau:CPV}.
In order to search for $CP$ violation in the angular distribution, 
we define the $CP$ asymmetry observable as the difference between 
the mean value of the product of the decay angles 
$\cos\beta \cos\phi$ in the $K^0_S \pi^{\pm}$ system
 for $\tau^-$ and  $\tau^+$:
\[
A^{\rm CP} = \langle \cos \beta \cdot \cos \psi \rangle_{\tau^-} -
             \langle \cos \beta \cdot \cos \psi \rangle_{\tau^+},
\]
where $\beta$ ($\psi$) is the angle between the direction of the $K^0_S$  
($\tau$) and the direction of the  $e^+e^-$ CM system measured 
in the $K^0_S \pi^\pm$ rest frame.

\begin{figure}[htb]
\parbox{\halftext}{
\centerline{\includegraphics[width=6.5cm]{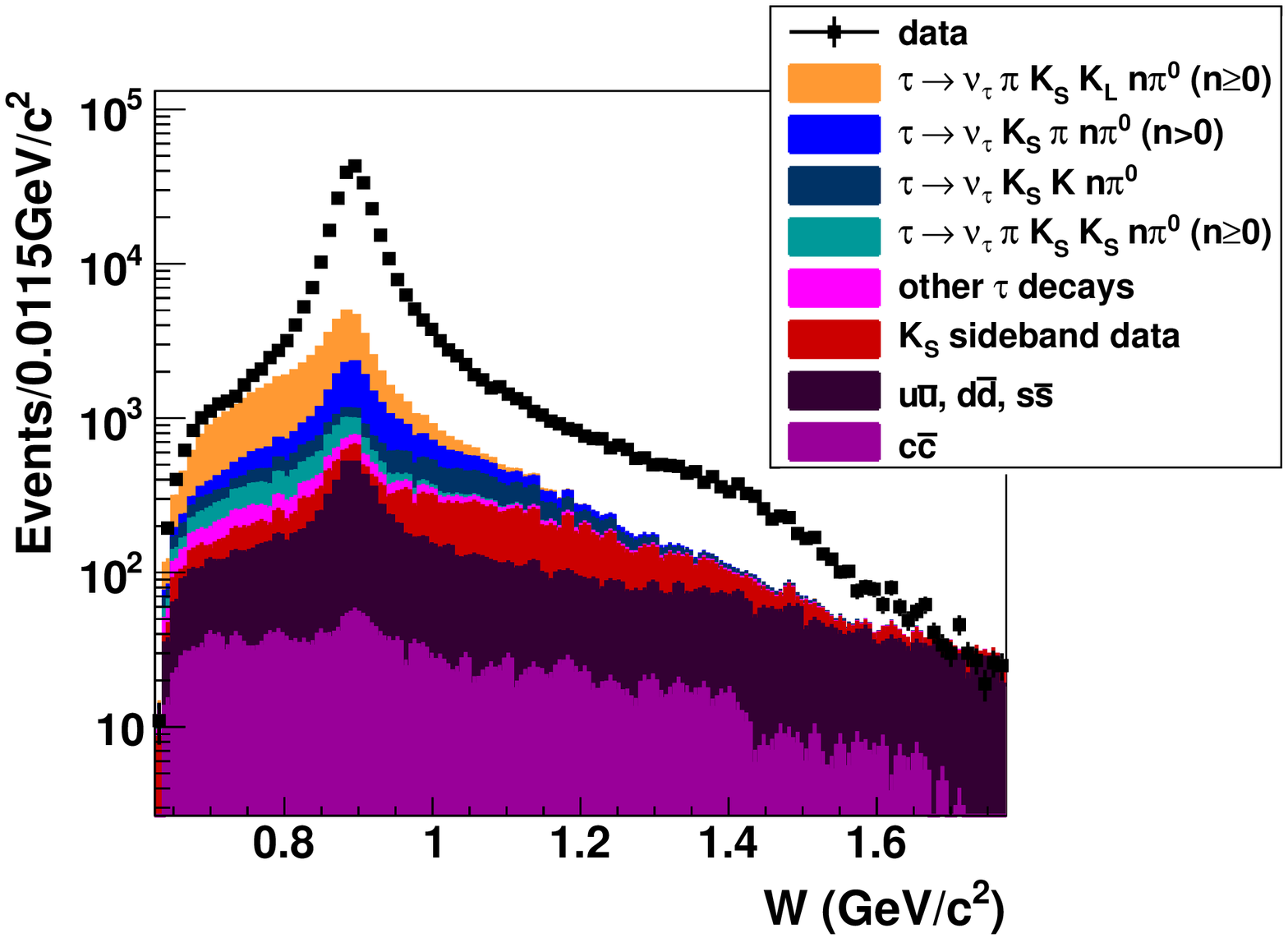}}
\caption{Invariant mass spectrum of the $K^0_S \pi^\pm$ system in 
$\tau \to K^0_S \pi^\pm \nu_{\tau}$ candidates~\cite{tau:CPV}.}
\label{fig:Kspi_1}
}
\hfill
\parbox{\halftext}{
\centerline{\includegraphics[width=6.5cm]{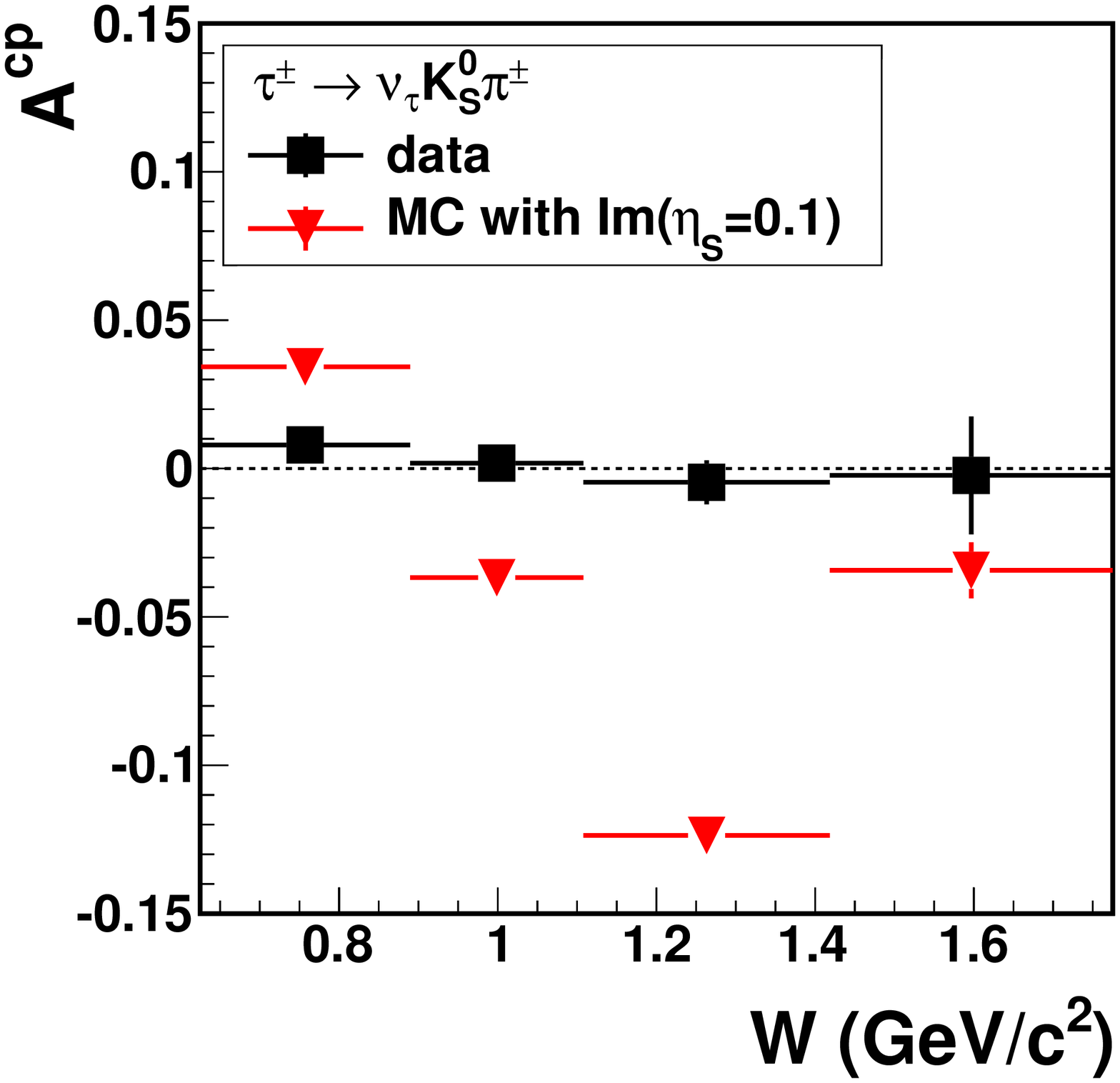}}
\caption{Measured $CP$-violating asymmetry $A^{\rm CP}$  as a function
of the $K^0_S \pi^\pm$ invariant
mass $W$ after subtraction of background (black squares)~\cite{tau:CPV}.
The inverted red triangles show the expected asymmetry
when $\Im (\eta_S )=0.1$.  Note that the previous CLEO result~\cite{tau:CLEO2002}
corresponds to $\Im(\eta_S) \le 0.19$.}
\label{fig:Kspi_2}}
\end{figure}

We obtain $3.2\times10^5$ $\tau^\pm \to K^0_S \pi^\pm \nu_\tau$ 
candidates. The $K^0_S \pi^\pm$ invariant mass distribution shown 
in Fig.~\ref{fig:Kspi_1} 
clearly indicates  that, in addition to the  $K^{*}(890)$ resonance, other 
resonant contributions are also needed to explain the full spectrum 
(see Sect.\ref{section:Kpi} for more details). 
The measured $CP$ asymmetry $A^{\rm CP}$ is shown in  Fig.~\ref{fig:Kspi_2} 
as a function of
$K^0_S \pi^\pm$ invariant mass after correcting for known detector effects.
The result indicates that there is no 
$CP$ asymmetry at the 1\% level.
Then we obtain the upper limit for the $CP$-violating scalar coupling constant 
$\eta_s$~\cite{tau:CPV}      
to be
\[
| {\rm Im}(\eta_s) | < (0.012 - 0.026),
\]
at the 90\% C.L. 
Our study achieved ten times higher sensitivity than the previous 
CLEO results shown by the inverted red triangles in Fig.~\ref{fig:Kspi_2}.

\subsubsection{Tau electric dipole moment}

If an elementary particle has a non-zero electric dipole moment (EDM), 
this is a clear indication of violation of the T-reversal symmetry 
and  thus  violation of $CP$ invariance according to the CPT theorem. 
The current limit for the $\tau$ EDM ($d_{\tau}$) 
is several orders of magnitude less restrictive than that for
the electron, muon, or  neutron. Measurement of $d_{\tau}$
is difficult  because of $\tau$'s short lifetime.
 However, improvements in sensitivity are interesting  both theoretically
and experimentally. As explained below, one can measure the $\tau$ EDM
 by using the correlation of decay product momenta
in the process $e^+e^- \to \tau^+\tau^-$.

The matrix element for the process $e^+e^- \to \tau^+\tau^-$,
is given by the sum of the SM term   ${\cal M}^2_{\rm SM}$, 
the EDM term $ |d_\tau|^2 {\cal M}^2_{d^2}$, and
the interference between them:
\[
{\cal M}^2 = {\cal M}^2_{\rm SM} + {\rm Re}(d_\tau) {\cal M}^2_{\rm Re}
+ {\rm Im}(d_\tau) {\cal M}^2_{\rm Im}+ |d_\tau|^2 {\cal M}^2_{d^2},
\]
where ${\rm Re}(d_\tau)$ ( ${\rm Im}(d_\tau)$ ) is the real (imaginary)  part
of the EDM. 
These interference terms  ${\cal M}^2_{\rm Re/Im}$ contain the following 
combination of spin-momentum
correlations:
\[
{\cal M}^2_{\rm Re} \propto (\bm{S}_+ \times \bm{S}_-)\cdot\hat{\bm{k}}, \quad 
                            (\bm{S}_+ \times \bm{S}_-)\cdot\hat{\bm{p}}, 
\]
\[
{\cal M}^2_{\rm Im} \propto (\bm{S}_+ - \bm{S}_-)\cdot\hat{\bm{k}}, \quad
                            (\bm{S}_+ - \bm{S}_-)\cdot\hat{\bm{p}}, 
\]
where $\bm{S}_\pm$ is a $\tau^\pm$ spin vector, and 
$\hat{\bm{k}}$ and $\hat{\bm{p}}$
are the unit vectors of the $\tau^-$ and  $e^-$  momenta in the CM 
system, respectively. 
These terms are $CP$-odd  since they change sign under a
$CP$ transformation.

One could evaluate the value of the matrix elements  if the values of  
 $\bm{S}_\pm$ and $\hat{\bm{k}}$ could be measured on an
 event-by-event basis from the $\tau$-decay products. Although
 one cannot know them completely due to
 missing neutrinos from $\tau$ decays,
 one can obtain the most probable values of $\bm{S}_\pm$ and $\hat{\bm{k}}$  
 by calculating approximate averages from 
 measurements of the momenta of $\tau$ decay products.
 In the analysis, we employ the method of optimal observables~\cite{tau:oo}.
\begin{wrapfigure}{r}{6.1cm}
  \centerline{\includegraphics[width=0.45\textwidth]
                                {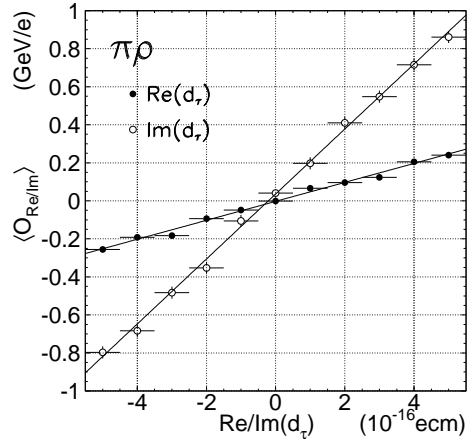}}
                              \caption{Correlation between the $\tau$ EDM 
and the optimal observable obtained by  MC simulation for 
$\tau^+\tau^- \to (\pi \nu_{\tau})(\rho \nu_{\tau})$~\cite{tau:edm}.
    Black dots and circles indicate the relations for the real and imaginary 
    parts, respectively.}
\label{fig:TauEDM_1}
\end{wrapfigure}
In this method, the observables 
${\cal O}_{\rm Re}$ and  ${\cal O}_{\rm Im}$ are 
\[
{\cal O}_{\rm Re} = \frac{{\cal M}^2_{\rm Re}}{{\cal M}^2_{\rm SM}},~~~
{\cal O}_{\rm Im} = \frac{{\cal M}^2_{\rm Im}}{{\cal M}^2_{\rm SM}},
\]
evaluated using the most probable values of $\bm{S}_\pm$ and $\hat{\bm{k}}$. 
The means of ${\cal O}_{\rm Re}$, ${\cal O}_{\rm Im}$ are proportional
to the EDM value and have maximal sensitivity.
The relation between the mean values and the EDM $d^{\tau}$
is shown in Fig.~\ref{fig:TauEDM_1}
for the $\tau^+\tau^- \to (\pi \nu_{\tau})(\rho \nu_{\tau})$ mode.

\begin{figure}
  \centerline{
    \includegraphics[height=5cm]{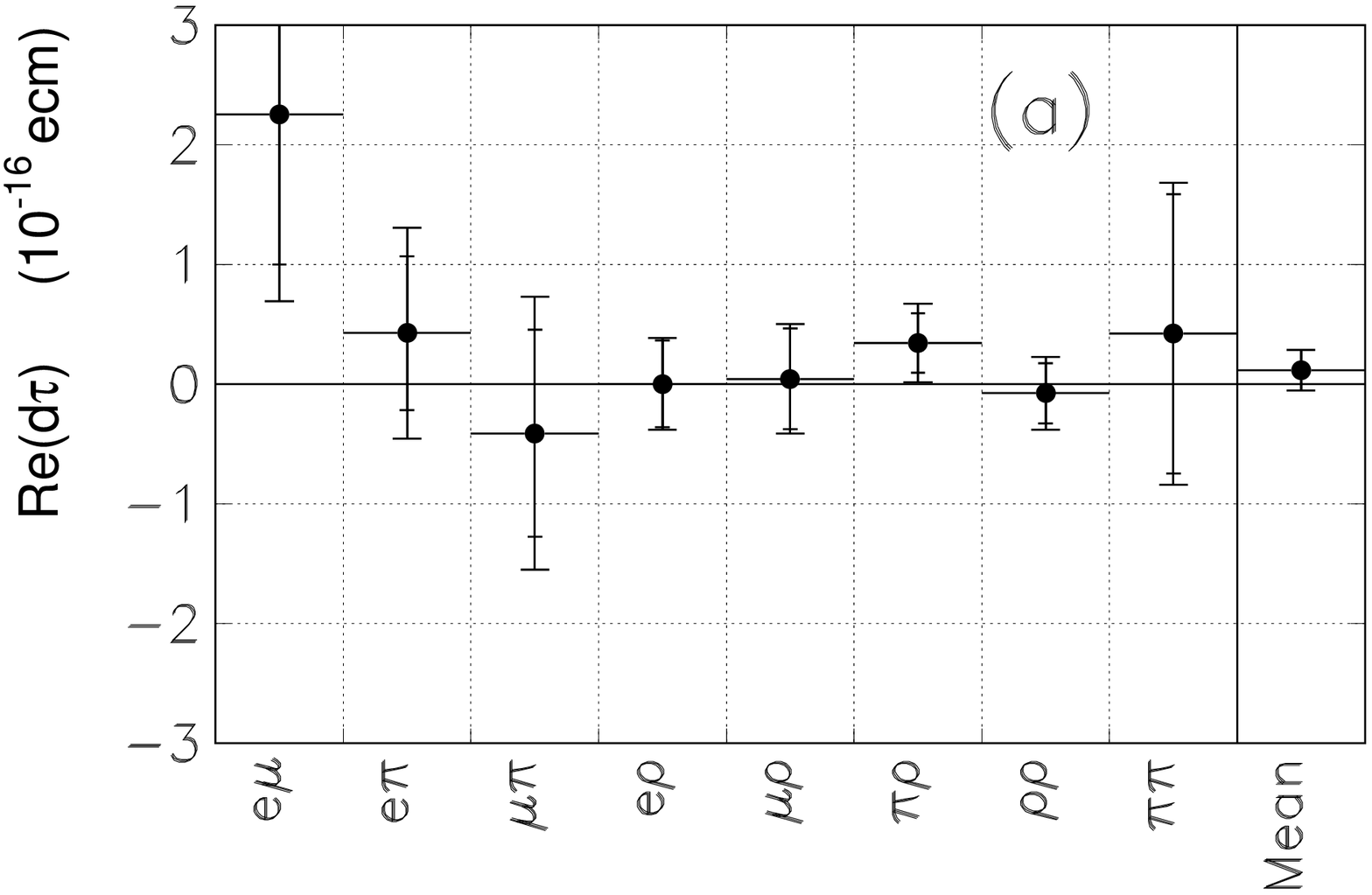}
    \includegraphics[height=5cm]{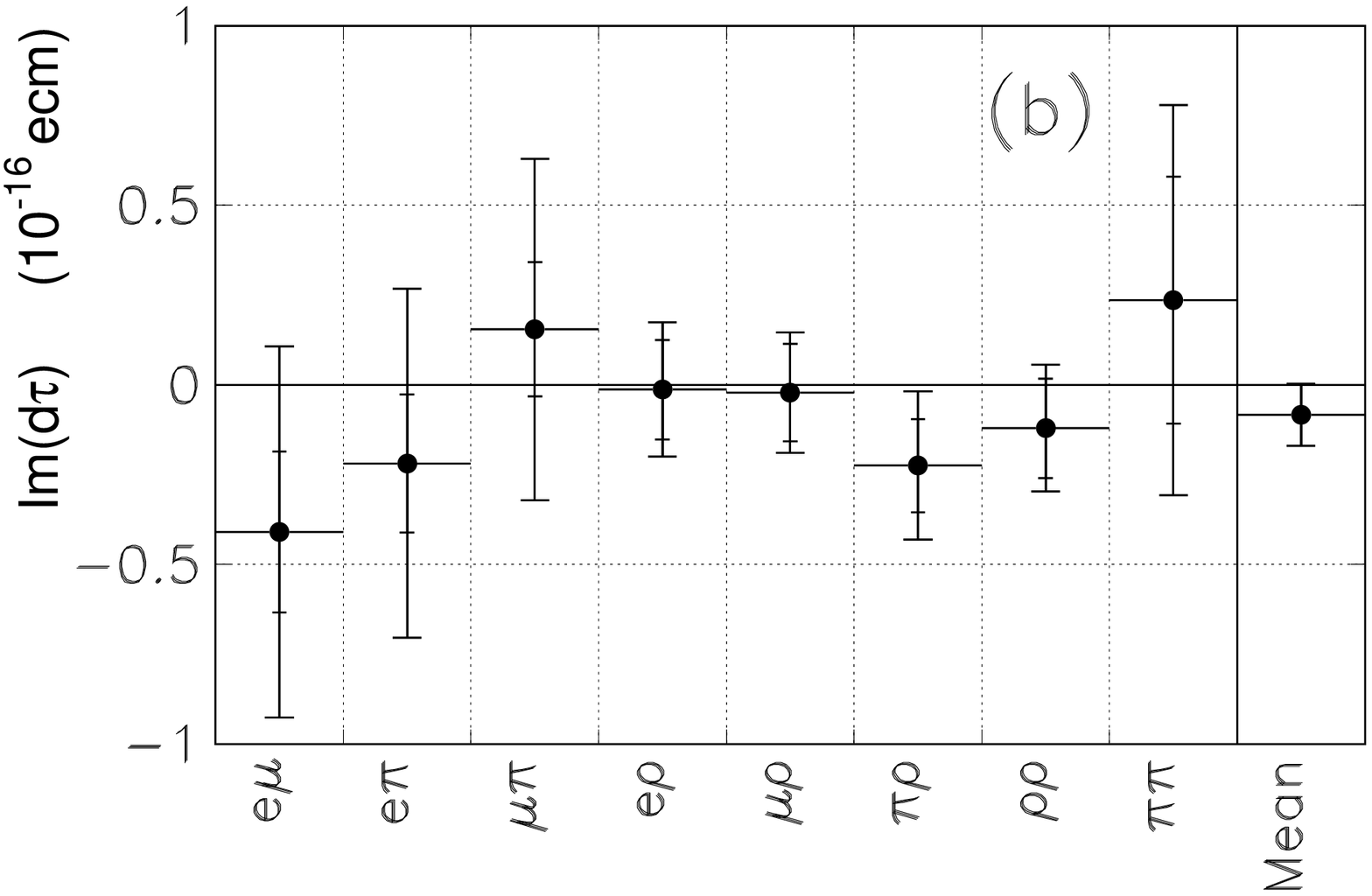}
  }
  \caption{Results on the $\tau$ EDM for 8 modes and the weighted mean
  for the (a) real and (b) imaginary parts.}
  \label{fig:TauEDM_fig6}
\end{figure}

We carried out the EDM analysis with a 29.5 fb$^{-1}$ data sample collected by
the Belle detector~\cite{tau:edm}.
In order to obtain the maximal sensitivity, we measured the EDM in 8 modes, 
$\tau^+\tau^- \to (e\nu_{e}\nu_{\tau})(\mu\nu_{\mu}\nu_{\tau})$,
$(e\nu_{e}\nu_{\tau})(\pi\nu_{\tau})$, 
$(\mu\nu_{\mu}\nu_{\tau})(\pi\nu_{\tau})$, $(e\nu_{e}\nu_{\tau})(\rho\nu_{\tau})$,
$(\mu\nu_{\mu}\nu_{\tau})(\rho\nu_{\tau})$, 
$(\pi\nu_{\tau})(\pi\nu_{\tau})$, $(\pi\nu_{\tau})(\rho\nu_{\tau})$, and
$(\rho\nu_{\tau})(\rho\nu_{\tau})$.
The values of EDM obtained from the mean values of the optimal observables 
are shown in Fig.~\ref{fig:TauEDM_fig6}. 
All results are consistent with zero EDM.

We obtain mean values for  ${\rm Re}(d_\tau)$ and   ${\rm Im}(d_\tau)$
by taking  the weighted mean of 8 modes  to be
\[
{\rm Re}(d_\tau) = (1.15 \pm 1.70) \times 10^{-17} e{\rm cm},~~~
{\rm Im}(d_\tau) = (-0.83 \pm 0.86) \times 10^{-17} e{\rm cm}.
\]
The  95\% C.L. intervals are 
\[
-2.2 \times 10^{-17} < {\rm Re}(d_\tau) < 4.5 \times 10^{-17} e{\rm cm},~~~
-2.5 \times 10^{-17} < {\rm Im}(d_\tau) < 0.8 \times 10^{-17} e{\rm cm}.
\]
These limits are ten times more restrictive  than previous experiments.

\subsection{Precision measurements}

\subsubsection{Tau lepton mass}

A precise measurement of the $\tau$ lepton mass is very important to test 
electroweak theory and lepton universality, 
since the decay width is proportional to the $\tau$ lepton mass to the fifth
power. For a long time the world average for the tau mass was dominated by a
single precise measurement carried out at the 
$e^+e^- $ threshold  by the BES experiment in 1996.
 
\begin{figure}[htb]
\parbox{\halftext}{
\centerline{\includegraphics[width=6.5cm]{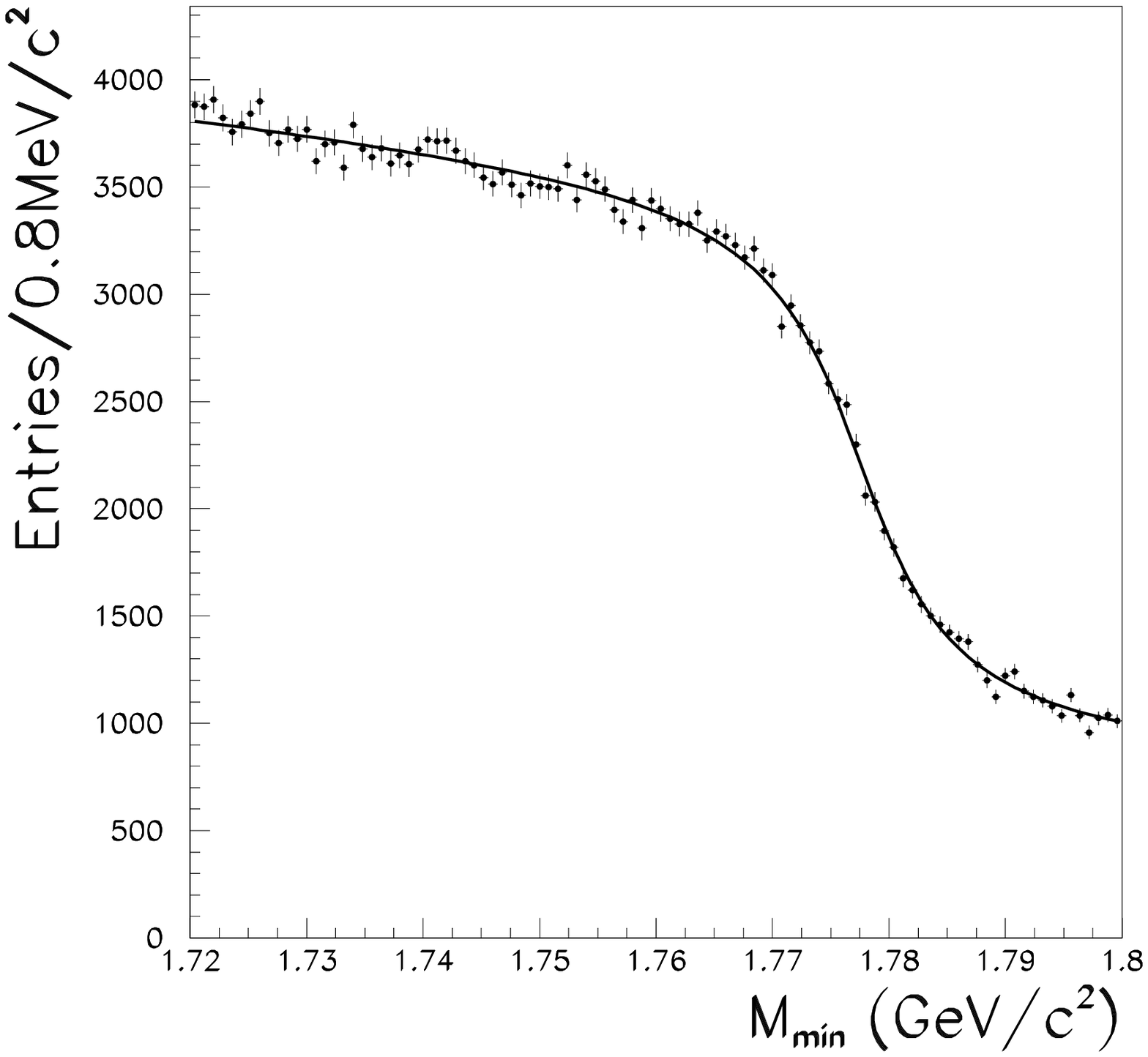}}
\caption{Pseudomass distribution $M_{\rm min}$ for 
    $\tau^\pm \to 3 \pi^\pm \nu_{\tau}$ candidates\cite{tau:mass}.}
\label{fig:TauMass_1}
}
\hfill
\parbox{\halftext}{
\centerline{\includegraphics[width=6.5cm]{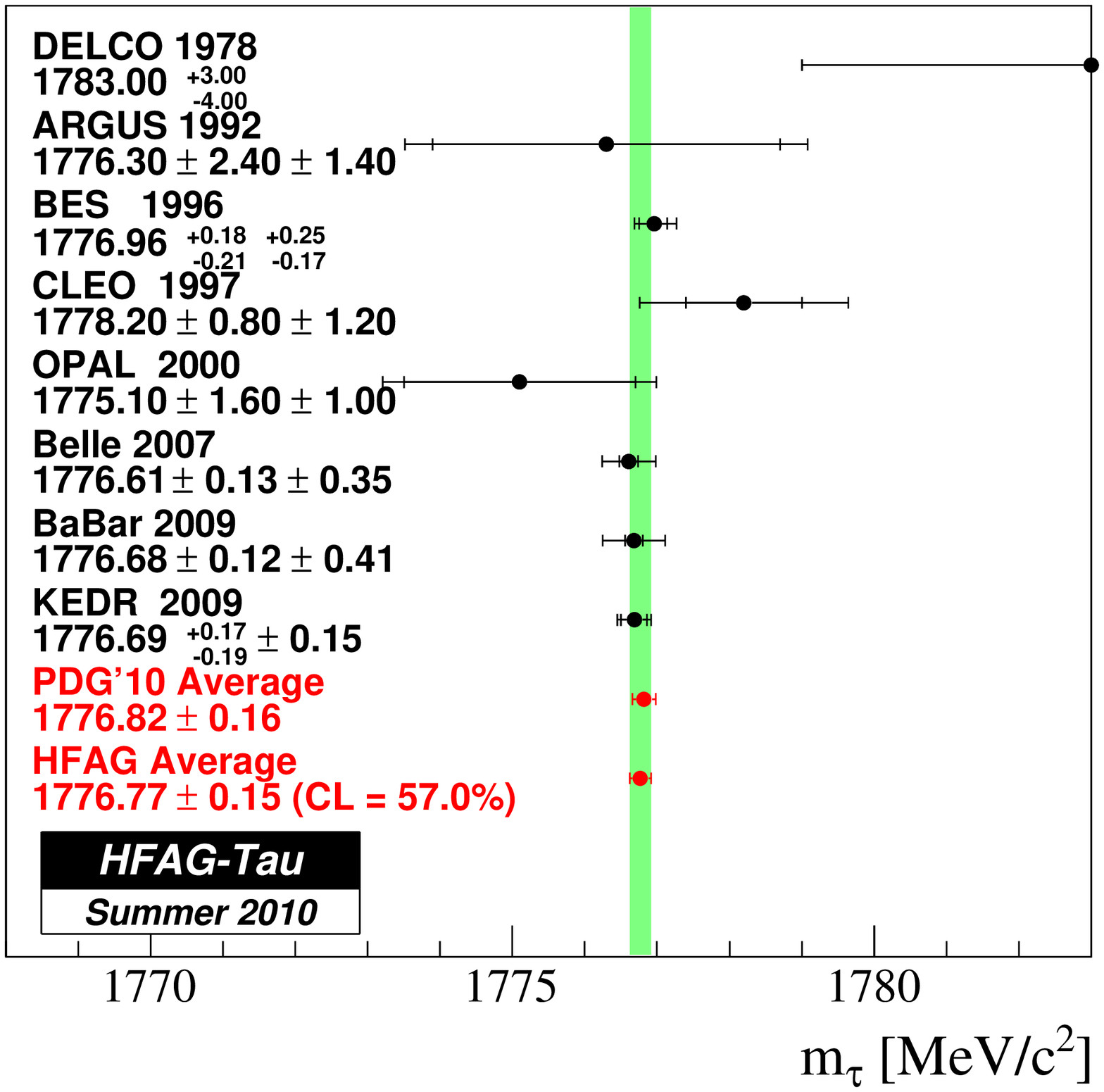}}
\caption{Summary of the $\tau$ mass measurements.}
\label{fig:TauMass_2}}
\end{figure}
 
Belle measured~\cite{tau:mass} the $\tau$ mass by using a pseudomass 
method and showed that a precision similar to that obtained
in the threshold region can be obtained with completely 
different systematic errors.
Figure~\ref{fig:TauMass_1} shows the pseudomass distribution 
obtained by  Belle, where
a few hundred thousand 
$\tau^{\pm}\to \pi^{\pm}\pi^-\pi^+ \nu_{\tau}$ events are used. 
The pseudomass is defined by
$$
  M_{\rm min}= \sqrt{ M_{x}^2 + 2 (E_{\rm beam} - E_{x}) (E_{x} - P_{x})},
$$
where $M_{x}, P_{x}, E_{x}$ are the mass, absolute momentum, and energy  
of the $3\pi$ system, respectively. We obtain
$$
  m_{\tau} = (1776.61 \pm 0.13 ({\rm stat}) 
\pm 0.35 ({\rm syst}) {\rm MeV/}c^{2}
 $$ 
for the $\tau$ mass and
 $$
   |m_{\tau^{+}} - m_{\tau^{-}}|/m_{\tau} < 2.8 
\times 10 ^{-4} \quad {\rm at\ 90\%\  C.L.},
 $$
 the most stringent limit for the relative mass difference between 
positive and negative $\tau$ leptons.

Measurements with a similar precision were subsequently carried out
by BaBar
with the pseudomass method~\cite{tau:babarmass} and by KEDR with the 
threshold-scan method~\cite{tau:kedrmass}. 
The current status of $\tau$ mass measurements is summarized in 
Fig.~\ref{fig:TauMass_2}. 

\subsubsection{Spectral function in 
$\tau^{\pm}\to \pi^{\pm}\pi^{0}\nu_{\tau} $ decay}

Among the decay channels of the $\tau$ lepton,
$\tau^{\pm}\to \pi^{\pm}\pi^{0}\nu_{\tau} $ has the largest branching fraction.
From the conservation of vector current (CVC), 
the $\pi^-\pi^0$ mass
spectrum can be related to the cross section for the process 
$e^+e^-\to \pi^+\pi^-$
and thus can be used to improve the theoretical error on the anomalous magnetic
moment of the muon $a_{\mu} = (g_{\mu} -2 )/2$.

\begin{figure}[htb]
\parbox{\halftext}{
\centerline{\includegraphics[width=6.5cm]{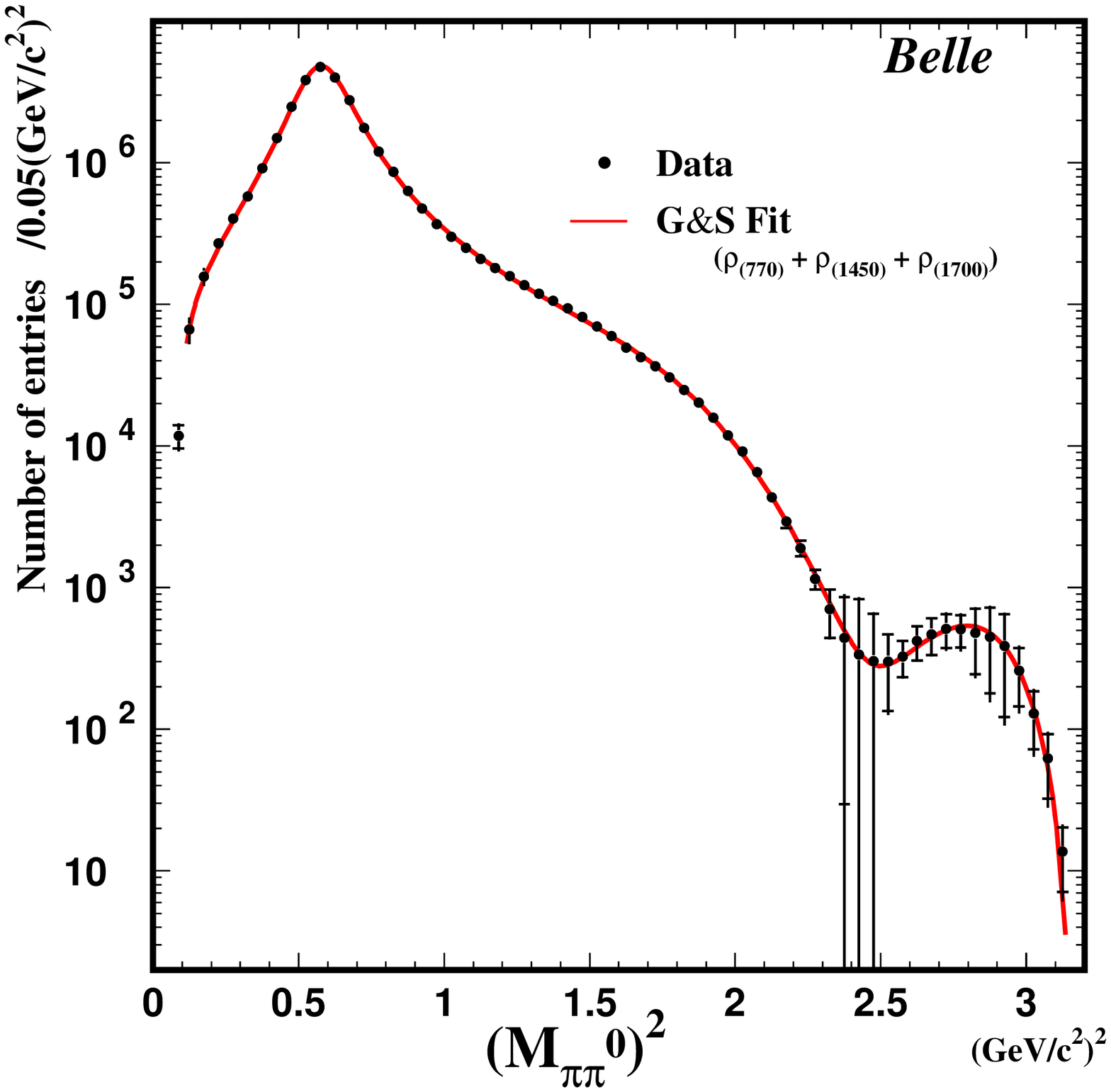}}
\caption{Unfolded $\pi^{\pm}\pi^{0}$ mass spectrum for 
 $\tau^{\pm}\to \pi^{\pm}\pi^{0}\nu_{\tau} $. Solid circles are the data
 and the solid curve is a fit. The error bars include both statistical 
and systematic uncertainties\cite{tau:2pi}.}
\label{fig:Tau2pi_1}
}
\hfill
\parbox{\halftext}{
\centerline{\includegraphics[width=6.5cm]{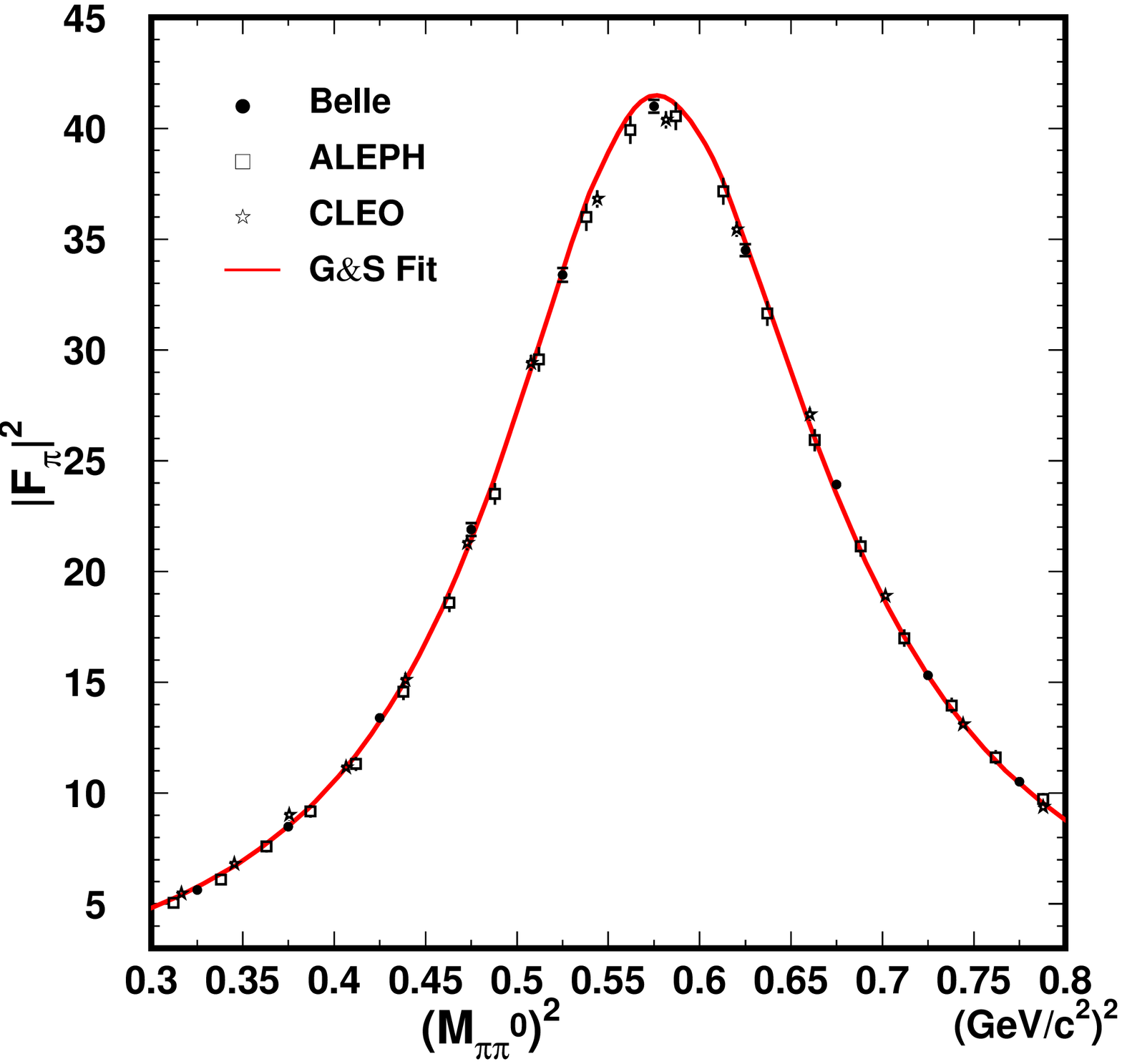}}
\caption{Pion form factor $|F_{\pi}(s)|^{2}$ in the $\rho(770)$
region extracted from the mass spectra in Fig.~\ref{fig:Tau2pi_1}.}
\label{fig:Tau2pi_2}}
\end{figure}

Using a sample of 5430000 $\tau^{\pm}\to \pi^{\pm}\pi^{0}\nu_{\tau} $,
Belle measures the branching fraction and the $\pi\pi^{0}$ mass 
spectrum~\cite{tau:2pi},
which are important for  obtaining the theoretical value of $g_{\mu}-2$.
After unfolding using the singular-value-decomposition 
method~\cite{Hocker:1995kb}, the $\pi\pi^{0}$ mass spectrum obtained is 
 shown in Fig.~\ref{fig:Tau2pi_1},  where  the shapes for 
 $\rho(770)$, $\rho(1450)$, and $\rho(1700)$ resonances and their 
interference pattern  
 are measured very precisely.  Figure~\ref{fig:Tau2pi_2} is the 
pion form factor
 in the $\rho(770)$ region obtained from the mass spectra 
in Fig.~\ref{fig:Tau2pi_1}. 
 The measured branching fraction is
 $$
 {\cal B}( \tau^{\pm}\to \pi^{\pm}\pi^{0}\nu_{\tau})= (25.24\pm 0.04({\rm stat}) \pm 0.40({\rm syst})) \%.
 $$
 
It is known that there is a significant difference in the
value of $a_{\mu}^{2\pi}$ obtained from $e^+e^-$ and $\tau$ data.
A lengthy discussion is ongoing about a possible source of this
difference\cite{tau:davier2010}.
Belle $\tau$ data are in very good agreement with the recent measurement 
of the $e^+e^- \to \pi^+\pi^-$ cross section from initial-state 
radiation (ISR) data by BaBar\cite{tau:ISRbabar}.
In addition, it has been pointed by 
F.~Jegerlehner~\cite{tau:Jegerlehner1,tau:Jegerlehner2}
that $\gamma-\rho$ interference, which is present only in $e^+e^-$ 
and does not contribute in the $\tau$ decay, plays an important role.
After taking this interference
effect into account, the discrepancy between the $e^+e^-$ and
$\tau$ data disappears; i.e.
the hadronic term  of $(g_{\mu}- 2)$ from the $e^+e^-$ data 
is $a_{\mu}^{\rm had} [e] = 690.8(4.7) \times 10^{-10}$,  while
including the $\tau$ data it becomes
$a_{\mu}^{\rm had} [e,\; \tau] = 691.0(4.7) \times 10^{-10}$\cite{tau:Jegerlehner1}.
Note that without the $\rho-\gamma$ interference correction,
$a_{\mu}^{\rm had} [e,\;\tau]$ was
$a_{\mu}^{\rm had} [e,\;\tau]= 696.6(4.7) \times 10^{-10}$.

The resulting difference   between theory and  
experiment for $a_{\mu}$
is greater than $3\sigma$, which strengthens the difference further.
Recently there have been efforts to evaluate $a_{\mu}^{\rm had}$
in lattice QCD~\cite{Aubin:2006xv,Boyle:2011hu}. The reported values
scatter in the range from $a_{\mu}^{\rm had}=641\times
10^{-10}$ to $748\times 10^{-10} $ with an error of $(30-64)\times 10^{-10}$.
The error is one order of magnitude larger than that obtained so far from
$e^+e^-$ and/or $\tau$ data.

\subsubsection{Observation of decays with three kaons}
\label{section:Kpi}

Using a data sample of 401~fb$^{-1}$ corresponding to 
$3.6 \times 10^8~\tau^+\tau^-$ pairs, Belle reported the first observation
of decays with three charged kaons in the final state~\cite{tau:3K}.
We select events in which a $K^+K^-$ pair comes from the
$\phi$ meson and, after taking into account a serious peaking background 
from $\tau^- \to \phi \pi^- \nu_\tau$, report the
branching fraction, ${\cal B}(\tau^- \to \phi K^- \nu_\tau)=
(4.05 \pm 0.25 \pm 0.26) \times 10^{-5}$.
In addition, we observe  $\tau^- \to \phi \pi^- \nu_\tau$ and 
 $\tau^- \to \phi \pi^- \pi^0 \nu_\tau$ decays, which is a serious
peaking background for the three kaon process. Later BaBar confirmed
the existence of this decay with a branching fraction consistent with 
ours~\cite{tau:babar3h}. 
 
\subsubsection{Study of  $\tau^-  \to K_S \pi^- \nu_\tau$}

A data sample of 351~fb$^{-1}$ has been used to study the 
$K_S \pi^-\nu_\tau$ final state~\cite{tau:Kpi}. As a result of the
analysis, 53110 lepton-tagged signal events have been selected.
The measured branching fraction, 
${\cal B}(\tau^-  \to K_S \pi^- \nu_\tau)=(0.404 \pm 0.002 \pm 0.013)\%$, 
is the most precise of all the published measurements. Although the Belle 
result is consistent with the other results within errors, the
central value is somewhat lower than all of them. 
\begin{wrapfigure}{r}{6.1cm}
  \centerline{\includegraphics[width=0.45\textwidth]
                                {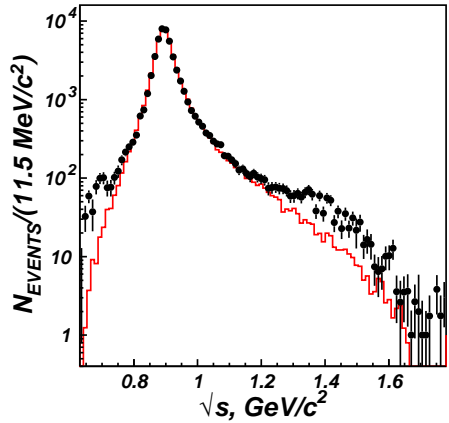}}
\caption{
The $K_S\pi$ mass distribution. Points with
error bars are data while the histogram shows the fitted result for the 
spectrum expected in a model incorporating only 
$K^*(892)$. The background is already subtracted.}
\label{fig:TauKpi}
\end{wrapfigure}
An analysis of the $K_S\pi^-$ invariant mass
spectrum shown in Fig.~\ref{fig:TauKpi} reveals the dominant
contribution from $K^*(892)^-$ with additional contributions of higher
states at 1400~MeV. A satisfactory fit is obtained only if the
existence of a broad scalar state, $K^*_0(800)$, is assumed. For the
first time the  $K^*(892)^-$ mass and width have been measured in 
$\tau$ decay: $M = (895.47 \pm 0.20 \pm 0.44 \pm 0.59)$~MeV, 
$\Gamma=(46.2 \pm 0.6 \pm 1.0 \pm 0.7)$~MeV, 
where the third uncertainty is from the model. The  $K^*(892)^-$ mass
is significantly higher than the world average value based 
on various hadronic experiments and is much closer to the world
average for the neutral $K^*(892)$. 

\subsubsection{Measurement of hadronic $\tau$ decays 
with an $\eta$ meson}

Using a data sample of 490~fb$^{-1}$ we have studied hadronic 
$\tau$  decay modes involving an $\eta$ meson. Candidate
 $\eta$ mesons are reconstructed from their decays 
into $\gamma\gamma$ and $\pi^+\pi^-\pi^0$~\cite{tau:etainc}.
Table~\ref{tau:tabeta}
lists the measured branching fractions or the  upper limits. 
In all cases the number of observed events is significantly higher 
and the results are more precise than previous 
measurements by CLEO~\cite{tau:cl1,tau:cl2,tau:cl3} and 
ALEPH~\cite{tau:aleph}. For the $K^-\eta\eta\nu_\tau$ decay mode, our result
is the first measurement. For $\pi^-\pi^0\eta\nu_\tau$ the invariant 
mass spectrum and the branching fraction are consistent with a prediction
based on the conserved vector current (CVC) theorem~\cite{tau:ei91}.

\begin{table}
\begin{center}
\caption{The branching fractions of various decay modes with 
an $\eta$ meson. The upper limits are at the 90\% C.L.}
\label{tau:tabeta}
\begin{tabular}{lc}
\hline\hline
Decay mode & ${\cal B}$ \\
\hline
$K^-\eta\nu_\tau, 10^{-4}$ & $1.58 \pm 0.05 \pm 0.09$ \\
$\pi^-\pi^0\eta\nu_\tau,  10^{-3}$ & $1.35 \pm 0.03 \pm 0.07$ \\
$K^-\pi^0\eta\nu_\tau, 10^{-5}$ & $4.6 \pm 1.1 \pm 0.4$ \\
$K^0_S\pi^-\eta\nu_\tau,  10^{-5}$ & $4.4 \pm 0.7 \pm 0.3$ \\
$K^{*-}\eta\nu_\tau, 10^{-4}$ &  $1.34 \pm 0.12 \pm 0.09$ \\
$K^-K^0_S\eta\nu_\tau,  10^{-6}$ & $<4.5  $ \\
$K^0_S\pi^-\pi^0\eta\nu_\tau,  10^{-5}$ & $<2.5 $ \\
$K^-\eta\eta\nu_\tau,  10^{-6}$ & $<3.0$ \\
$\pi^-\eta\eta\nu_\tau, 10^{-6}$ & $<7.4$ \\
$(K^-\pi^0\eta\nu_\tau)_{\rm nonres}$, $10^{-5}$ & $<3.5$ \\
\hline\hline
\end{tabular}
\end{center}
\end{table} 

\subsubsection{Decays with three hadrons in the final state}

With a data sample of 666~fb$^{-1}$ Belle has also studied various
decay modes of the $\tau$ lepton  with three hadrons in the final 
state~\cite{tau:3h}.
The results of this analysis for the branching fractions of various
three-prong modes are listed in Table~\ref{tau:tabcompare} together with
recent results from BaBar~\cite{tau:babar3h}. 
Note that, 
for the $\pi^-\pi^+\pi^-\nu_\tau$ and $K^-\pi^+\pi^-\nu_\tau$ modes, 
the branching fractions listed do not include any $K^0$ contribution,  
while the result for $K^-K^+K^-\nu_\tau$ includes $\phi K^-\nu_\tau$.

\begin{table}[htp]
\begin{center}
\caption{
Comparison of the branching fractions of three hadron decay modes 
from Belle and BaBar. 
}
\begin{tabular}{lcc}
\hline\hline
Decay mode & BaBar & Belle \\
\hline
 $\pi^-\pi^+\pi^-\nu_\tau$, \%   & $8.83 \pm 0.01 \pm 0.13$      & 
$8.42 \pm 0.00 ^{+0.26} _{-0.25}$       \\ 
 $K^-\pi^+\pi^-\nu_\tau$, \%    & $0.273 \pm 0.002 \pm 0.009$   & 
$0.330 \pm 0.001 ^{+0.016} _{-0.017}$           \\ 
$K^-K^+\pi^-\nu_\tau$, \%     & $0.1346 \pm 0.0010 \pm 0.0036$ & 
$0.155 \pm 0.001 ^{+0.006} _{-0.005}$           \\ 
$K^-K^+K^-\nu_\tau$, $10^{-5}$ & $1.58 \pm 0.13 \pm 0.12$    & 
$3.29 \pm 0.17 ^{+0.19} _{-0.20}$               \\ 
\hline\hline
\end{tabular}
\label{tau:tabcompare}
\end{center}
\end{table}

\begin{figure}[!ht]
\begin{center}
\includegraphics[height=8.5cm]{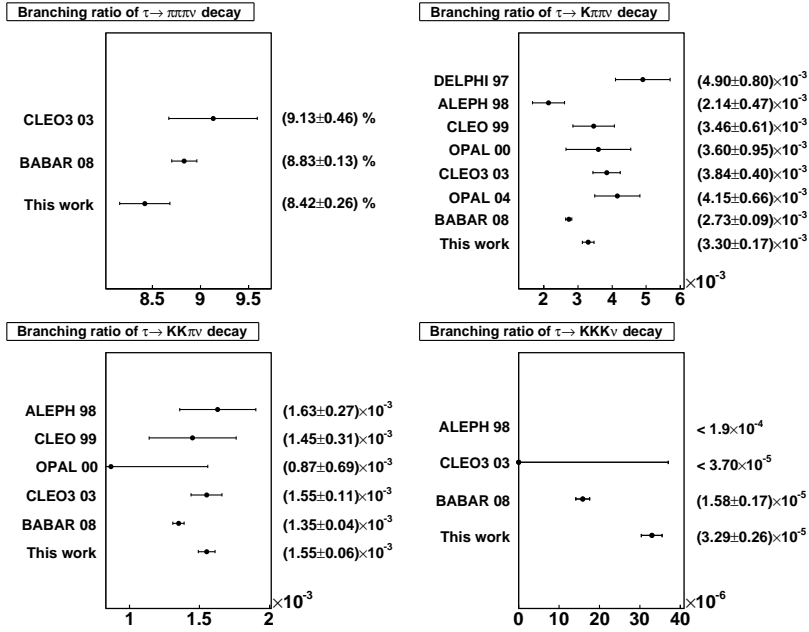}
\end{center}
\caption{Summary of the branching fraction measurements
for three-prong $\tau$ decays.}
\label{fig:tauhistory3h}
\end{figure}

In Fig.~\ref{fig:tauhistory3h}, our results are compared with 
the previous measurements. For all modes studied, the precision of 
the branching fraction measurements
for both BaBar~\cite{tau:babar3h} and Belle is significantly better 
than previous results.
The accuracy of our results is comparable to that of BaBar, but the 
central values show striking differences in all channels other than 
$\pi^-\pi^+\pi^-\nu_\tau$. For this mode, our result is $1.4\sigma$
lower than that of BaBar, while for the other modes the branching fractions
obtained by Belle are higher by $3.0\sigma$, $3.0\sigma$, and 
$5.4\sigma$ than those of BaBar for the $K^-\pi^+\pi^-\nu_\tau$, 
 $K^-K^+\pi^-\nu_\tau$, and $K^-K^+K^-\nu_\tau$ modes, respectively.

\subsubsection{Summary of precise measurements}

These measurements, as well as additional measurements of missing modes, 
are very important for obtaining separately the inclusive branching 
fractions of vector, axial-vector, and strange decay modes and corresponding
spectral functions. 

For the rare decay modes with branching fractions of less than $10^{-2}$,
there is a significant improvement compared to the previous experiments.

\newcommand{\dzero}{D^0}
\newcommand{\dzerobar}{\overline{D}{}^0}
\newcommand{\zdmix}{$D^0$\---$\overline{D}{}^0$~}
\newcommand{\taudzero}{\tau_{D^0}}
\newcommand*{\bfrac}[2]{\genfrac{}{}{0pt}{}{#1}{#2}}
\section{$D^0$ mixing and $CPV$}
\label{chap_charm}
The neutral $D$ meson system is one of the four flavored neutral 
particle\---antiparticle systems that can exhibit oscillations. 
Particle\---antiparticle mixing causes an initial (at time $t=0$) 
pure $D^0$ meson state to evolve in time to a linear
combination of $D^0$ and $\overline{D}{}^0$ states:
\begin{equation}
|D^0(t)\rangle=\left[ |D^0\rangle \cosh\left(\frac{ix+y}{2}\Gamma t\right)+\frac{q}{p}|\bar{D}^0\rangle \sinh\left(\frac{ix+y}{2}\Gamma t\right)\right]
\times e^{-(i m-\frac{\Gamma}{2})t},
\end{equation}
where the two parameters that describe the $D^0$ \--- $\overline{D}^0$ mixing, $x$ and $y$, are defined as
the mass and width difference of the two mass eigenstates $|D_{1,2}\rangle = p|D^0\rangle\pm q|\overline{D}{}^0\rangle$:
\begin{equation}
x =  \frac{m_1-m_2}{\Gamma},~~~~~y = \frac{\Gamma_1-\Gamma_2}{2\Gamma},~~~~~\Gamma = \frac{\Gamma_1+\Gamma_2}{2},
\end{equation}
and $\Gamma$ is the mean decay width. The coefficients $p$ and $q$
are complex, satisfying the normalization condition $|p|^2+|q|^2=1$.
The time-dependent decay rates for $\dzero\to f$ (favored) and $\dzero\to\overline{f}$ (suppressed) decays are given by:
\begin{eqnarray}
 \Gamma(D^0(t)\to f) & = & |{\cal A}_f|^2e^{-\Gamma t}\left( \frac{1+|\lambda_f|^2}{2}\cosh(y\Gamma t)-\Re[\lambda_f]\sinh(y\Gamma t)\right.\nonumber\\
& & \left. +\frac{1-|\lambda_f|^2}{2}\cos(x\Gamma t)+\Im[\lambda_f]\sin(x\Gamma t)\right),\label{eq_th_dofdr}\\
 \Gamma(D^0(t)\to \overline{f}) & = & |\overline{\cal A}_{\overline{f}}|^2\left|\frac{q}{p}\right|^2e^{-\Gamma t}\left( \frac{1+|\lambda_{\overline{f}}^{-1}|^2}{2}\cosh(y\Gamma t)-\Re[\lambda_{\overline{f}}^{-1}]\sinh(y\Gamma t)\right.\nonumber\\
& & \left. -\frac{1-|\lambda_{\overline{f}}^{-1}|^2}{2}\cos(x\Gamma t)-\Im[\lambda_{\overline{f}}^{-1}]\sin(x\Gamma t)\right),\label{eq_th_dofbdr}
\end{eqnarray}
where $\lambda_f = \frac{q}{p}\frac{\overline{\cal A}_f}{{\cal A}_f}$ and $\lambda_{\overline{f}}\equiv\frac{q}{p}\frac{\overline{\cal A}_{\overline{f}}}{{\cal A}_{\overline{f}}}$.
The time evolution of neutral $D$ meson decays is exponential with lifetime $\tau_{\dzero}=1/\Gamma$, 
modulated by the mixing parameters $x$ and $y$ (see the expressions
above). 
Time-dependent measurements of $\dzero$ 
and $\dzerobar$ decays thus enable us to measure the mixing parameters $x$ and $y$. 
Since the dependence on $x$ and $y$ depends on the final state, 
different decay modes exhibit different sensitivities to 
the parameters $x$ and $y$.

\begin{figure}[t]
\begin{center}
 \includegraphics[width=0.45\textwidth]{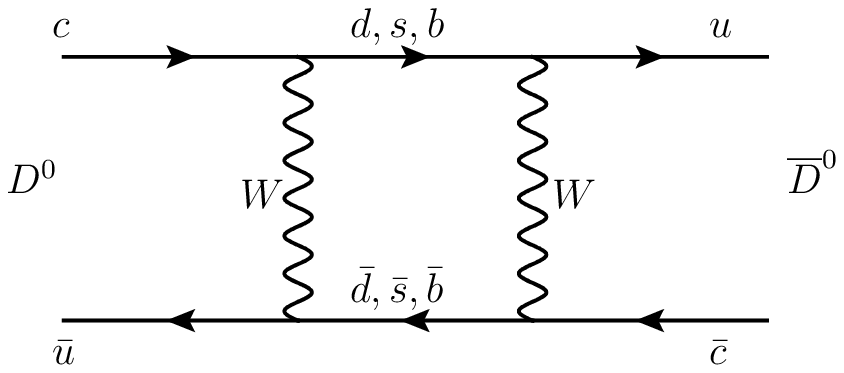}\hspace{0.09\textwidth}\raisebox{1cm}{\includegraphics[width=0.45\textwidth]{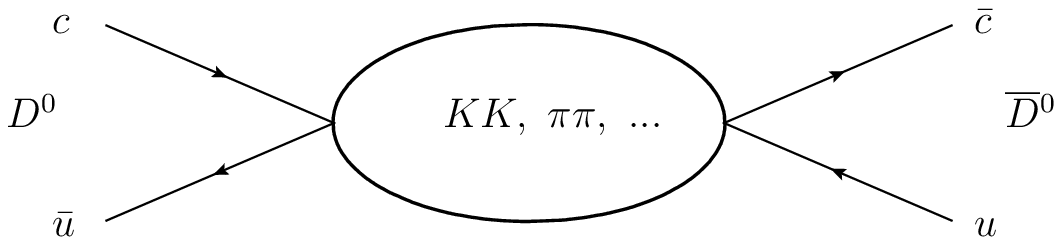}}
\end{center}
 \caption{Short distance (left) and long distance (right) contributions to \zdmix mixing in the SM.}
 \label{fig_th_boxD0}
\end{figure}

Out of the four flavored neutral meson systems, 
the neutral $D$ meson system is the only one in 
which down-type quarks are involved 
in the box diagram loop (see Fig.~\ref{fig_th_boxD0}). 
The neutral pion is its own antiparticle and 
the top quark decays before it forms a hadron and therefore cannot oscillate. 
Hence studies of charm mixing offer a
unique probe of NP via flavor changing neutral 
currents in the down-type quark sector.
In the SM, mixing in the neutral $D$ meson system 
can proceed through a double weak boson exchange 
(short distance contributions) represented by box diagrams, 
or through intermediate states that are accessible to 
both $\dzero$ and $\dzerobar$ (long distance effects), as 
represented in Fig.~\ref{fig_th_boxD0}. 
The potentially large long distance contributions are non-perturbative
and therefore difficult to estimate, so the predictions for the mixing parameters $x$ and $y$
within the SM span several orders of magnitude between $10^{-8}$ and $10^{-2}$~\cite{Nelson:1999fg,Falk:2001hx}. 
Due to large uncertainties in the SM mixing predictions,
 it is difficult to identify NP contributions (a clear hint would be if $x$ 
is found to be much larger than $y$); however, 
measurements can still provide useful and competitive constraints on many NP models, as will be discussed later.

The study of $CP$ violation in decays of charmed hadrons also holds the potential for uncovering NP. 
In the SM, direct $CP$ violation can occur in singly Cabibbo suppressed (SCS; $c\to du\overline{d}$, $c\to su\overline{s}$) 
decays, but not in Cabibbo favored (CF; $c\to su\overline{d}$) or doubly Cabibbo suppressed (DCS; $c\to du\overline{s}$) 
decays. This is due to the fact that the 
final state particles in SCS decays contain at 
least one quark\---antiquark pair of the same flavor, 
which makes a contribution 
from penguin-type or box amplitudes induced by virtual $b$-quarks possible 
in addition to the tree amplitudes. However, the contribution of these 
second order amplitudes is strongly suppressed by the small
combination of CKM matrix elements $V_{cb}V_{ub}^{\ast}$.
Therefore, in processes involving 
 charmed hadrons, mainly the first two generations of quarks 
are involved. From the Wolfenstein parameterization of the 
CKM matrix \cite{bib_Wolfenstein} one can see that the elements related to the 
first two generations of the quarks are nearly real. Of course,  
by using the unitarity of the matrix one can still parameterize and
estimate the 
imaginary part of those elements. For example, examining the phase 
difference between the decays $D^0(\bar{D}^0)\to K^+K^-$, one finds 
that it is $2 \arg(V_{cs}V_{us}^\ast)$, which can be expressed using 
the unitarity and the Wolfenstein parameterization as 
$2A^2\lambda^4\eta\approx 10^{-3}$. Hence the expected $CPV$ 
asymmetries in the charm sector are of the order of $10^{-3}$, which is small compared 
to the measured $CP$ asymmetries in the bottom sector.
Recently, with the experimental precision reaching the per mille 
level, some authors \cite{Zupan} have argued that the asymmetries could be 
much larger than naively expected. Nevertheless, at the 
current level of experimental sensitivity, the measurements of the 
$CPV$ in the charm sector are mainly a search for a significant effect, 
which would point to so-far unknown NP processes.  

\subsection{Experimental techniques of $\dzero$\---$\dzerobar$ mixing and $CP$ symmetry violation}

Often, the flavor of initially produced neutral $D$ mesons needs to be tagged in order 
to identify \zdmix transitions and $CP$ violation. The flavor is tagged by requiring that 
neutral $D$ mesons originate from $D^{\ast +}\to \dzero\pi^+$ decays, where the charge 
of the pion accompanying $\dzero$ tags the flavor of the neutral $D$ meson at production. 
Another common property of the measurements described below is the selection of 
$D$ meson candidates based on the CM momentum, typically $p^\ast>2.5$~GeV/$c$ for data 
taken at the $\Upsilon(4S)$ resonance. 
This requirement completely removes charmed mesons arising 
from possibly $CP$-violating $B$ meson decays that have a
 displaced production vertex. Hence the Belle charm samples
consist entirely of $e^+ e^- \to  c \bar{c}$ continuum data.

The most precise constraints on the mixing parameters 
$x$ and $y$ are obtained using the time dependence 
of $\dzero$ decays. In time-dependent measurements, 
the $\dzero$ decay time is determined according to
$t=m_{\dzero}(\vec{L}\cdot\vec{p}_{\dzero})/|\vec{p}_{\dzero}|^2$ , 
where $\vec{L}$ is the vector joining the $\dzero$'s 
production and decay vertices, 
and $\vec{p}_{\dzero}$ and $m_{\dzero}$ are its momentum and nominal mass. 
The reconstructed tracks of $\dzero$ decay products 
are refitted to a common vertex to determine the $\dzero$ decay point, 
and then the $\dzero$'s production point is 
determined from the kinematic fit of the $\dzero$ momentum vector 
with the beam spot profile. 
The decay-time uncertainty $\sigma_t$ is evaluated 
event-by-event from the covariance matrices of the production and decay
vertices. Typically, for decays with two charged tracks in the
final state, $\langle\sigma_t\rangle\sim \taudzero/2$. 
Candidates with badly reconstructed decay time (with large $\sigma_t$) are
excluded from the analysis.

The mixing parameters are extracted by performing a fit to the decay-time distribution using the following PDF:
\begin{equation}
 {\cal P}(t) = \int_{-\infty}^{+\infty} \Gamma_{\rm sig}(t'; x, y)R_{\rm sig}(t-t')dt'+{\cal P}_{\rm bkg}(t),
\end{equation}
where the signal contribution is a 
convolution of the (final state dependent) 
time-dependent decay rate ($\Gamma_{\rm sig}$) and the 
detector resolution function ($R_{\rm sig}$). To reduce 
the systematic uncertainties related to the parameterization of the
resolution function, kinematically similar decays (from high
statistics control samples) are usually used 
to determine the resolution function
parameters directly from data. 
The background ${\cal P}_{\rm bkg}(t)$ is parameterized 
using an exponential (to describe
the background candidates originating from mis-reconstructed charm
decays) and a $\delta$ function (to describe
random combinations of final state particles), 
each convolved with its corresponding resolution function. 
The parameters of the background PDF are determined
using events populating the sideband region 
in the invariant mass of $D^0$ candidates.

Experimental determinations of $CPV$ can be divided into those 
using the decay time distribution of certain decays to determine the 
unknown parameters and those using the decay time-integrated 
methods. The unknown $CPV$ parameters often follow from the
parameterization below: 
\begin{equation}
\left|\frac{\bar{A}_{\bar{f}}}{A_f}\right|^2 \equiv 1+A_D^f~~,
~~~~~~\left|\frac{q}{p}\right|^2 \equiv 1+A_M~~, 
~~~~~~\Im\bigl[\frac{q}{p}\frac{\bar{A}_f}{A_f}\bigr]\equiv
\left| \frac{q}{p}\frac{\bar{A}_f}{A_f}\right| \sin\phi~~.
\label{D_cpv_eq1}
\end{equation}
$A_D^f\ne 0$ is the asymmetry from $CPV$ in decays to a final state $f$, 
$A_M\ne 0$ is from $CPV$ in the mixing, and 
$\sin\phi \ne 0$ is a manifestation of  
$CPV$ in the interference between decays with and without mixing. 

While in charged $D$ meson processes only the $CPV$ in decays 
is present, neutral charmed mesons may include contributions 
from all three types of violation. 

All measurements are performed blindly, i.e. 
the selection criteria are determined 
using samples of simulated events 
or data events that are statistically independent from 
those used to perform the measurement, in order to avoid possible biases. 

\subsection{Time-dependent measurements of $\dzero$\---$\dzerobar$ mixing and $CP$ violation}

\subsubsection{Decays to $CP$ eigenstates}

\begin{figure}[t]
\begin{center}
 \includegraphics[width=0.65\textwidth]{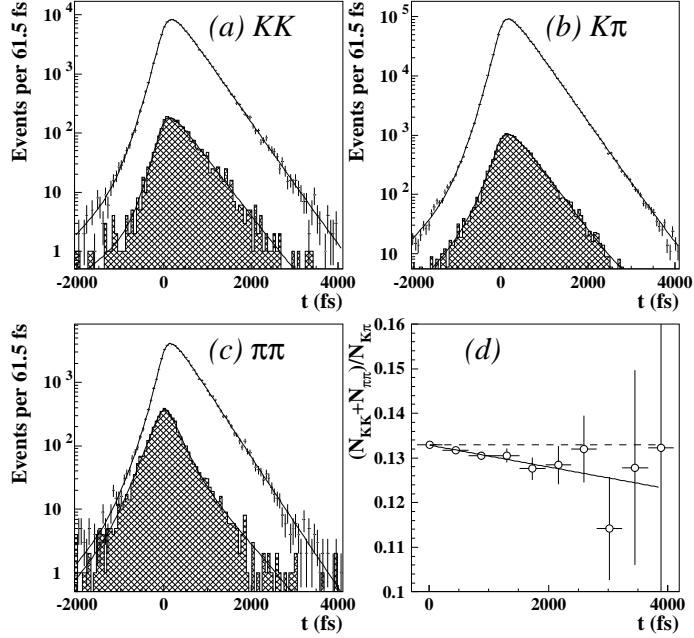}
\end{center}
 \caption{Results of the simultaneous fit to decay-time distributions
of (a) $\dzero\to K^+K^-$, (b) $\dzero\to K^-\pi^+$, and (c) $\dzero\to\pi^+\pi^-$ decays. 
The cross-hatched area represents background contributions, the shape of which was fitted using $\dzero$ invariant mass sideband
events. (d) Ratio of decay-time distributions between $\dzero\to K^+K^-$, $\pi^+\pi^-$, and $\dzero\to K^-\pi^+$ decays. 
The solid line is a fit to the data points.}
 \label{fig_belle_ycp}
\end{figure}

Belle found the first evidence for \zdmix mixing~\cite{Staric:2007dt} in a data sample of $540$~fb$^{-1}$ 
using the ratios of lifetimes extracted from a 
sample of $\dzero$ mesons produced through the process $D^{\ast +}\to \dzero\pi^+$, which decay to $K^-\pi^+$, $K^-K^+$, or 
$\pi^-\pi^+$. The time-dependent decay rates of the CF mode, $K^-\pi^+$, and the SCS modes $h^-h^+$ ($h=K$ or $\pi$) 
are obtained from the time-dependent decay rates given in the previous section:
\begin{eqnarray}
 \Gamma(\dzero(t)\to K^-\pi^+,\dzerobar(t)\to K^+\pi^-) & \propto & e^{-t/\taudzero}\\
\Gamma(\dzero(t),\dzerobar(t)\to h^+h^-) & \propto & e^{-(1+y_{CP})t/\taudzero},
\end{eqnarray}
where we assume that $x,y\ll1$ and $|\overline{\cal A}_f/{\cal A}_f|=1$ 
($|\overline{\cal A}_f/{\cal A}_f|\ll1$) for $\dzero$ meson decays to $h^-h^+$ ($K^-\pi^+$). For 
$\dzero\to h^+h^-$ decays, linear 
terms in $xt$ and $yt$ are the first terms in the Taylor expansion
of the exponential function above.
The lifetime difference between decays to the $CP$ eigenstates $h^-h^+$ and $CP$-mixed state $K^-\pi^+$, $y_{CP}$, 
is defined as
\begin{equation}
 y_{CP}\equiv\frac{\tau_{K^{\mp}\pi^{\pm}}}{\tau_{h^+h^-}}-1=y\cos\phi-\frac{1}{2}A_M x\sin\phi.\label{eq_th_ycp}
\end{equation}
The lifetimes $\tau_{K\pi}$ and $\tau_{hh}$ are the 
effective lifetimes extracted from samples of $\dzero$ 
mesons decaying to the $CP$ mixed final state $K^-\pi^+$, 
and $CP$ even final states $K^-K^+$ and $\pi^-\pi^+$. 
If $|q/p|=1$ and $\phi=\arg(q/p)=0(\pi)$, 
$CP$ symmetry in mixing and interference 
between mixing and decay is conserved, 
and hence the parameter $y_{CP}$ corresponds to the mixing parameter $y$. 
In these time-dependent measurements of 
neutral $D$ mesons decaying to $CP$ eigenstates,  
 indirect $CP$ violation is also probed by comparing the lifetimes 
of $\dzero$ and $\dzerobar$ mesons decaying to $CP$ eigenstates:
\begin{equation}
A_{\Gamma} =
\frac{\tau_{h^+h^-}^{\dzerobar}-\tau_{h^+h^-}^{\dzero}}
{\tau_{h^+h^-}^{\dzerobar}+\tau_{h^+h^-}^{\dzero}}
= 
\frac{1}{2}A_My \cos\phi - x \sin\phi.
\end{equation}
By performing a simultaneous fit to the decay-time distributions 
of around 0.15 (1.2) million reconstructed tagged $\dzero$ decays with purity
above 90\% to $h^-h^+$ ($K^-\pi^+$),  
Belle found $y_{CP} = (1.13\pm0.32\pm0.25)\%$ and $A_{\Gamma}=(0.01\pm0.30\pm0.15)\%$.
Figure~\ref{fig_belle_ycp} shows the decay-time distributions 
with fit results superimposed as well as the decay-time dependent ratio 
of $\dzero$ decays to $CP$-even eigenstates $K^-K^+$ and $\pi^-\pi^+$ to 
the $CP$ mixed final state $K^-\pi^+$, as measured by Belle 
\cite{Staric:2007dt}. In case of $y_{CP}=0$, this ratio would be constant, 
which is inconsistent with Belle's data at $3.2\sigma$.
No evidence for indirect $CP$ violation is found. 

\subsubsection{Decays to hadronic wrong sign decays}

Belle also performed a search for neutral $D$ meson mixing 
and $CP$ violation in a time-dependent study of 
DCS $\dzero\to K^+\pi^-$ decays\cite{Zhang:2006dp}
based on $400$~fb$^{-1}$ of data. 
These decays (also referred to as wrong sign decays) 
can proceed both through mixing followed by a CF decay, 
$\dzero\to\dzerobar\to K^+\pi^-$, or 
directly through a DCS decay such as $\dzero\to K^+\pi^-$. 
To distinguish the two processes, an analysis of the decay time 
distribution is performed. The most general form 
(e.g. allowing for
direct $CP$ violation in DCS decays, 
mixing and interference between mixing and decay) 
for the time-dependent decay 
rates of the two-body wrong sign decays $D^0\to K^+\pi^-$ or
$\dzerobar\to K^-\pi^+$
to second order in $x$ and $y$ is given by:
\begin{eqnarray}
 \Gamma\left(\bfrac{D^0(t)\to K^+\pi^-}{\dzerobar(t)\to K^-\pi^+}\right) & \propto & e^{-t/\taudzero} \left(R_D(1\pm A_D\right).\nonumber\\
& & \left. +\sqrt{R_D(1\pm A_D)}\left[\frac{1\pm A_M}{1\mp A_M}\right]^{1/4}(y'\cos\phi\mp x'\sin\phi)\frac{t}{\taudzero}\right.\nonumber\\
& & \left. +\frac{1}{4}\left[\frac{1\pm A_M}{1\mp A_M}\right]^{1/2}(x'^2+y'^2)\frac{t^2}{\taudzero^2} \right),\label{eq_th_wsdo}
\end{eqnarray}
where $R_D$ is the ratio of DCS to CF decay rates, and 
the parameters $x'$ and $y'$ are rotated mixing parameters, which are
rotated by an unknown strong phase difference between the DCS and CF amplitudes, $\delta_{K\pi}$: 
$x'=x\cos\delta_{K\pi}+y\sin\delta_{K\pi}$ and $y'=y\cos\delta_{K\pi}-x\sin\delta_{K\pi}$. 
The three terms in the time-dependent decay rates of wrong sign decays are due to the DCS amplitude, the interference between the DCS and CF amplitudes, 
and the CF amplitude, respectively. 
In addition to the wrong sign decays, 
the Cabibbo favored (or right sign) $D^0\to K^-\pi^+$ decays 
are reconstructed in order
to determine the resolution function parameters, 
as well as the distribution of wrong sign signal events in $D^0$ invariant mass
and mass difference distributions, 
which are fitted to extract the number of 
correctly reconstructed wrong sign decays.

\begin{figure}[t]
\centering
 \includegraphics[width=0.55\textwidth]{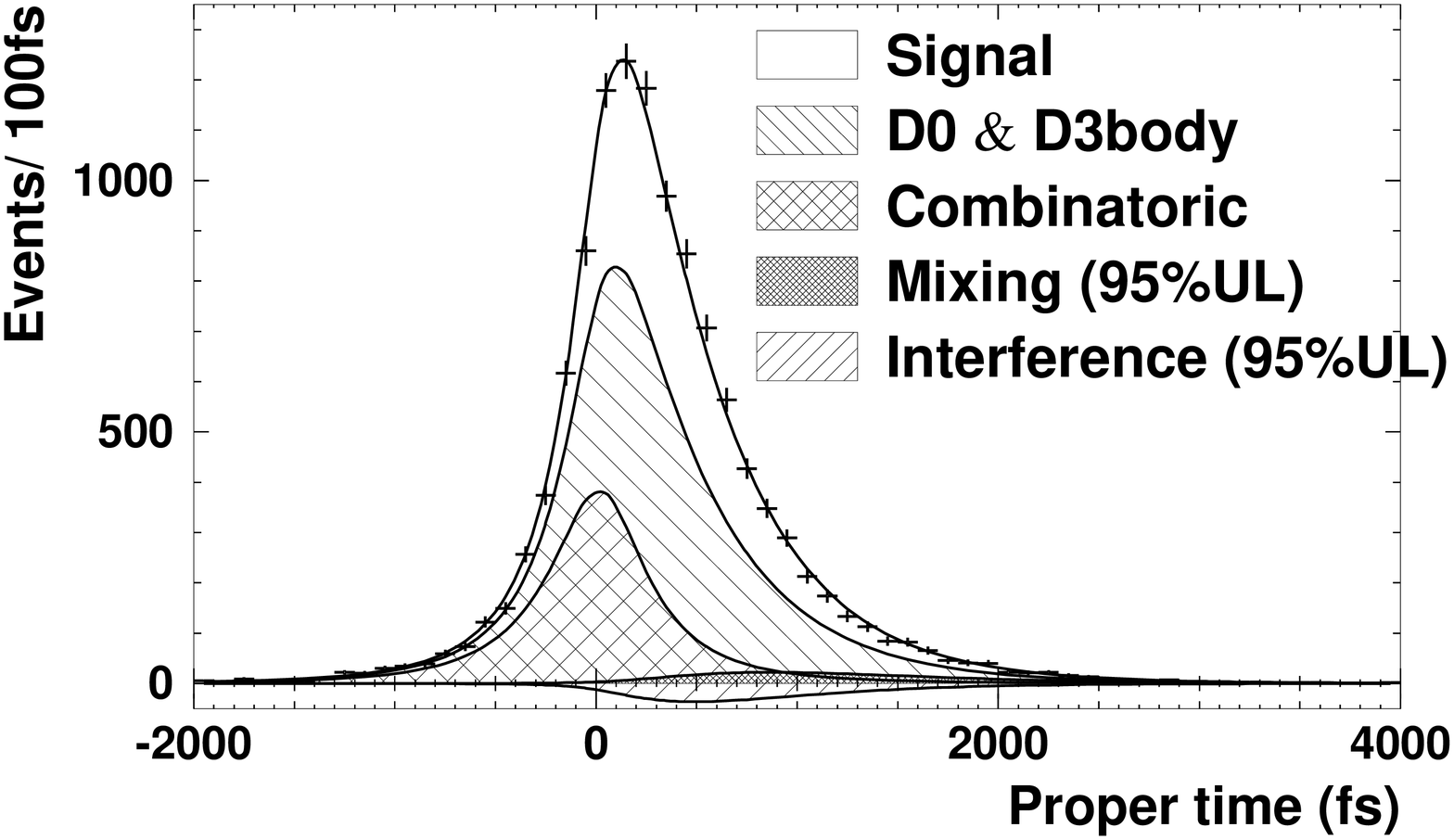}
 ~~~\includegraphics[width=0.35\textwidth]{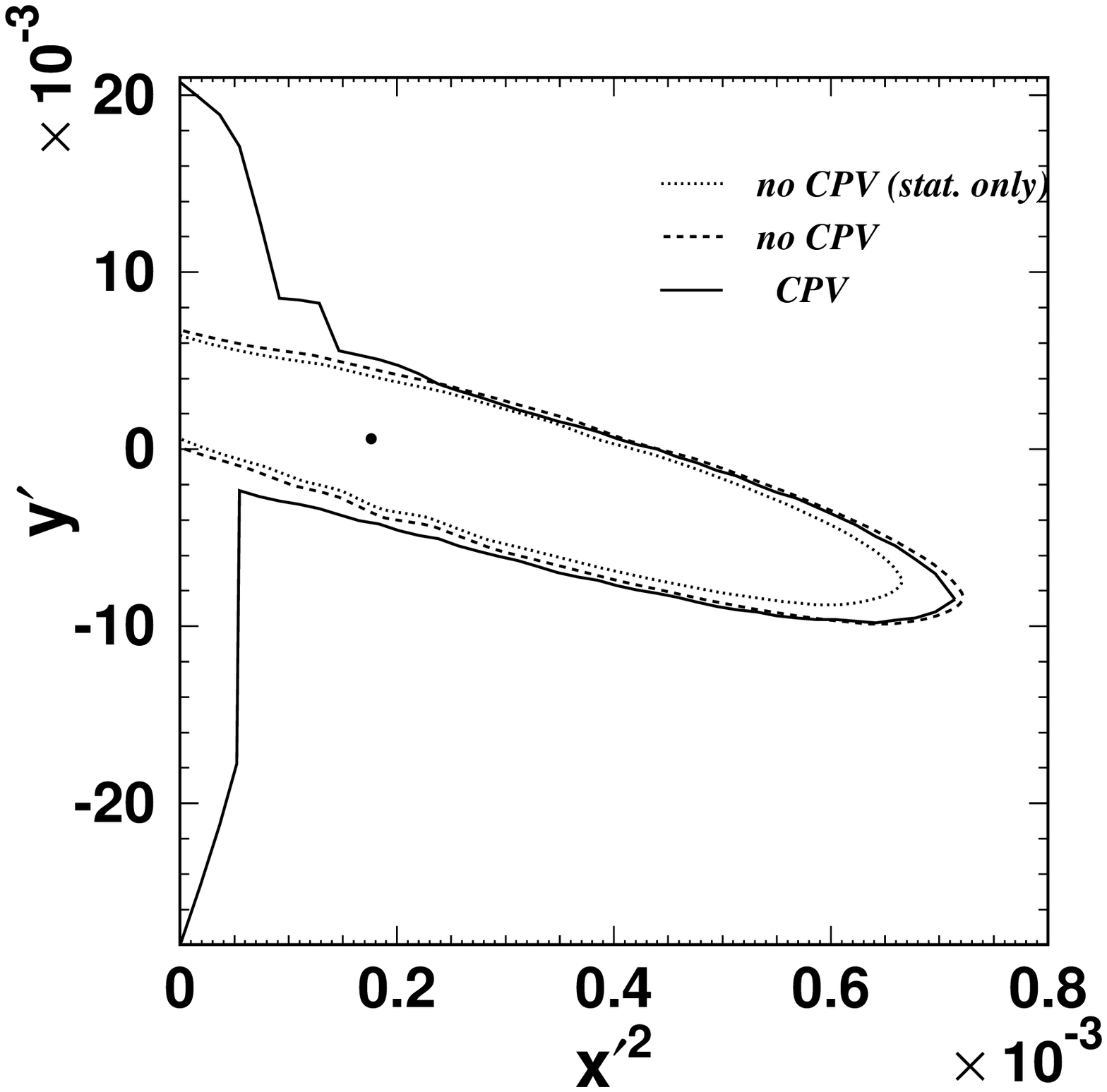}
 \caption{(Left) The decay-time distribution for wrong sign (WS) events. Superimposed
on the data (points with error bars) are projections of
the decay-time fit when no $CPV$ is assumed. The mixing and
interference terms are shown at the 95\% C.L. upper
limits. (Right) 95\% C.L. regions for $(x'^2,y')$. The point is the best fit
result assuming $CP$ conservation. The dotted (dashed) curve is
the statistical (statistical and systematic) contour for no $CPV$.
The solid curve is the statistical and systematic contour in the
$CPV$-allowed case.}
\label{fig:kpiws:mix}
\end{figure}

From a fit to the decay-time distribution of around $4\times 10^3$ signal wrong sign decays (and with purity around 50\%) Belle
found $x'^2=(0.18^{+0.21}_{-0.23})\times10^{-3}$ 
and $y'=(0.6^{+0.4}_{-3.9})\times 10^{-3}$ 
assuming no $CP$ violation
(setting $A_D=A_M=\phi=0$ in Eq. \ref{eq_th_wsdo}). The errors in
$x'^{2}$ and $y'$   include both statistical and systematic 
uncertainties. 
A projection of this fit superimposed 
on the data is shown in Fig.~\ref{fig:kpiws:mix} and
the non-mixing point ($x'^2=y'=0$) is found to be excluded at 95\%
C.L. 
In a second fit, 
$CP$-violating parameters are allowed to float and 
no evidence for either direct
or indirect $CPV$ is found. Belle obtains 
the following 95\% C.L. intervals for 
$CP$-violating parameters: $A_D\in(-76,107)\times10^{-3}$ 
and $A_M\in(-995,1000)\times 10^{-3}$. 

\subsubsection{Self-conjugated three-body decays}

Several intermediate resonances can contribute to a hadronic three-body decay 
of a neutral $D$ meson. For example, 
$\dzero\to K^0_S\pi^+\pi^-$ decays can proceed 
via $D^0\to K^{\ast -}\pi^+$ (CF amplitude), 
$\dzero\to K^0_S\rho^0$ (SCS amplitude and $CP$ eigenstate), 
$\dzero\to K^{\ast +}\pi^-$ (DCS amplitude), and others. 
In the isobar model, the instantaneous amplitudes for $\dzero$ and $\dzerobar$ decays to the three-body 
final state $f$ are parameterized as a sum of Breit\---Wigner
resonances and a constant nonresonant term 
(in the case of no direct $CP$ violation, e.g., 
there is no difference between 
amplitudes and phases in $\dzero$ and $\dzerobar$ decays):
\begin{eqnarray}
 {\cal A}_f(s_+,s_-) & = & \sum_r a_re^{i\phi_r}{\cal A}_r(s_+,s_-) + a_{\rm NR}e^{i\phi_{\rm NR}},\\
\overline{\cal A}_f(s_+,s_-) & = & \sum_r a_re^{i\phi_r}{\cal A}_r(s_-,s_+) + a_{\rm NR}e^{i\phi_{\rm NR}},
\end{eqnarray}
where $\sqrt{s_{\pm}}$ is the invariant mass of a 
pair of final state particles (e.g. $K^0_S\pi^{\pm}$), 
and the sum runs over possible intermediate resonances $r$. 
The time-dependent decay rate for $\dzero$ decays is thus given 
by (the corresponding expression for $\dzerobar$ decays is 
obtained by multiplying the equation below by $|p/q|^2$):
\begin{eqnarray}
  \frac{d\Gamma(D^0\to f)}{ds_+ds_-dt} & \propto &
  |{\cal A}_1(s_+,s_-)|^2e^{-\frac{t}{\tau}(1+y)} + 
  |{\cal A}_2(s_+,s_-)|^2e^{-\frac{t}{\tau}(1-y)} \nonumber\\
  & & + 2\Re[{\cal A}_1(s_+,s_-){\cal A}_2^{\ast}(s_+,s_-)]\cos{\left(x\frac{t}{\tau}\right)}
  e^{-\frac{t}{\tau}} \nonumber\\
  & & + 2\Im[{\cal A}_1(s_+,s_-){\cal A}_2^{\ast}(s_+,s_-)]\sin{\left(x\frac{t}{\tau}\right)}
  e^{-\frac{t}{\tau}}, \label{eq_mD0}
\end{eqnarray}
where ${\cal A}_{1,2}(s_+,s_-) = \frac{1}{2}\left({\cal A}_f(s_+,s_-)\pm\frac{q}{p}\overline{\cal A}_f(s_+,s_-)\right)$. 
Different regions in the $s_+ - s_-$ plane (also called the Dalitz plot) exhibit different forms of time 
dependence, as can be seen from the above decay rate; 
therefore, the time-dependent Dalitz plot analysis  of neutral $D$ meson decays to a self-conjugated three-body final state 
enables us to measure the $x$ and $y$ parameters simultaneously. In 
the case where the analysis is performed separately for 
$\dzero$ and $\dzerobar$ samples, indirect $CP$ violation can be
probed by measuring the amplitude and phase of $q/p$. 
This method of measuring the mixing parameters $x$ and $y$ was 
pioneered by CLEO in $\dzero\to K^0_S\pi^+\pi^-$ decays \cite{Asner:2005sz}, 
and was applied by Belle to 
$\dzero\to K^0_S\pi^+\pi^-$ decays \cite{Abe:2007rd} 
using a data sample of 540~fb$^{-1}$ in which around 530 000
signal events are reconstructed with a purity of around 95\%. 

\begin{figure}[t]
\begin{center}
\includegraphics[width=0.32\textwidth]{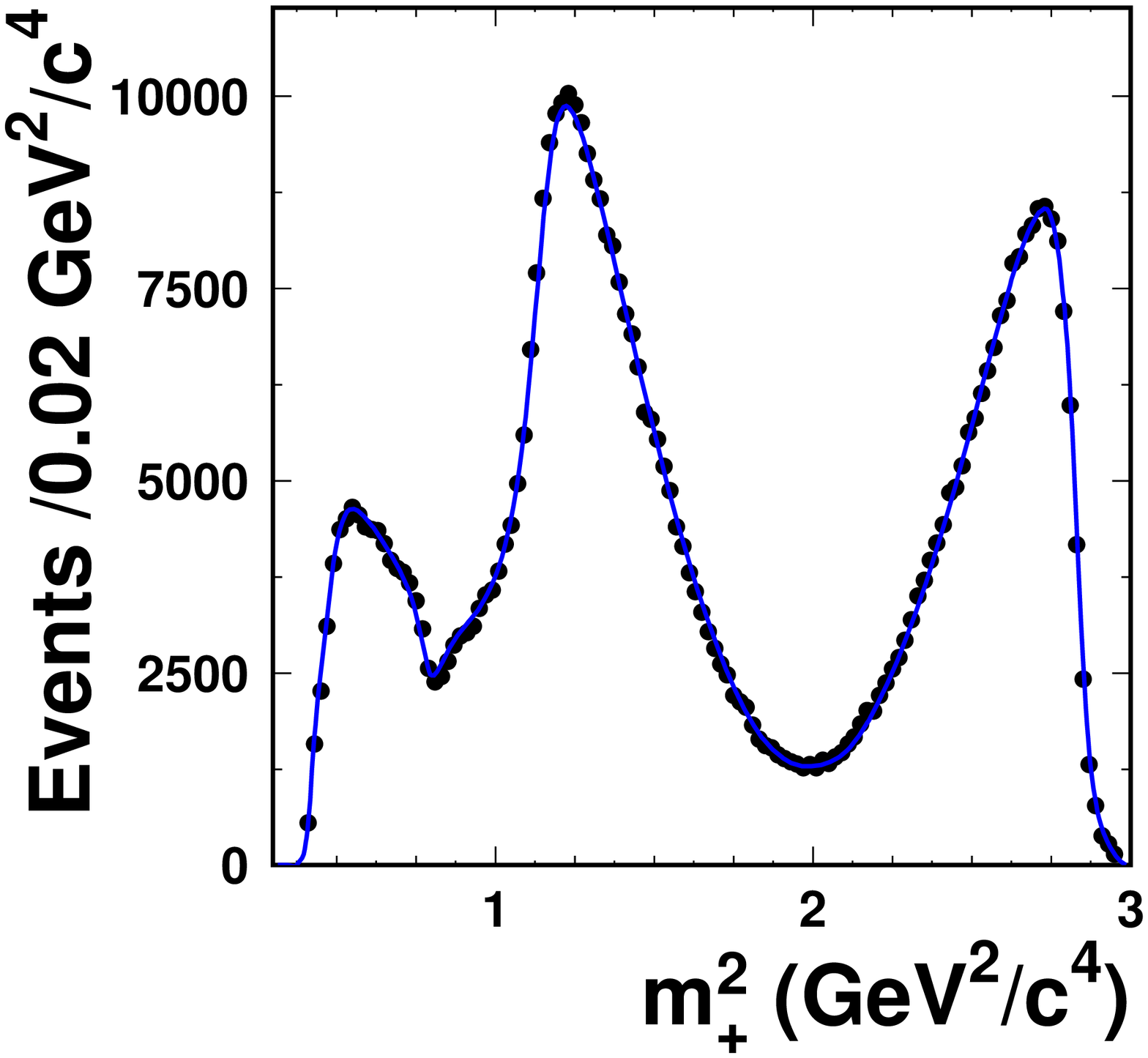}~
\includegraphics[width=0.32\textwidth]{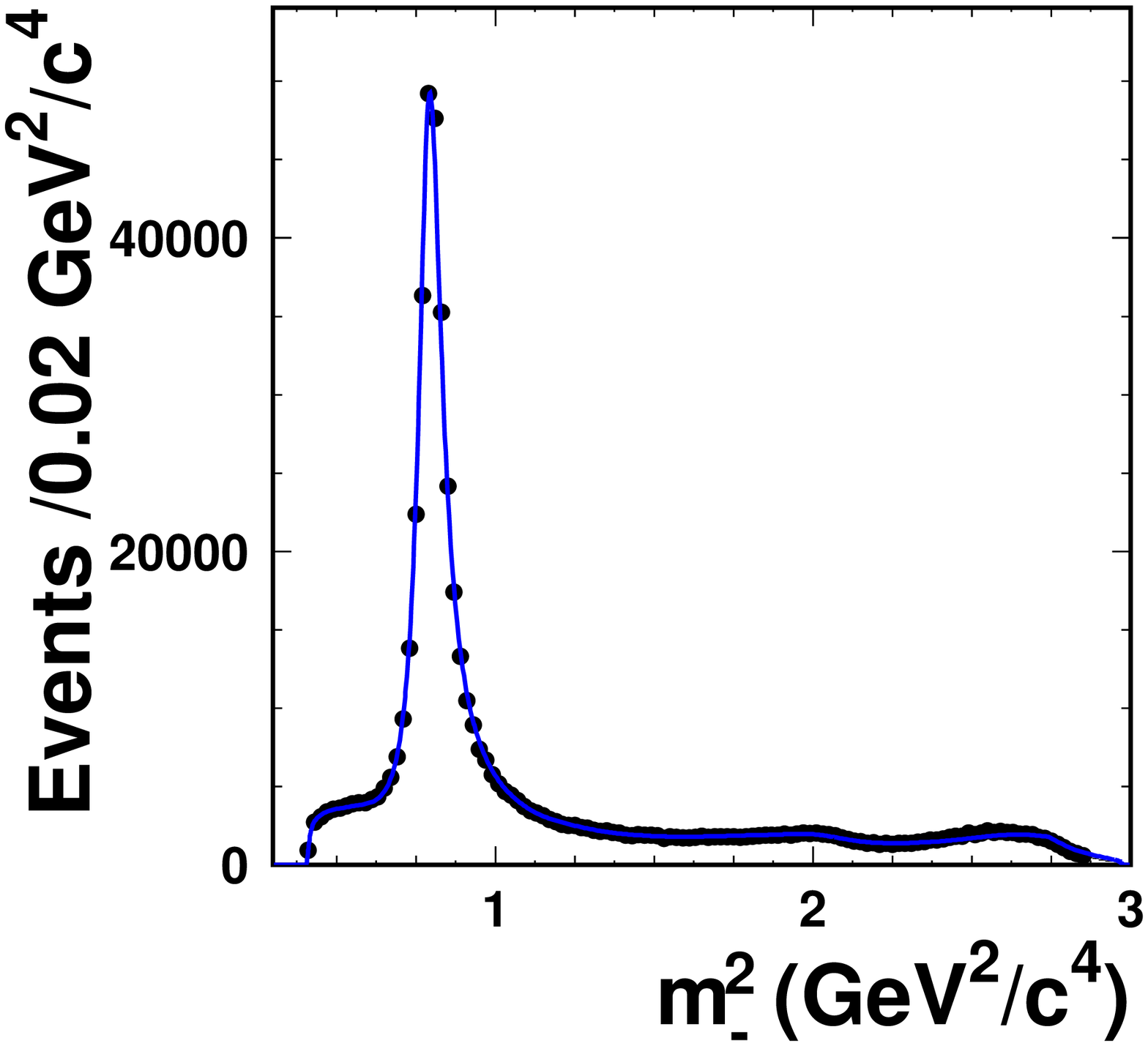}~
\includegraphics[width=0.32\textwidth]{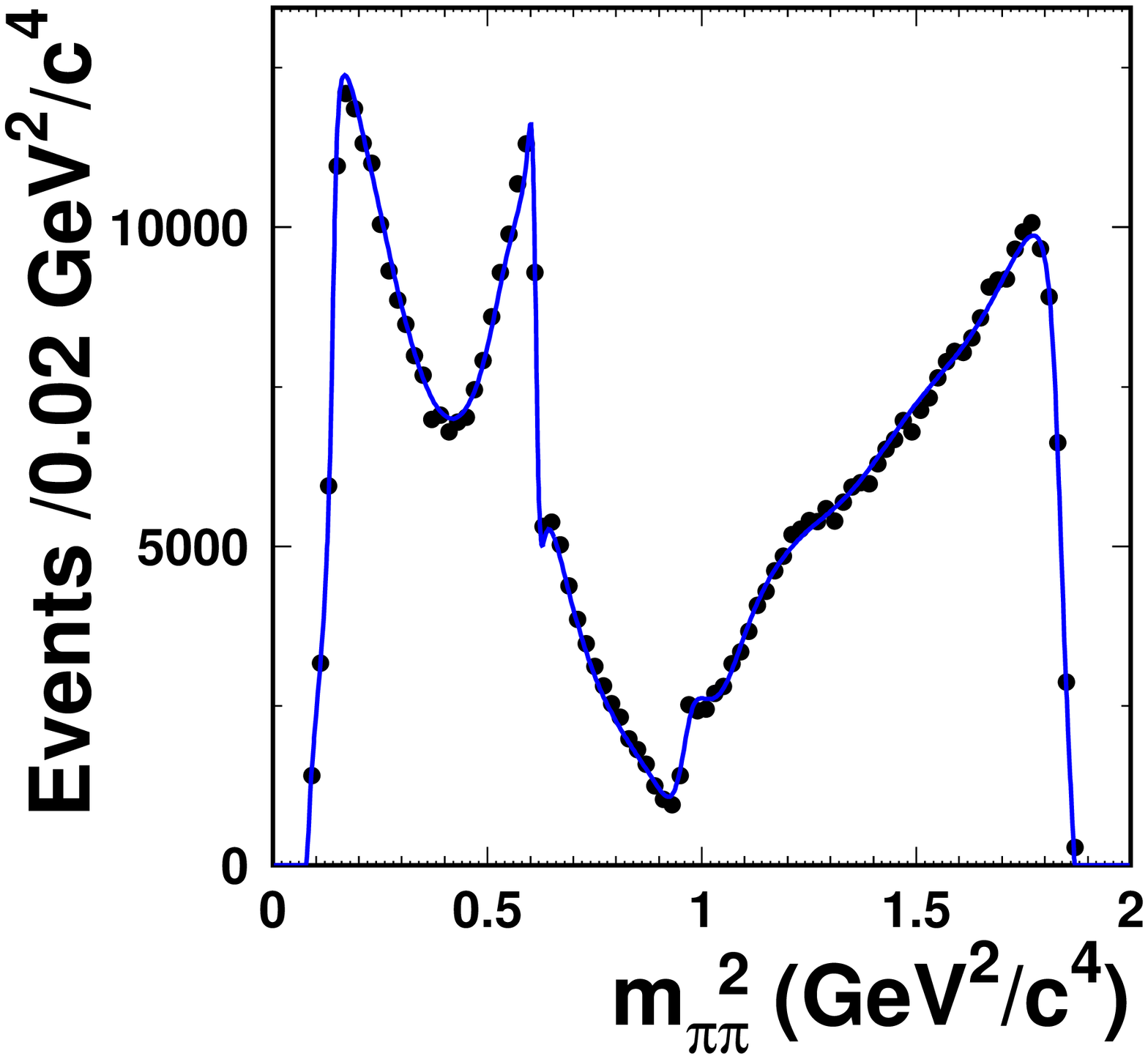}~
\end{center}
 \caption{Projections of the Dalitz distribution (points with error bars) 
and the fit result (curve) for $\dzero\to K^0_S\pi^+\pi^-$ decays 
\cite{Abe:2007rd}. Here, $m^{2}_{\pm}$ corresponds to $m^{2}(K^0_S\pi^{\pm})$ for $\dzero$ decays and to
$m^{2}(K^0_S\pi^{\mp})$ for $\dzerobar$ decays.}
 \label{fig:k0spipi}
\end{figure}

The decay amplitude is not a priori known and has to be extracted
from the data. This is done by first performing a time-integrated 
Dalitz plot analysis in which a model for the decay amplitude 
(${\cal A}(s_+,s_-)$) that describes the observed decay kinematics 
best is obtained. Belle finds that
a good description of the Dalitz plot 
is obtained when 18 quasi-two-body resonances and a nonresonant term are used
(see Fig.~\ref{fig:k0spipi} for the results of the fit). 
Once the decay amplitude composition is determined, 
a time-dependent Dalitz analysis is performed 
to determine the mixing parameters. In a fit with conserved
$CP$ symmetry ($|q/p|=1$ and $\phi=0$) 
the mixing parameters are found to be: 
$x=(0.80\pm0.29{}^{+0.09}_{-0.07}{}^{+0.10}_{-0.14})\%$
and $y=(0.33\pm0.24{}^{+0.08}_{-0.12}{}^{+0.06}_{-0.08})\%$, 
excluding the non-mixing point with 95\% C.L. The errors quoted are 
the statistical, systematic error arising from
experimental sources (e.g. modeling of background, resolution function, etc.) and the systematic error arising from the decay model 
(determined by using alternative models with different parameterizations, excluding resonances with small contributions, etc.). In a fit
allowing for $CPV$, the $|q/p|$ and $\phi$ parameters 
are found to be consistent with no $CP$ violation: 
$|q/p|=0.86{}^{+0.30}_{-0.29}{}^{+0.06}_{-0.03}\pm0.08$ and $\phi=(-14{}^{+16}_{-18}{}^{+5}_{-3}{}^{+2}_{-4})^{\circ}$.

Large fractions of $\dzero\to K^0_SK^+K^-$ decays
proceed via $\dzero\to K^0_S\phi$ ($CP$-odd) and $\dzero\to K^0_Sa_0(980)$ ($CP$-even) decays. 
Belle took advantage of this fact and performed a measurement of
the $y_{CP}$ mixing parameter integrated over the Dalitz plot
using an untagged sample of 
$\dzero\to K^0_SK^+K^-$ decays \cite{Zupanc:2009sy}.
We measure the effective lifetimes of $\dzero$ mesons, $\tau_{\rm ON,OFF}$, in two different regions of $K^+K^-$
invariant mass (at the $\phi$ peak (ON) and in $\phi$ sidebands
(OFF)), which are given by 
$\tau_{\rm ON,OFF}=(1+(1-2f_{\rm ON,OFF})y_{CP})\taudzero$, 
where $f_{\rm ON,OFF}$ is the $CP$-even fraction in the ON or OFF region 
calculated using the decay model obtained by BaBar
\cite{Aubert:2008bd}. 
The obtained value of $y_{CP}=(0.11\pm0.61\pm0.52)\%$ is consistent with 
$y_{CP}$ obtained in $D^0\to hh$ decays. 

\subsection{World average and constraints on new physics models}

Various measurements of \zdmix mixing performed in different decay
modes can be combined to obtain the world average values of $x$ and $y$. 
The Charm subgroup of the Heavy Flavor Averaging Group has done this by performing a global $\chi^2$ fit from measurements of 
relevant observables \cite{hfag} performed by the Belle\footnote{
In the world average fit, HFAG also includes Belle's time-integrated
measurement of the mixing rate $R_{M}=(x^2+y^2)/$ in $D^0\to
K^+\ell\nu_{\ell}$ \cite{Bitenc:2008bk}, which is not described in
detail in this report.}, 
BaBar, CDF, LHCb, CLEO-c, Focus, and FNAL E791 experiments. 
The world average values are found to be
\begin{eqnarray}
 x & = & (0.63\pm{}^{+0.19}_{-0.20})\%,\\
 y & = & (0.75\pm0.12)\%,
\end{eqnarray}
and
\begin{eqnarray}
 \left|\frac{q}{p}\right| & = & (0.88\pm{}^{+0.18}_{-0.16}),\\
 \phi & = & (-10.1\pm{}^{+9.5}_{-8.9})^{\circ}.
\end{eqnarray}
The non-mixing point, $(x,y)=(0,0)$, is excluded at the 10.2 
standard deviation level while the $|q/p|$ and $\phi$ values 
are consistent with
conservation of $CP$ symmetry 
in mixing and interference between mixing and decay.

Golowich et al. \cite{Golowich:2007ka} 
studied the implications of existing \zdmix measurements on many NP
models. In many scenarios they found strong constraints 
that surpass those from other search techniques 
and provide an important test of flavor changing neutral currents 
in the up-quark sector. 
One simple extension to the SM that they studied is the addition of 
a fourth family of fermions. The obtained constraint on the CKM 
mixing parameters $V_{cb'}V_{ub'}^{\ast}$ ($b'$ is the down-quark 
of the fourth generation) is an order of magnitude more restrictive 
than those obtained from unitarity considerations of the CKM matrix.

\subsection{Time-integrated measurements of $CPV$ in charm}

In the time-integrated measurements one usually determines the asymmetry 
of the partial decay widths, 
\begin{equation}
A_{CP}^f\equiv \frac{\Gamma(D\to f)-\Gamma(\bar{D}\to \bar{f})}
{\Gamma(D\to f)+\Gamma(\bar{D}\to \bar{f})}~.
\label{D_cpv_eq2}
\end{equation}
The measured asymmetry 
\begin{equation}
A_\textrm{rec}^f=\frac{N(D^+\to f)-N(D^-\to \bar{f})}
{N(D^+\to f)+N(D^-\to \bar{f})}~~,
\label{D_cpv_eq3}
\end{equation}
where $N$ denotes the number of detected decays, 
receives a contribution from several non-$CP$ 
violating sources, the detector-induced asymmetries due to a possible 
asymmetry in the acceptance of positively and negatively charged pions 
and kaons, or the different acceptances for neutral kaons and 
their antiparticles. In addition, the physical forward\---backward asymmetry 
in the process $e^+ e^-\to c\bar{c}$ affects the measured asymmetry, as we will 
see in the following. All these effects must be carefully determined using  
control data samples in order 
to achieve an accuracy at the level of ${\cal{O}}(10^{-3})$, 
the level of uncertainty of $A_{CP}$ measurements 
in various final states reached by the 
Belle experiment. The existing MC simulation tools 
cannot be used for corrections at this level of accuracy. 

Currently the best sensitivity on $A_{CP}$ at Belle has been achieved in the 
decays $D^+\to \pi^+K_S^0$. This decay mode is a mixture of CF ($D^+\to \pi^+\bar{K}^0$) and 
DCS ($D^+\to \pi^+K^0$) decay. If NP processes with unknown $CP$-violating phases would contribute, the 
$CPV$ in the decays may be significantly different from zero. 
The measured asymmetry in these decays 
can be written as 
\begin{equation}
A_\textrm{rec}^{K_S \pi^+} = A_{CP}^{K_S \pi^+} + A^{\pi^{+}}_\epsilon(p_{\pi^+},\cos\theta_{\pi^+}) + A_{FB}(\cos\theta^\ast)~,
\label{D_cpv_eq4}
\end{equation}
where $A_{CP}^{K_S \pi^+}$ is the physical 
$CPV$ asymmetry, $A^{\pi^{+}}_\epsilon$ the 
detector-induced asymmetry between 
the $\pi^+$ and $\pi^-$ reconstruction efficiencies, 
and $A_{FB}$ the contribution of the forward\---backward asymmetry. 
The latter is an odd function 
of the $D$ meson polar angle in the CM $\cos\theta^\ast$ (see e.g. 
Ref.~\cite{afb}), while the first term is independent 
of any kinematic variables. The detector-induced asymmetry depends on the 
momentum and the polar angle of the charged track in the laboratory frame. 
In bins of these variables the measured asymmetry can be corrected for 
$A^{\pi^{+}}_\epsilon$ using samples of 
$D^0\to K^-\pi^+\pi^0$ and $D^+\to K^-\pi^+\pi^+$ decays.
The measured asymmetries for these decays are 
\begin{eqnarray}
\nonumber
A_\textrm{rec}^{K\pi\pi} = A_{FB} + A^{K^{-}}_\epsilon + A^{\pi_1^{+}}_\epsilon + A^{\pi_2^{+}}_\epsilon \\
A_\textrm{rec}^{K\pi\pi^0} = A_{FB} + A^{K^{-}}_\epsilon + A^{\pi_1^{+}}_\epsilon ~~,
\label{D_cpv_eq5}
\end{eqnarray}
assuming negligible $CP$ violation in the Cabibbo favored $D$ meson 
decays and the universality of the forward\---backward asymmetry 
for different types of charmed mesons \footnote{Within the SM only
 Cabibbo suppressed decays 
of charmed mesons have two possible amplitudes with 
different weak and strong phases---the tree and the 
penguin amplitude---which is a necessary condition for non-zero
$CPV$ in decays.}. 
By inspecting Eqs.~(\ref{D_cpv_eq5}) one finds that in the difference 
of the measured asymmetries in $D^0\to K^-\pi^+\pi^0$ 
and $D^+\to K^-\pi^+\pi^+$ some of the detector-induced 
asymmetries and the forward\---backward contribution 
cancel and hence one can determine $A^{\pi^{+}}_\epsilon$. 
In turn, $A^{\pi^{+}}_\epsilon$ is then used to correct
$A_\textrm{rec}^{K_S \pi^+} $ 
in bins of the charged pion momentum and polar angle, 
and to extract $A_{CP}^{K_S \pi^+}$\cite{dcpv1}. 
However, in the $D^+\to h^+K_S^0$ decay modes,  
one needs additional corrections due to the 
presence of a neutral kaon in the final state. 
In such $D^+$ meson decay modes 
either a $K^0$ or $\bar{K}^0$ is produced, which 
interact differently in the detector material. However, 
in the final state a $K_S^0$ is reconstructed, and hence
this affects the value of the asymmetry. 
A separate dedicated study \cite{Ks-study} was performed and the 
appropriate correction factor applied to the asymmetry. 
Furthermore, because of the $CP$ 
violation in the neutral kaon system, the asymmetry expected 
in this final state with  
$K_S^0$ is $A^{K_S}=(-0.332\pm 0.006)\%$. The Belle result is given 
in Table~\ref{D_cpv_tab1} and is in good agreement with the expectation 
due to $CP$ violation in the neutral kaon system. 

Belle searched extensively for non-zero time-integrated $CP$
asymmetries in a number of other decay modes 
and achieved the best sensitivity in many of these. 
The results (see Table~\ref{D_cpv_tab1}) are consistent with 
no $CPV$ at levels varying from ${\cal{O}}(10^{-2})$ to ${\cal{O}}(10^{-3})$. 
 
\begin{table}
  \caption{Measured time-integrated $CPV$ asymmetries in the $D$ meson system.}
  \label{D_cpv_tab1}
  \begin{tabular}{lcclc}
\hline\hline
Decay mode                 & $\cal L$ (fb$^{-1}$) & $A_{CP}$ (\%) & Comment  & Ref.\\\hline
$D^0\to K^0_S\pi^0$         &      & $-0.28 \pm 0.19 \pm 0.10$   &          &              \\ 
$D^0\to K^0_S\eta$          & 791  & $+0.54 \pm 0.51 \pm 0.16$   &          &  \cite{dcpv4}\\
$D^0\to K^0_S\eta'$         &      & $+0.98 \pm 0.67 \pm 0.14$   &          &              \\ \hline
$D^0\to\pi^+\pi^-$          & 540  & $+0.43 \pm 0.52 \pm 0.12$   &          & \cite{Staric:2008rx}\\
$D^0\to K^+K^-$             &      & $-0.43 \pm 0.30 \pm 0.11$   &          &                     \\\hline 
$D^0\to \pi^+\pi^-\pi^0$    & 532  & $+0.43 \pm 1.30$            &          & \cite{Arinstein:2008zh} \\\hline
$D^0\to K^+\pi^-\pi^0$      & 281  & $-0.6\phantom{0} \pm 5.3\phantom{0}$ & & \cite{Tian:2005ik} \\
$D^0\to K^+\pi^-\pi^+\pi^-$ &      & $-1.8\phantom{0} \pm 4.4\phantom{0}$ & &  \\\hline
$D^+\to K_S^0{\pi^+}$       & 977  & $-0.363\pm 0.094\pm 0.067$  & signif. asymmetry due to $K_S^0$ & \cite{dcpv1}\\\hline
$D^+\to \phi\pi^+$         & 955   & $+0.51 \pm 0.28 \pm 0.05$   & universality of $A_{FB}$ in $D_s^+$ & \cite{dcpv3}\\
                           &      &                             & and $D^+$ decays to $\pi^+\phi$ tested& \\ \hline
$D^+\to \eta\pi^+$         & 791   & $+1.74 \pm 1.13 \pm 0.19$   & $D^+\to K^+\eta^{(\prime)}$ & \cite{dcpv2}\\
$D^+\to \eta'\pi^+$        & 791   & $-0.12 \pm 1.12 \pm 0.17$   & also observed & \\\hline
$D^+\to K^0_S K^+$         & 673   & $-0.16\pm0.58\pm0.25$ & & \cite{dcpv5}\\\hline
$D_s^+\to K^0_S\pi^+$      & 673   & $+5.45\pm2.50\pm0.33$ & & \cite{dcpv5}\\
$D_s^+\to K^0_S K^+$       &       & $+0.12\pm0.36\pm0.22$ & & \cite{dcpv5}\\
\hline\hline
\end{tabular}
\end{table}

\subsection{Conclusions}

With the world's largest sample of recorded charmed hadron decays Belle has experimentally observed
mixing phenomena in the last remaining neutral meson system, $D^0$. The mixing 
parameters in this system are nowadays becoming a precision measurement, 
with world average values \cite{hfag} of 
$x=(0.63\pm{}^{+0.19}_{-0.20})\%$ and $y=(0.75\pm0.12)\%$. 
Further measurements and advances in 
theoretical predictions are required to determine
 whether the observed values are consistent with the SM 
or receive contributions from NP. Furthermore, 
an extensive search for $CPV$ in the charm sector 
was carried out. The measurement methods that were developed
allowed for the observation of a significant 
$CPV$ asymmetry in decay modes with a neutral kaon in the final
state and sensitivities to possible time-integrated $CP$ asymmetries at the per mille level 
in a variety of decay modes. No significant 
indirect $CP$ violation has been observed so far.

\def\ra{\!\rightarrow\!}
\def\cp{$CP$\/}

\def\mevm{~MeV/$c^2$\/}
\def\mevp{~MeV/$c$\/}
\def\meve{~MeV}
\def\gevm{~GeV/$c^2$\/}
\def\gevp{~GeV/$c$\/}
\def\geve{~GeV}

\def\kbar{\overline{K}{}^{\,0}}
\def\kkbar{$K^0$-$\kbar$}
\def\dbar{\overline{D}{}^{\,0}}
\def\ddbar{$D^0$-$\dbar$}
\def\bbar{\overline{B}{}^{\,0}}
\def\bbbar{$B^0$-$\bbar$}

\def\bsdsds{$B^0_s\ra D^{(*)+}_s D^{(*)-}_s$}
\def\bsdsstdsst{$B^0_s\ra D^{*+}_s D^{*-}_s$}
\def\bsdsstds{$B^0_s\ra D^{*\pm}_s D^{\mp}_s$}
\def\bsdspi{$B^0_s\ra D^{(*)-}_s \pi^+$}
\def\bdsd{$B^0\ra D^{(*)+}_s D^-$}
\def\bsjks{$B^0_s\ra J/\psi\,K^0_S$}
\def\bdjks{$B^0_d\ra J/\psi\,K^0_S$}
\def\bs{B^{}_s}
\def\bsst{B^{*}_s}
\def\bsbar{\overline{B}{}^{}_s}
\def\bsbarst{\overline{B}{}^{\,*}_s}
\def\fl{f^{}_L}
\def\jmm{$J/\psi\ra\mu^+\mu^-$}
\def\jee{$J/\psi\ra e^+ e^-$}
\def\fkk{$\phi\ra K^+ K^-$}
\def\kspp{$K^0_S\ra\pi^+\pi^-$}
\def\mbc{M^{}_{\rm bc}}
\def\de{\Delta E}
\def\qq{$q\bar{q}$}
\def\mm{$\mu^+\mu^-$}
\def\ee{$e^+e^-$}

\def\dms{\Delta m^{}_s}
\def\dgs{\Delta\Gamma^{}_s}
\def\dgcp{\Delta\Gamma^{CP}_s}
\def\gs{\Gamma^{}_s}
\def\kstz{\overline{K}{}^{\,*0}}
\def\kstp{K^{*+}}
\def\qqbar{$q\bar{q}$}

\section{$B$ physics at the $\Upsilon$(5$S$)}
\label{chap_bs}

The $\Upsilon$(10860), ($M=10 876\pm 11$~MeV/$c^2$,  $\Gamma=55\pm 28$~MeV)\cite{pdg2012}, is generally interpreted as the $\Upsilon$(5$S$), the fourth excitation of the vector bound state of $b\bar b$, and is just above $B_s^*\bar B_s^*$ threshold.
The Belle experiment collected a total of 121.4~fb$^{-1}$ at the $\Upsilon$(10860) peak energy and a total of 27.6~fb$^{-1}$ at off-peak CM energies nearby, between 10.683 and 11.021~GeV.
The on-resonance data sample corresponds to 37 million resonance events and includes 7.1 million $B_s$ events.
These data were analyzed to pursue investigations of $B_s$ meson properties, hadronization to $B_q$ and $B_s$ events ($q$ is a $u$- or $d$-quark), energy dependence of various types of events, and possible new bottomonia and bottomonium-like states.
Published on-peak results are based on two subsets, 1.86~fb$^{-1}$ and 23.6~fb$^{-1}$ (including the 1.86~fb$^{-1}$), as well as the full set of 121.4~fb$^{-1}$, which will be referred to as sets 2FB, 24FB, and 121FB, respectively.

The $e^+e^-\to \Upsilon$(10860) is an excellent venue for studying several aspects of $B_s$ decay; given clean, efficiently triggered events with precisely known CM energy, collected by a well-understood detector, the Belle experiment has been uniquely positioned to measure absolute branching fractions, access modes that include photons in the final state, and do comparative studies of $B$ and $B_s$ mesons with minimal systematic uncertainties.

\subsection{$B_s^{(*)}$ masses: method of full reconstruction}
\label{sec:BsMass}

\begin{figure}[ht!]
\begin{center}
{\includegraphics[width=5.2cm]{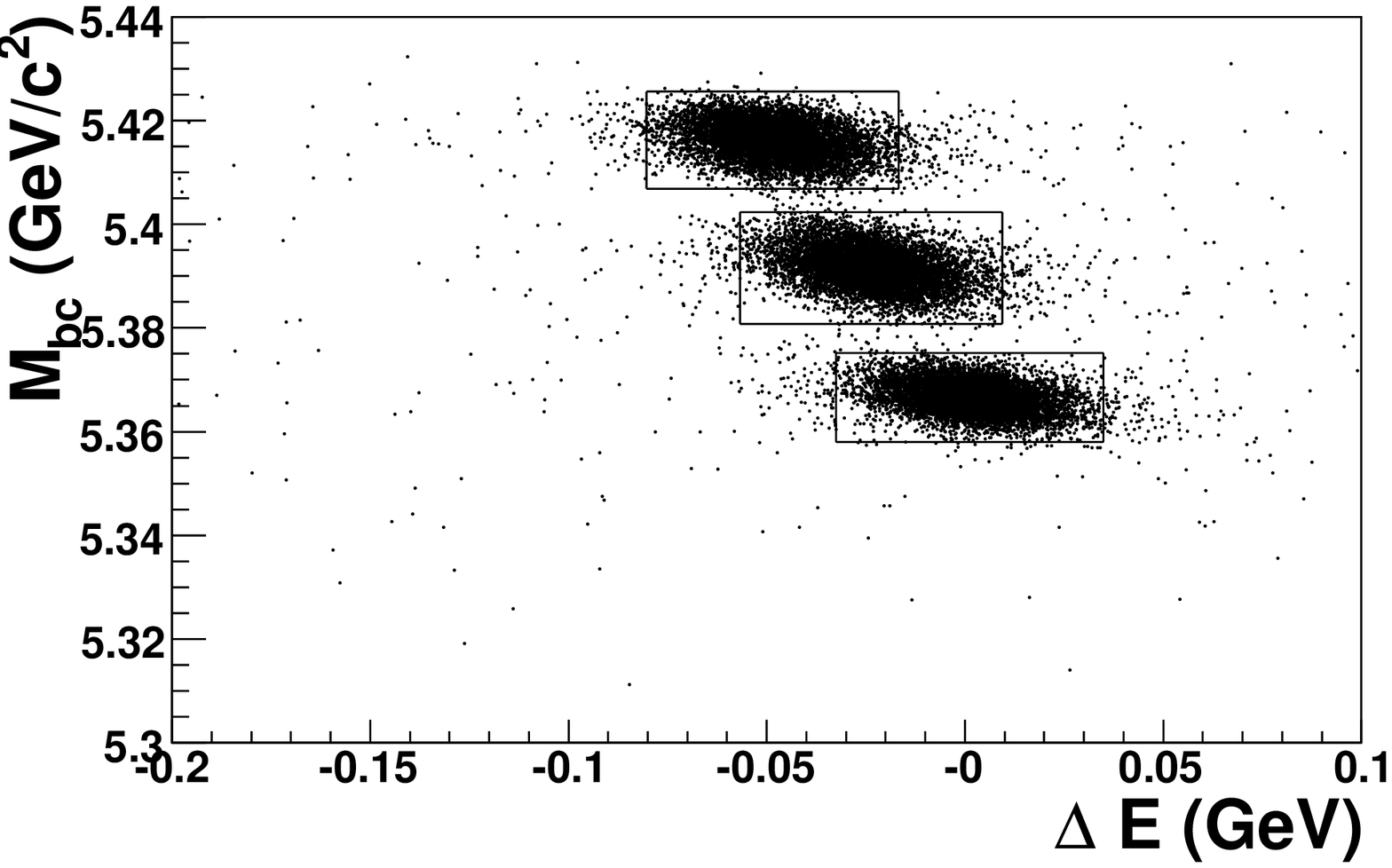}}
{\includegraphics[width=5.2cm]{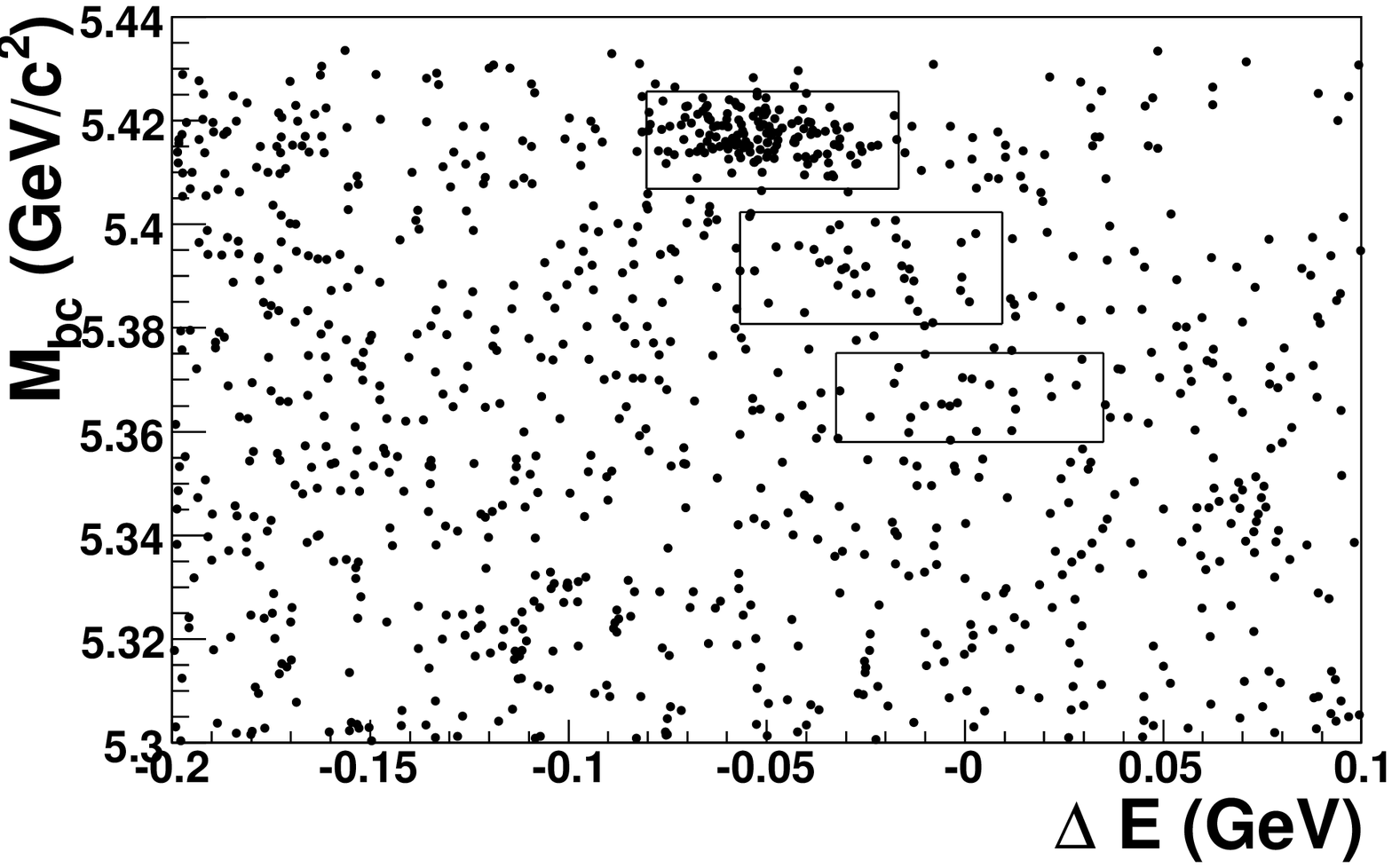}}
\end{center} 
\caption{Illustration of the full reconstruction  method,  $B_s\to D_s\pi$.
Distributions in $\Delta$E and $M_{\rm bc}$ of candidates, (left) Monte Carlo simulation, (right) data, 24FB set.
Also shown are  signal regions for  $B_s^*\bar B_s^*$(upper signal box), $B_s^*\bar B_s$(middle box), and $B_s\bar B_s$(lower box) events~\cite{Dspi}.
\label{fig:Dspi}}
\end{figure}

At the energy of the $\Upsilon$(10860), three types of $B_s$ events are allowed: $B_s\bar B_s$, $B_s^*\bar B_s^*$, and $B_s\bar B_s^*$ (and $\bar B_s B_s^*$) events.
Each is an exclusive 2-body decay, so the energy of the daughter $B_s^{(*)}$ in the collision CM frame is fully constrained.
The method of ``full reconstruction,'' where all decay products are detected and measured, was used with great success for $B_q$ at the $\Upsilon$(4$S$).
The reconstruction of $B_s$ in $B_s\bar B_s$ events is analogous:
each $B_s$ carries energy equal to the beam energy (in the collision CM system), so upon reconstructing a candidate, the quantity $\Delta E$  accumulates at $\Delta E =0$ GeV and $M_{\rm bc}$ at the true $B_s$ mass, $m_{B_s}$.
In the decay $B_s^*\to  B_s\gamma$, the photon carries away essentially all of the released energy, which is equal to the mass difference, $\delta M\approx 50$~MeV/$c^2$.
In a $B_s^*\bar B_s$ event, the $\bar B_s$ ($B_s^* $) carries energy $\sim E_{\rm beam}-\delta Mc^2/2$ ($\sim E_{\rm beam}+\delta Mc^2/2$).  
The daughter $B_s$ from $B_s^* \to B_s\gamma$ carries energy $\sim E_{\rm beam}-\delta Mc^2/2$.
Thus, for both of these $B_s$'s, one can expect reconstructed decays to accumulate around $\Delta E = -\delta Mc^2/2$ and $M_{\rm bc}= m_{B_s}+\delta M/2$.
Carrying the process another step further, both $B_s$'s in $B_s^*\bar B_s^*$ events accumulate at $\Delta E = -\delta Mc^2$ and $M_{\rm bc}= m_{B_s}+\delta M=m_{B_s^*}$.
Given the Belle detector's momentum resolution, these three event types accumulate in well-separated regions of  $\Delta E$ and $M_{\rm bc}$, as shown in Fig.~\ref{fig:Dspi} for $B_s\to D_s^-\pi^+$ candidates, signal MC simulations, and data~\cite{Dspi}.

As can be seen from Fig.~\ref{fig:Dspi}, the $B_s^*\bar B_s^*$ events dominate in the data. 
As explained above, the $M_{\rm bc}$ distribution peaks at $m_{B_s}$ (with very minor corrections).
The $B_s^*-B_s$ mass difference is found from the mean $\Delta E$ of the candidates; the $B_s$ candidate mass reconstructed as $M_{\rm bc}'=\sqrt{(E_{\rm beam}^*+\langle\Delta E\rangle)^2-(p_{\rm cand}^*)^2}$ accumulates at the $B_s$ mass.
The modes $\bar B_s\to D_s^+\pi^-$\{$D_s\to \phi (\to K^+K^-)\pi^-$, $K^{*0}(\to K^+K^-)K^-$,  $K_S(\to \pi^+\pi^-)K^-$\} were reconstructed for this measurement. 
$B_s^*\bar B_s^*$ candidates are  selected by requiring $-0.08<\Delta E<-0.02$~GeV.
From the 24FB data set we measure\cite{Dspi}
\begin{eqnarray*}
m_{B_s^*}&=& 5416.4\pm0.4\pm0.5
\ {\rm MeV}/c^2\\
m_{B_s}&=& 5364.4\pm1.3\pm0.7\ {\rm MeV}/c^2
\end{eqnarray*}

\subsection{Event composition at the $\Upsilon(10860)$ peak}
To study production and decay rates of $B_s$, their abundance and properties 
in $\Upsilon$(10860) events are needed.
This evaluation proceeds in three steps.
First, we measure the hadronic $b\bar b$ cross section~\cite{5Sinclusive}.
We then  find the fraction of $b\bar b$ events containing $B_s$~\cite{5Sinclusive}.
Finally, we measure the relative rates to the three possible event types~\cite{Dspi}.

\subsubsection{${\sigma(e^+e^-\to b\bar b)}$}

\begin{figure}[htb]
\parbox{\halftext}{
\centerline{\includegraphics[width=5.5cm]{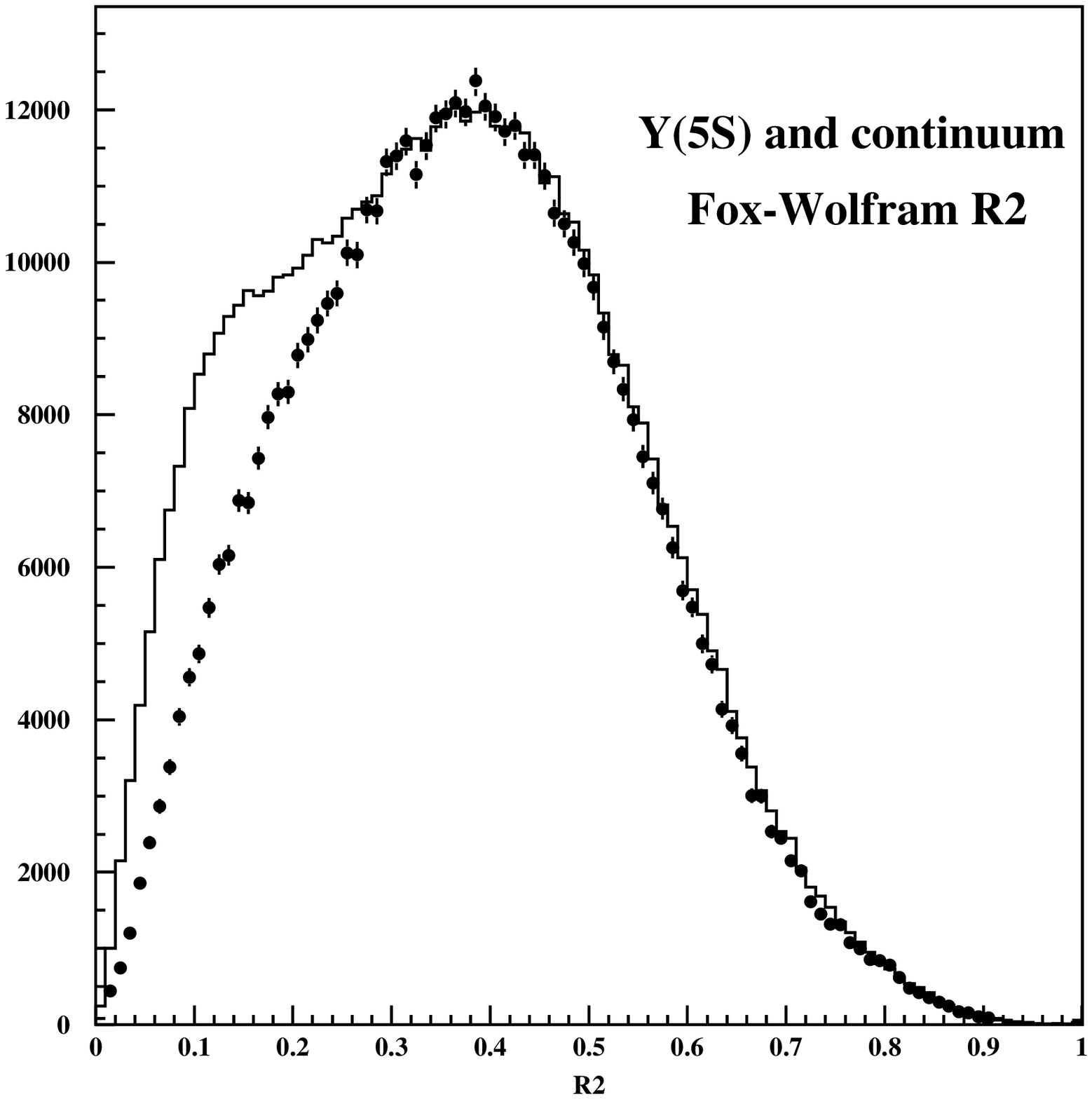}}
\caption{
Distribution in $R_2$, (histogram) data set 2FB, at the $\Upsilon$(10860) and (points) continuum below $\Upsilon$(4$S$), scaled.
}
\label{fig:5S_profile_a}
}
\hfill
\parbox{\halftext}{
\centerline{\includegraphics[width=5.5cm]{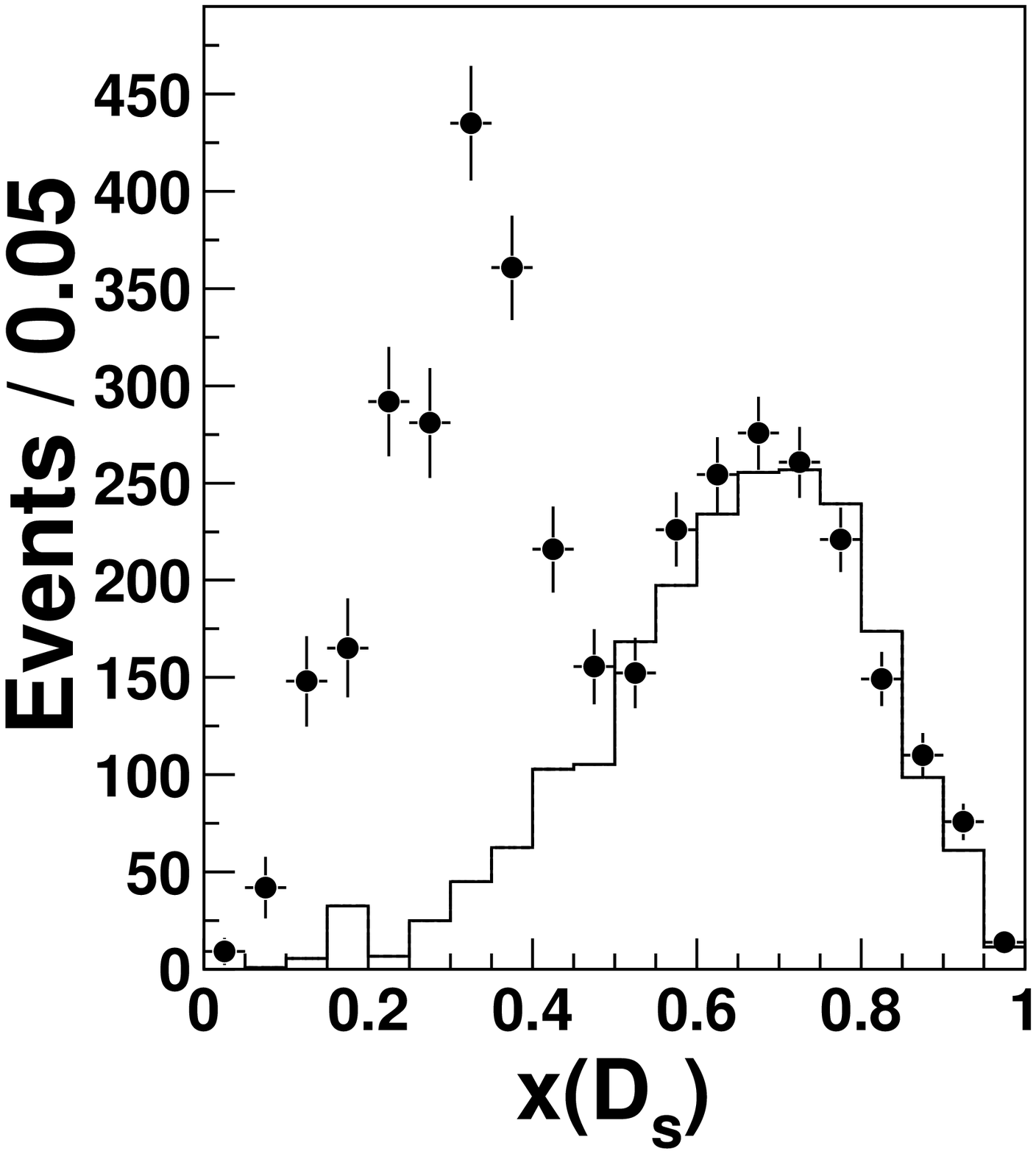}}
\caption{
Distribution of $D_s$ in $x\equiv p_{D_s}/\sqrt{E_{\rm beam}^2-m_{D_s}^2}$, (points) $\Upsilon$(10860) and (histogram) scaled continuum~\cite{5Sinclusive}.
}
\label{fig:5S_profile_b}}
\end{figure}

As is the case at the $\Upsilon$(4$S$), $b\bar b$ events at the $\Upsilon$(10860) (where ``$b\bar b$'' includes both resonance and $b\bar b$ continuum events, which are indistinguishable) are readily distinguished statistically from the continuum of lighter quarks $e^+e^-\to q\bar q$ ($q=u,d,s,c$) via their distribution in $R_2$, the ratio of the second and zeroth Fox\---Wolfram moments~\cite{FW}, a measure of ``jettiness'' that tends to be lower for the more isotropic $b\bar b$ events.
The $R_2$ distribution for the 2FB data and a scaled continuum sample are shown in Fig.~\ref{fig:5S_profile_a}.
We found $\sigma_{e^+ e^-\to b\bar b}=(3.01\pm0.02\pm 0.16)\times 10^2$~pb\cite{5Sinclusive}, which constitutes $\approx$10\% of the total hadronic cross section.
This value was averaged with the corresponding result from CLEO\cite{cleo_5Sb} to obtain the PDG average of $\sigma_b=(3.02\pm 0.14)\times 10^2$~pb~\cite{pdg2012}.

\subsubsection{${\sigma(e^+e^-\to B_s\bar B_s)}$/${\sigma(e^+e^-\to b\bar b)}$}
The fraction ($f_s$) of $b\bar b$ events that hadronize to $B_s$  ($B_s^{(*)} \bar B_s^{(*)}$) may be determined through measurement of the inclusive rate ${\mathcal B}(\Upsilon(10860)\to D_s X)\equiv{\mathcal B}_{\Upsilon}$.
The $B_s$ decays predominantly via the spectator mechanism, as do the lighter $B$ mesons, and as such we can assume a direct correspondence between $B\to DX$ and $B_s\to D_sX$ for a large fraction of the decays. 
Based on our understanding of the mechanisms of $B$ decay and the measured branching fractions ${\cal B}(B\to DX)$ and  ${\cal B}(B\to D_sX)$, a reasonable estimate may be made\cite{cleo_5S}: 
$
{\cal B}(B_s\to D_sX)=(92\pm 11)\%.
$
The inclusive rate of $\Upsilon$(5$S$)$\to D_sX$ is an average over $B_s$, $B_d$, and $B_u$, weighted by abundance:
\begin{eqnarray}
\frac{{\cal B}(\Upsilon(5S)\to D_sX)}{2}&=&f_s\cdot{\cal B}(B_s\to D_s X)\nonumber\\
&&+(1-f_s)\frac{{\cal B}(\Upsilon(4S)\to D_s X)}{2}
\label{eq:inclDs}
\end{eqnarray}
where $f_s$ is the fraction of $B_s$ and we assume that $B_d$ and $B_u$ are produced equally and that non-$B$ production is negligible.
The distributions of $D_s$ in normalized momentum $x\equiv p_{D_s}/\sqrt{E_{\rm beam}^2-m_{D_s}^2}$ for $\Upsilon$(5$S$) and scaled continuum data are shown in Fig.~\ref{fig:5S_profile_b}.
The measured value, ${\cal B}(\Upsilon(5S)\to D_sX)/2=(22.6\pm 1.2\pm 2.8)\%$ for the 2FB data set, is fed into Eq.~(\ref{eq:inclDs}) and solved to obtain\cite{5Sinclusive}
$f_s=(16.4\pm 1.4\pm 4.1)\%$,
which corresponds to $(4.95\pm 1.31)\times 10^4\ B_s$~events/fb$^{-1}$.
The same analysis may be performed for $D^0$ to obtain an independent value of $f_s$, albeit with larger uncertainties;  ${\mathcal B}(B_s\to D^0X)\ll {\mathcal B}(B_q\to D^0X)$.
The results are combined to obtain\cite{5Sinclusive}
$f_s=(18.0\pm 1.3\pm 3.2)\%.$
The Belle result is averaged with the corresponding CLEO result\cite{cleo_5Sb} to obtain the PDG average\cite{pdg2012} 
\begin{eqnarray*}
f_s=(19.5^{+3.0}_{-2.3})\%.
\end{eqnarray*}
The same method applied to the 121FB set yields 
\begin{eqnarray*}
f_s=(17.1\pm 3.0)\%.
\end{eqnarray*}


\subsubsection{$B_s^*\bar B_s^*:B_s^*\bar B_s:B_sB_s$}		

As described in Sect.~\ref{sec:BsMass}, reconstructed $B_s$ signals from the three event types are well separated in $\Delta E$ and $M_{\rm bc}$.
These three modes account for 100\% of $B_s$ events, so the fraction comprised by each is derived from a simultaneous fit to $\Delta E$ and $M_{\rm bc}$ that yields all three signals.
For this measurement we use $B_s\to D_s^-\pi^+$, the mode with the greatest statistical significance.
To date, statistically significant signals have been observed in the $B_s^{*} \bar B_s^{*}$ and $B_s^*\bar B_s + B_s\bar B_s^*$ channels in the 24FB data set, from which we obtain\cite{Dspi} 
\begin{eqnarray*}
f_{B_s^*B_s^*}
&\equiv& \frac{\sigma(e^+e^-\to B_s^*\bar B_s^*)}{\sigma(e^+e^-\to B_s^{(*)}\bar B_s^{(*)})}=(90.1^{+3.8}_{-4.0}\pm 0.2)\% \\
f_{B_s^*B_s}&\equiv& \frac{\sigma(e^+e^-\to B_s^*\bar B_s + B_s\bar B_s^*)}{\sigma(e^+e^-\to B_s^{(*)}\bar B_s^{(*)})} =(7.3^{+3.3}_{-3.0}\pm 0.1)\% 
\end{eqnarray*}
The value $f_{B_s^*B_s^*}=(87.0\pm 1.7)\%$ from the 121FB data set (unpublished)~\cite{remi_full} is used in evaluating branching fractions from the 121FB set.

\subsubsection{$B^{(*)}\bar B^{(*)}(\pi)(\pi)$}		
\begin{figure}[h!]
\begin{center}
{\includegraphics[width=5.2cm]{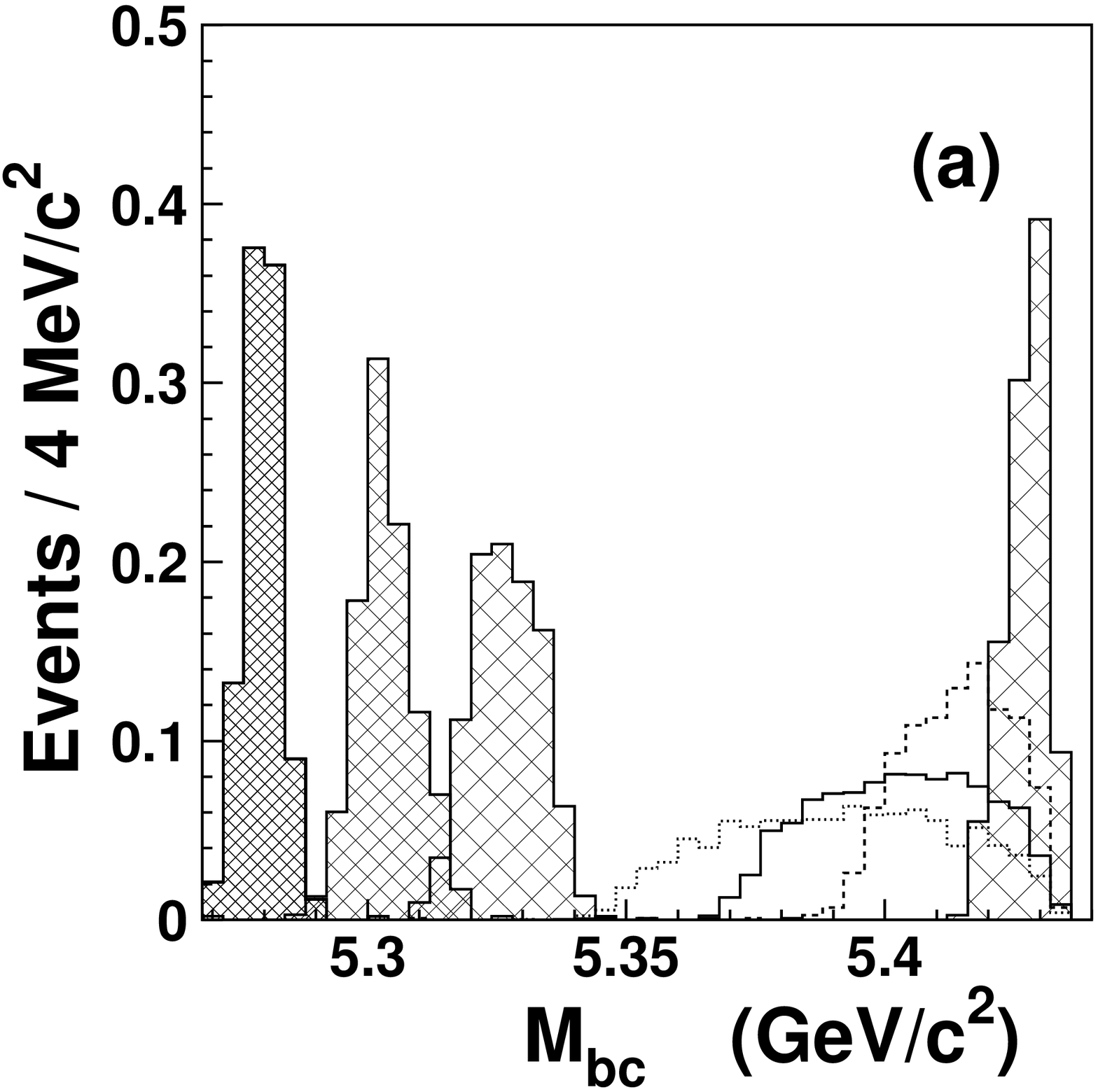}}
{\includegraphics[width=5.2cm]{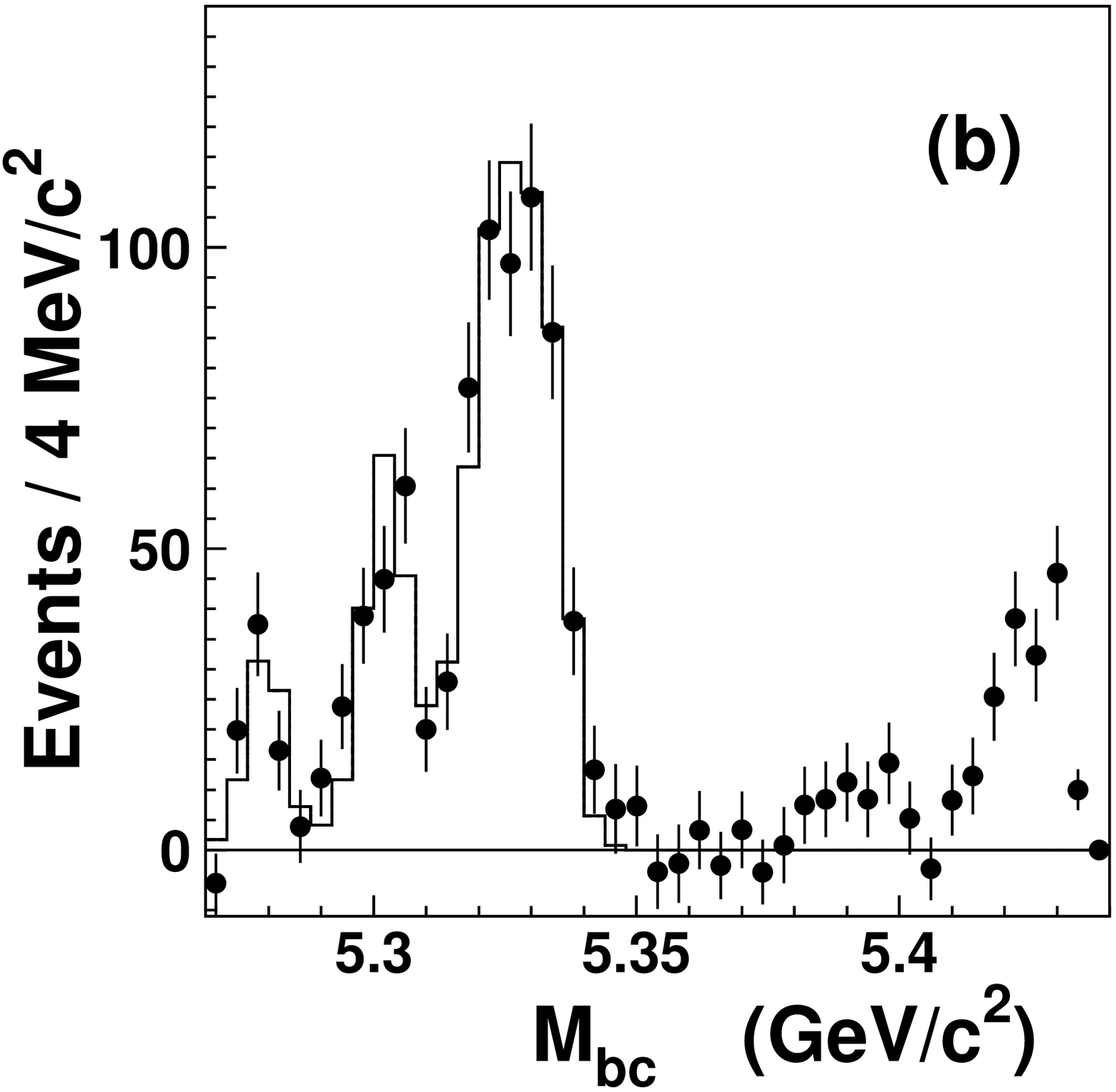}}
\end{center}
\caption{(a) Distributions in $M_{\rm bc}$ for reconstructed 
$B^0 \to D^- \pi^+$, for 
$B\bar{B}$, $B\bar{B}^\ast+B^\ast\bar{B}$, $B^\ast\bar{B}^\ast$, and
$B\bar{B}\,\pi \pi$ channels (cross-hatched histograms, left to right)
and for the three-body
channels $B\bar{B}^\ast\,\pi+B^\ast\bar{B}\,\pi$ (plain histogram),
$B\bar{B}\,\pi$ (dotted), and $B^\ast\bar{B}^\ast\,\pi$ (dashed).
The distributions are normalized to unity. 
(b) $M_{\rm bc}$ distribution in data after background subtraction.
The sum of the five studied $B$ decays (points with error bars) 
and results of the
fit (histogram) used to extract the two-body channel fractions are shown.
}  
\label{fig:B-recon}
\end{figure}

The well-tuned methods of $B$ reconstruction at the $\Upsilon$(4$S$) (Sect.~\ref{sec:FRB2taunu}) have been applied to study the more complicated assortment of $B$ events at the $\Upsilon$(10860)\cite{BBpi}.
The following final states that include non-strange $B$ mesons are energetically allowed:
$B_q^{(*)} \bar B_q^{(*)}$, $B_q \bar B_q^{(*)}\pi$, $B_q \bar B_q\pi\pi$.
The relative rates can improve our understanding of hadronization dynamics.
Neutral and charged $B$'s are reconstructed in the following modes and submodes:
$B^+\to J/\psi K^+,\ \bar D^0\pi^+$; 
$B^0\to J/\psi K^{*0},\ D^-\pi^+$;
$J/\psi \to e^+e^-,\ \mu^+\mu^-$;
$K^{*0}\to K^+\pi^-$;
$\bar D^0\to K^+\pi^-,\ K^+\pi^+\pi^-\pi^-$;
$D^-\to K^+\pi^-\pi^-$.
As with the fully reconstructed $B_s$, the signal events populate the $(\Delta E,M_{\rm bc})$ plane in clusters depending on the type of event.
Figure~\ref{fig:B-recon}(a) shows the projections in $M_{\rm bc}$ of the distributions for the various event types.
The distribution of candidates in data, after  background subtraction, are shown in Fig.~\ref{fig:B-recon}(b).
While the distributions for events containing additional pions overlap each other, it is clear from the data that their contribution is relatively small and that the majority of the rate is due to  two-body events, $B^{(*)}\bar B^{(*)}$.
It is also noted that there is an accumulation of events in the region of high $M_{\rm bc}$, where $B\bar B\pi\pi$ events would accumulate, according to the MC simulation.
The fractions of $b\bar b$ events fragmenting to $B\bar B$, $B^*\bar B$, and $B^*\bar B^*$ are measured to be $(5.5^{+1.0}_{-0.9}\pm 0.4)\%$, $(13.7\pm 1.3\pm 1.1)\%$, and $(37.5^{+2.1}_{-1.9}\pm 3.0)\%$, respectively.
The events where $M_{\rm bc}$ is above the two-body limit are grouped together as ``large $M_{\rm bc}$'' and found to comprise $(17.5^{+1.8}_{-1.6}\pm 1.3)\%$.

\begin{figure}[h!]
\begin{center}
{\includegraphics[width=5.2cm]{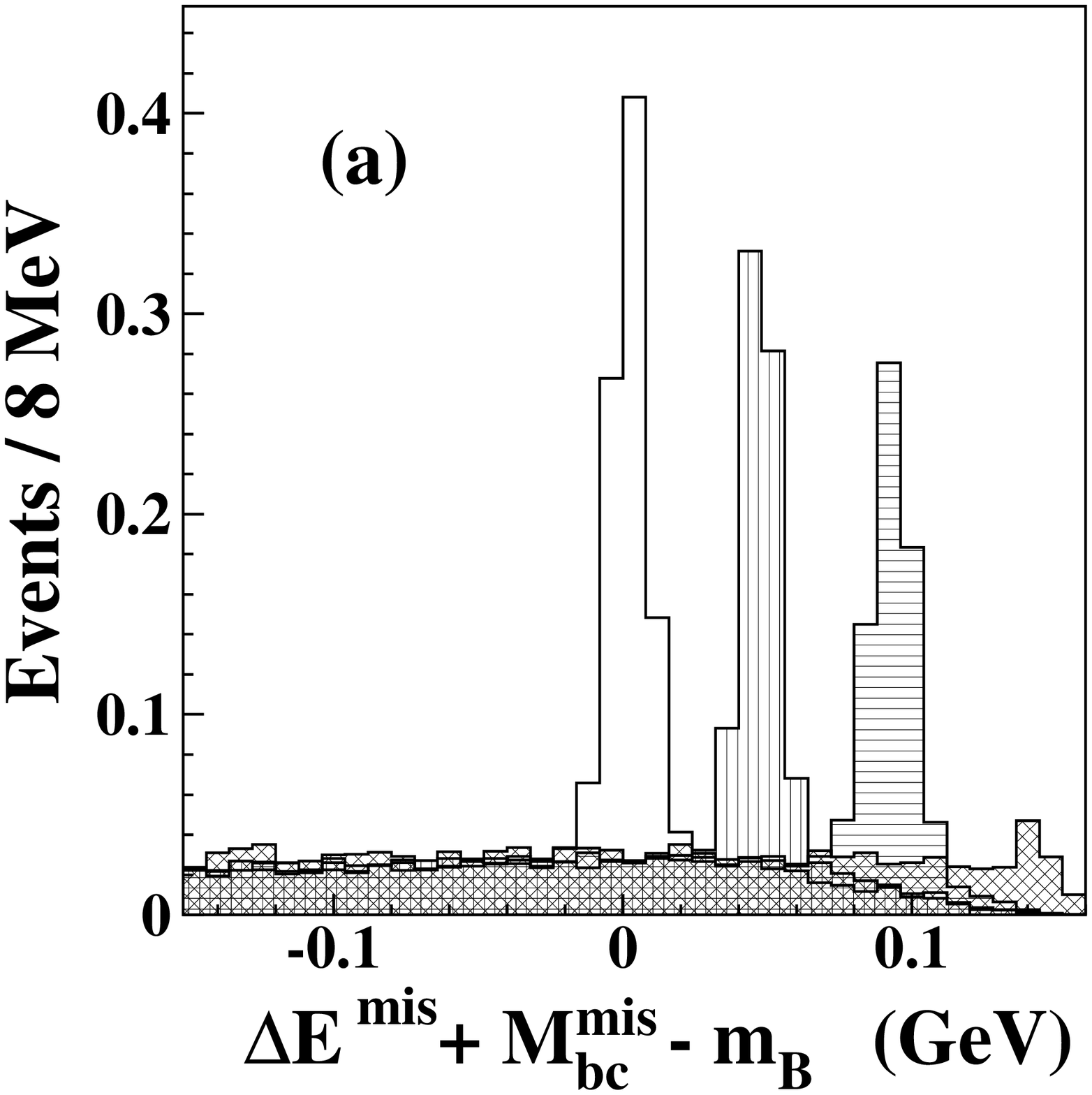}}
{\includegraphics[width=5.2cm]{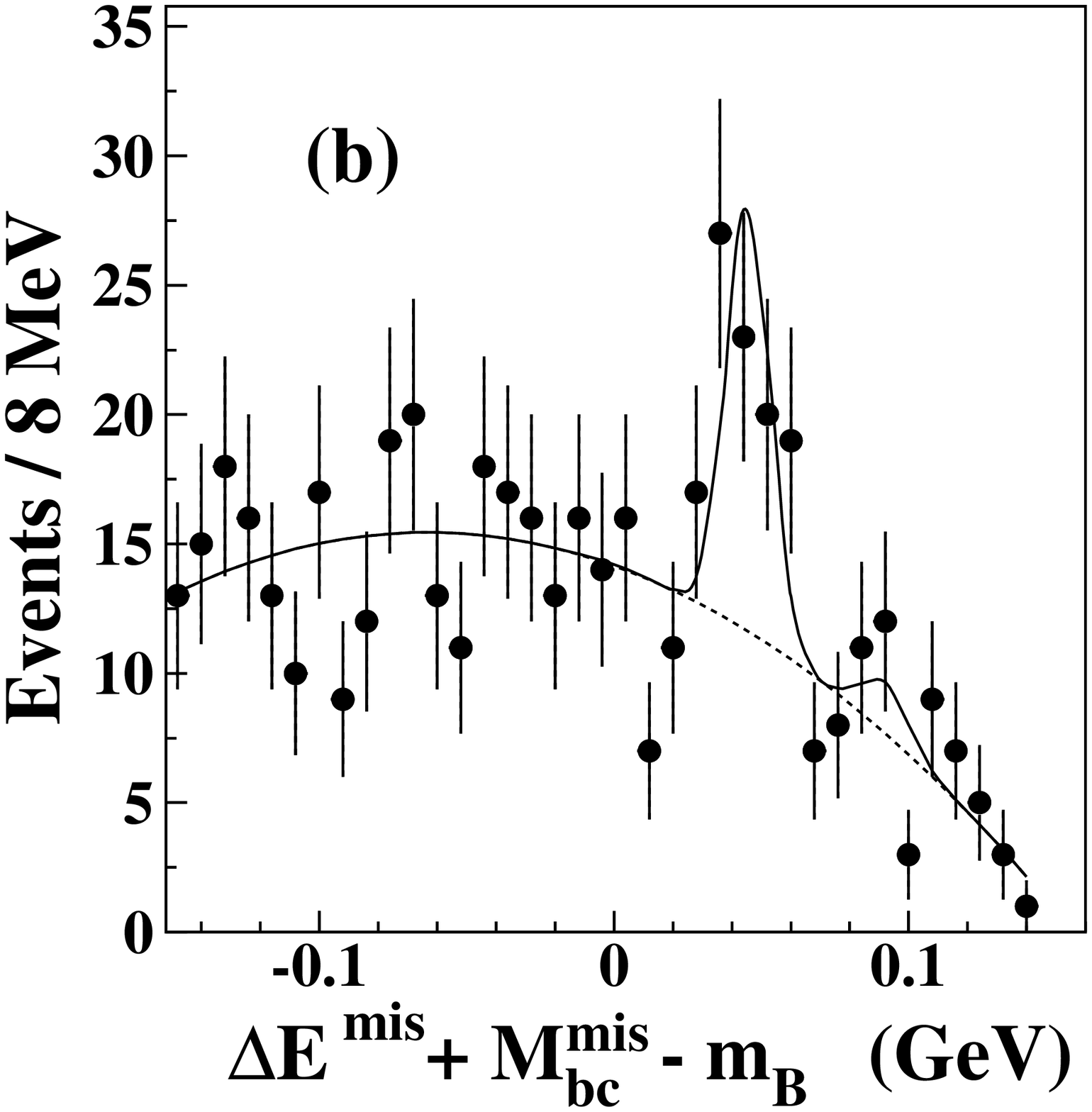}}
\end{center}
\caption{(a) The \mbox{$\Delta E_{\rm miss}+M_{{\rm bc}, \rm miss}-m_B$}
distribution normalized per reconstructed $B$ meson
for MC simulated $B^+ \to J/\psi K^+$ decays
in the (peaks from left to right) $B\bar{B}\,\pi^+$,
$B\bar{B}^\ast\,\pi^+ +B^\ast\bar{B}\,\pi^+$,
$B^\ast\bar{B}^\ast\,\pi^+$, and $B\bar{B}\,\pi \pi$
channels.
(b) The \mbox{$\Delta E_{\rm miss}+M_{{\rm bc}, \rm miss}-m_B$} data
distribution for right-sign $B^{-/0}\,\pi^+$ combinations 
for all five studied $B$ modes. The curve shows the
result of the fit~\cite{BBpi}}.
\label{fig:Bpi-frecon}
\end{figure}

Events containing one or more additional pions may be identified by pairing reconstructed $B$'s with additional charged pions in the event and examining the residual, or {\it missing}, event energy and momentum, $E_{\rm miss}$ and $\vec{P}_{\rm miss}$, which by inference are carried by the opposing $B^{(*)}$ and up to one  additional pion.
From these we reconstruct $\Delta E_{\rm miss}$ and $M_{{\rm bc}, \rm miss}$.
Projections onto \mbox{$\Delta E_{\rm miss}+M_{{\rm bc}, \rm miss}-m_B$} for various simulated event types are shown in Fig.~\ref{fig:Bpi-frecon}(a).
The corresponding distribution in data, with the fit result, is shown in Fig.~\ref{fig:Bpi-frecon}(b).
The fractions of $b\bar b$ events hadronizing to three-body modes $B\bar{B}\pi$, $B\bar{B}^\ast\pi$, and $B^\ast\bar{B}^\ast\pi$ are found to be  $(0.0 \pm 1.2 \pm 0.3)\%$, $(7.3\,^{+2.3}_{-2.1} \pm 0.8)\%$, and $(1.0\,^{+1.4}_{-1.3} \pm 0.4)\%$, respectively.
Paradoxically, no evidence for $B\bar B\pi\pi$ was observed, so this channel does not account for the remaining $(9.2\,^{+3.0}_{-2.8} \pm 1.0)\%$ of the ``large $M_{\rm bc}$'' contribution observed in  $\Upsilon$(10860)$\to BX$.
The residual is quantitatively consistent with initial state radiation, $e^+e^-\to e^+e^-\gamma,\ e^+e^-\to b\bar b$, where about half the $b\bar b$ form the $\Upsilon$(4$S$) resonance~\cite{BBpi}.

\subsection{$B_s$ decays}

To a large degree, the general properties of the $B_s$ meson parallel those of the non-strange $B$ mesons.
Like its lighter cousins, the $B_s$ is expected to decay predominantly by a spectator process, where the lighter valence quark has no role in the weak interaction, and its spectator-dominated properties such as lifetime are expected to be similar.
This is serendipitous at the $B$-factory, allowing many of the techniques developed for analysis of $B$'s at the $\Upsilon$(4$S$) to be applied to $B_s$ at the $\Upsilon$(5$S$).
Furthermore, close correspondences between the hadronic final states in spectator decays of $B_s$ and $B_d$ allow for sensitive tests of quark\---hadron duality and of hadronic models that may reduce theoretical uncertainties  limiting precision CKM tests in $B$ physics.

Being electrically neutral, the $B_s$ experiences mixing and may thus address  questions of interest regarding $CP$ violation and roles for physics beyond the Standard Model.
Notably, $B_s$ experiences a much higher rate of mixing than $B_d$, and very little $CP$ violation in the SM.

All branching fractions described in this section are listed in Table~\ref{tab:BsDecays}.
The branching fractions measured in the 24FB set are evaluated using $f_{B_s^*B_s^*}=(90.1^{+3.8}_{-4.0})\%$, $f_s=(19.5^{+3.0}_{-2.3})\%$, and $\sigma_{e^+e^-\to b\bar b}=0.302\pm 0.014$~nb (a weighted average from \cite{5Sinclusive,cleo_5Sb}).
The results based on the 121FB set have used $N_{B_s^{(*)}\bar B_s^{(*)}}= (7.1 \pm 1.3) \times 10^6={\cal L}\times \sigma_{e^+e^-\to b\bar b}\times f_s$ and $f_{B_s^*B_s^*}=(87.1\pm 3.0)\%$.

\subsubsection{Modes with single $D_s$}

\begin{figure}[ht]
\begin{center}
\includegraphics[width=5.2cm]{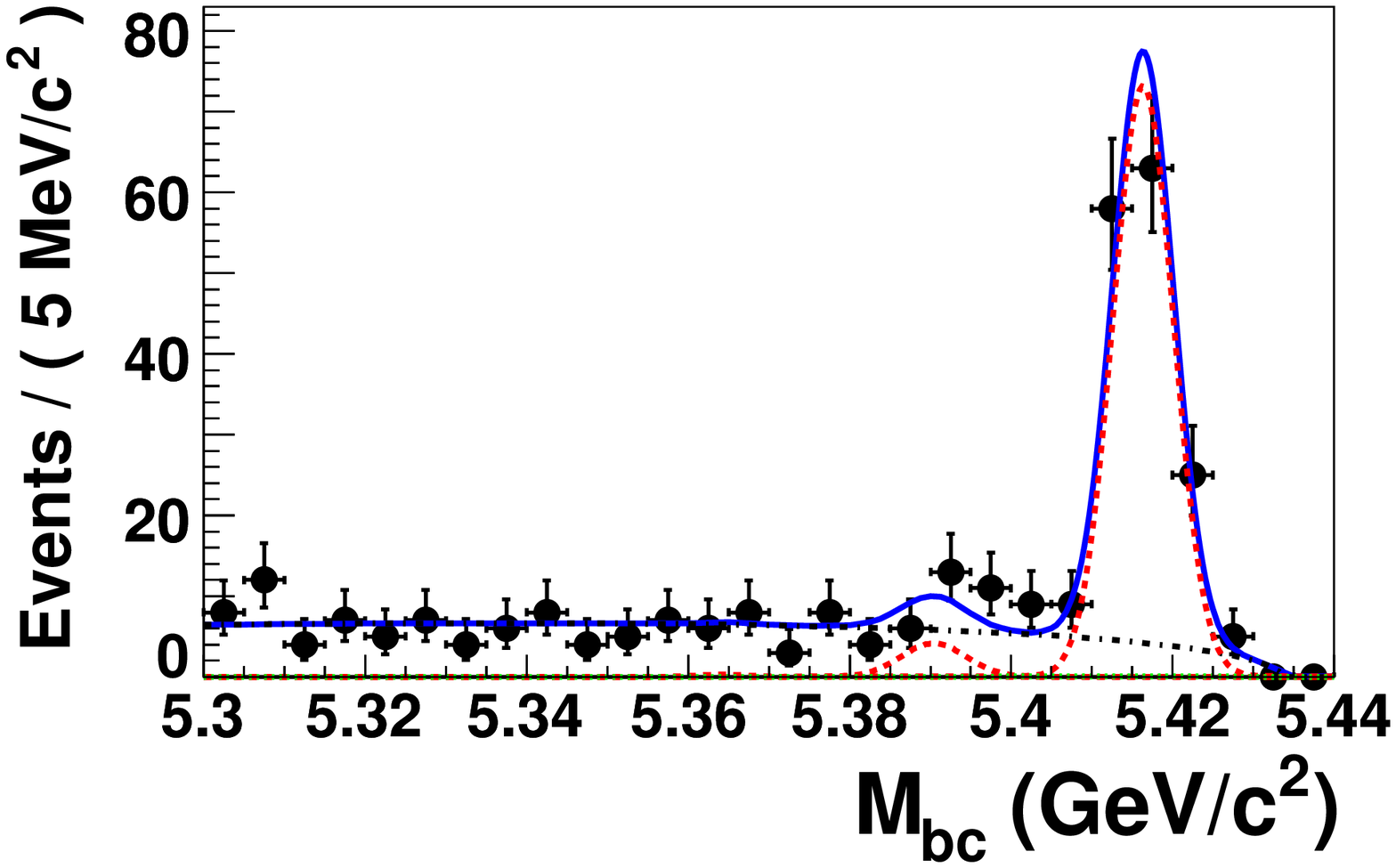}
\includegraphics[width=5.2cm]{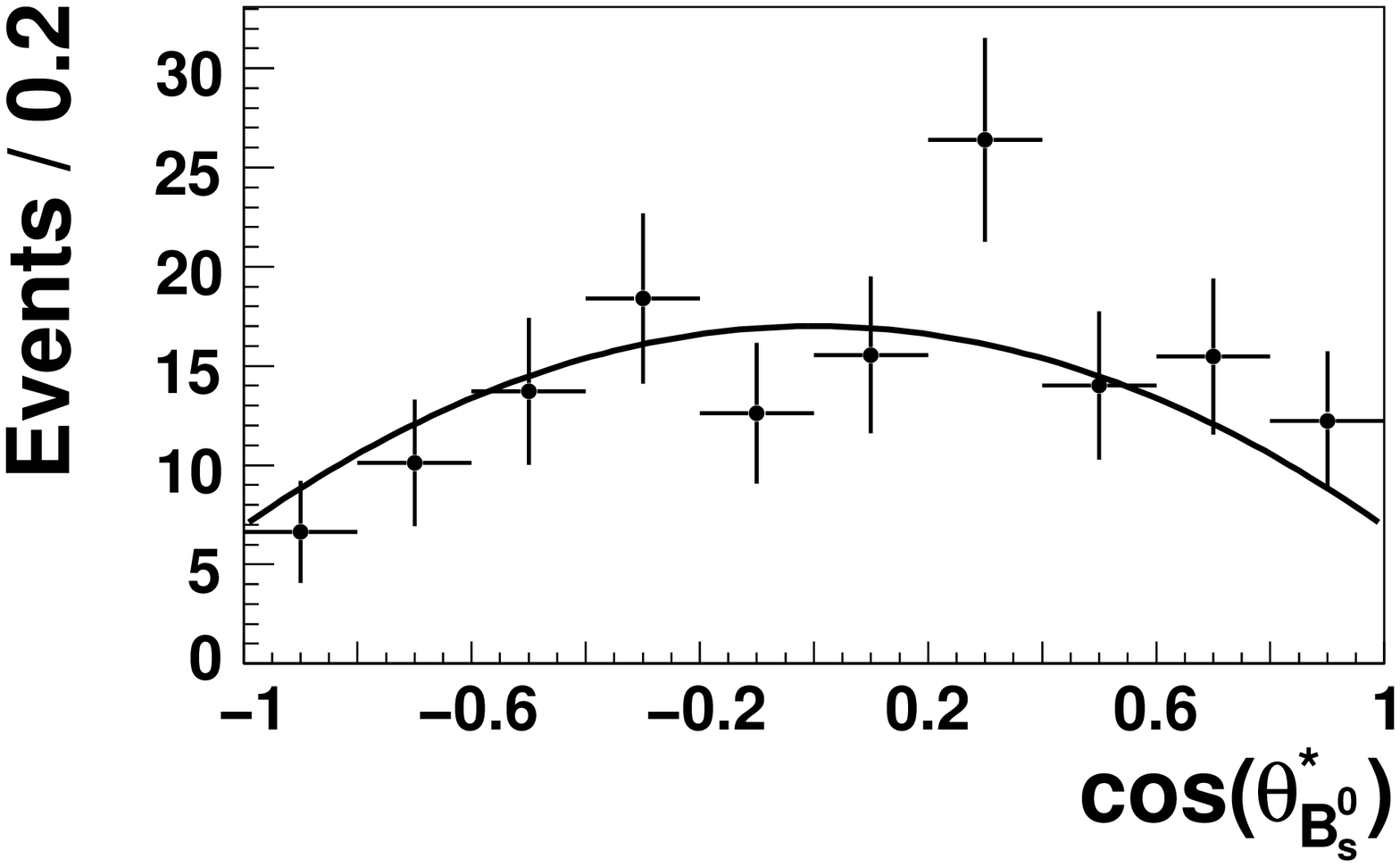}
\end{center} 
\caption{(Left) $M_{\rm bc}$ distribution of $B_s\to D_s^{-}\pi^+$ candidates with
          $\Delta E$ in the $B_s^* \bar B_s^*$ signal region $[-80,-17]$~MeV, 24FB data set~\cite{Dspi}.
The different fitted components are shown with dashed curves for the signal, dotted curves for the $B_s\to D_s^{*-}\pi^+$ background, and dash-dotted curves for the continuum.
(Right) Fitted distribution of the cosine of the angle between the $B_s$ momentum and the beam axis in the CM frame for the $\Upsilon(5S)\to B_s^*\bar B_s^*$ signal.
\label{fig:Dssth}}
\end{figure}

The decays $B_s\to D_s^{(*)-}h^+$, where $h$ is a light non-strange meson, proceed dominantly via a CKM-favored spectator process.
The $D_s$ are reconstructed in the modes
$\phi (\to K^+K^-)\pi^-$, $K^{*0}(\to K^+K^-)K^-$, and $K_S(\to \pi^+\pi^-)K^-$ and $\rho^\pm$ in $\pi^\pm\pi^0$. 
As described in Sect.~\ref{sec:BsMass}, the signal is extracted by fitting the distributions in $\Delta E$ and $M_{\rm bc}$ (and decay angles, 
for $B_s\to D_s^{*-}\rho^+$). 
Shown in Fig.~\ref{fig:Dssth}(left) is the projection into $M_{\rm bc}$ for 
$B_s\to D_s^{-}\pi^+$ candidates in the 24FB set.
The branching fraction for modes other than $D_s^{-}\pi^+$ are obtained using only the $B_s^*\bar B_s^*$ sample and the value of $f_{B_s^*B_s^*}$ measured with $D_s^{-}\pi^+$~\cite{Dspi,Dssth}.

The distribution of the angle between the $B_s$ momentum and the beam axis in the CM frame, $\theta_{B_s}^*$, is of theoretical interest \cite{PRD_55_272} and is presented in Fig.~\ref{fig:Dssth}(right) for the signal events in the $B_s^*\bar B_s^*$ region. A fit of the distribution to $1+a\cos^2\theta_{B_s}^{*}$  returns $\chi^2/{\rm{n.d.f.}}=8.74/8$ with $a=-0.59^{+0.18}_{-0.16}$. {We naively expect $a=-0.27$ by summing over all the possible polarization states.}

For  $B_s\to D_s^{*-}\rho^+$, a pseudoscalar decay to two vectors, the distributions in the helicity angles $\theta_{D_s^{*-}}$ and $\theta_{\rho^+}$ depend on the relative contribution from the different helicity states, which depends on the detailed hadronization mechanism for the decay; for example, the factorization hypothesis predicts that longitudinal polarization dominates: $f_L\approx 88\%$~\cite{factorization}.
A four-dimensional fit yields $77.7^{+14.6}_{-13.3}$ ($7.4\sigma$) signal events
and  $f_L = 1.05{^{+0.08}_{-0.10}} {^{+0.03}_{-0.04}}$~\cite{Dssth}.

\begin{table}
	\centering
	\caption{\label{tab:BsDecays}  Branching fractions with statistical 
and systematic uncertainties.  A third uncertainty, due to $f_s$, is quoted 
where it is separated from other systematics. The data set analyzed is 
identified in the rightmost column.
	  }
	  \begin{tabular}{cccc}
	    	    \hline\hline
	    	    Mode & ${\cal B}(10^{-4})$  & Data set\\ \hline
	    	    Single-$D_s$ modes &&\\
	    $D_s^{-}\pi^+$&$36.7{^{+3.5}_{-3.3}}{^{+4.3}_{-4.2}}\pm 4.9$& 24FB\\
	    $D_s^{*-}\pi^+$&$24^{+5}_{-4}\pm 3\pm 4$& 24FB\\
	    $D_s^-\rho^+$&$85^{+13}_{-12}\pm 11\pm 13$& 24FB\\
	    $D_s^{*-}\rho^+$&$118^{+22}_{-20}\pm 17\pm 18$& 24FB\\
	    \hline
	    $c\bar c s\bar s$ modes && \\
	    $J/\psi\eta$ & $5.10\pm 0.50\pm 0.25^{+1.14}_{-0.79}$ & 121FB\\
	    $J/\psi\eta\prime$ & $3.71\pm 0.61\pm 0.18{^{+0.83}_{-0.57}}$ & 121FB\\
	    $J/\psi f_0(980)$ & $1.16{^{+0.31}_{-0.19}}{^{+0.15}_{-0.17}}{^{+0.26}_{-0.18}}$ & 121FB\\
	    $J/\psi f_0(1370)$ & $0.34{^{+0.11}_{-0.14}}{^{+0.03}_{-0.02}}{^{+0.08}_{-0.05}}$ & 121FB\\
	    $D_s^{*+}D_s^{*-}$ & $200\pm 30\,\pm 50$ &  121FB\\
	    $D_s^{*+}D_s^{-}+c.c.$ & $180\pm 20\,\pm 40$ &  121FB\\
	    $D_s^{+}D_s^{-}$ & $58^{+11}_{-9}\,\pm 13$ &  121FB\\
	    $D_s^{(*)+}D_s^{(*)-}$ sum & $430 \pm 40\,^{+60}_{-50}\,\pm 90$ &  121FB\\ \hline
	    $h\bar h$ modes && \\
	    $K^+K^-$ & $0.38^{+0.10}_{-0.09}\pm 0.05 \pm 0.05$ &  24FB\\
	    $K^0\bar K^0$ & $<0.66$ (90\% C.L.) &  24FB\\
	    $K^-\pi^++c.c.$ & $<0.26$ (90\% C.L.) &  24FB\\
	    $\pi^+\pi^-$ & $<0.12$ (90\% C.L.) &  24FB\\ \hline
	$\phi\gamma$ & $0.57 { ^{+0.18}_{-0.15}} { ^{+0.12}_{-0.11}}$ & 24FB \\
	$\gamma\gamma$ & $<870$ (90\% C.L.) &  24FB\\
	    \hline\hline
	    \vspace{5pt}
	  \end{tabular}
\end{table}

\subsubsection{Flavor-neutral channels}

An interesting characteristic of  the $B_s$ stems from the fact that it experiences an appreciable rate to the flavor-neutral combination $c\bar c s\bar s$, via a tree-level CKM-favored process. 
The massiveness of the participating quarks and proximity to mass threshold  argue for the applicability of predictions at the limit $m^{}_{(b,c)}\ra\infty$ with $(m^{}_b-2m^{}_c)\ra 0$ and $N^{}_c ({\rm number\ of\ colors})\ra\infty$, where the $c\bar c s\bar s$ final states are \cp-even and the $D^{*\pm}_s D^{\mp}_s$ and $D^{*+}_s D^{*-}_s$ modes (along with $D_s^+ D_s^-$) saturate the width difference $\Delta\Gamma^{CP}_s$ between the two $CP$-eigenstates~\cite{aleksan}. This parameter equals $\dgs/\cos\phi^{}_s$, 
where $\dgs$ is the decay width difference between the mass 
eigenstates, and $\phi^{}_s$ is 
the \cp-violating phase in $\bs$\---$\bsbar$ mixing~\cite{Dunietz}.
Thus the summed branching fraction 
${\cal B}(B_s^0\to D_s^{(*)+}D_s^{(*)-})$
gives a constraint in the $\dgs$\---$\phi_s$ 
parameter space. Both parameters can receive contributions 
from NP; see, e.g., Refs.~\cite{Nierste,new_physics}. 
Assuming negligible \cp\ violation ($\phi^{}_s\!\approx\!0$), the branching fraction is related to $\dgs$ via 
\begin{equation}
\dgs/\gs = 2{\cal B}/(1-{\cal B}). 
\label{eqn:DeltaGamma}
\end{equation}
The quantity of interest, the summed branching fraction ${\cal B}={\cal B}(B_s\to D_s^{(*)+}D_s^{(*)-}$, is more easily measured in $e^+e^-\to \Upsilon$(5$S$) than at a hadron machine because the decay $D_s^*\to D_s\gamma$ can be fully reconstructed. 

The final Belle result is based on the 121FB set~\cite{esen12}.
It includes the first measurement of the fraction of longitudinal polarization ($f_L$) 
of $B^0_s\ra D^{*+}_s D^{*-}_s$. The final states reconstructed consist of 
$D^+_sD^-_s$, $D^{*+}_sD^-_s\!+\!D^{*-}_sD^+_s$ ($\equiv D^{*\pm}_sD^\mp_s$), and 
$D^{*+}_sD^{*-}_s$, where 
$D^{*+}_s\ra D^+_s\gamma$,
$D^+_s\ra \phi\pi^+$,
$K^0_S\,K^+$,
$\kstz K^+$,
$\phi\rho^+$,
$K^0_S\,\kstp$, and 
$\kstz \kstp$,
$K^0_S\ra\pi^+\pi^-$,
$K^{*0}\ra K^+\pi^-$,
$K^{*+}\ra K^0_S\pi^+$,
$\phi\ra K^+ K^-$,
$\rho^+\ra\pi^+\pi^0$, and
$\pi^0\ra\gamma\gamma$~\cite{charge-conjugates}.

Events containing candidates satisfying 
$5.25\mbox{\gevm}<\mbc <5.45$\gevm\ and $-0.15\mbox{\geve}<\de <0.10${\geve} are selected.
Approximately half of the selected events have multiple \bsdsds\ candidates. These typically arise from photons produced via $\pi^0\ra\gamma\gamma$ that are wrongly assigned as $D_s^*$ daughters. For these events we select the candidate that minimizes a $\chi^2$ constructed from the reconstructed $D_s^+$ and (if present) $D_s^{*+}$ masses. 

Signal yields are measured by performing a two-dimensional unbinned maximum-likelihood fit to the $\mbc$\---$\de$ distributions. 
The combinatorial effects of analyzing multiple multi-body decays present a particular challenge in this analysis.
The signal PDFs have three components: correctly reconstructed (CR) decays; ``wrong combination'' (WC) decays in which a non-signal track or $\gamma$ is included in place of a true daughter track or $\gamma$; and ``cross-feed'' (CF) decays in which a $D^{*\pm}_s D^{\mp}_s$ ($D^{*+}_s D^{*-}_s$) is reconstructed as a $D^+_s D^-_s$ ($D^+_s D^-_s$ or $D^{*\pm}_s D^{\mp}_s$), or a $D^+_s D^-_s$ ($D^{*\pm}_s D^{\mp}_s$) is reconstructed as a $D^{*\pm}_s D^{\mp}_s$ or $D^{*+}_s D^{*-}_s$ ($D^{*+}_s D^{*-}_s$). In the former case, the $\gamma$ from $D^{*+}_s\ra D^+_s\gamma$ is lost and $\de$ is shifted down by 100\---150\meve; this is called ``CF-down.'' In the latter case, an extraneous $\gamma$ is included and $\de$ is shifted up by a similar amount; this is called ``CF-up.'' In both cases $\mbc$ remains almost unchanged.
The small contributions from $\bs\bsbar$ and $\bs\bsbarst$ events are fixed relative to $\bsst\bsbarst$ according to our measurement on $B_s^0 \to D^-_s \pi^+$ decays~\cite{remi_full}. The fitted signal yields from $\bsst\bsbarst$ only are used to determine the branching fractions.

\begin{figure}
\hbox{\centerline{
\includegraphics[width=4.5 cm,height=4.5 cm]{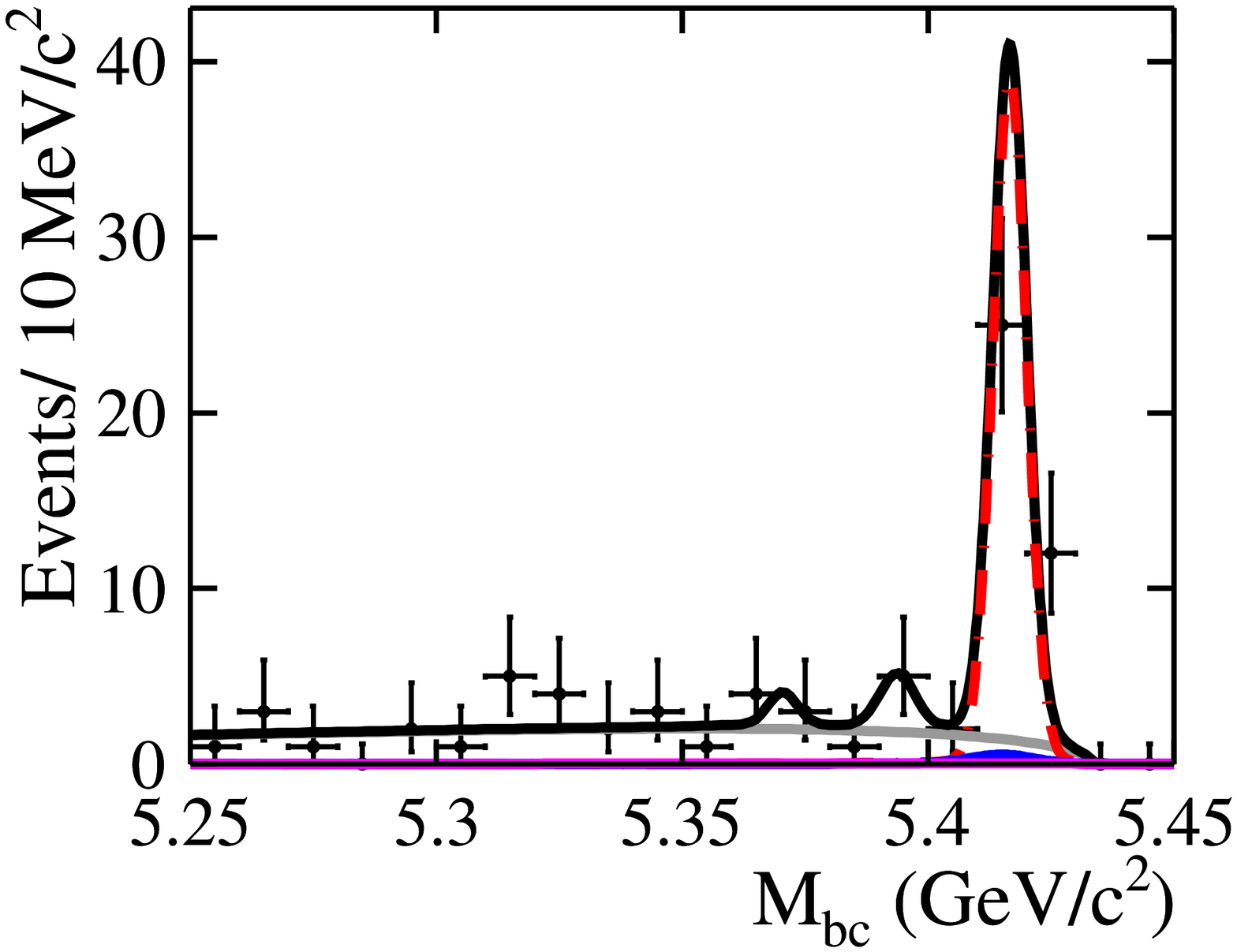}
\includegraphics[width=4.5 cm,height=4.5 cm]{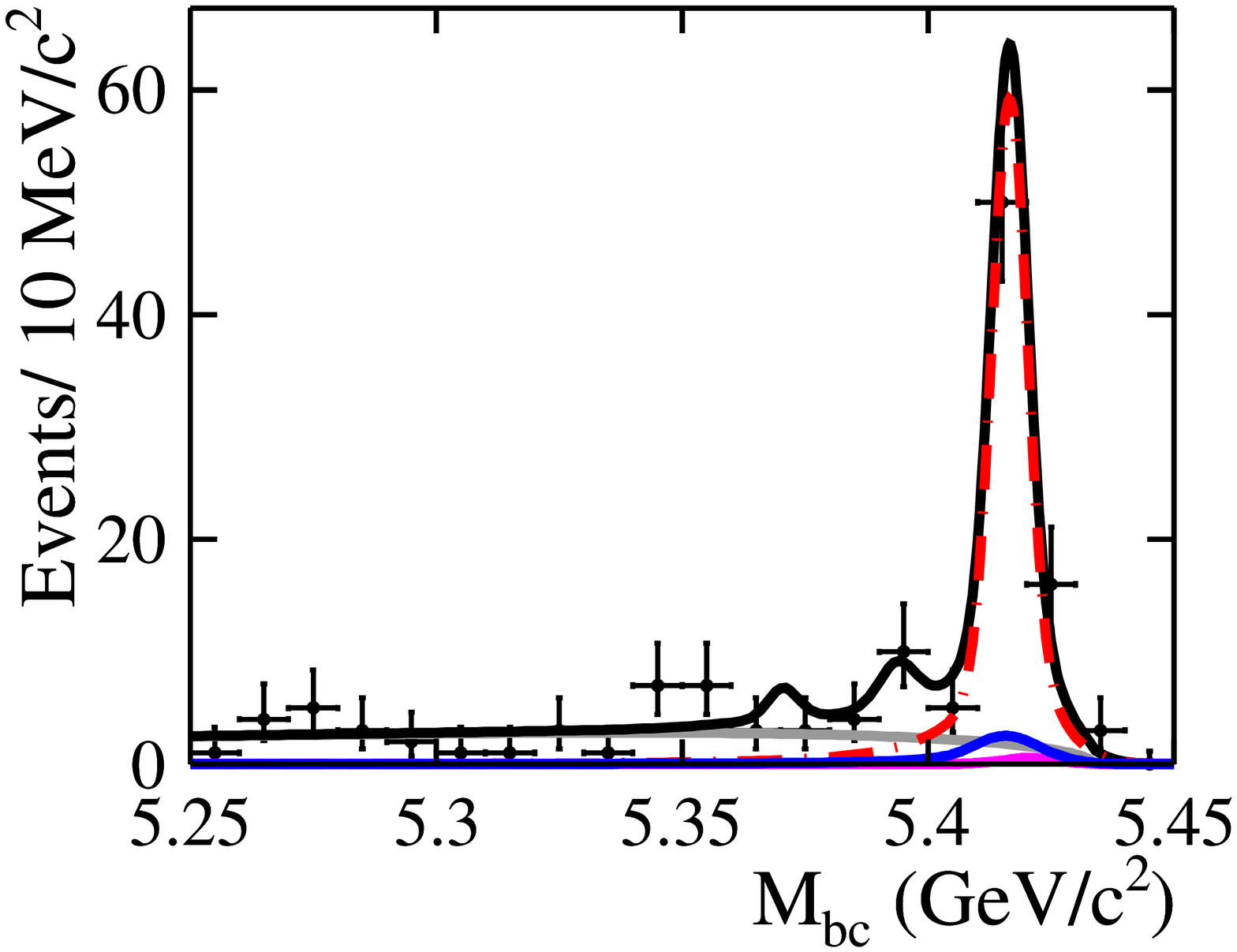}
\includegraphics[width=4.5 cm,height=4.5 cm]{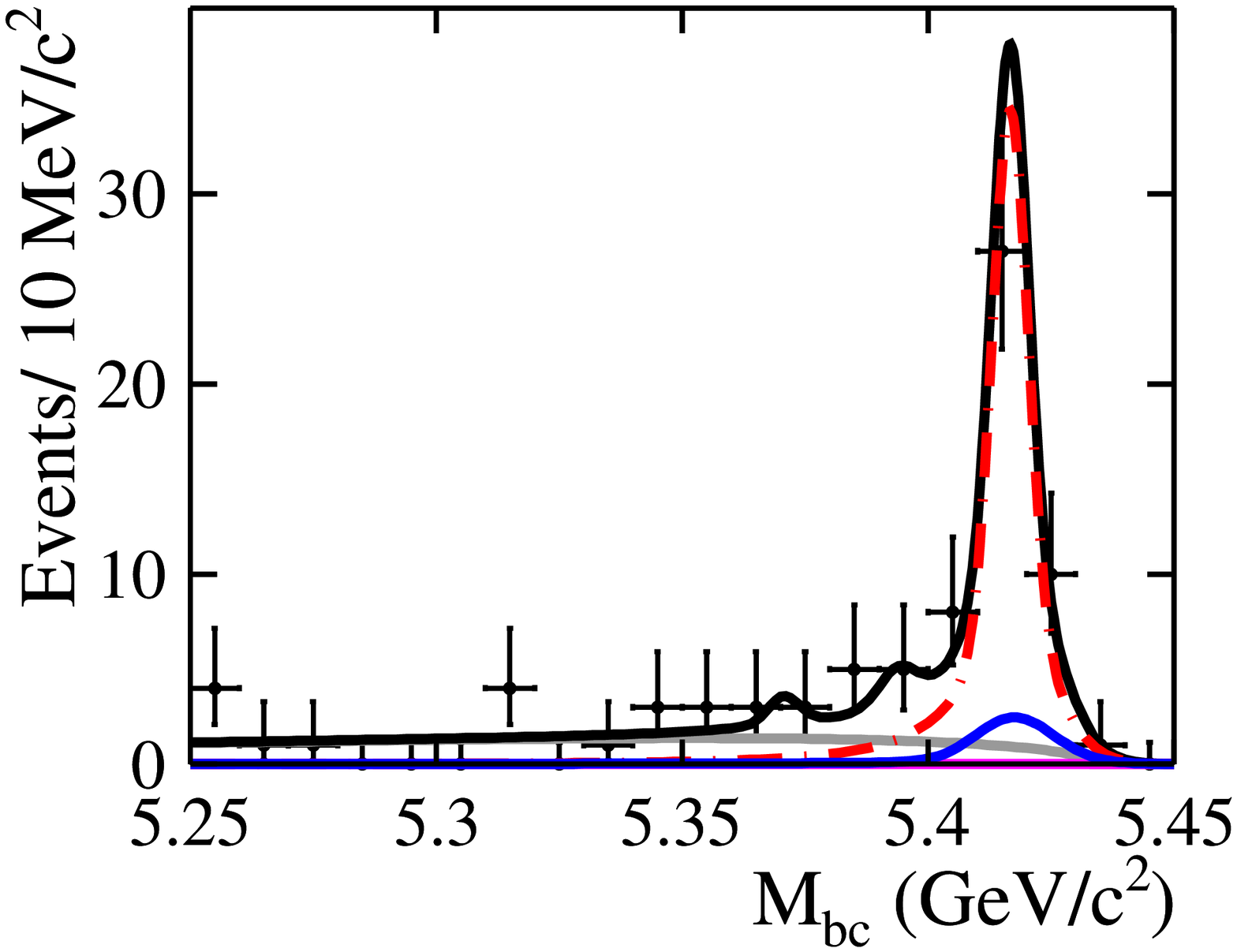}
}}
\caption{$\mbc$ projections and fit results, (left) $B^0_s\ra D^+_s D^-_s$, (center) $B^0_s\ra D^{*\pm}_s D^{\mp}_s$, (right) $B^0_s\ra D^{*+}_s D^{*-}_s$. The red dashed curves show CR+WC signal; the blue and purple solid curves show CF; the gray solid curves show background; and the black solid curves show the total.}
\label{fig:fit_results}
\end{figure}

The projections of the fit are shown in Fig.~\ref{fig:fit_results}. The branching fraction for channel $i$ is calculated as ${\cal B}^{}_i = Y^{}_i/(\varepsilon^i_{MC}\cdot N^{}_{\bs\bsbar} \cdot f^{}_{B^*_s\overline{B}{}^{\,*}_s}\cdot 2)$, where $Y^{}_i$ is the fitted CR yield, and $\varepsilon^i_{MC}$ is the MC signal efficiency with intermediate branching fractions~\cite{pdg2012} included. 
The statistical significance is calculated as $\sqrt{-2\ln(\mathcal{L}_0 / \mathcal{L}_{\mathrm{max}})}$, where $\mathcal{L}_0$ and $\mathcal{L}_{\mathrm{max}}$ are the values of the likelihood function when the signal yield $Y^{}_i$ is fixed to zero and when it is floated, respectively. We include systematic uncertainties (discussed below) in the significance by smearing the likelihood function by a Gaussian having a width equal to the total systematic error related to the signal yield.

Inserting the total ${\cal B}$ from Table~\ref{tab:BsDecays} into 
Eq.~\ref{eqn:DeltaGamma} gives
\begin{eqnarray}
\dgs/\gs & = & 0.090\pm0.009 \,\pm0.023\,,
\label{eqn:dg_result}
\end{eqnarray}
where the first error is statistical and the second is systematic. 
This result has precision similar to that of other recent 
measurements~\cite{LHCb_deltagamma,CDF_deltagamma}. The central 
value is consistent with, but lower than, the theoretical 
prediction~\cite{Nierste}; the difference may be due to the 
unknown \cp-odd component in $B^0_s\ra D^{*+}_sD^{*-}_s$, and 
contributions from three-body final states. 
With more data these unknowns can be measured.
The former is estimated to be only 6\% for analogous $B^0\ra D^{*+}D^{*-}_s$
decays~\cite{Rosner}, but the latter can be significant: 
Ref.~\cite{Hou} calculates
$\Delta\Gamma(B^{}_s\ra D^{(*)}_s D^{(*)} K^{(*)})/\Gamma^{}_s=
0.064\pm 0.047$.
This calculation predicts $\Delta\Gamma^{}_s/\Gamma^{}_s$ from 
$D^{(*)+}_s D^{(*)-}_s$ alone to be $0.102\pm 0.030$, which 
agrees well with our result.

To measure $f^{}_L$, we perform an unbinned ML fit to the helicity angles $\theta^{}_1$ and $\theta^{}_2$, which are the angles between the daughter $\gamma$ momentum and the opposite of the $B^{}_s$ momentum in the $D^{*+}_s$ and $D^{*-}_s$ rest frames, respectively. The angular distribution is $\left(|A_+|^2 + |A_-|^2\right)\left(\cos^2\theta_1 +1\right) \left(\cos^2\theta_2 +1\right) + |A^{}_0|^2 4\sin^2\theta^{}_1\sin^2\theta^{}_2$, where $A^{}_+$, $A^{}_-$, and $A^{}_0$ are the three polarization amplitudes in the helicity basis. The fraction $f^{}_L = |A_0|^2/(|A_0|^2+|A_+|^2 + |A_-|^2)$.
We obtain\cite{esen12}
\begin{eqnarray}
f^{}_L & = & 0.06\,^{+0.18}_{-0.17}\,\pm0.03,
\label{eqn:helicity_result}
\end{eqnarray}
where the first error is statistical and the second is systematic. 
Reconstruction of $B_s$ decays to well-defined $CP$ final states are of interest 
for studies of $CP$ violation.
In the SM, mixing-mediated $CP$ violation occurs in neutral mesons due to the complex argument of the product of CKM matrix elements participating in the mixing ``box diagram.'' 
For $B_s$ the relevant product is $V_{\rm tb}^{*2}V_{\rm ts}^2$, which is real, so no significant asymmetry is expected.
Searches for $CP$ asymmetry in decays of $B_s$ thus present an opportunity to reveal NP.
Such measurements will require the reconstruction of a sizable sample of $CP$-defined final states. 

\begin{figure}[ht]
\begin{center}
{\includegraphics[width=6.5cm]{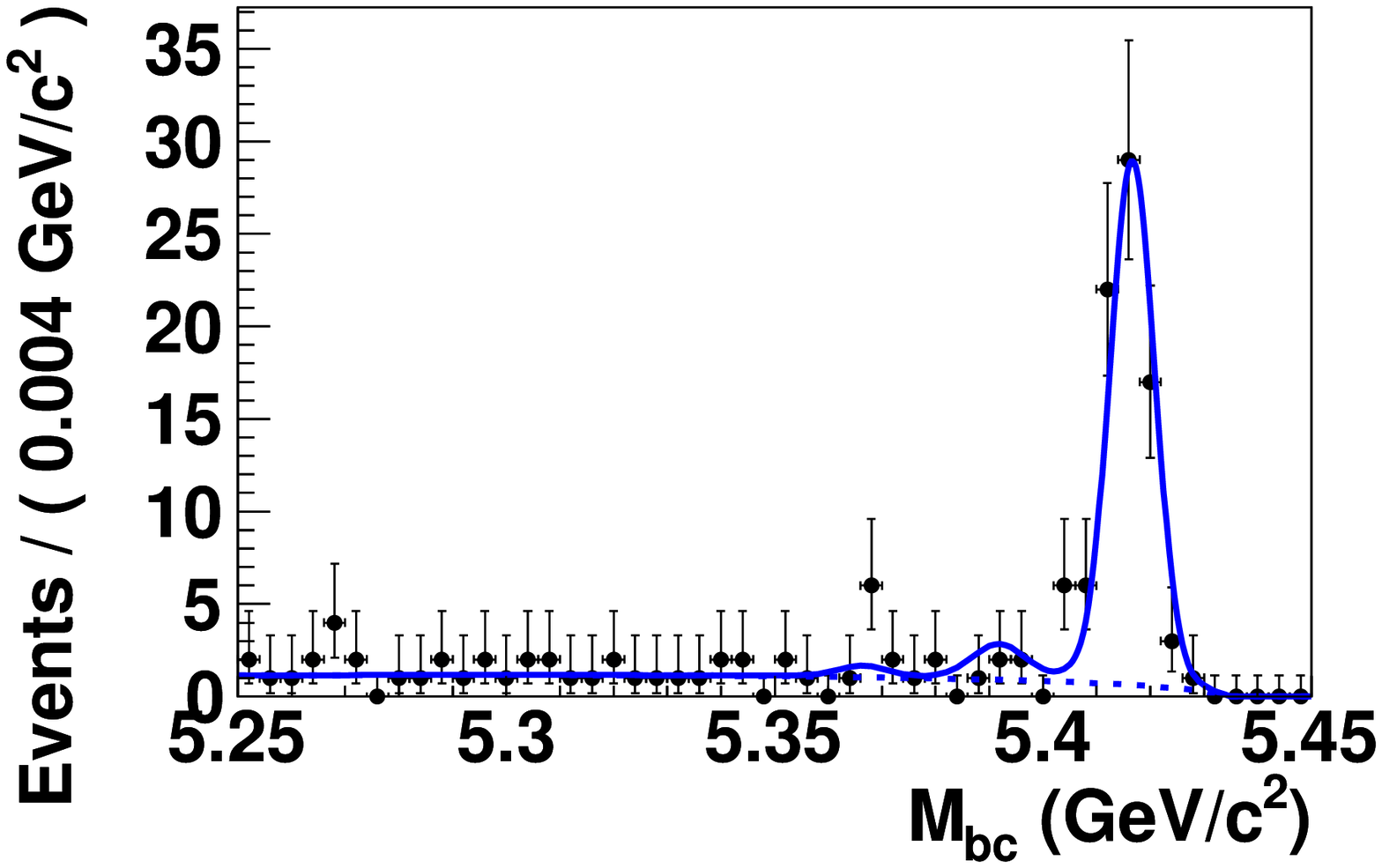}
\includegraphics[width=6.5cm]{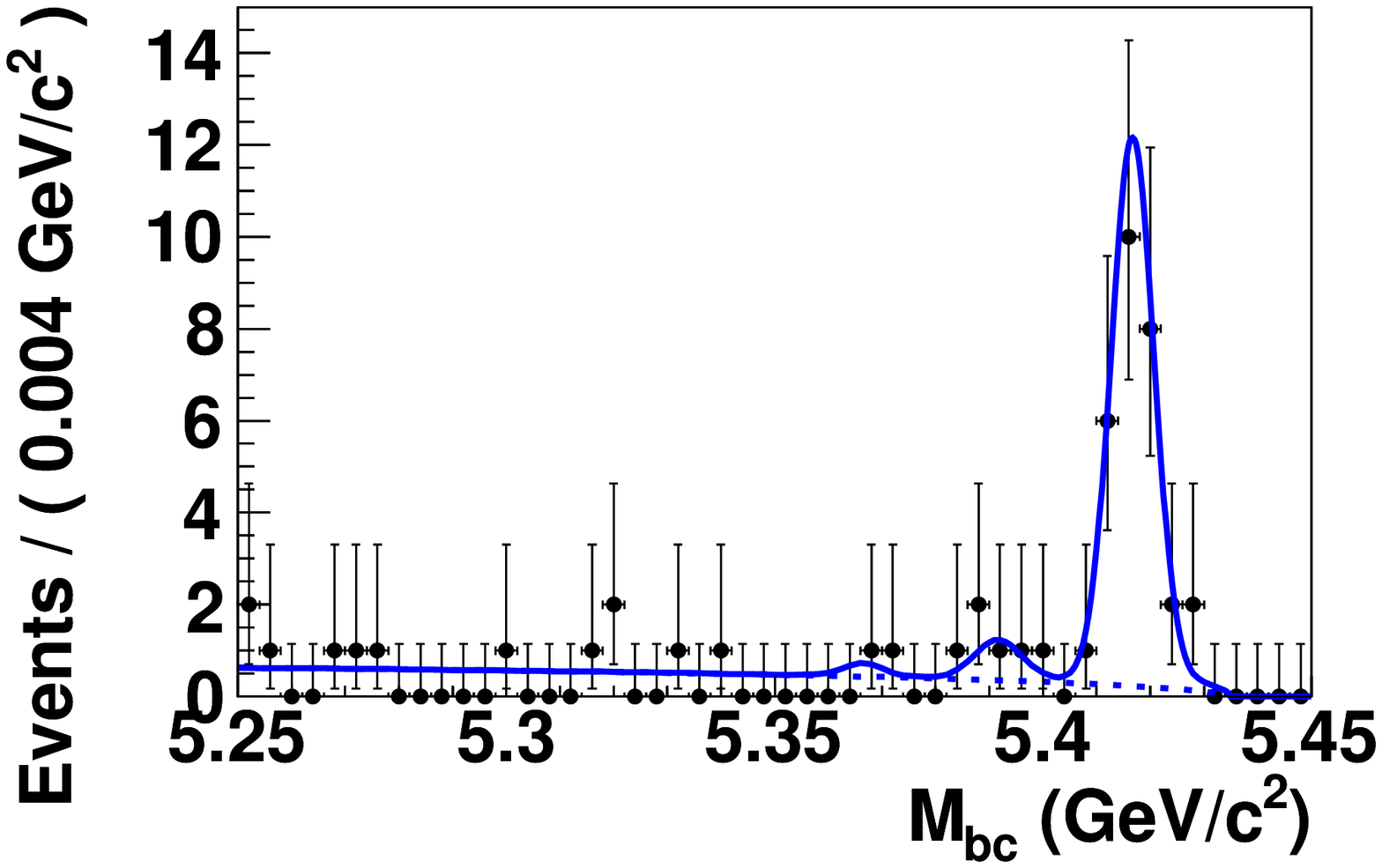}
}
\end{center} 
\caption{Projections in  $M_{\rm bc}$, based on the 121FB data set at $\Upsilon$(10860):  (left) $B_s\to J/\psi\eta(\gamma\gamma)$,  (right) $B_s\to J/\psi\eta(\pi^+\pi^-\pi^0)$.
Solid curves show projections of fit results. Backgrounds are represented by the blue dotted curves. Two small bumps around 5.37 and 5.39 GeV$/c^2$ are contributions from $B_s^0 \bar B_s^0$ and $B_s^*\bar B_s^0$ production channels, due to the overlap of the $\Delta E$ signal regions. 
\label{psi-eta}}
\end{figure}

The decays $B_s \to J/\psi\eta^{(\prime)}$ ($CP=+1$) proceed by the same process as $B\to J/\psi K^0$, so the branching fractions may be estimated based on the measured  branching fractions~\cite{pdg2012}, ${\cal B}(B_d^0\to J/\psi K^0) = 8.71\times 10^{-4}$: 
${\cal B}(B_s\to J/\psi \eta) \approx 3.5\times 10^{-4}$, and
${\cal B}(B_s\to J/\psi \eta^{\prime}) \approx 4.9\times 10^{-4}$.
The decays are reconstructed in the following modes:
$J/\psi \to e^+e^-,\ \mu^+\mu^-$; $\eta \to \gamma \gamma,$ $\pi^+\pi^-\pi^0$;
$\eta^{\prime}\to \eta\pi^+\pi^-$, $\rho^0\gamma$.
The signals are extracted via a 2-dimensional fit in $\Delta E$ and $M_{\rm bc}$.
Projections in $M_{\rm bc}$ are shown in Fig.~\ref{psi-eta}.

\begin{figure}[ht]
\begin{center}
{\includegraphics[width=8.0cm]{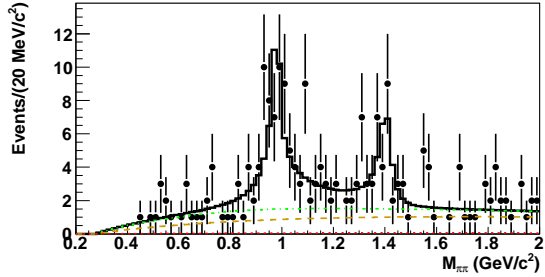}}
\caption{\label{fig:Jpsipipi} Pion pair mass distribution for $B_s\to J/\psi \pi^+\pi^-$ candidates in the 121FB set, for $ -79.7\;\mathrm{MeV}<\Delta E <
-19.7\;\mathrm{MeV}$.  The solid line represents the total
PDF.  The dash-dotted curve represents the
total background, the dashed curve shows other $J/\psi$ background, 
and the dotted curve the nonresonant component.  }
\end{center}
\end{figure}

\medskip
The same $b\to c \bar c s$ process can also produce the decay $B_s^0\to J/\psi f_0(980)$, another promising channel for $CP$ studies, with the clear advantage of being an all-charged final state with no angular analysis required because of the $J^P=0^+$ quantum numbers of the $f_0(980)$.
The mode was reconstructed as $B_s\to J/\psi \pi^+\pi^-$,\ \{$J/\psi\to \mu^+\mu^-,e^+e^-$\}, analyzing the 121FB set.
The fit to data include the $f_0(980)$ and another resonance in the $\pi\pi$ mass spectrum at $\sim1.4$~GeV$^2$, $f_X$ (Fig.~\ref{fig:Jpsipipi}).
The $f_X$ mass, measured at 
$1.405\pm0.015^{+0.001}_{-0.007}$ GeV$/c^2$, is consistent with that of the $f_0(1370)$.
The nonresonant yield is consistent with zero.

\medskip

We have also searched for the 2-body $CP$-eigenstate modes $B_s\to K^+K^-$, $K^0\bar K^0$, and $\pi^-\pi^+$, as well as the flavored mode $B_s\to K^-\pi^+$, in the 24FB data set~\cite{Bs2hh}.
The findings for $K^+K^-$ and $K^0\bar K^0$ were the first absolute branching fraction and first reported limit, respectively.

\subsubsection{Radiative decays}

Radiative penguin decays, which produce a photon via a one-loop Feynman diagram, are a promising venue to search for physics beyond the SM because particles at mass scales not yet directly accessible at accelerators can contribute to such loop effects. 
The $B_s \to \phi \gamma$ mode is a radiative process described within the SM by a $\bar{b} \to \bar{s} \gamma$ penguin diagram; it is the counterpart of the $B \to K^*(892) \gamma$ decay. 
In the SM, the $B_s \to \phi \gamma$ branching fraction has been computed with 30 \% uncertainty to be about $40 \times 10^{-6}$~\cite{bs2phigam-sm1,bs2phigam-sm2}. 
This channel was first observed at Belle, with $\phi$ reconstructed in the mode $K^+K^-$~\cite{bs2phigam} .
For photon selection, major sources of background in the signal region included $\pi^0\to \gamma\gamma$ and $\eta\to\gamma\gamma$ as well as calorimeter hits that were out of time with the  beam crossing.
Based on the 24FB set, we reported ${\cal B}(B_s \to \phi \gamma) = (57 { ^{+18}_{-15}(\rm stat) } { ^{+12}_{-11} (\rm syst) } ) \times 10^{-6} $, which is in agreement with both the SM predictions and with extrapolations from measured $B^{+}\to K^*(892)^{+} \gamma$ and $B^{0}\to K^*(892)^{0} \gamma$ decay branching fractions~\cite{bs2phigam}.

\subsubsection{Modes suppressed in the Standard Model}
\begin{figure}[htb]
\parbox{\halftext}{
\centerline{\includegraphics[width=5.5cm]{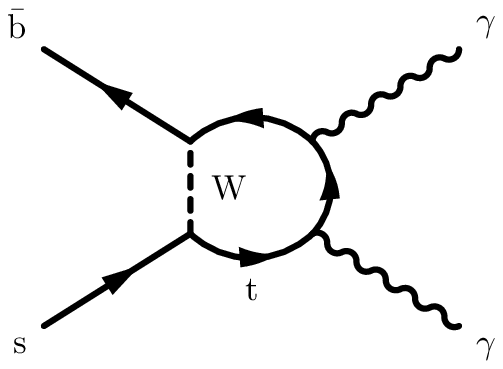}}
\caption{
Diagram describing the dominant processes for $B_s\to \gamma\gamma$.
}
\label{figure:feynman}
}
\hfill
\parbox{\halftext}{
\centerline{\includegraphics[width=5.5cm]{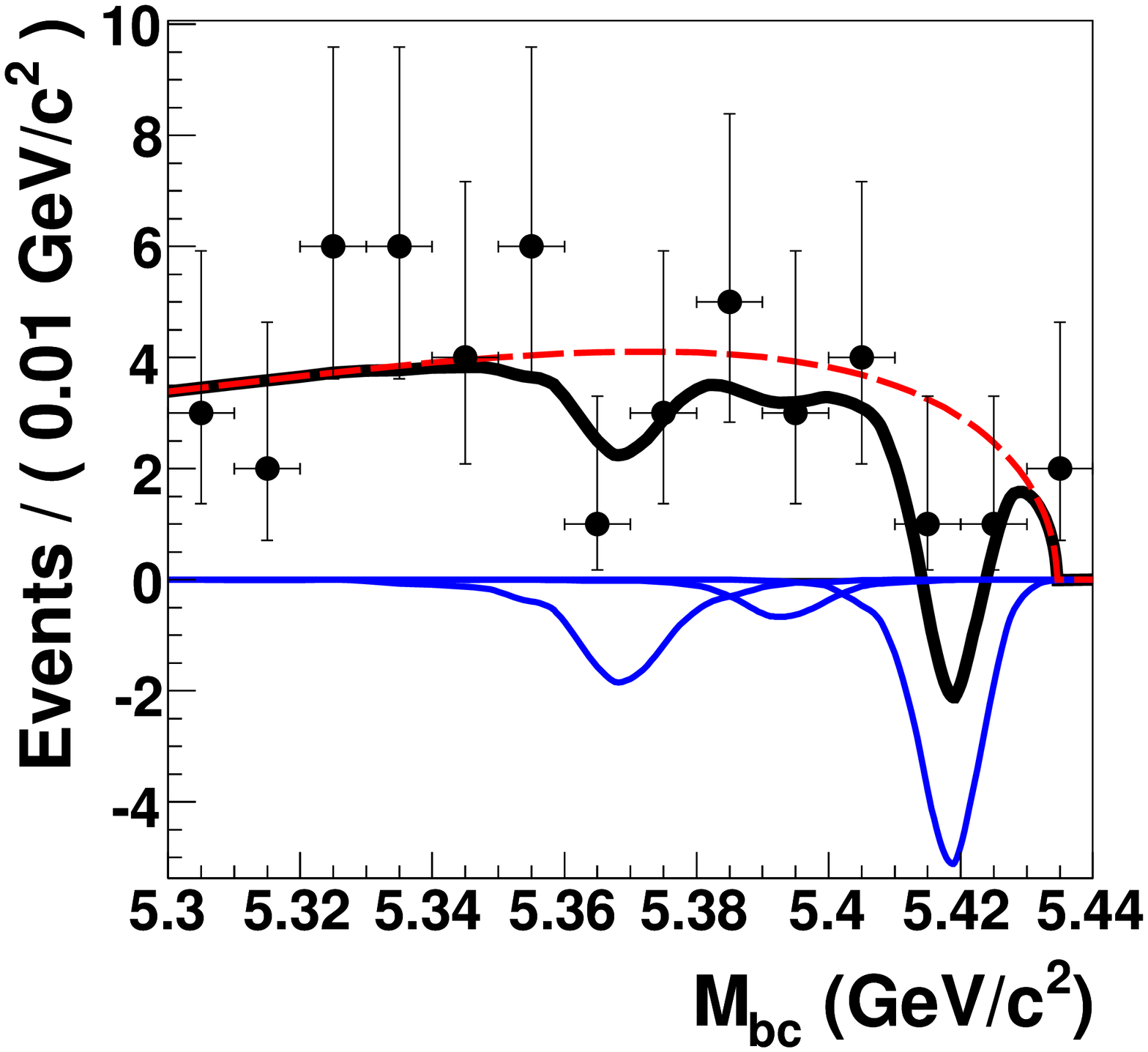}}
\caption{
$M_{\rm bc}$ projection and fit for the $B_s\to \gamma\gamma$ search 24FB data set.
}
\label{figure:mbc_gamgam}}
\end{figure}

The $B_s \to \gamma \gamma$ mode is described in the SM by a penguin annihilation diagram (Fig.~\ref{figure:feynman}), and its branching fraction has been calculated to be in the range (0.5\---1.0)$ \times 10^{-6}$~\cite{bs2gamgam-sm1,bs2gamgam-sm2,bs2gamgam-sm3}.
Belle has searched for this mode in the 24FB data set~\cite{bs2phigam}.
No significant signal was observed (Fig.~\ref{figure:mbc_gamgam}), and a 90\%C.L. upper limit of ${\cal B}(B_s \to \gamma \gamma) < 8.7 \times 10^{-6}$ was obtained.  This limit significantly improves on the previously reported one and is only an order of magnitude larger than the SM prediction, providing the possibility of observing this decay at a future Super $B$-factory~\cite{superbelle,superb}.

\subsection{Measurement of $\sin 2\phi_1$}

\begin{figure}[h]
\includegraphics[width=5.cm]{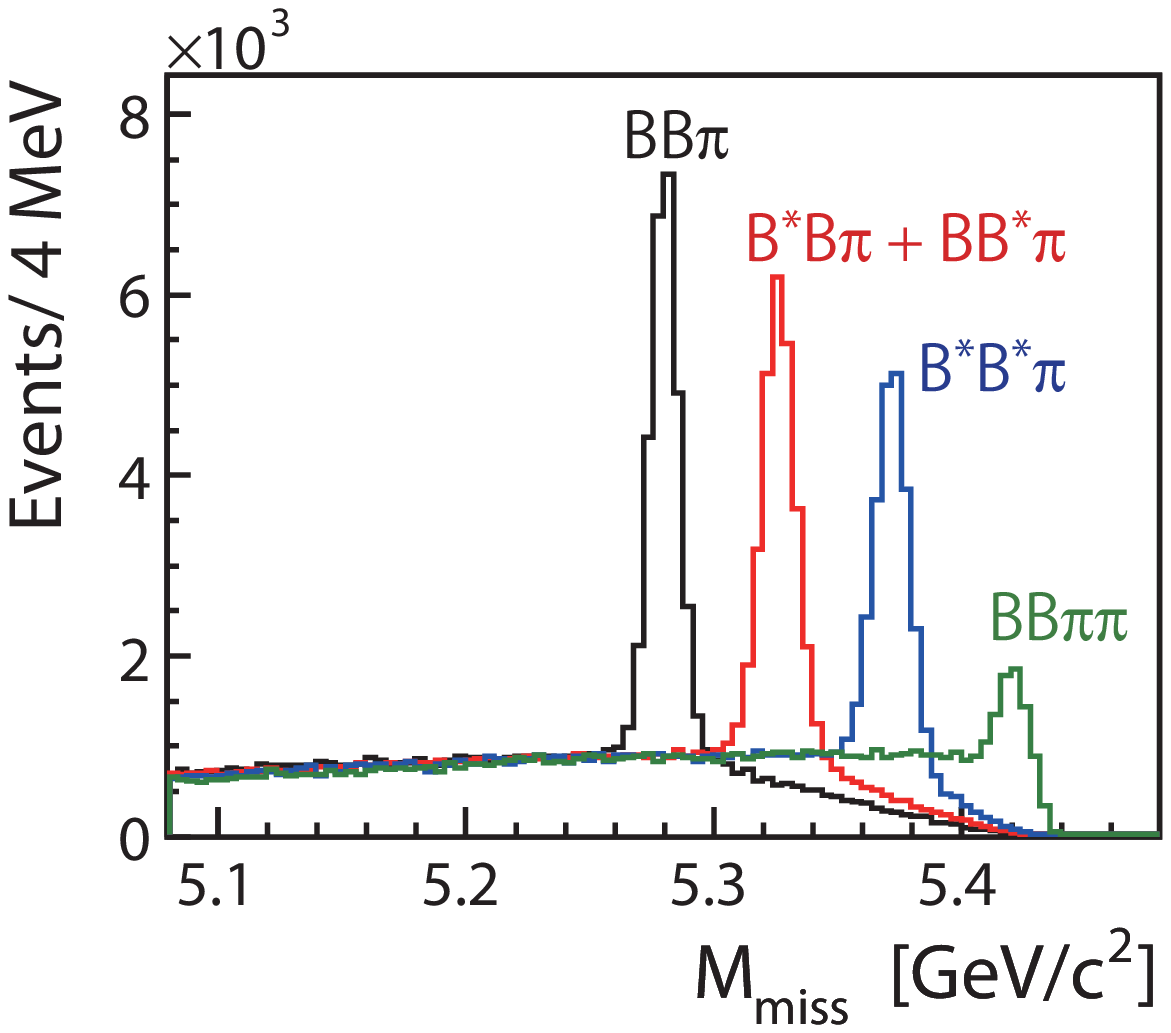}
\includegraphics[width=5.cm]{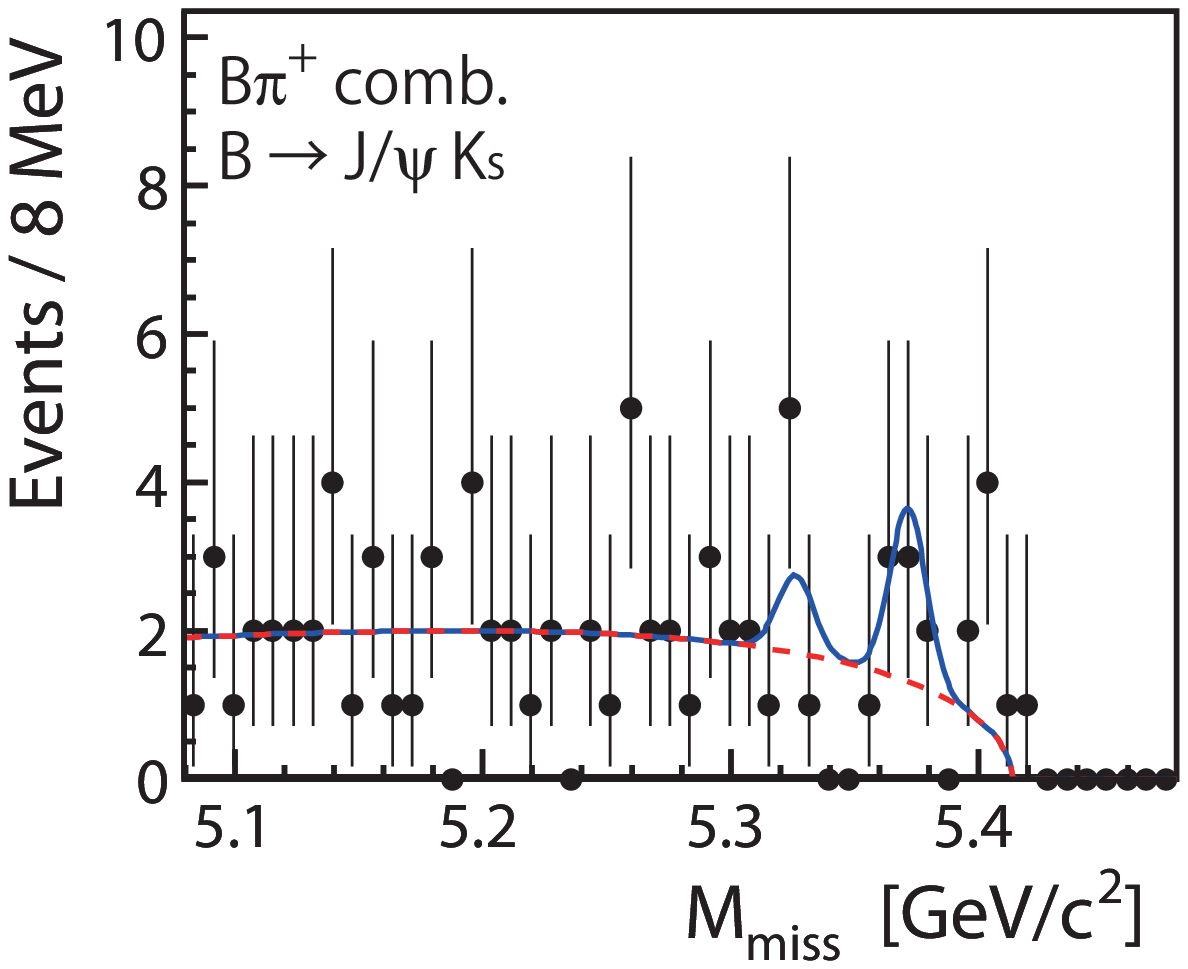}
\includegraphics[width=5.cm]{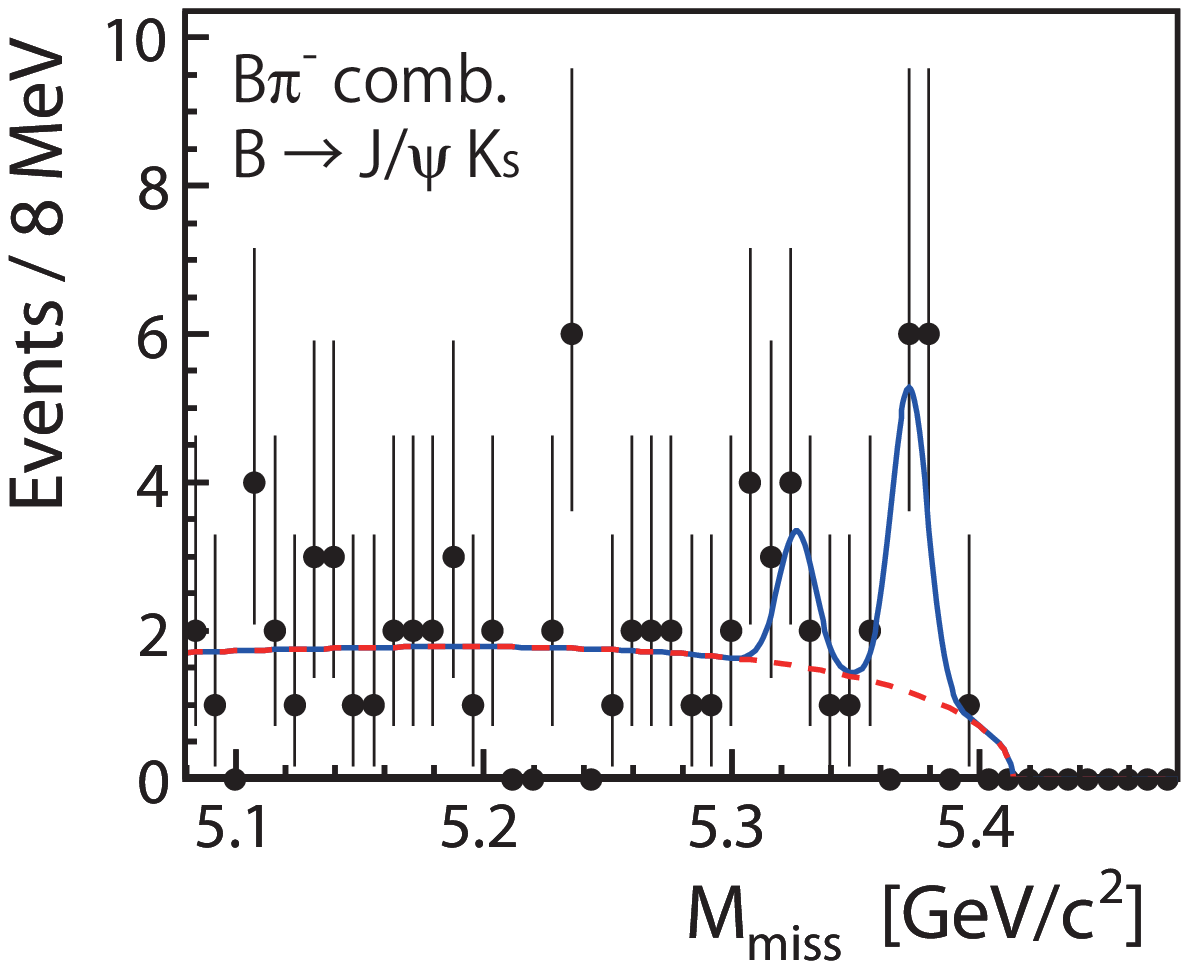}
\caption{Distributions in  $M_{\rm miss}$ of tagged $B^0\pi$ candidates for (left) simulated $B\bar B\pi$, $B\bar B^*\pi$, and $B^*\bar B^*\pi$ events and 121.4~fb$^{-1}$ of data, (center) $B^0\pi^+$  and (right) $B^0\pi^-$.
\label{fig:BBpi}}
\end{figure}

The method of full $B$ reconstruction, used to study the assortment of $B$ events at the $\Upsilon$(5$S$)~\cite{BBpi},
has been applied to a novel tag to measure $\sin 2\phi_1$~\cite{Bpisin2phi1}.
Three-body final states \mbox{$B^{(*)0}\{\to B^0(\gamma)\} B^{(*)-}\pi^+$ }($+c.c.$) are identified through full reconstruction of a neutral $B$ in a $CP$-eigenstate and a charged pion.
The event residue, consisting of a charged $B$ and up to two photons, is characterized through ``missing mass,'' calculated through energy and momentum conservation:
\begin{eqnarray*}
E_{\rm miss}=E_{\rm beam}-E_{B^0\pi};\ \  \vec{p}_{\rm miss}=-\vec{p}_{B^0\pi} ;\ \  MM(B^0\pi)=M_{\rm miss}=\sqrt{E_{\rm miss}^2-\vec{p}_{\rm miss}^2}.
\end{eqnarray*}
The missing mass distributions are well separated for  $B\bar B\pi\pi$, $B\bar B\pi$, $B\bar B^*\pi$, and $B^*\bar B^*\pi$ events, as can be seen in Fig.~\ref{fig:BBpi}(left).
The sign of the charged pion tags the initial flavor of the neutral $B$ and enables a {\it time-independent} measurement of $CP$ asymmetry, which is related to $\sin 2\phi_1$ as:
\begin{eqnarray*}
A_{BB\pi}\equiv \frac{N_{BB\pi^-}-N_{BB\pi^+}}{N_{BB\pi^-}+N_{BB\pi^+}}
=\frac{{\cal S}x+{\cal A}}{1+x^2}
\end{eqnarray*}
where ${\cal S}= -\eta_{CP}{\rm sin} 2\phi_1$ ($\eta_{CP}$ is the $CP$-eigenvalue of the $B^0$ mode), $x=\Delta m/\Gamma$, and ${\cal A}$, a measure of direct $CP$ violation, is zero in the SM.

Neutral $B$'s are reconstructed in the following modes and submodes:
$B^0\to J/\psi K_S;
\ J/\psi \to e^+e^-,\ \mu^+\mu^-$.
Figure~\ref{fig:BBpi} shows the distributions in $M_{\rm miss}$ for (center) $B^0\pi^+$  and (right) $B^0\pi^-$  combinations, respectively, where the fits yield a total of $21.5\pm 6.8$ events.
The asymmetry is found to be $A_{BB\pi}=0.28\pm 0.28$, corresponding to $\sin 2\phi_1 = 0.57\pm 0.58\pm 0.06$.
This result establishes a new time-independent method of measuring $\sin 2\phi_1$.  The value is consistent with measurements in $\Upsilon$(4$S$) data.

\section{New resonances}
\label{chap_resonances}
\subsection{Charmonium physics }

In $e^+e^-$ collisions at CM energies near $\sqrt{s}\simeq 10.58$~GeV, 
there are a number of ways to produce final states that contain a $c\bar{c}$ 
quark pair.  These include: i) $B$-meson decays, in which
$b\to c\bar{c} s$ is a favored transition; ii) $\gamma\gamma$ fusion,
which is proportional to the square of the quark charge and, thus, favors
production of $c\bar{c}$ and $u\bar{u}$ pairs over $s\bar{s}$ and $d\bar{d}$
pairs; iii) near-threshold $s$-channel $c\bar{c}$ production via initial-state
radiation; and iv) $c\bar{c}$ associated production with $J/\psi$ mesons in 
$e^+e^-$ annihilation, which Belle found to be the dominant mechanism for
$J/\psi$ productions in $e^+e^-$ annihilation near
$\sqrt{s}=10$~GeV. Belle exploited 
all four of these processes to make a series of interesting discoveries
related to the spectroscopy and interactions of $c\bar{c}$ charmonium mesons.

\subsubsection{First observation of the $\eta_c(2S)$}
\begin{figure}
\begin{minipage}[t]{75mm}
\includegraphics[height=0.6\textwidth,width=0.9\textwidth]{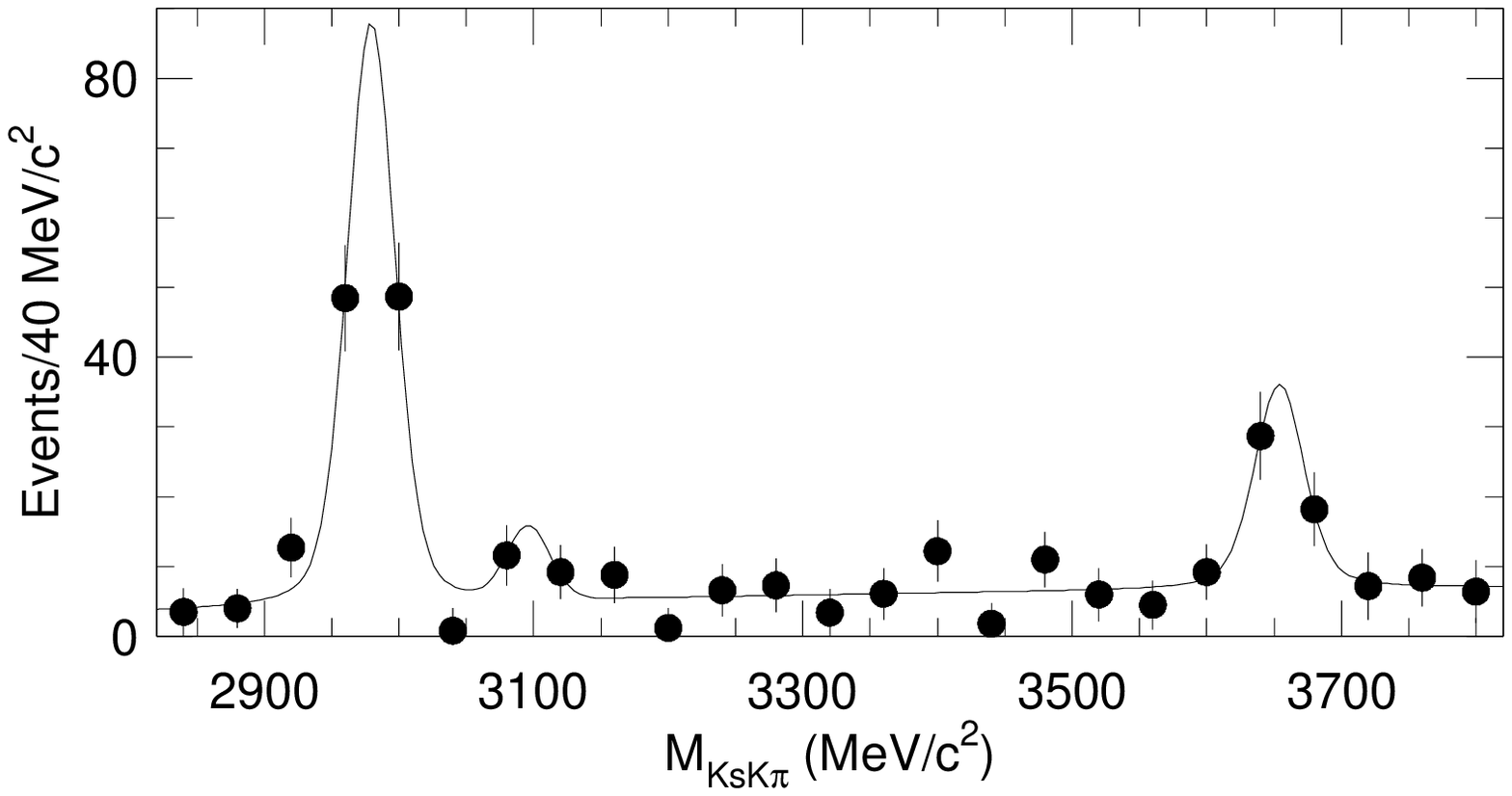}
\end{minipage}
\begin{minipage}[t]{75mm}
  \includegraphics[height=0.6\textwidth,width=0.9\textwidth]{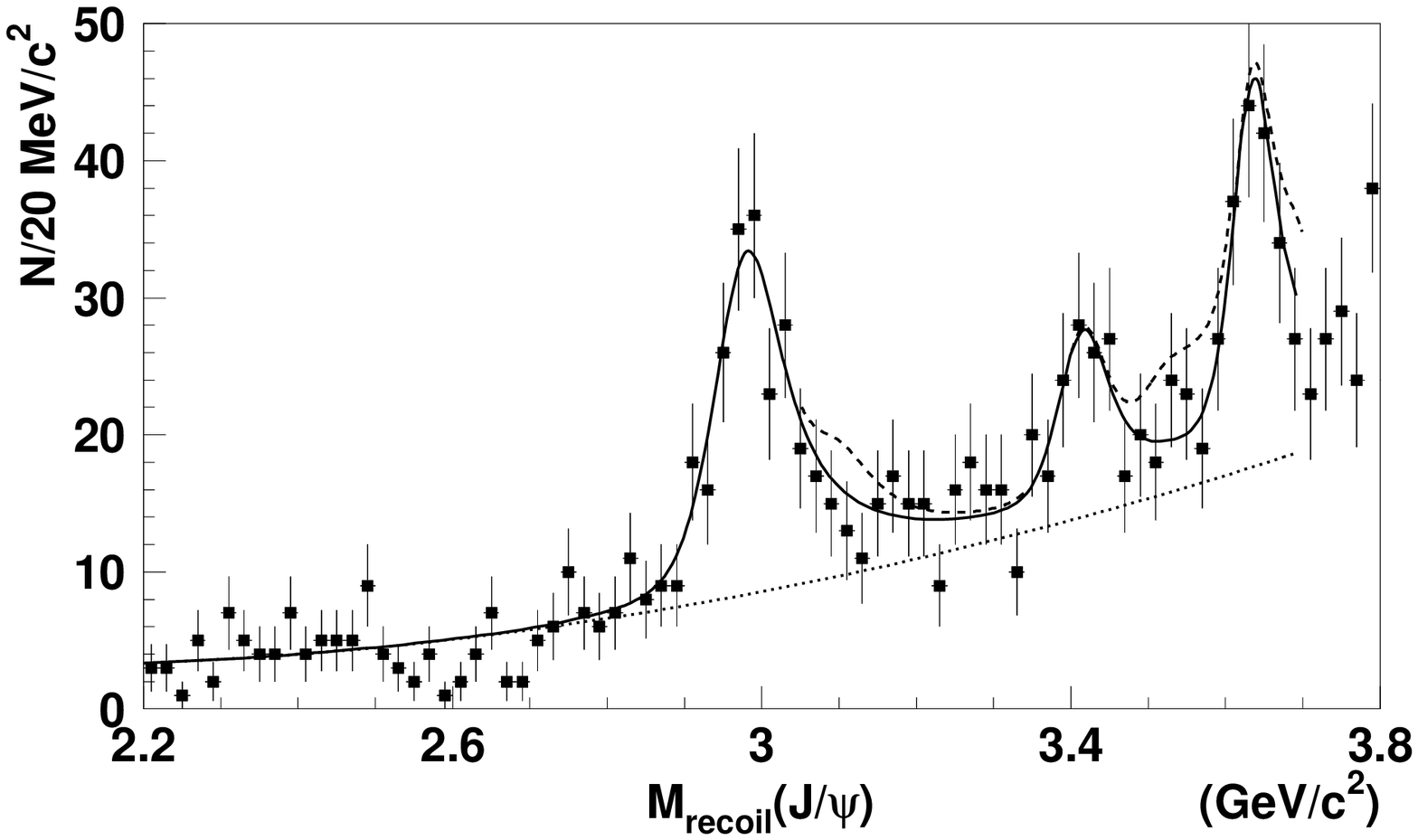}
\end{minipage}
\caption{Left: The $K_S K^{\pm}\pi^{\mp}$ mass distribution from
$B\to K K_S K^{\pm}\pi^{\mp}$ decays~\cite{choi_etac2s}. The large peak on 
the left is the $\eta_c$; 
the smaller peaks on the right are the $J/psi$ (around 3.1~GeV) and 
$\eta_c(2S)$ signals.
Right: The $J/\psi$ recoil mass spectrum in inclusive
$e^+e^-\to J/\psi \, X$ processes~\cite{pakhlov_etac2s}. 
A fit with $\eta_c$, $\chi_{c0}$, and $\eta_c(2S)$ contributions is 
shown as a solid curve. 
The dashed curve in the figure corresponds to the case where 
the contributions of the $J/\psi$, $\chi_{c1}$, $\chi_{c2}$, and 
$\psi(2S)$ are set at their 90\% C.L. upper limit values. The dotted
curve is the background function. }
\label{fig:etac2s}
\end{figure}

Prior to 2002, the only ``positive'' observation of the $\eta_c(2S)$, the first
radial excitation of the charmonium ground state meson, 
the $\eta_c$, was a peak in  the $\gamma$ energy 
spectrum from exclusive $\psi(2S)\to \gamma X$
decays reported by the Crystal Ball Experiment~\cite{Xball_etac2s}.  However,
this result was somewhat suspicious 
since the hyperfine $\psi(2S)$-$\eta_c(2S)$ mass
splitting inferred from the measured mass value,
$\Delta M_{\rm hfs}(2S)=92\pm 5$~MeV, 
is substantially higher than the theoretical expectation of  
$\Delta M^{\rm theory}_{\rm hfs}(2S)\simeq 58\pm 8$~MeV; see, e.g., 
Ref.~\cite{badalian}.
In 2002, Belle reported the observation of a higher-mass $\eta_c(2S)$ candidate
in the $\eta_c(2S)\to K_S K^{\pm}\pi^{\mp}$ mass distribution produced via the
$B\to K \eta_c(2S)$, $\eta_c(2S) \to K_SK^{\pm}\pi^{\mp}$ decay chain (see 
Fig.~\ref{fig:etac2s}(left))~\cite{choi_etac2s}.  Belle subsequently
observed a signal at the same mass 
in the $J/\psi$ recoil mass spectrum for inclusive $e^+e^-\to J/\psi \, X$
processes~\cite{pakhlov_etac2s}, shown in the right-hand panel of
Fig.~\ref{fig:etac2s}.

The original Belle $\eta_c(2S)$ signal has since been confirmed by a number of
reports, including a higher statistics Belle study of $B\to K
K_SK^{\pm}\pi^{\mp}$ decays~\cite{anna_etac2s}.  The current PDG
world-average hyperfine splitting value, 
$\Delta M^{\rm PDG}_{\rm hfs}(2S)=49\pm 4$~MeV~\cite{pdg2012}, is close
to theoretical expectations and inconsistent with 
the Crystal Ball result, which is now
generally thought to have been incorrect. 
\subsubsection{The $X(3872)$}
\begin{figure}
\begin{minipage}[t]{75mm}
\includegraphics[height=0.65\textwidth,width=0.9\textwidth]{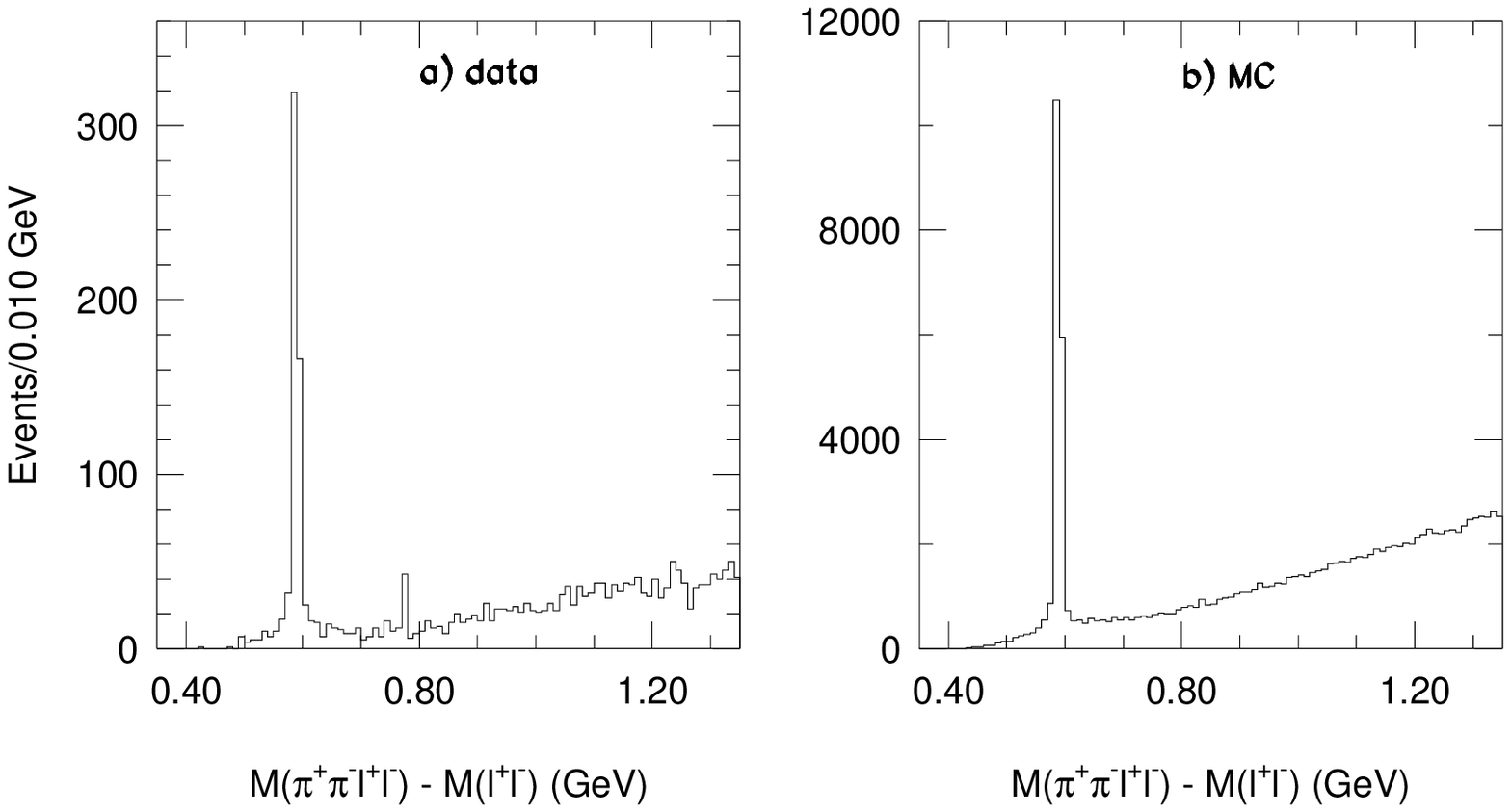}
\end{minipage}
\begin{minipage}[t]{75mm}
  \includegraphics[height=0.6\textwidth,width=0.9\textwidth]{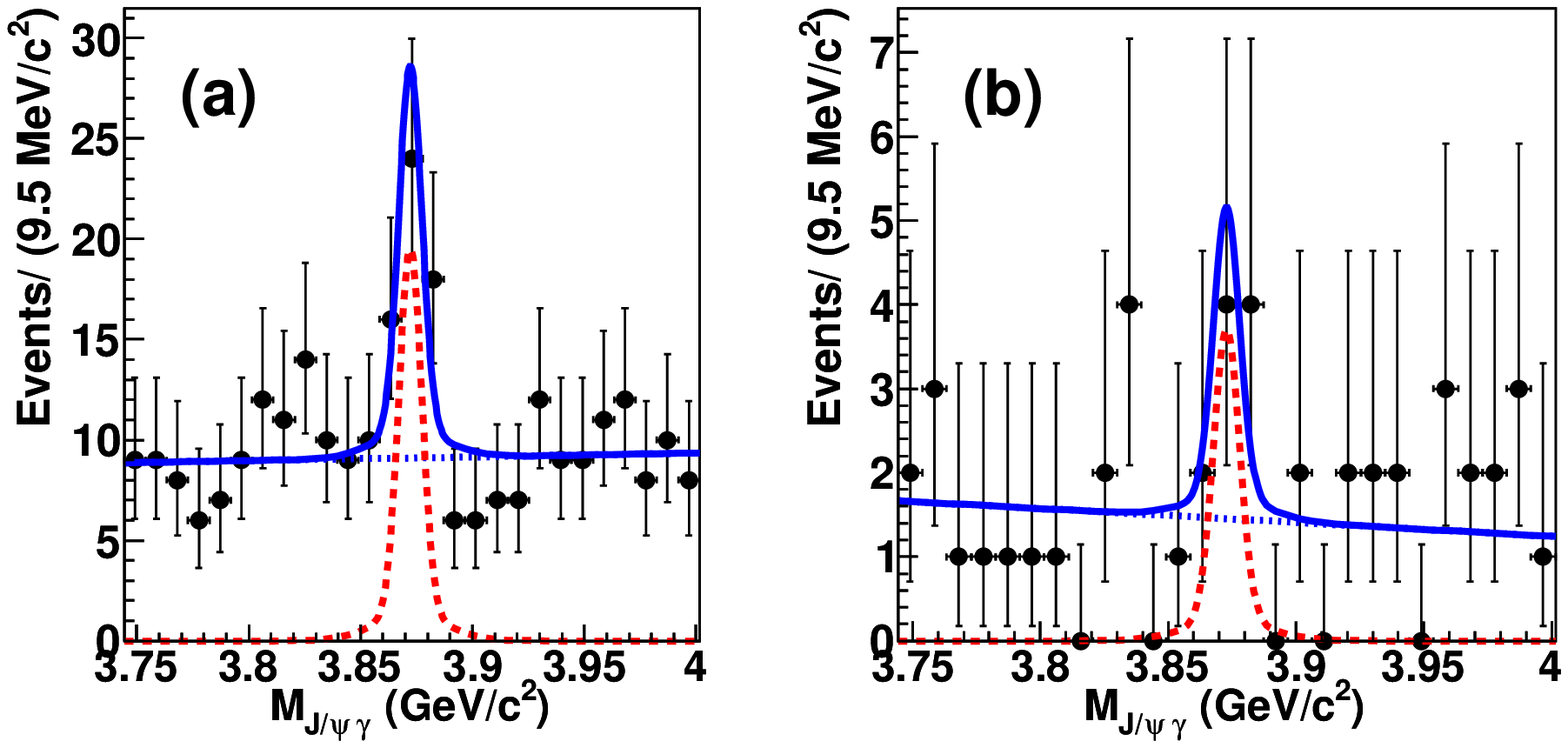}
\end{minipage}
\caption{Left: $\Delta M = M(\pi^+\pi^- \ell^+\ell^-)-M(\ell^+\ell^-)$
distributions for $B \to K\pi^+\pi^- J/\psi$, $J/\psi \to \ell^+\ell^-$ decays
for a) data and b) inclusive $B\to J\psi X$ 
MC~\cite{choi_x3872}.   The peak near 
$\Delta M\simeq 0.6$~GeV is due 
to $\psi^{\prime}\to \pi^+\pi^- J/\psi$ decays; the peak 
at $\Delta M\simeq 0.75$~GeV in the data, which does not show up in the MC,
is due to $X(3872)\to \pi^+\pi^- J/\psi$. 
 Right: $M(\gamma J/\psi)$ distributions for
 a) $B^+ \to K^+\gamma J/\psi$ and 
 b) $B^0 \to K^0 \gamma J/\psi$ decays~\cite{belle_gamjp}.
 }
\label{fig:x3872}
\end{figure}

The $X(3872)$ was first observed by Belle~\cite{choi_x3872} 
as a small narrow peak
in the $\pi^+\pi^- J/\psi$ invariant 
mass spectrum from $B\to K \pi^+\pi^- J/\psi$
decays shown in the leftmost panel of Fig.~\ref{fig:x3872}. It was subsequently
confirmed by CDF, D0, and BaBar~\cite{cdf_x3872}.  Other $X(3872)$ decay
modes that have been identified include the radiative decay, 
$X(3872)\to \gamma J/\psi$~\cite{belle_gamjp,babar_gamjp}, which establishes
its charge conjugation parity as $C=+1$, subthreshold decays to
$\omega J\psi$~\cite{x3872_omegajp},
and the decay to open charm,
$X(3872)\to D^{*0}\bar{D}^{0}$~\cite{aushev_ddstar,x_2_ddstar}.  
The Belle signals for $X(3872)\to\gamma J/\psi$ are shown in the right panel of
Fig.~\ref{fig:x3872}.  Angular correlation studies by CDF~\cite{cdf_angles}
and Belle~\cite{belle_angles} indicate a preferred quantum number assignment
of $J^{PC}=1^{++}$, although $2^{-+}$ cannot be ruled out.
The only available $1^{++}$ $c\bar{c}$ charmonium assignment 
for the $X(3872)$ is the $\chi^{\prime}_{c1}$.  
However, the 3872~MeV mass value is significantly lower than
the expected $\chi^{\prime}_{c1}$ mass of 3905~MeV, 
a value that is pegged to the
measured $3929\pm 6$~MeV mass of its $J=2$ multiplet partner, 
the $\chi^{\prime}_{c2}$, which
was discovered by Belle in 2006 (see below). A $\chi^{\prime}_{c1}$ mass of
3872~MeV would imply that the mass splitting 
for the radially excited $\chi_{cJ}(2P)$
multiplet is larger than that for the $\chi_{cJ}(1P)$ multiplet, contrary to
expectations from potential models and lattice QCD~\cite{davies}.   
There are similar problems for the $J^{PC}=2^{-+}$ assignment, 
for which the only available $c\bar{c}$ level
is the $\eta_{c2}$, the $^1D_2$ state. 
In this case, the 3872~MeV mass value is too
high compared to the expected value of 3837~MeV, 
an expectation that is tightly
constrained by the measured mass of its $^3D_1$ multiplet partner, the well
established $\psi(3770)$.  

\noindent
The lack of a natural charmonium assignment 
and the close proximity of the $X(3872)$ mass,
$3871.68 \pm 0.17$~MeV,~\cite{pdg2012} 
to the $D^{*0}\bar{D}^{0}$ mass threshold, 
$3871.94\pm 0.35$~MeV,~\cite{pdg2012} has led to speculations 
that the $X(3872)$ is a loosely bound 
$D^{*0}\bar{D}^{0}$ molecule-like structure; see, e.g., Ref.~\cite{oset},
although other interpretations have been proposed; see, e.g.,
Refs.~\cite{ktchao}.

\subsubsection{The $Y(3940)$}
\begin{figure}
\begin{minipage}[t]{75mm}
\includegraphics[height=0.65\textwidth,width=0.9\textwidth]{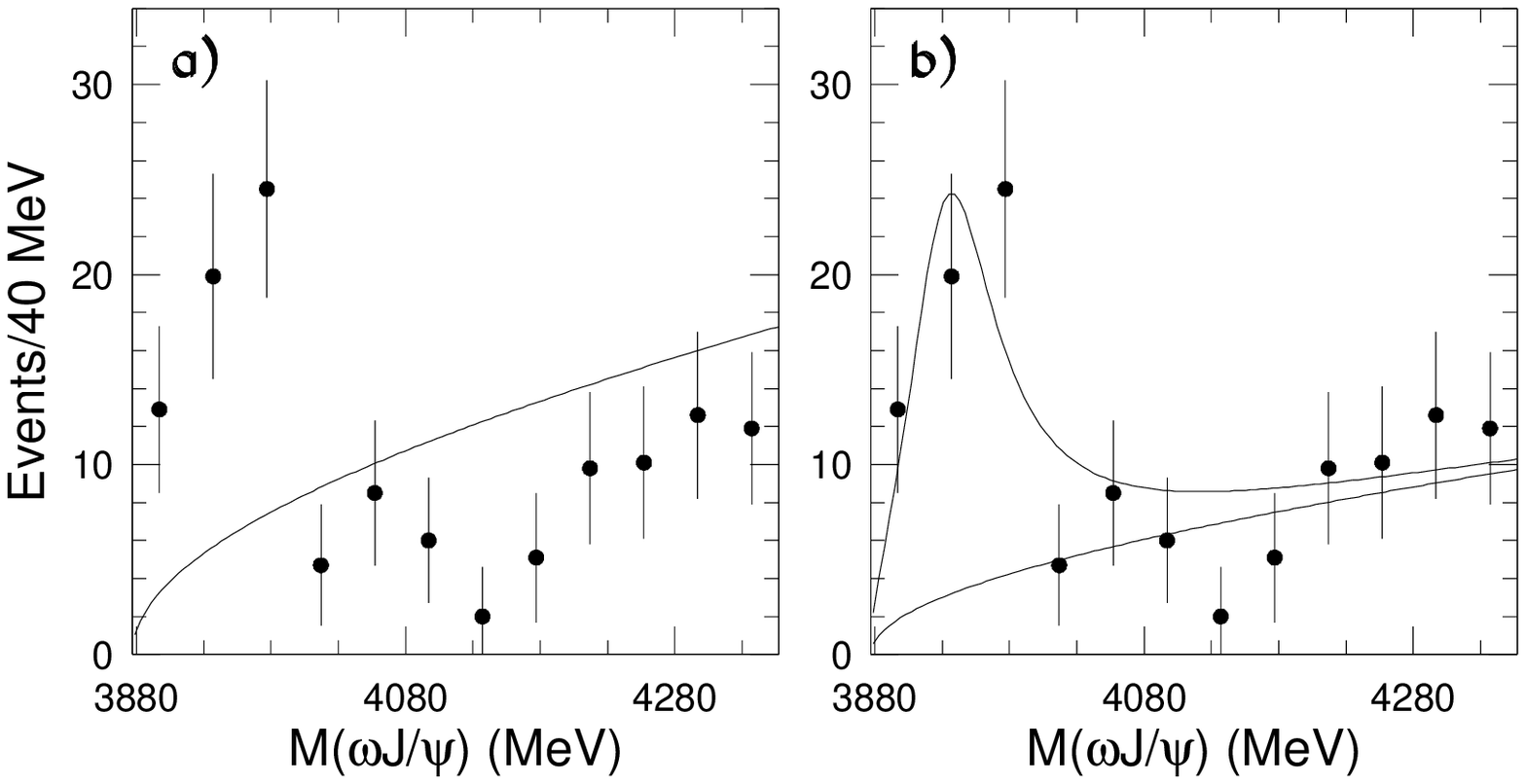}
\end{minipage}
\begin{minipage}[t]{75mm}
  \includegraphics[height=0.6\textwidth,width=0.9\textwidth]{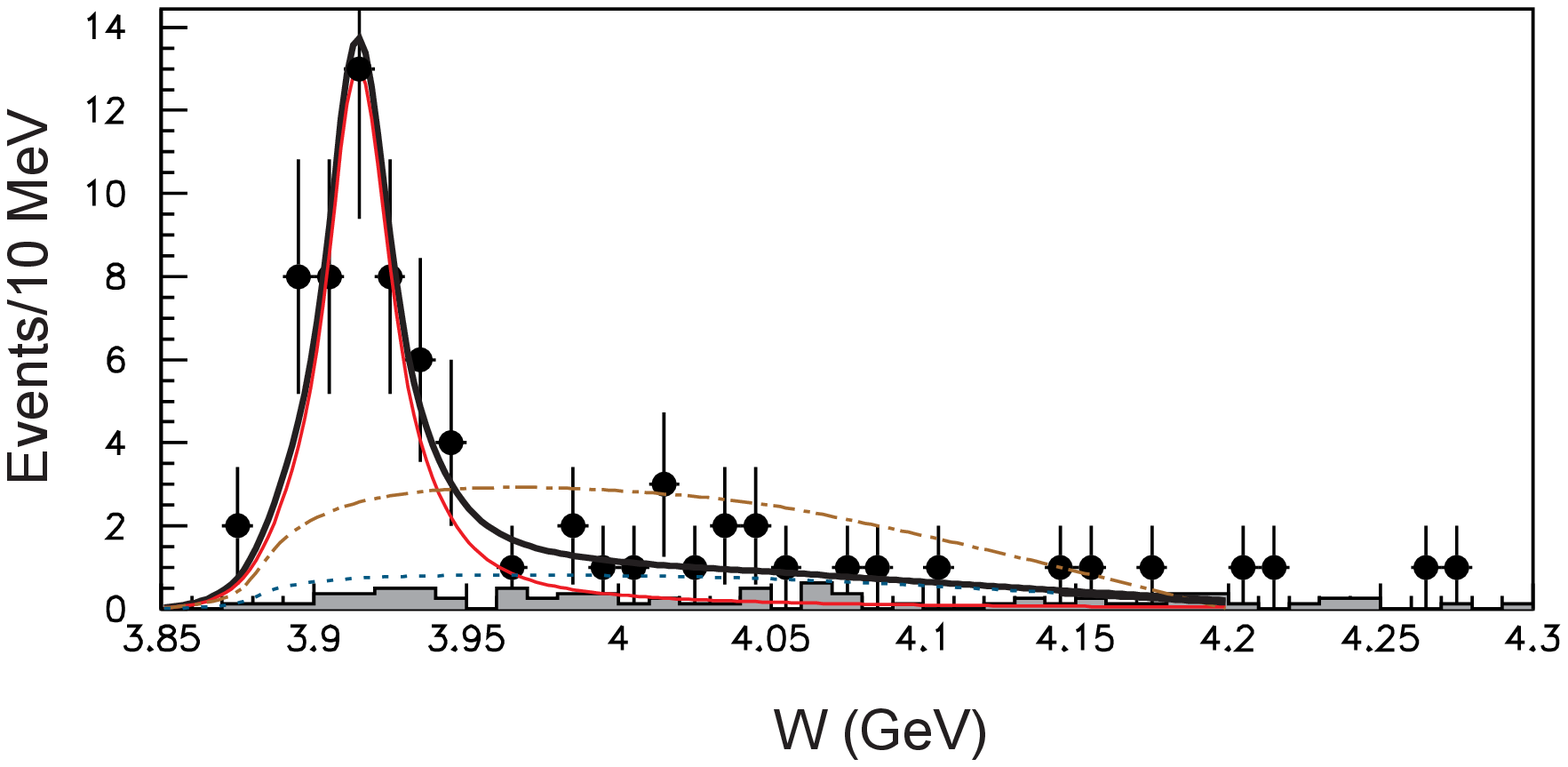}
\end{minipage}
\caption{Left: The points with error bars show the $M(\omega J/\psi)$
distribution for $B \to K \omega J/\psi$  decays.  The curve in 
a) shows results of a fit to a phase-space-like threshold function.
The curve in b) shows the results of a fit 
with a Breit\---Wigner resonance function
included~\cite{choi_y3940}. 
Right: The $\omega J/\psi$ invariant mass distributions for
the two-photon fusion process 
$\gamma\gamma \to \omega J/\psi$~\cite{uehara_y3940}.
The bold solid curve shows results of a fit including a resonance 
(thinner solid curve) and the dot-dashed curve shows
a fit to a phase-space-only distribution; 
the histogram shows $J/\psi$ sideband data. }
\label{fig:y3940}
\end{figure}

The $Y(3940)$ was first observed by Belle as the near-threshold peak 
in the $\omega J/\psi$
invariant mass distribution in 
$B\to K \omega J/\psi$ decays~\cite{choi_y3940}, as shown 
in the left panel of Fig.~\ref{fig:y3940}.  
This observation was subsequently confirmed by
BaBar~\cite{babar_y3940}.  The Belle experiment reported a similar peak in
the near-threshold $\omega J/\psi$ mass 
distribution produced in the two-photon process
$\gamma\gamma\to \omega J/\psi$~\cite{uehara_y3940} 
(see Fig.~\ref{fig:y3940} (right)).
Although the mass of the $Y(3940)$ is well 
above the open-charm threshold, decays to
$D\bar{D}$~\cite{jolanta_ds2710,babar_ds2710}  
and $D^*\bar{D}$~\cite{aushev_ddstar} have not been seen;
in the latter case, a 90\% C.L. upper limit of 
${\mathcal B}(Y(3940)\to D^*\bar{D})<1.4 {\mathcal B}(Y(3940)\to
\omega J/\psi)$ has been established.  
This limit and the rate of production in two-photon processes, 
implies that the partial width to $\omega J/\psi$ is large, 
namely $\Gamma(Y(3940)\to \omega J/\psi) > 1$~MeV,
which is very large for charmonium.

\noindent
Belle's $\gamma\gamma \to Y(3940)\to\omega J/\psi$ observation was confirmed
by BaBar, which also included results of an angular analysis that
favors a $J^{PC} = 0^{++}$ quantum number assignment~\cite{babar_ggomegajp}. 
The only available $0^{++}$ $c\bar{c}$ assignment is the
$\chi^{\prime}_{c0}$, for which the mass value is somewhat high, 
but, perhaps, acceptable.  The
$\chi^{\prime}_{c0}\to D^*\bar{D}$ decay mode is forbidden by parity, but 
$\chi^{\prime}_{c0}\to D\bar{D}$ is allowed and expected to be a strongly
favored mode~\cite{barnes}, so the lack of any prominent signal for it
is a mystery~\cite{jolanta_ds2710}.

\subsubsection{The $Z(3930)$ candidate for the 
$\chi^{\prime}_{c2}$ charmonium state}

\begin{figure}
\begin{minipage}[t]{75mm}
\includegraphics[height=0.5\textwidth,width=0.9\textwidth]{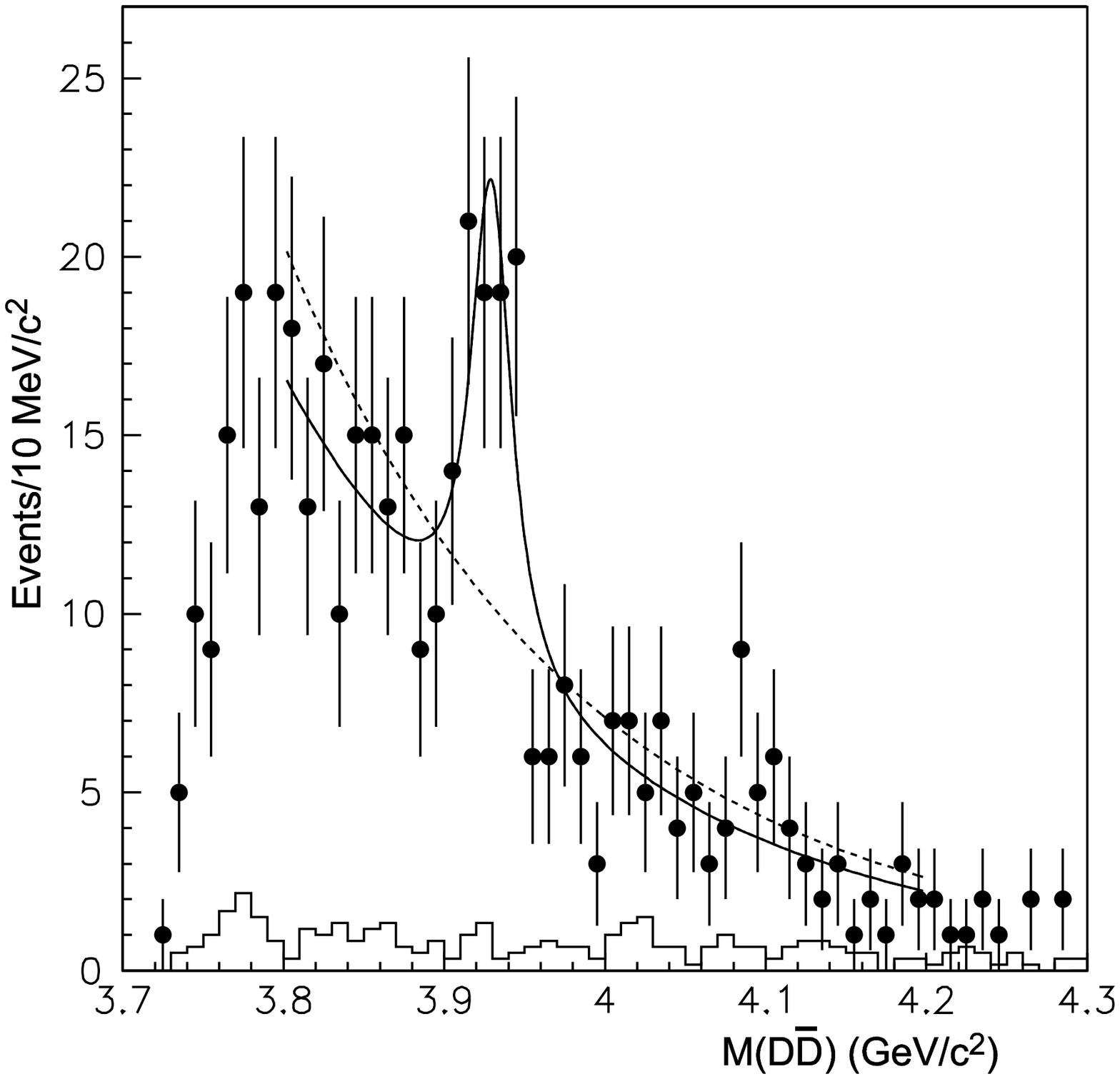}
\end{minipage}
\begin{minipage}[t]{75mm}
  \includegraphics[height=0.5\textwidth,width=0.9\textwidth]{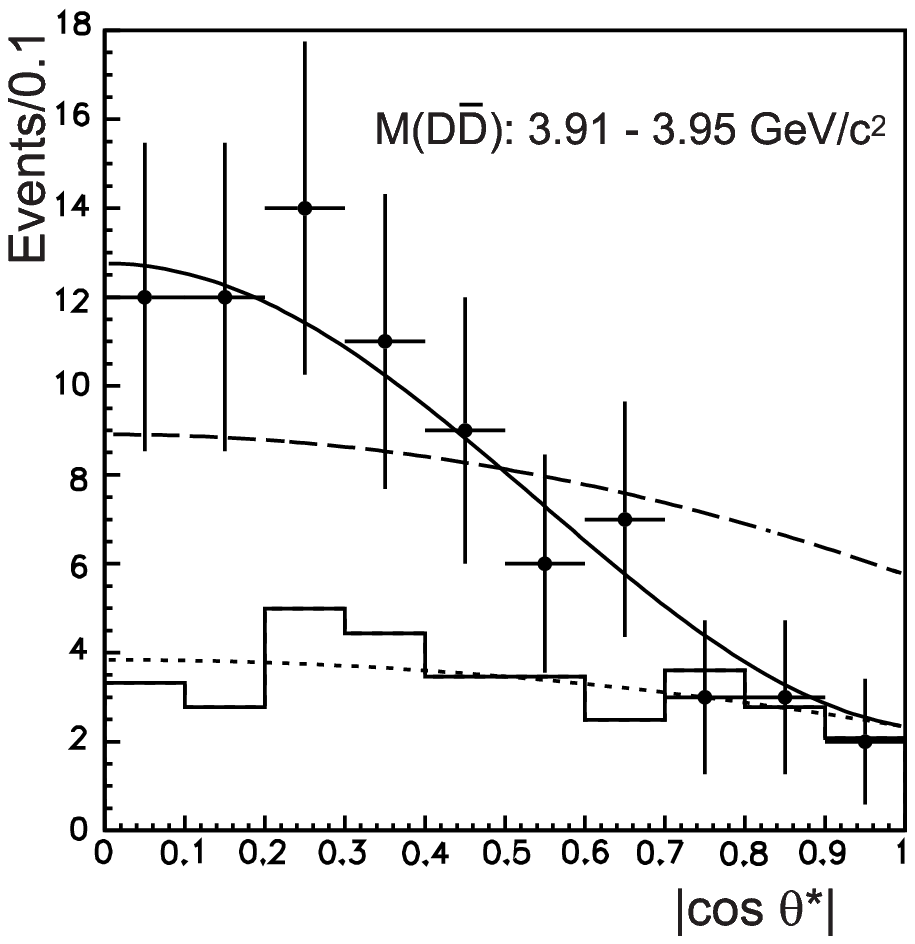}
\end{minipage}
\caption{Left: Invariant mass distributions for $D\bar{D}$
pairs produced via the $\gamma\gamma\to D\bar{D}$ two-photon process.  
The curves show fits to the data with (solid)
and without a resonance term (dashed)~\cite{uehara_z3930}.
Right: 
The yield of events with $3.91<M(D\bar{D})<3.95$~GeV  {\it versus}
$|\cos\theta^*|$. 
The curves are expectations for $J=2$ (solid) and $J=0$ (dashed);
the histogram shows the $M(D\bar{D})$ sideband yield~\cite{uehara_z3930}.
 }
\label{fig:z3930}
\end{figure}

The left panel of Fig.~\ref{fig:z3930} 
shows the $D\bar{D}$ invariant mass distribution
for the process $\gamma\gamma\to D\bar{D}$ 
measured by Belle~\cite{uehara_z3930}, where
a strong peak near 3930~MeV is evident.  
The right panel shows the $|\cos\theta^*|$ 
distribution for events in the $\pm 20$~MeV mass interval centered at 3930~MeV,
where $\theta^*$ is the CM angle 
between the $D$ meson direction and the beamline.
Small values of $|\cos\theta^*|$ are favored, 
which is consistent with expectations
for a $J=2$ resonance (shown in the figure as a solid curve).  
The mass, angular distribution, and the strong decay 
to $D\bar{D}$ are all consistent with expectations
for the $\chi^{\prime}_{c2}$, {\it i.e.,} 
the radially excited $2^{3}P_{2}$ charmonium
state.   

\subsubsection{The $X(3940)$}

\begin{figure}
\begin{minipage}[t]{75mm}
\includegraphics[height=0.6\textwidth,width=0.9\textwidth]{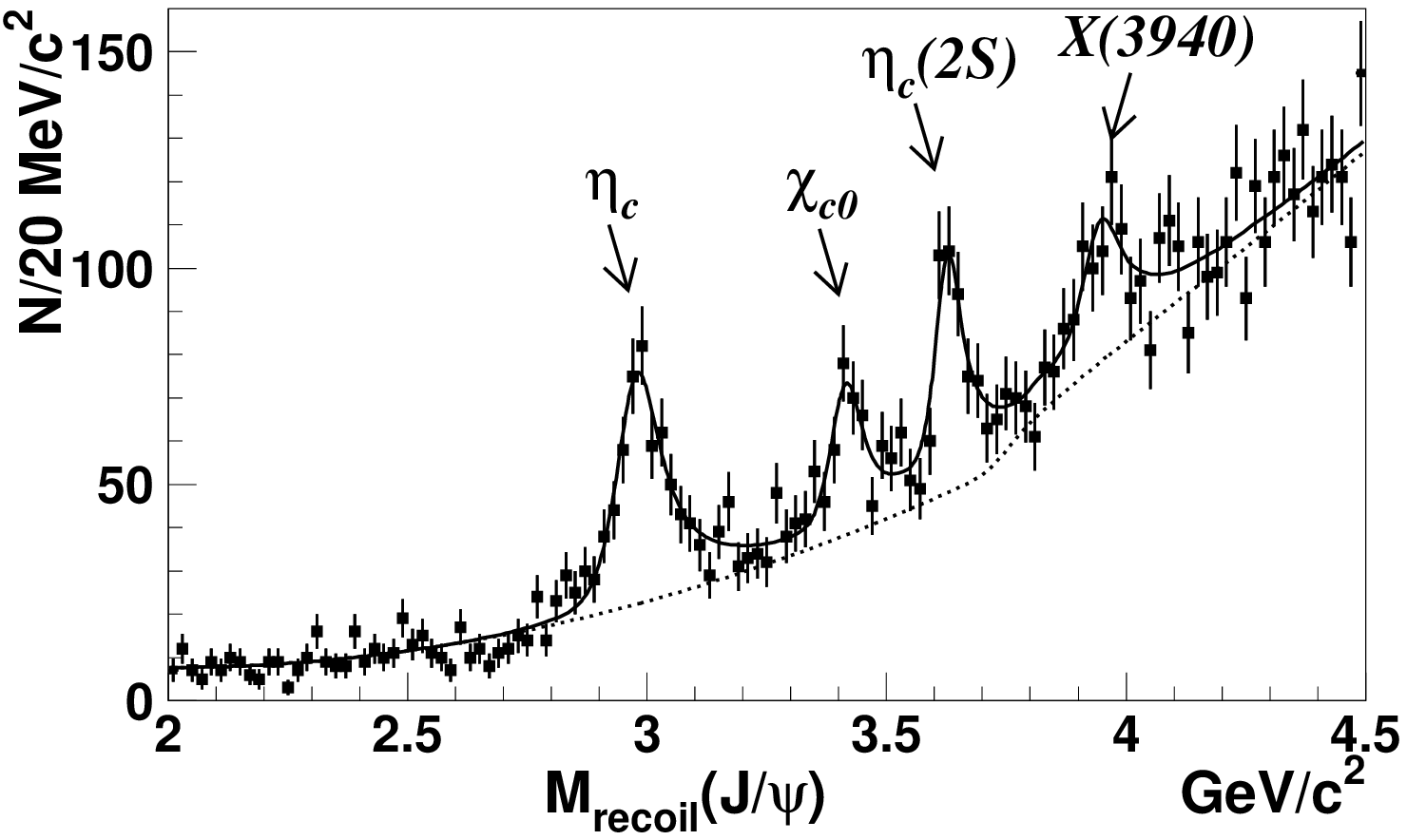}
\end{minipage}
\begin{minipage}[t]{75mm}
  \includegraphics[height=0.6\textwidth,width=0.9\textwidth]{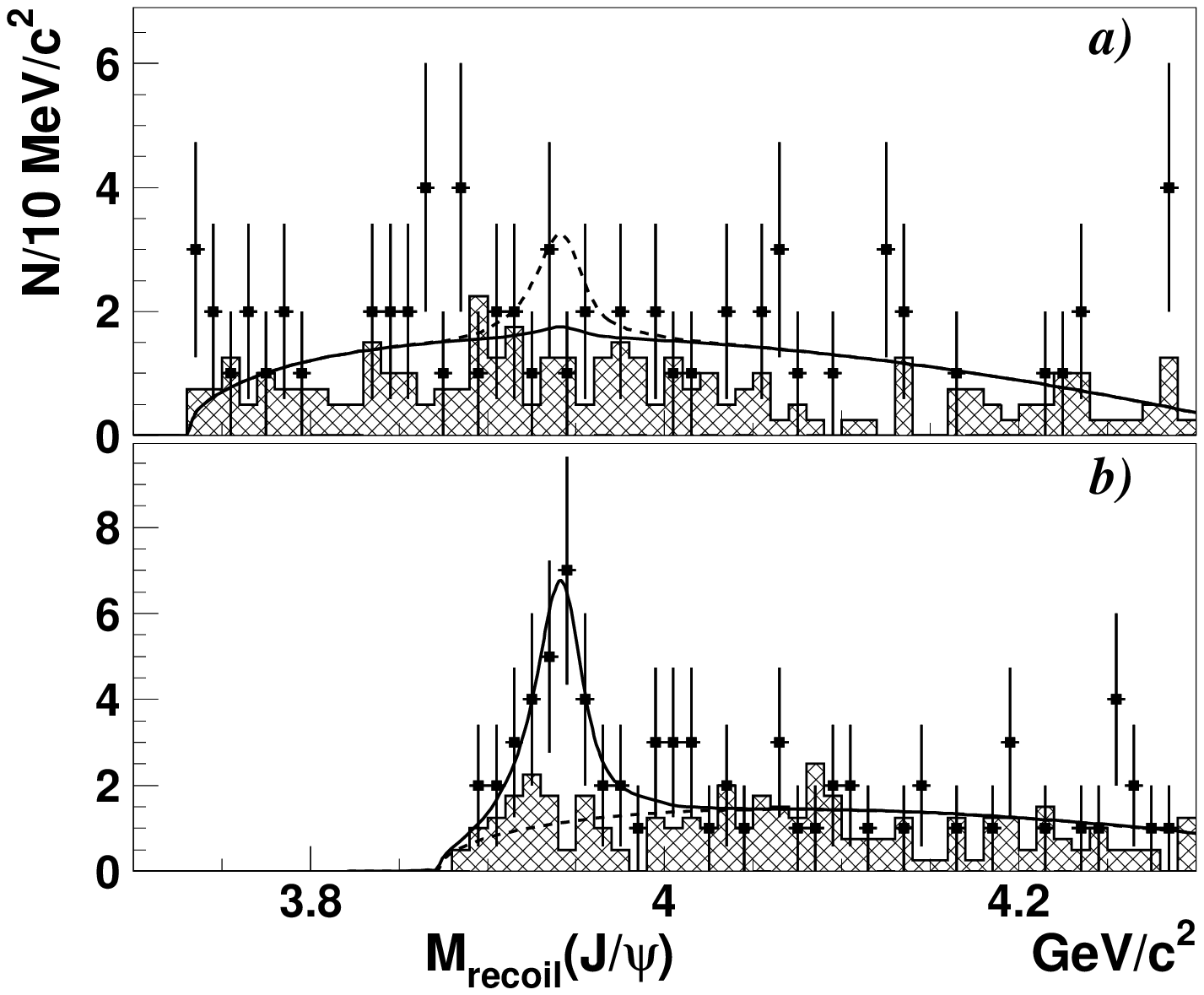}
\end{minipage}
\caption{Left: The distribution of masses recoiling from the $J/\psi$ in
inclusive $e^+e^-\to J/\psi \, X$ reactions~\cite{pakhlov_x3940}. 
The solid curve
shows the result of a fit that includes $\eta_c$, $\chi_{c0}$, $\eta_c(2S)$, and
$X(3940)$ resonance terms as well as
 a smooth background function that has a
step at the $D\bar{D}$ threshold (dotted curve).
Right: 
The a) $D\bar{D}$ and b) $D^*\bar{D}$ invariant mass distributions
from exclusive $e^+e^-\to J/\psi D^{(*)}\bar{D}$ annihilation~\cite{pakhlov_x3940}.
The curves are fits that include possible resonance terms and the histograms
are backgrounds determined from the $D$-meson sidebands. 
The dashed curves show: 
a) the 90\% C.L. upper limit on the signal; b) the background function.
 }
\label{fig:x3940}
\end{figure}

Belle discovered a third meson state with mass near 3940~MeV, the $X(3940)$,
produced in association with a $J/\psi$ in $e^+e^-$ annihilation. The left
panel of Fig.~\ref{fig:x3940} shows the distribution of masses 
recoiling from the $J/\psi$ in inclusive 
$e^+e^-\to J/\psi \, X$ reactions~\cite{pakhlov_x3940}.
With a partial reconstruction technique, Belle was able to isolate samples
of exclusive $e^+e^-\to J/\psi D\bar{D}$ and $J/\psi D^*\bar{D}$ events.
The $D\bar{D}$ and $D^*\bar{D}$ invariant mass distributions for these
samples are shown in the right panels of Fig.~\ref{fig:x3940}.  There
is no sign of the $X(3940)$ in the $ D\bar{D}$ events, but there is a distinct
signal for $X(3940)\to D^*\bar{D}$. 

\noindent
To address the question of
whether or not the $X(3940)$ is the same state as the $Y(3940)$,
a search~\cite{pakhlov_x3940} was made for $e^+e^-\to J/\psi \omega J/\psi$.  
No signal for $X(3940)\to\omega J/\psi$ was seen
and a 90\% C.L. lower limit 
${\mathcal B}(X(3940)\to D^*\bar{D})>1.7 {\mathcal B}(X(3940)\to
\omega J/\psi)$ was established, 
which is inconsistent with the corresponding upper limit for the $Y(3940)$
discussed  above.  This implies that the $Y(3940)$, produced in $B$ decays
and decaying to $\omega J/\psi$, and the $X(3940)$, produced in
association with a $J/\psi$ and decaying to $D\bar{D}^*$, are distinct
states.  The only $c\bar{c}$ assignment available for the $X(3940)$ is
the $\eta_c(3S)$, for which decays to $D^*\bar{D}$ are expected to be
dominant and decays to $D\bar{D}$ are forbidden by parity.  However,
the $^3S_1$ triplet partner state of the $\eta_c(3S)$ is the well established
$\psi(4040)$, with a mass of $4040\pm 4$~MeV~\cite{pdg2012}.  Assigning
the $X(3940)$ as the $\eta_c(3S)$ would mean
$\Delta M_{hfs}(3S) = 98 \pm 8$~MeV, {\it i.e.,} twice as large
as $\Delta M_{hfs}(2S)$ (see above) and in strong disagreement with
theoretical expectations.

\subsubsection{Anomalous $J^{PC}=1^{--}$ states 
seen in initial-state-radiation processes}

\begin{figure}
\begin{minipage}[t]{75mm}
\includegraphics[height=0.6\textwidth,angle=0,width=0.8\textwidth]{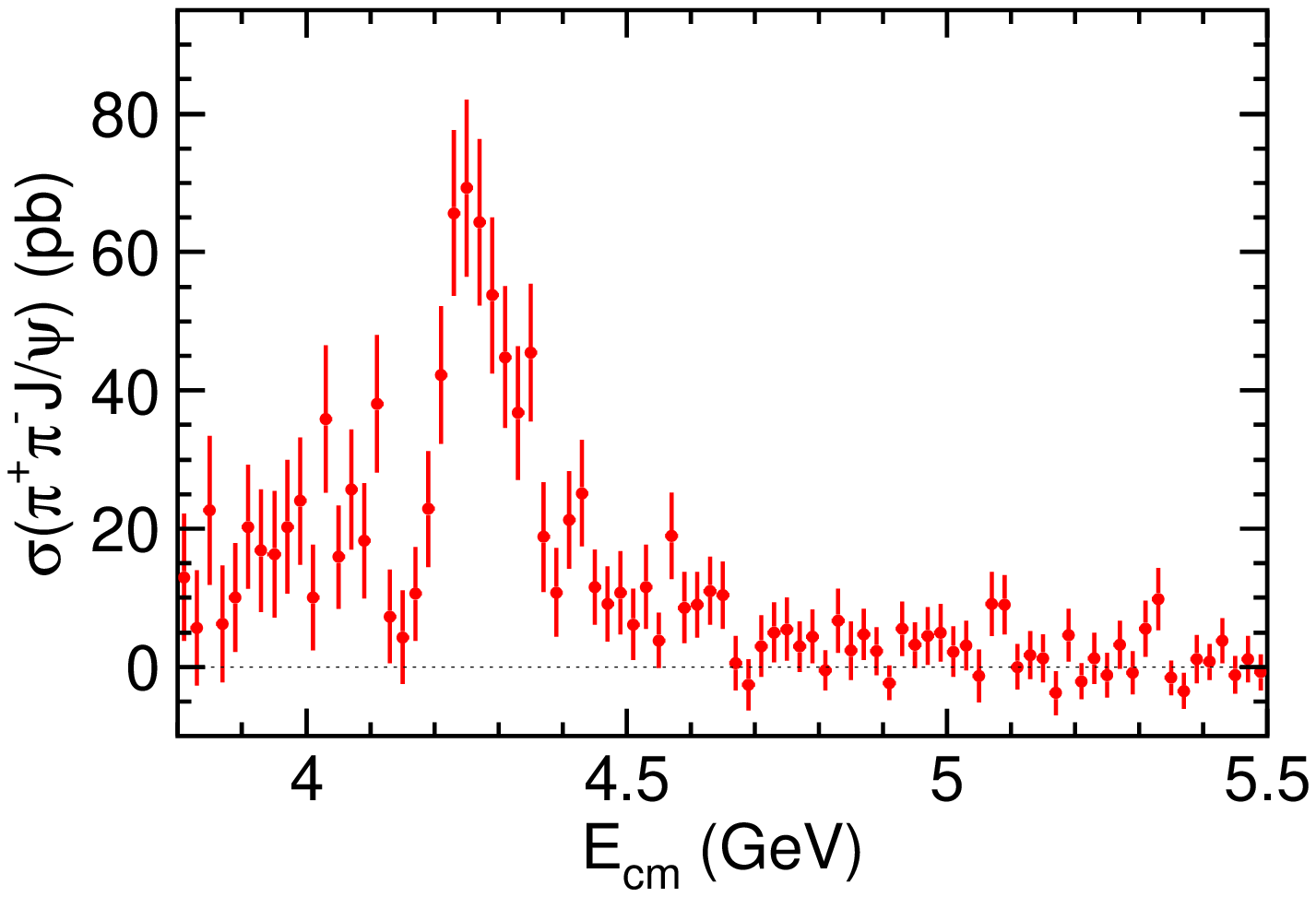}
\end{minipage}
\begin{minipage}[t]{75mm}
  \includegraphics[height=0.55\textwidth,width=0.8\textwidth]{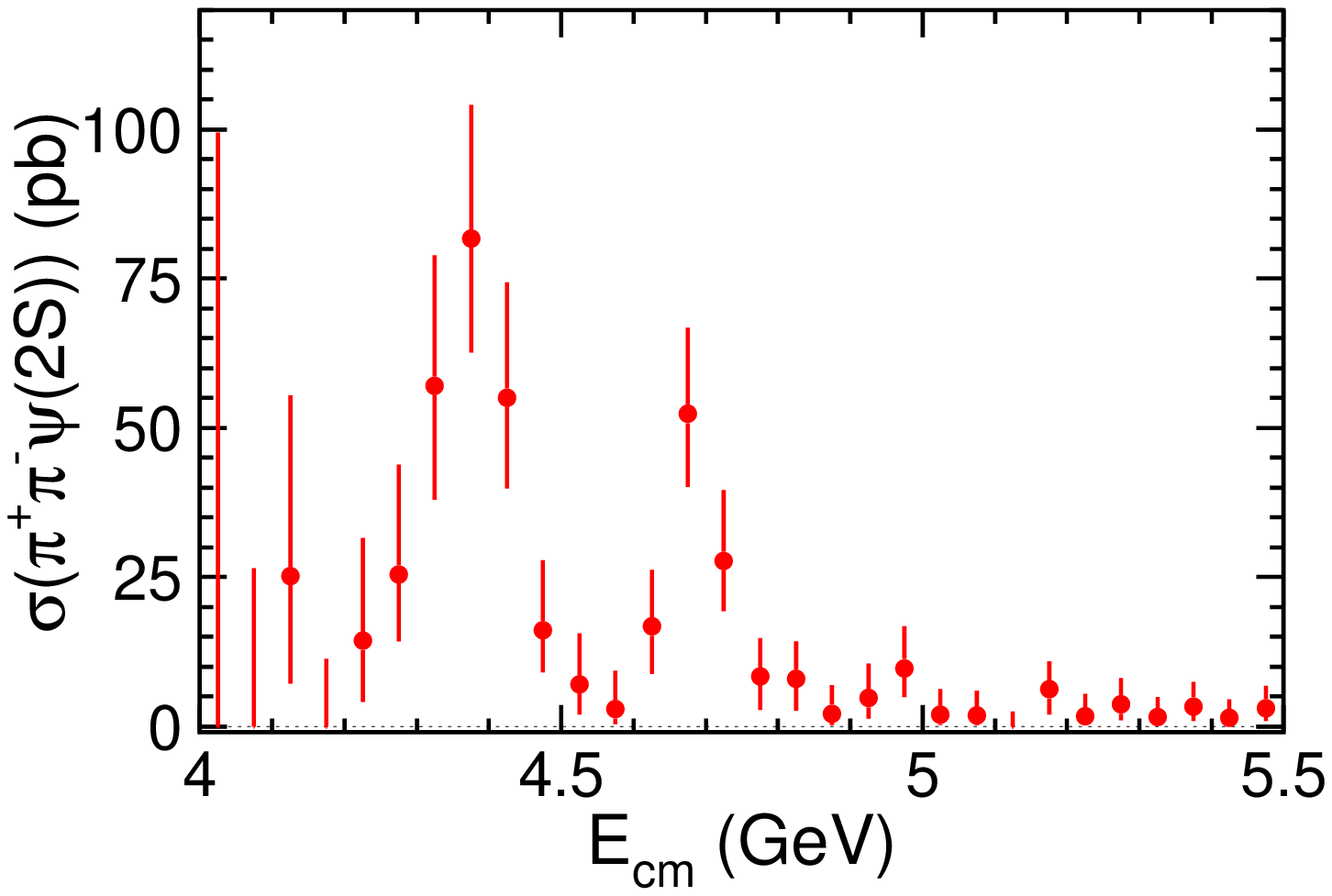}
\end{minipage}
\caption{
 The cross sections for left:
$e^+e^-\to \pi^+ \pi^- J/\psi$~\cite{yuan_y4260} and
right:
$e^+e^-\to \pi^+ \pi^- \psi(2S)$~\cite{wang_y4360}.
}
\label{fig:y4260}
\end{figure}

In 2005, BaBar reported the discovery of a striking $\pi^+\pi^- J/\psi$ peak
near 4260~MeV in the initial-state-radiation process
$e^+e^-\to \gamma_{isr}\pi^+\pi^- J/\psi$~\cite{babar_y4260}. This observation
was subsequently confirmed 
by CLEO~\cite{cleo_y4260} and Belle~\cite{yuan_y4260}.
The cross section for $e^+e^-\to \pi^+\pi^- J/\psi$ from the Belle paper is
shown in the left panel of Fig.~\ref{fig:y4260}, where a prominent signal
for the $Y(4260)$ with a peak cross section of $\sim 70$~pb is evident.
Curiously, the total cross section for $e^+e^-$ annihilation into
open charmed mesons shows no sign of a peak at 4260~MeV; the total
cross section for open charm at the $Y(4260)$ peak is about
3~pb~\cite{bes_R},  which, taken together with the measured natural width 
$\Gamma_{\rm tot}[Y(4260)]=95\pm 14$~MeV, implies a 90\% C.L. lower limit on the
partial width $\Gamma(Y(4260)\to \pi^+\pi^- J/\psi)>1.6$~MeV~\cite{xhmo_y4260}.
This is much larger than values that are typical for $1^{--}$ charmonium
states ({\it e.g.}, $\Gamma(\psi(3770)\to \pi^+\pi^- J/\psi)=53\pm 8$~keV).  

BaBar also reported a similar peak in the $\pi^+\pi^-\psi(2S)$ cross section
at 4325~MeV~\cite{babar_y4360}.  With higher statistics, Belle
confirmed this (now called the $Y(4360)$), and found a second, higher
mass peak, the $Y(4660)$~\cite{wang_y4360} (see the right panel of
Fig.~\ref{fig:y4260}).  Here too, there are no evident accompanying
structures in the open charm cross sections near these masses.  Another
peculiar feature is that, with the currently available statistics,
there are no signs of the $Y(4260)$ in the $\pi^+\pi^-\psi(2S)$ channel
or of the $Y(4360)$ or $Y(4660)$ in the $\pi^+\pi^- J/\psi$ channel.

\subsubsection{The electrically charged  {\it $Z^-$} charmonium-like 
meson candidates}

\begin{figure}[htb]
\begin{minipage}[t]{75mm}
\includegraphics[height=0.3\textwidth,angle=0,width=0.8\textwidth]{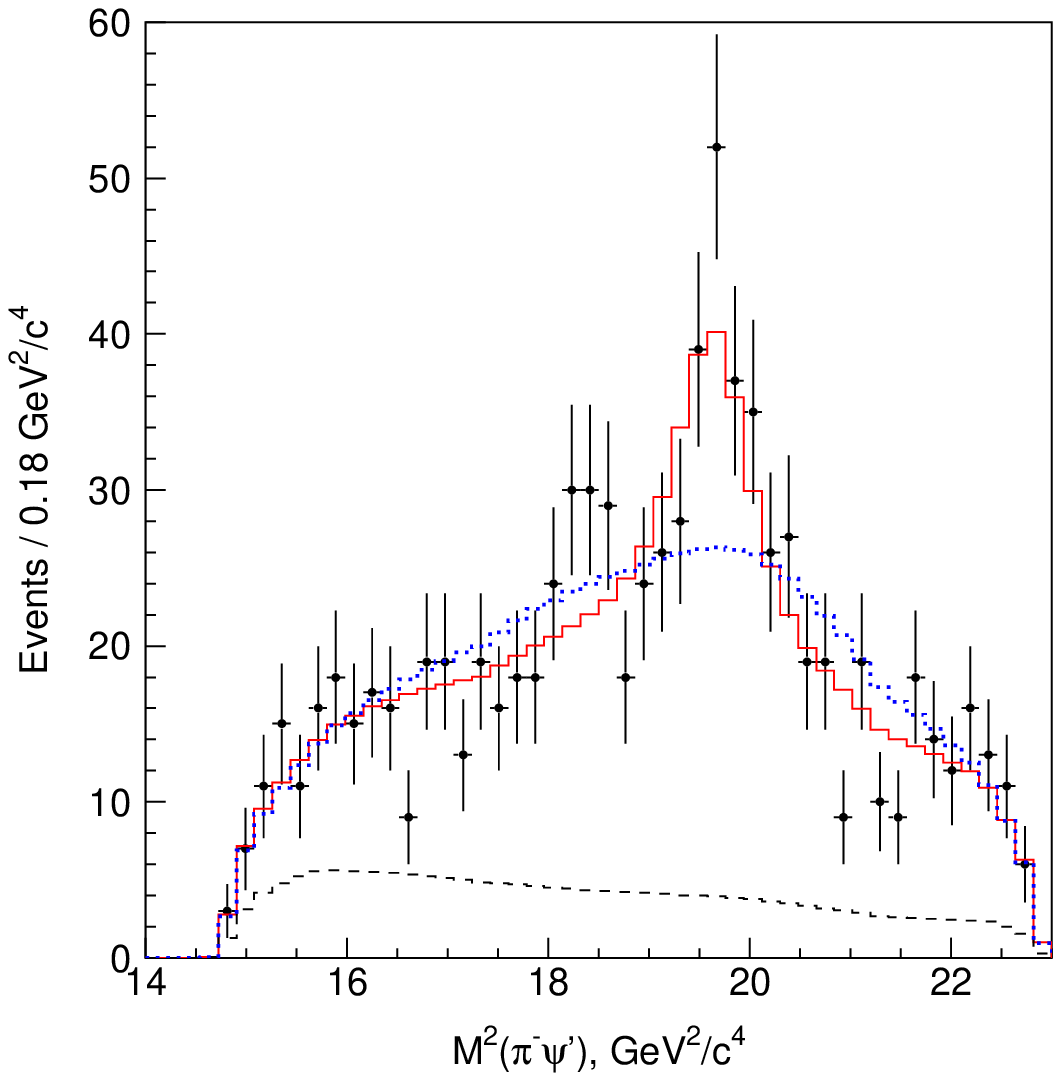}
\end{minipage}
\begin{minipage}[t]{75mm}
  \includegraphics[height=0.75\textwidth,width=0.8\textwidth]{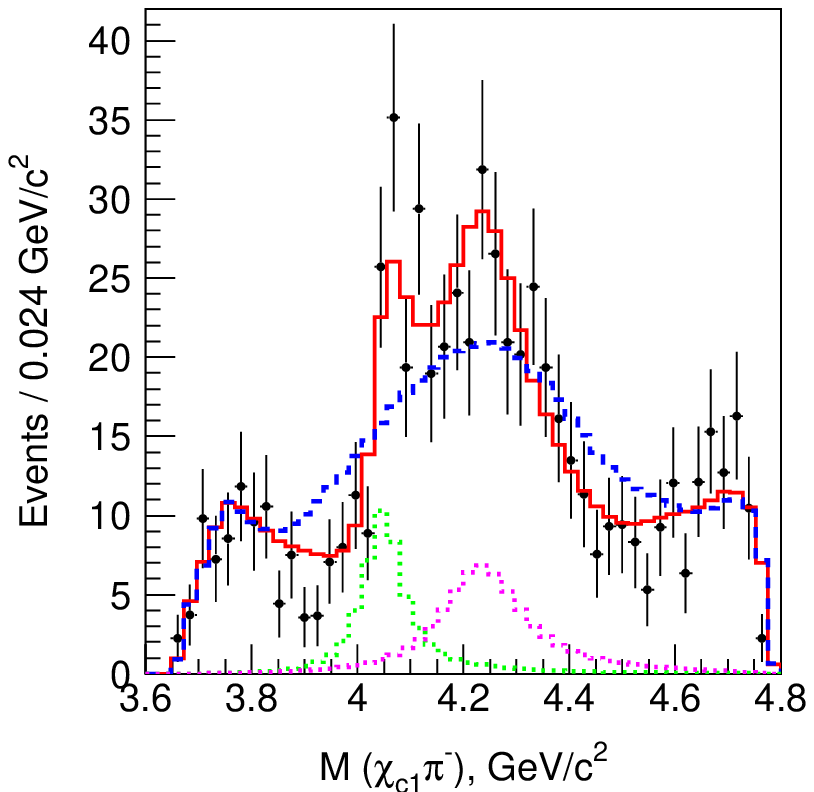}
\end{minipage}

\caption{Left: The
$M^2_{\psi^{\prime}\pi^-}$ projection of the Dalitz plot with the
$K^\ast$ bands removed is shown as data points~\cite{belle_z4430_2}. 
The histograms show the corresponding projections of the Dalitz-plot fits with 
(red solid) and without (blue dotted) a $Z^{-}\to\psi^{\prime}\pi^-$ 
resonance term. 
The dashed histogram is the background.
Right: The data points show the
$M_{\chi_{c1}\pi^-}$ projection of the Dalitz plot with the $K^\ast$
bands removed. The histograms show the corresponding projections of
the fits with (red solid) and without (blue dotted) two $Z^{-} \to \chi_{c1}\pi^-$ 
resonance terms, the dotted histograms represent the contribution of the two 
$\chi_{c1}\pi^-$ resonances.}
\label{fig:z4430}
\end{figure}

In 2008, Belle reported peaks in the
$\psi^{\prime}\pi^-$ and $\chi_{c1}\pi^-$ invariant
mass distributions in $B\to \psi^{\prime}\pi^-K$
(Fig.~\ref{fig:z4430}~(left)~\cite{belle_z4430_1,belle_z4430_2} 
and $B\to\chi_{c1}\pi^- K$ (Fig.~\ref{fig:z4430}~(right)~\cite{belle_z14050},
respectively. If these peaks are meson resonances, they would necessarily
have a minimal quark content of $c\bar{c}d\bar{u}$ and be unmistakably
exotic.  Although in both cases the peaks have greater than 
5$\sigma$ statistical significance, the experimental situation remains
uncertain since none of these peaks have yet been confirmed by other
experiments. Analyses by BaBar of the same channels
neither confirm nor contradict the Belle
claims~\cite{babar_z4430}.

\subsubsection{Studies of open charmed hadron pair-production 
via initial-state-radiation}

The observation of the $Y(4260)$ motivated a Belle program of
measurements of exclusive $e^+ e^-$ cross sections for 
charmed hadron pairs near threshold. Belle presented the 
first measurements of exclusive cross sections for the 
production of charmed-hadron pairs in electron\---positron 
annihilation in the vicinity of the threshold for open-charm 
production performed at CM energies near the $\Upsilon(4S)$
resonance using the initial-state-radiation process.
The continuous energy spectrum of this radiation allows
investigating the production of charmonium with quantum
numbers $J^{PC} = 1^{- -}$ over the whole energy range. 
The electromagnetic suppression of hard photon radiation 
is compensated by an enormous integrated luminosity collected 
at the $B$-factories, and selection criteria specific for 
the ISR processes provide high efficiency at considerable 
suppression of the background. Taken together, these factors 
resulted in measurements that are competitive in precision
with the CLEOc~\cite{cleo_y4260} and BESII~\cite{bes_psi} 
experimental data in which charmed-hadron 
cross sections were measured using $e^+ e^-$ energy scans
with and without electromagnetic suppression.

\begin{figure}
\includegraphics[height=0.7\textwidth,width=1.0\textwidth]{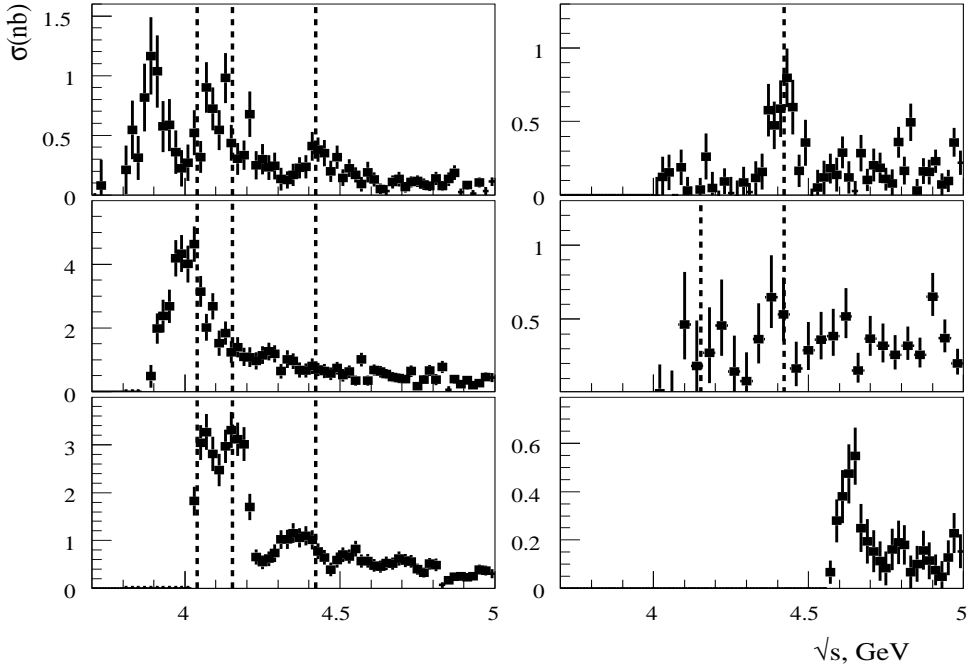}
\caption{Exclusive cross sections for charmed-hadron pair production
measured in Belle~\cite{Galina_1,Galina_2, Galina_3,Galina_4,Galina_5}.
Left: $D\bar{D}$, $D=D^0$~or~$D^+$ (upper); $D^+ D^{*-}$ (middle); $D^{*+}D^{*-}$ (lower). Right: $D^0D^-\pi^+$ (upper); $D^0 D^{*-}\pi^+$ (middle); $\Lambda_c^+ \Lambda_c^-$ (lower).
The vertical dashed lines indicate the mass values of the established
$1^{--}$ charmonium states: $\psi(4040)$, $\psi(4160)$, and $\psi(4415)$.}
\label{fig:galina}
\end{figure}

The exclusive $e^+ e^-$ cross sections to $D \overline{D}$ 
($D$ = $D^0$ or $D^+$), $D^+ D^{*-}$, $D^{*+} D^{*-}$, 
$D^0 D^- \pi^+$, and $D^0 D^{*-} \pi^+$ final state using 
ISR~\cite{Galina_1,Galina_2,Galina_3,Galina_4}, shown in 
Fig.~\ref{fig:galina}, have no evident peaks that can be
associated with any of the above-mentioned $Y$ states,
contrary to expectations for conventional $J^{PC}=1^{--}$
charmonium states with such large masses and total widths.

In 2008, the Belle collaboration reported the observation 
of a significant near-threshold peak, called the $X(4630)$, in 
the $e^+ e^- \to \Lambda_c^+ \Lambda_c^-$ exclusive cross section
shown in the lower right-hand panel of Fig.~\ref{fig:galina}
\cite{Galina_5}. It remains unclear whether or not this observed peak
is a resonance. In particular,  peaks near the baryon\---antibaryon pair
mass threshold are observed in many processes, including three-body baryon
decays of $B$-mesons\cite{Wang:2003yi}. The mass and width
of the $X(4630)$ peak determined under the assumption that the 
$X(4630)$ is due to a resonance are 
$M = (4634 \pm 10)$ MeV/$c^2$ and 
$\Gamma = (92 \pm 40)$ MeV. These values agree 
within errors with the mass and the full width of the
$Y(4660)$ peak seen in the $Y(4660)\to\pi^+\pi^-\psi'$
decay channel\cite{wang_y4360} as mentioned above. 
Such a coincidence (including quantum numbers) may not 
be accidental, although the possibility that the 
$X(4630)$ and $Y(4660)$ peaks have different origins
cannot be ruled out.   Among possible conventional interpretations,
it has been suggested that the $X(4630)$ is the $\psi(5S)$ or
$\psi(6S)$ $1^{--}$ charmonium state\cite{psi5s}, or a threshold effect
caused by the presence of the $\psi(3D)$ state with mass
slightly below the $\Lambda_c^+ \Lambda_c^-$ threshold.

\subsubsection{Summary}

This section has highlighted only a fraction of the charmonium and 
charmonium-related results from Belle.
In addition to the observations described above, Belle reported a number
of other observations related to charmonium.
A near-threshold peak was found in the $D^*\bar{D}^*$
$e^+e^- \to J/\psi D^*\bar{D}^*$ annihilation
process\cite{pakhlov_x4160}.  A Belle search for the $Y(4140)$ ---a 
$\phi J/\psi$ resonance reported by CDF\cite{cdf_y4140} ---in the 
$\phi J/\psi$ mass distribution produced via the $\gamma\gamma\to \phi J/\psi$
two-photon process found no evidence for the $Y(4140)$ but, instead, uncovered
a 3.2$\sigma$ significant peak at higher mass that was dubbed the $X(4350)$~\cite{shen_x4350}.
Belle cross section measurements of exclusive processes of the type
$e^+e^-\to J/\psi \eta_c$~\cite{pakhlov_etac2s} and
$e^+e^- \to J/\psi D^{(*)}\bar{D}^{(*)}$~\cite{pakhlov_jpccbar}
found order-of-magnitude disagreements with NRQCD predictions~\cite{nrqcd} 
and have had a profound impact on
subsequent developments in the theory; see, e.g., Ref.~\cite{nrqcd_qwg}.
A recent study of the $\gamma \chi_{c1}$ mass distribution in the $B$-meson
decay process $B\to K\gamma \chi_{c1}$ found 
strong evidence for the long-sought-for
$\psi_{c2}$, the $^3D_2$ charmonium state~\cite{bhardwaj_psi2c}.

In the original physics program planned for Belle outlined in
the Belle Letter of Intent~\cite{belle_loi}, no mention was made of charmonium
physics or searches for non-conventional, multi-quark meson states.  Somewhat
unexpectedly, thanks in part to the huge data samples provided by the KEKB
collider, Belle turned out to be a powerful instrument for both 
conventional charmonium physics, and for 
uncovering a new class of charmonium-like
states that have yet to be understood~\cite{godfrey}.

\subsection{Bottomonium(-like) states}
As described in the previous section, most of the new charmonium 
states discovered in recent years at the $B$-factories do not seem 
to have a simple $c\bar{c}$ structure. Although the masses of these 
states are above the corresponding thresholds for decay into a pair of 
open charm mesons, they decay readily into $J/\psi$  or $\psi(2S)$ and pions,
which is unusual for $c\bar{c}$ states. In addition, their masses and decay 
modes are not in agreement with the predictions of potential models, 
which, in general, describe $c\bar{c}$ states very well. For
these reasons, some of these charmonium-like states
 are probably more complex than simple quark\---antiquark
states and are candidates for exotic objects 
such as hybrid, molecular, or tetraquark states.
Recently, Belle has made a series of exciting discoveries of
new states in the bottomonium sector using its unique data sample
taken around the $\Upsilon(5S)$ resonance.
\begin{figure}
\centerline{\includegraphics[width=13cm,height=7cm]{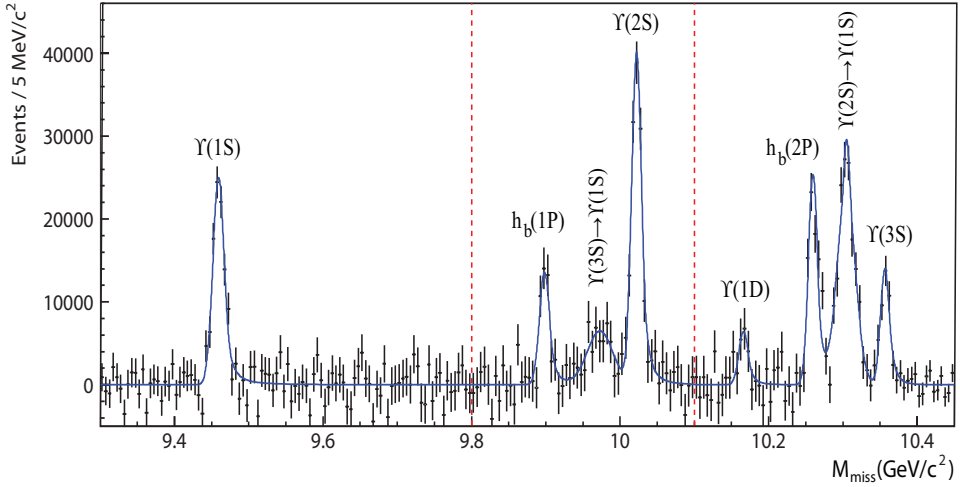}}
\caption{ The inclusive $M_{\rm miss}$ spectrum with 
the combinatorial background subtracted (points
with error bars) and the signal component of the fit function overlaid (smooth 
curve). The vertical lines indicate boundaries of the fit regions.
The expected high-statistics reference signals, 
$\Upsilon(5S) \to \Upsilon(nS) \pi^+ \pi^-$ where $n=1,2,3$, as well as
the newly observed $h_b(1P)$ and $h_b(2P)$ states are seen.}
\label{fig:mmiss}
\end{figure}

Bottomonium refers to bound states of $b \bar{b}$ quarks and
is considered an excellent laboratory to study QCD at low energy. 
The spin-singlet states $h_b(nP)$ and $\eta_b(nS)$ alone provide
information concerning the spin\---spin (or hyperfine) interaction in 
bottomonium. Measurements of the $h_b(nP)$ masses would provide unique
access to the $P$-wave hyperfine splitting, $\Delta M_{hfs}(nP) \equiv 
  \; <M(n^3P_J)> - \;M(n^1 P_1)$, the difference between the 
spin-weighted average mass of the $P$-wave triplet states 
($\chi_{bJ}(nP)$ or $n^3 P_J$) and that of corresponding $h_b(nP)$, 
or $n^1 P_1$. 
These splittings are predicted to be close to zero. 
Recently, CLEO observed the process 
$e^+e^- \to h_c(1P)\pi^+ \pi^-$ at a rate comparable to that for 
$e^+e^- \to J/\psi \pi^+ \pi^-$ in data taken at the $\psi(4160)$ resonance.
Such a large rate was unexpected because the production
of $h_c(1P)$ requires a $c$-quark spin-flip, while production of 
$J/\psi$ does not. 
Belle previously reported anomalously 
high rates for $e^+e^- \to \Upsilon(nS) \pi^+ \pi^-$ 
($n$ = 1, 2, 3) at energies near the $\Upsilon(5S)$ mass~\cite{y_belle}. 
If the $\Upsilon(nS)$ signals are attributed entirely to 
$\Upsilon(5S)$ decays, the measured partial decay widths 
$\Gamma [\Upsilon(5S) \to \Upsilon(nS) \pi^+ \pi^-] \sim$~0.5~MeV
are about two orders of magnitude larger than typical widths
for di-pion transitions among the four lower $\Upsilon(nS)$ states.
Using the large data sample collected at energies near the 
$\Upsilon(5S)$ resonance and motivated
by the suggestive CLEO result, Belle decided to investigate the 
missing $h_b(mP)$ singlet bottomonium states~\cite{hb_belle}.

We do not expect the $h_b(mP)$ states to have a large dominant 
exclusive decay mode, which would allow their reconstruction with
high efficiency. Instead, they are reconstructed inclusively using 
the missing mass (recoil mass) of the $\pi^+ \pi^-$ pair. 
The $\pi^+ \pi^-$ missing mass is defined 
as $M^2_{\rm miss} \equiv (P_{\Upsilon(5S)} - P_{\pi^+ \pi^-})^2$, where
$P_{\Upsilon(5S)}$ is the 4-momentum of the $\Upsilon(5S)$ determined
from the beam momenta and $P_{\pi^+ \pi^-}$ is the 4-momentum 
of the $\pi^+ \pi^-$ system. 
The $\pi^+ \pi^-$ transitions between $\Upsilon(nS)$ states 
provide high-statistics reference signals as shown in 
Fig.~\ref{fig:mmiss}. The $h_b(nP)$ states are also very clearly,
and for the first time, observed here. 
The measured masses of the $h_b(1P)$ and $h_b(2P)$, $M = (9898.3 \pm 1.1 
{}^{+1.0}_{-1.1})$~MeV/$c^2$ and $M = (10259.8 \pm 0.6 {}^{+1.4}_{-1.0})$~MeV/$c^2$
respectively, correspond to hyperfine splittings that are consistent
with zero. The processes $\Upsilon(5S) \to h_b(mP) 
\pi^+ \pi^-$, which require a heavy-quark spin flip, are then
found to have rates that are comparable
to those for the heavy-quark spin conserving transitions 
$\Upsilon(5S) \to \Upsilon(nS) \pi^+ \pi^-$.
These observations differ from a priori theoretical expectations
and strongly suggest that exotic mechanisms contribute
to $\Upsilon(5S)$ decays.

\begin{figure}[htb]
\parbox{\halftext}{
\centerline{\includegraphics[width=6.5cm,height=5.5cm]{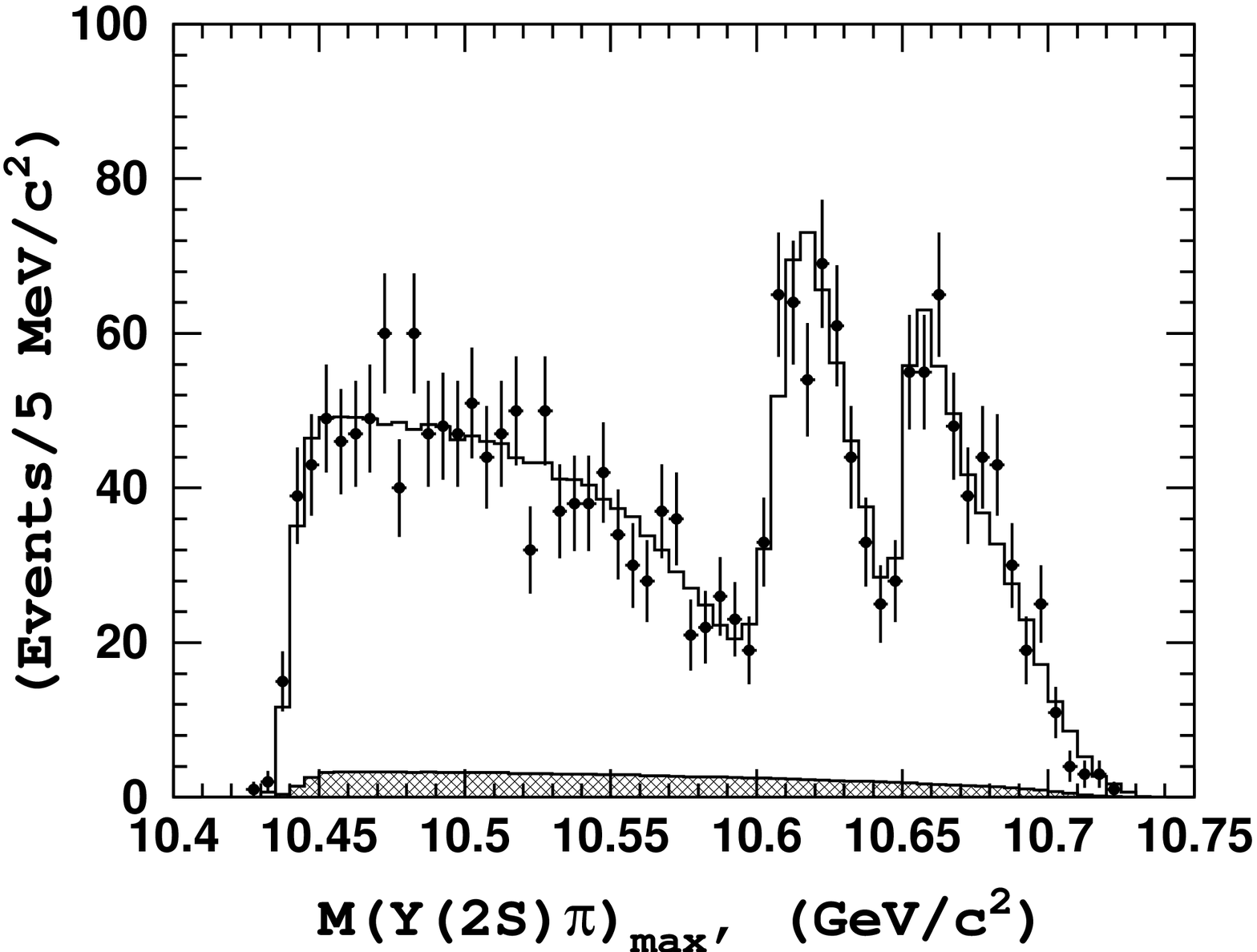}}
\caption{Comparison of fit results (open histograms) with experimental
data (points with error bars) for events in the $\Upsilon(2S)$ signal
regions. $M(\Upsilon(2S)\pi)_{\rm max}$ is the maximum invariant
mass of the two $\Upsilon(2S)\pi$ combinations. 
The hatched histogram shows the background component.}
\label{fig:mass_1}
}
\hfill
\parbox{\halftext}{
\centerline{\includegraphics[width=6.5cm,height=5.5cm]{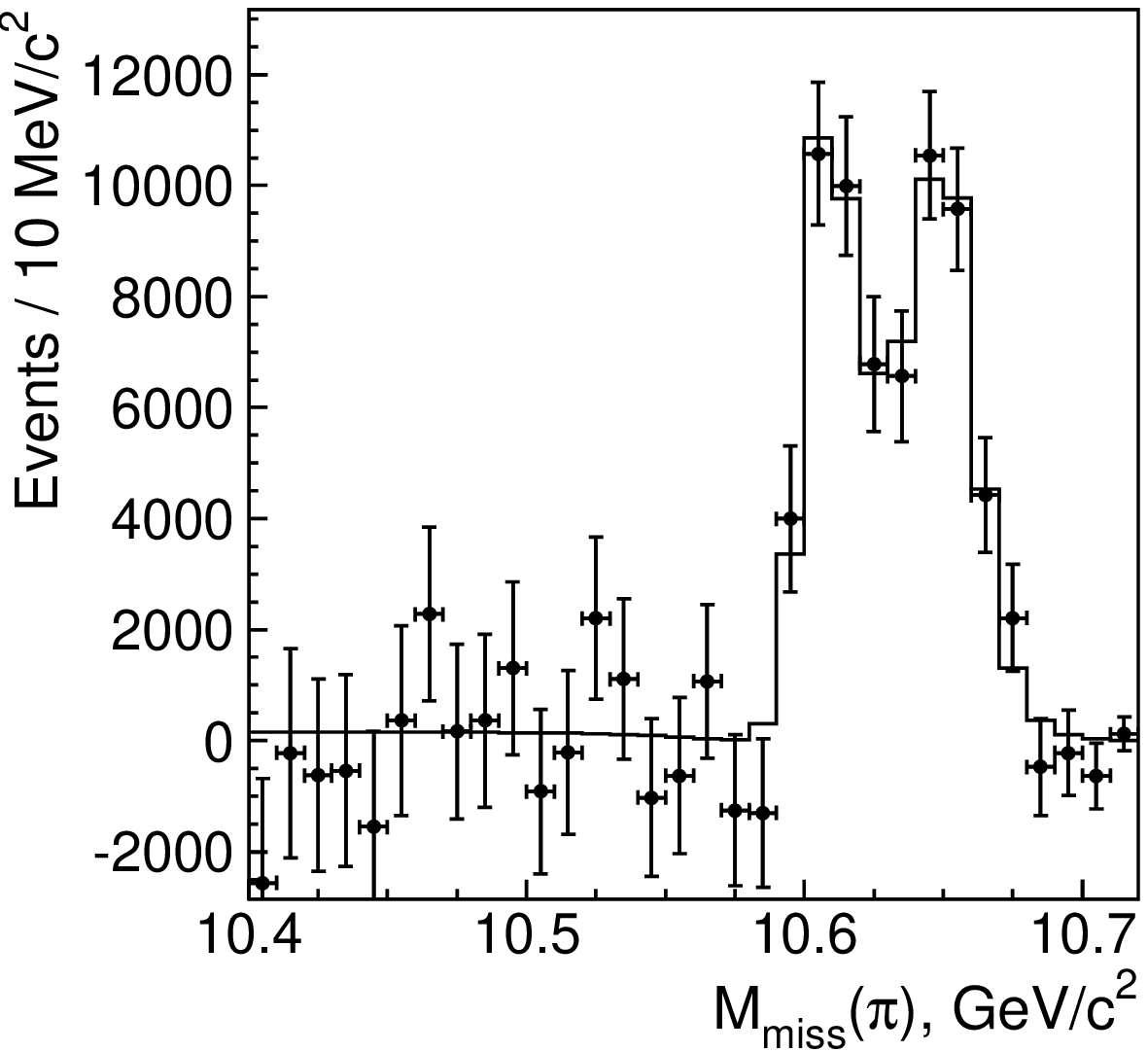}}
\caption{Comparison of fit results for events in the $h_b(1P)$ signal region. 
The $Z_b(10610)$ and $Z_b(10650)$ are clearly observed in both
cases; the result of the fit is represented by the histogram.
}
\label{fig:mass_2}}
\end{figure}

%
%
To understand the $\Upsilon(nS)$ and $h_b(mP)$ production mechanism 
at the $\Upsilon(5S)$ resonance, it is necessary to study in 
detail the resonant structure of the $\Upsilon(5S) \to \Upsilon(nS)
\pi^+ \pi^-$ and $\Upsilon(5S) \to h_b(mP) \pi^+ \pi^-$ 
transitions~\cite{zb_belle}.
In the case of $\Upsilon(5S) \to \Upsilon(nS)\pi^+ \pi^-$, the
$\Upsilon(nS)$ is reconstructed in the $\mu^+ \mu^-$ channel 
and one examines the $\pi^\pm \Upsilon(nS)$ mass spectra.
This is illustrated for the $\Upsilon(2S)$ case in Fig.~\ref{fig:mass_1}. 
Two charged bottomonium-like resonances, the $Z_b(10610)$ and
$Z_b(10650)$, are observed (Table~\ref{table:1}). 
A similar structure is found (Fig.~\ref{fig:mass_2}) 
for the $h_b(mP) \pi^+ \pi^-$ decay, where this time the appropriate
observable is $M_{\rm miss}(\pi^\mp)$, the missing mass of the opposite 
sign pion as the decays are reconstructed inclusively using the missing
mass of the $\pi^+ \pi^-$ pair. Production of the $Z_b$'s saturates
the $\Upsilon(5S) \to h_b(mP) \pi^+ \pi^-$ transitions and accounts 
for the high inclusive $h_b(mS)$ production rate.
All channels yield consistent results and weighted averages over 
all five channels give $M = 10607.2 \pm 2.0$~MeV/$c^2$,
$\Gamma = 18.4 \pm 2.4$~MeV for the $Z_b(10610)$ and 
$M = 10652.2 \pm 1.5$~MeV/$c^2$,
$\Gamma = 11.5 \pm 2.2$~MeV for the $Z_b(10650)$, where statistical
and systematic errors are added in quadrature. The $Z_b(10610)$
production rate is similar to that of the $Z_b(10650)$ for each of the 
five decay channels. Analyses of charged pion angular distributions 
favor the $J^P = 1^+$ spin-parity assignment for both the $Z_b(10610)$
and $Z_b(10650)$. 

\begin{table}
\caption{Comparison of results on $Z_b(10610)$ and $Z_b(10650)$ parameters
(mass and width in MeV, relative normalization and phase in degrees) 
obtained from $\Upsilon(5S) \to \Upsilon(nS) \pi^+ \pi^-$ ($n = 1, 2, 3$)
and $\Upsilon(5S) \to h_b(mP) \pi^+ \pi^-$ ($m = 1, 2$) analyses.}
\label{table:1}
\begin{center}
\begin{tabular}{lccccc} \hline \hline
Final state & $\Upsilon(1S) \pi^+ \pi^-$ & $\Upsilon(2S) \pi^+ \pi^-$ & 
$\Upsilon(3S) \pi^+ \pi^-$ & $h_b(1P) \pi^+ \pi^-$ & $h_b(2P) \pi^+ \pi^-$ \\
\hline
$M[Z_b(10610)]$ &  $10611 \pm 4 \pm 3$ & $10609 \pm 2 \pm 3$ &
$10608 \pm 2 \pm 3$ & $10605 \pm 2 {}^{+3}_{-1}$ & $10599 {}^{+6}_{-3} {}^{+5}_{-4}$ \\
$\Gamma[Z_b(10610)]$ & $22.3 \pm 7.7 {}^{+3.0}_{-4.0}$ & 
$24.2 \pm 3.1 {}^{+2.0}_{-3.0}$ & $17.6 \pm 3.0 \pm 3.0$ & 
$11.4 {}^{+4.5}_{-3.9} {}^{+2.1}_{-1.2}$ & $13 {}^{+10}_{-8} {}^{+9}_{-7}$ \\
$M[Z_b(10650)]$ &  $10657 \pm 6 \pm 3$ & $10651 \pm 2 \pm 3$ &
$10652 \pm 1 \pm 2$ & $10654 \pm 3 {}^{+1}_{-2}$ & $10654 {}^{+2}_{-3} {}^{+3}_{-2}$ \\
$\Gamma[Z_b(10650)]$ & $16.3 \pm 9.8 {}^{+6.0}_{-2.0}$ & 
$13.3 \pm 3.3 {}^{+4.0}_{-3.0}$ & $8.4 \pm 2.0 \pm 2.0$ & 
$20.9 {}^{+5.4}_{-4.7} {}^{+2.1}_{-5.7}$ & $19 \pm 7 {}^{+11}_{-7}$ \\
Rel. norm. & $0.57 \pm 0.21 {}^{+0.19}_{-0.04}$ & $0.86 \pm 0.11 {}^{+0.04}_{-0.10}$ &
$0.96 \pm 0.14 {}^{+0.08}_{-0.05}$ & $1.39 \pm 0.37 {}^{+0.05}_{-0.15}$ &
$1.6 {}^{+0.6}_{-0.4} {}^{+0.4}_{-0.6}$ \\
Rel. phase & $58 \pm 43 {}^{+4}_{-9}$ & $-13 \pm 13 {}^{+17}_{-8}$ &
$-9 \pm 19 {}^{+11}_{-26}$ & $187 {}^{+44}_{-57} {}^{+3}_{-12}$ & 
$181 {}^{+65}_{-105} {}^{+74}_{-109}$ \\
\hline
\end{tabular}
\end{center}
\end{table}

These states defy a standard bottomonium assignment.
In principle, a bottomonium particle's electric charge is zero; 
therefore, the minimal quark content of the $Z_b(10610)$ and $Z_b(10650)$ 
is a four-quark combination. 
Theoretical interpretations of these hidden-bottom meson 
resonances were proposed immediately after their observation.
The proximity (within a few MeV/$c^2$) of the 
measured masses of these unexpected new states to 
the open beauty thresholds,
$B {\bar{B}}^*$ (10604.6 MeV/$c^2$) and $B^* {\bar{B}}^*$ 
(10650.2 MeV/$c^2$), suggests a ``molecular'' nature of these
new states, which can in turn explain most of their observed properties.
In the case of a molecule, it would be natural to expect that 
$Z_b^0 (10610)$ and $Z_b^0 (10650)$ states to decay respectively to 
$B {\bar{B}}^*$ and $B^* {\bar{B}}^*$ final states at substantial
rates. Recently, Belle reported preliminary results
on the analysis of three-body $\Upsilon(5S) \to B B^* \pi$ 
($B^+ {\bar{B}}^{*0} \pi^-$, $B^- B^{*0} \pi^+$, $B^0 B^{*-} \pi^+$
and ${\bar{B}}^0 B^{*+} \pi^-$) and $\Upsilon(5S) \to B^* B^* \pi$
($B^{*+} {\bar{B}}^{*0} \pi^-$ and $B^{*-} B^{*0} \pi^+$) including an
observation of the 
$\Upsilon(5S) \to Z_b^\pm(10610)\pi^\mp \to [B \bar{B}^*]^\pm \pi^\mp$
and $\Upsilon(5S) \to Z_b^\pm(10650)\pi^\mp \to [B^* \bar{B}^*]^\pm \pi^\mp$
decays as intermediate channels.
Evidence (with a significance of 4.9$\sigma$) 
for a neutral $Z_b^0(10610)$ decaying to $\Upsilon(2S) \pi^0$ has 
been also obtained by Belle in a Dalitz plot 
analysis of $\Upsilon(5S) \to \Upsilon(2S) \pi^0 \pi^0$ 
using their full $\Upsilon(5S)$ data sample~\cite{zb0_belle}. Its 
measured mass, $M(Z_b^0 (10610)) = 10609 {}^{+8}_{-6} \pm 6$~MeV/$c^2$,
is consistent with the mass of the corresponding charged state, the
$Z_b^\pm (10610)$.

The $Z_b$ states have also been interpreted as cusps at the $B^* {\bar{B}}$ 
and $B^* {\bar{B}}^*$ thresholds and as tetraquark states.

After observing that the decay $\Upsilon(5S) \to h_b(nP)\pi^+ \pi^-$
proceeds via the $Z_b$ intermediate resonances, Belle~\cite{etab_belle} 
exploited this
information to look for the $\eta_b(1, 2S)$ resonances in the processes
$e^+ e^- \to h_b(nP)\pi^+ \pi^-$, $h_b(nP) \to \eta_b(mS)\gamma$. 
%
\begin{figure}
  \centerline{\includegraphics[width=0.45\textwidth]
                                {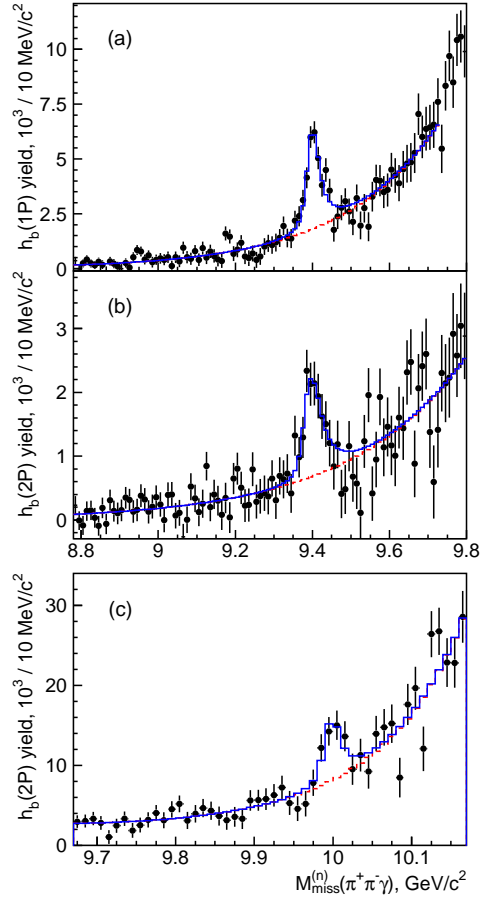}}
                              \caption{
The $h_b(1P)$ yield versus $M_{\rm miss}^{(1)}(\pi^+ \pi^- \gamma)$
(a), and $h_b(2P)$ yield versus $M_{\rm miss}^{(2)}(\pi^+ \pi^- \gamma)$
in the $\eta_b(1S)$ region (b) and in the $\eta_b(2S)$ region (c).
The solid (dashed) histogram presents the fit result (background component
of the fit function).
}
\label{fig:metab}
\end{figure}
%
Here only the $\pi^+$, $\pi^-$, and $\gamma$ are reconstructed and 
the requirement 10.59 GeV/$c^2 < M_{\rm miss}(\pi^\pm) <$ 10.67 GeV/$c^2$ 
helps to reduce the background significantly. The 
$M_{\rm miss}(\pi^+ \pi^-)$ spectra are fitted for different
$M_{\rm miss}^{(n)}(\pi^+ \pi^- \gamma)$ bins to measure the $h_b(nP)$
yield. The $h_b(nP)$ yield peaks at $M_{\rm miss}^{(n)}(\pi^+ \pi^- \gamma)$ 
values corresponding to $m_{\eta_b(mS)}$ due to the 
$h_b(nP) \to \eta_b (mS) \gamma$ transitions (Fig.~\ref{fig:metab}).

The $h_b(1P) \to \eta_b(1S) \gamma$ and $h_b(2P) \to \eta_b(1S) \gamma$
transitions are observed for the first time and first evidence for the
$\eta_b(2S)$ is obtained using the $h_b(2P) \to \eta_b(2S) \gamma$ transition. 
The mass and width parameters of the $\eta_b(1S)$ and $\eta_b(2S)$ are
measured to be $m_{\eta_b(1S)} = (9402.4 \pm 1.5 \pm 1.8)$ MeV/$c^2$, 
$m_{\eta_b(2S)} = (9999.0 \pm 3.5 {}^{+2.8}_{-1.9})$ MeV/$c^2$, and 
$\Gamma_{\eta_b(1S)} = (10.8 {}^{+4.0}_{-3.7} {}^{+4.5}_{-2.0})$ MeV.
Our value of the $\eta_b(1S)$ mass is about 11 MeV higher than the
previous world average and the hyperfine splittings are
$57.9 \pm 2.3$ MeV and $24.3^{+4.0}_{-4.5}$ MeV for the $1S$ and $2S$ states, 
respectively, consistent with theoretical predictions.

\subsection{Others}
%
In addition to $c \bar{c}$ and $b\bar{b}$ states, Belle has 
also studied charmed mesons and baryons. They are copiously 
produced at KEKB either directly in $e^+e^-$ collisions or 
as products of $B$ meson decays.
At the 10.53 GeV CM energy, the cross section for prompt $c \bar{c}$ 
pair production exceeds that of $b \bar{b}$, assuring large samples of ground 
and excited charmed states hadronizing from the produced $c\bar{c}$ quarks. 
Charm hadrons are usually studied inclusively, however such an approach 
often suffers from large background. 
Charm production in $B$ decays is governed 
by the Cabibbo-favoured $b\to c$ transition. The restricted kinematics of 
$\Upsilon(4S) \to B \bar{B}$ production enables selection 
of clean $B$ samples. The
fixed spin of the parent $B$ 
constrains possible quantum numbers of daughter particles, 
simplifying spin-parity determinations. However, charmed states with
high spin and highly excited charm states are suppressed in $B$ decays. 
 
\subsubsection{Charmed mesons}

The spectra of quark-antiquark systems 
are predicted using potential models, which 
attempt to model QCD features 
by describing the interquark potential~\cite{Qq-models}. 
Charmed mesons, having $c\bar{u}$, $c\bar{d}$ or $c\bar{s}$ quark content, are 
heavy-light systems for which the models employ Heavy Quark Symmetry (HQS). 
In the limit of an infinitely heavy-quark mass, 
heavy-light mesons become similar 
to a hydrogen atom; which gives many theoretical simplifications.
 However, since the $c$ quark mass is finite, HQS is only an approximate 
symmetry. An important consequence of its breaking 
is the $D_{(s)}-D_{(s)}^{*}$ splitting. 
The orbitally excited $P$-wave multiplet ($L=1$), denoted $D_{(s)}^{* *}$, 
is expected to consist of a broad $J^P=(0^+, 1^+)$ doublet having total 
light-quark angular momentum $j_{q}=\frac{1}{2}$ and a narrow $(1^+, 2^+)$ 
doublet with $j_{q}=\frac{3}{2}$.

Before the advent of the $B$-Factories, in addition to the 
ground state $D_{(s)}$ and $D_{(s)}^{*}$ 
mesons, only the narrow $D_{(s)}^{* *}$ doublets were established: 
$(D_{1}(2420), D^{*}_{2}(2460))$ and $(D_{s1}(2536),
D_{s2}^*(2573))$; the broad ones remained missing. 
Discovery of two narrow and unexpected states, the 
$D_{s0}^{*}(2317)^{+}$ and $D_{s1}(2460)^{+}$, began a renaissance in charm 
spectroscopy~\cite{dsj-obs}. They were found in the 
$D_s^{+} \pi^0$ and $D_s^{* +} \pi^0$ final states, respectively, 
and were produced inclusively in the $c \bar{c}$ continuum. 
\begin{figure}[b]
\centerline{
\includegraphics[width=0.33\textwidth,height=4cm]{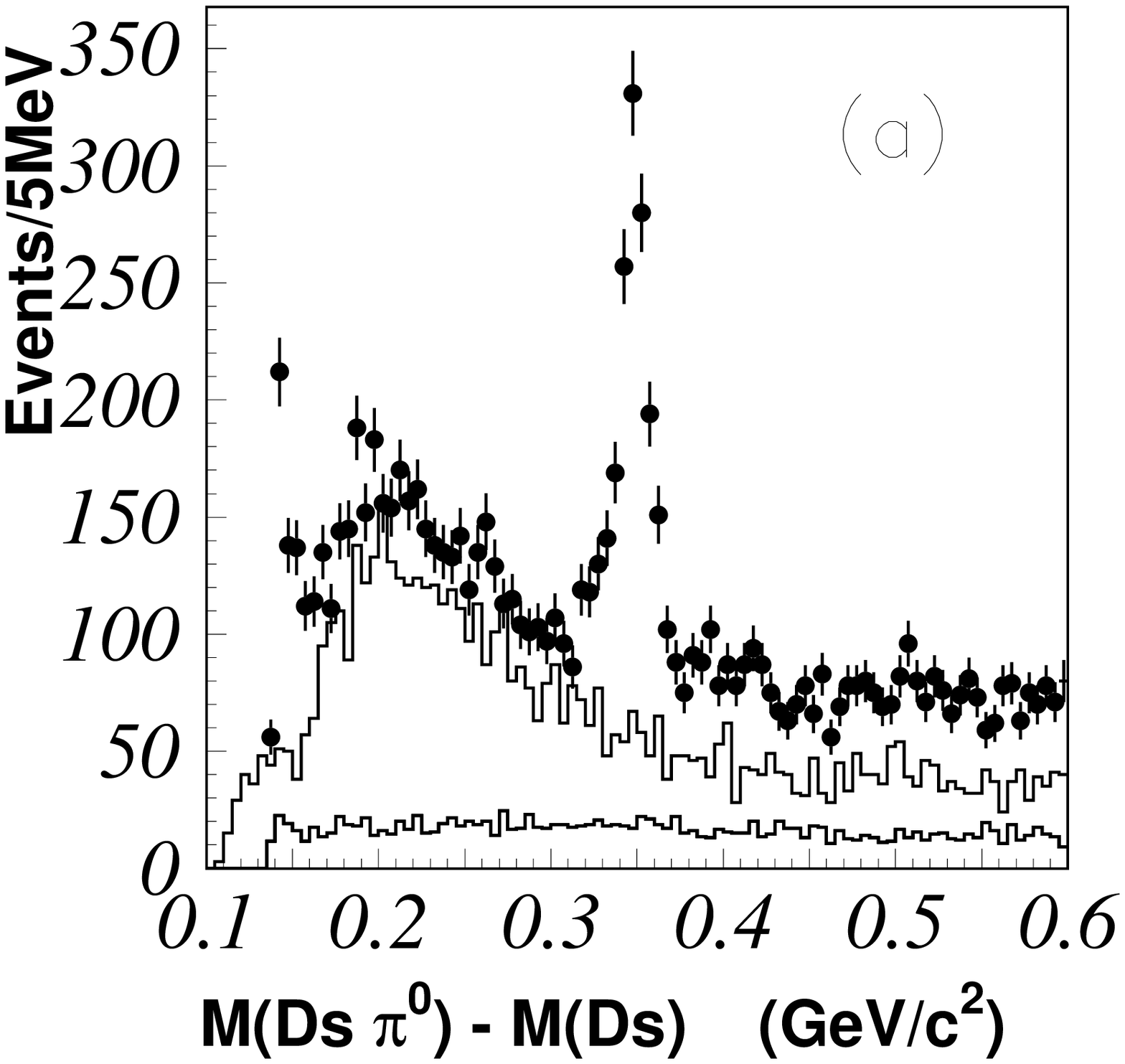}
\includegraphics[width=0.33\textwidth,height=4cm]{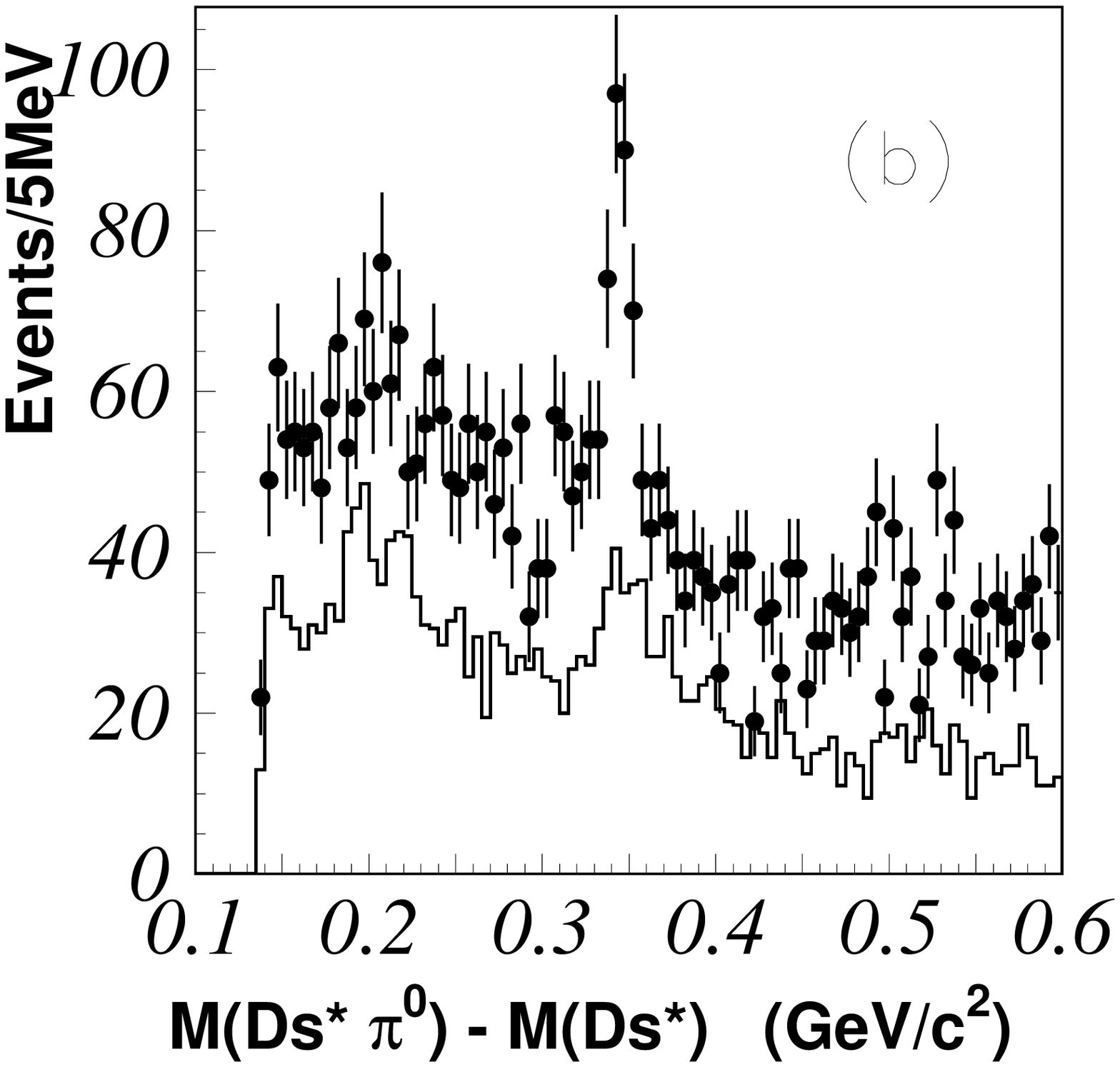}
\includegraphics[width=0.33\textwidth,height=4cm]{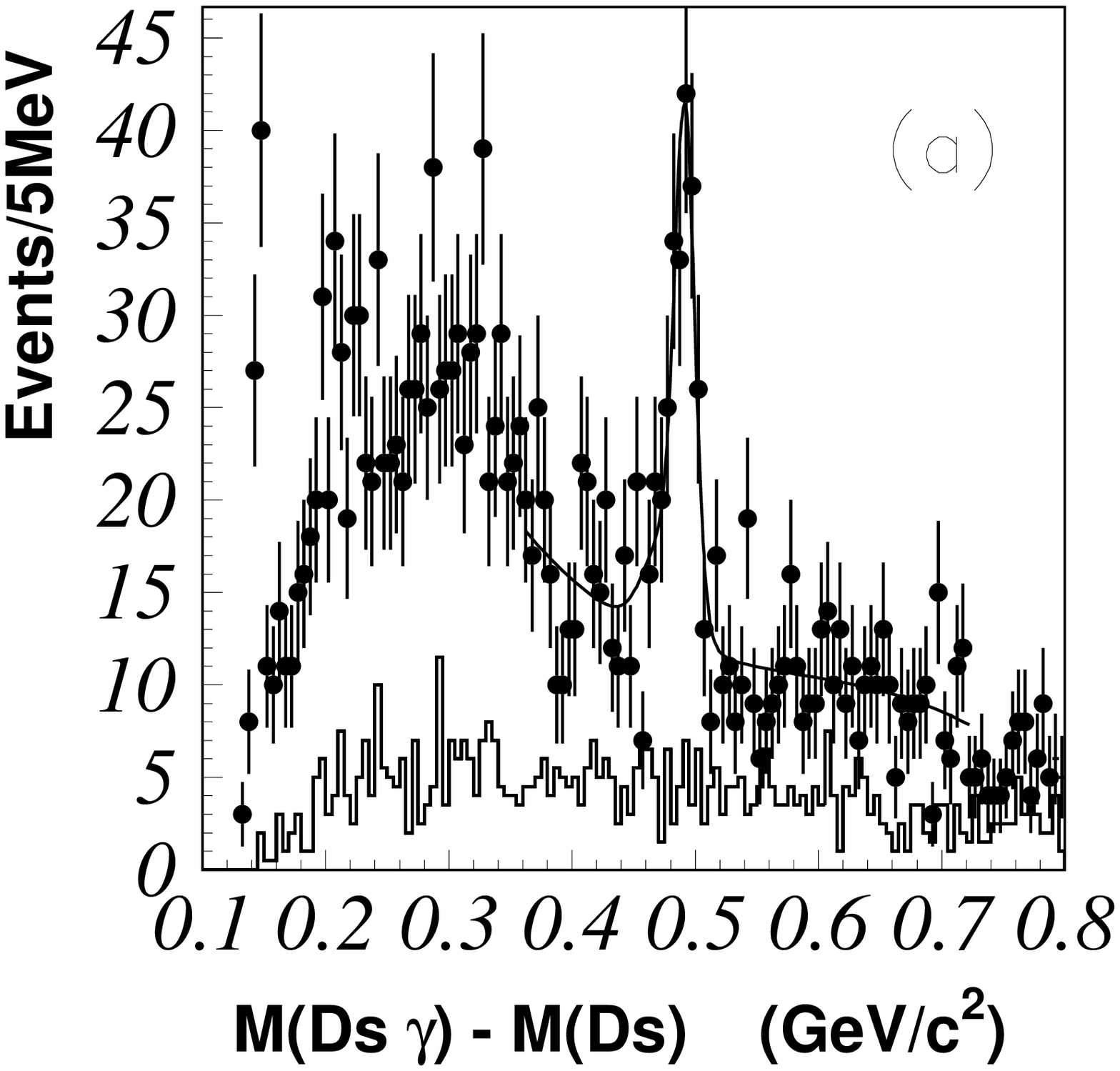}}
\caption{Distributions of $\Delta M(D_s \pi^0)$ (left), 
$\Delta M(D_s^{*} \pi^0)$ (middle), and $\Delta M(D_s \gamma)$ (right). 
Histograms show data from the
$D_s$ and/or $\pi^0$ sideband regions.}
\label{fig:dsj1}
\end{figure}  
Spectra of the
$\Delta M(D_s^{(*)} \pi^0) \equiv M(D_s^{(*)} \pi^0)-M(D_s^{(*)})$
mass difference
measured by Belle are shown in Fig.~\ref{fig:dsj1}; prominent peaks at 
$\Delta M(D_s \pi^0) \approx 350~\text{MeV}/\text{c}^2$ and 
$\Delta M(D_s^{*} \pi^0) \approx 350~\text{MeV}/\text{c}^2$, are 
the $D_{s0}^{*}(2317)$ and $D_{s1}(2460)$, respectively. 
Their masses and upper limits on their 
widths, measured from fits to the $\Delta M(D_s^{(*)} \pi^0)$, 
are summarized in Table~\ref{tab:dss_parameters1}. 
\begin{table}[t]
  \caption{Parameters of $D_{s0}^{*}(2317)$ and $D_{s1}(2460)$, 
   compared with PDG parameters of $j_{q}=\frac{3}{2}$ states.} 
  \label{tab:dss_parameters1}
  \begin{center}
\begin{tabular}{ccccc}
  \hline \hline
$J^P (j_q)$ & $D_{s}^{**}$ & Decay modes & Mass (MeV/$c^2$) & Width (MeV/$c^2$)     \\ 
  \hline
$  0^+ (\frac{1}{2})$ & $D_{s0}^*(2317)$ & $D_s   \pi$              & $2317.2 \pm 0.5 \pm 0.9$ & $<4.6$     \\
$  1^+ (\frac{1}{2})$ & $D_{s1}(2460)$   & $D_s^* \pi, D_s \gamma$  & $2456.5 \pm 1.3 \pm 1.3$ & $<5.5$     \\
$  1^+ (\frac{3}{2})$ & $D_{s1}(2536)$   & $D^* K$                  & $2535.3 \pm 0.2$         & $<2.3$     \\
$  2^+ (\frac{3}{2})$ & $D_{s2}^*(2573)$ & $D K$                    & $2572.6 \pm 0.9$         & $20 \pm 5$ \\
  \hline\hline
\end{tabular}
  \end{center}
\end{table}

Observation of 
radiative $D_s \gamma$ (Fig.~\ref{fig:dsj1}) and dipion $D_s \pi^+ \pi^-$ 
decays of the $D_{s1}(2460)$ ruled out a $J^P=0^{\pm}$ assignment. 
For the $D_{s0}^{*}(2317)$ 
no decay channel was found apart from the discovery mode. Such a decay pattern 
was consistent with spin-parity assignments of $0^+$ and $1^+$ for 
the $D_{s0}^{*}(2317)$ and $D_{s1}(2460)$ respectively, as expected for 
the $P$-wave $c\bar{s}$ doublet with $j_q=\frac{1}{2}$. 
However, the measured masses were much lower than 
predicted by potential models and, thus, decays 
to $D^{(*)}K$, expected to be dominant, were not permitted kinematically.  
Instead, the dominant decays into 
isospin-violating modes resulted in very small 
widths. All this triggered exotic interpretations of these mesons as
$D K$ molecules, multiquark states, 
mixtures of $P$-wave $c\bar{s}$ meson with a $c \bar{s}q\bar{q}$ 
tetraquark, or chiral partners of $D^{(*)}_{s}$~\cite{ds_interpret}.

\begin{figure}
\centerline{
\includegraphics[width=8.5cm,height=5.5cm]{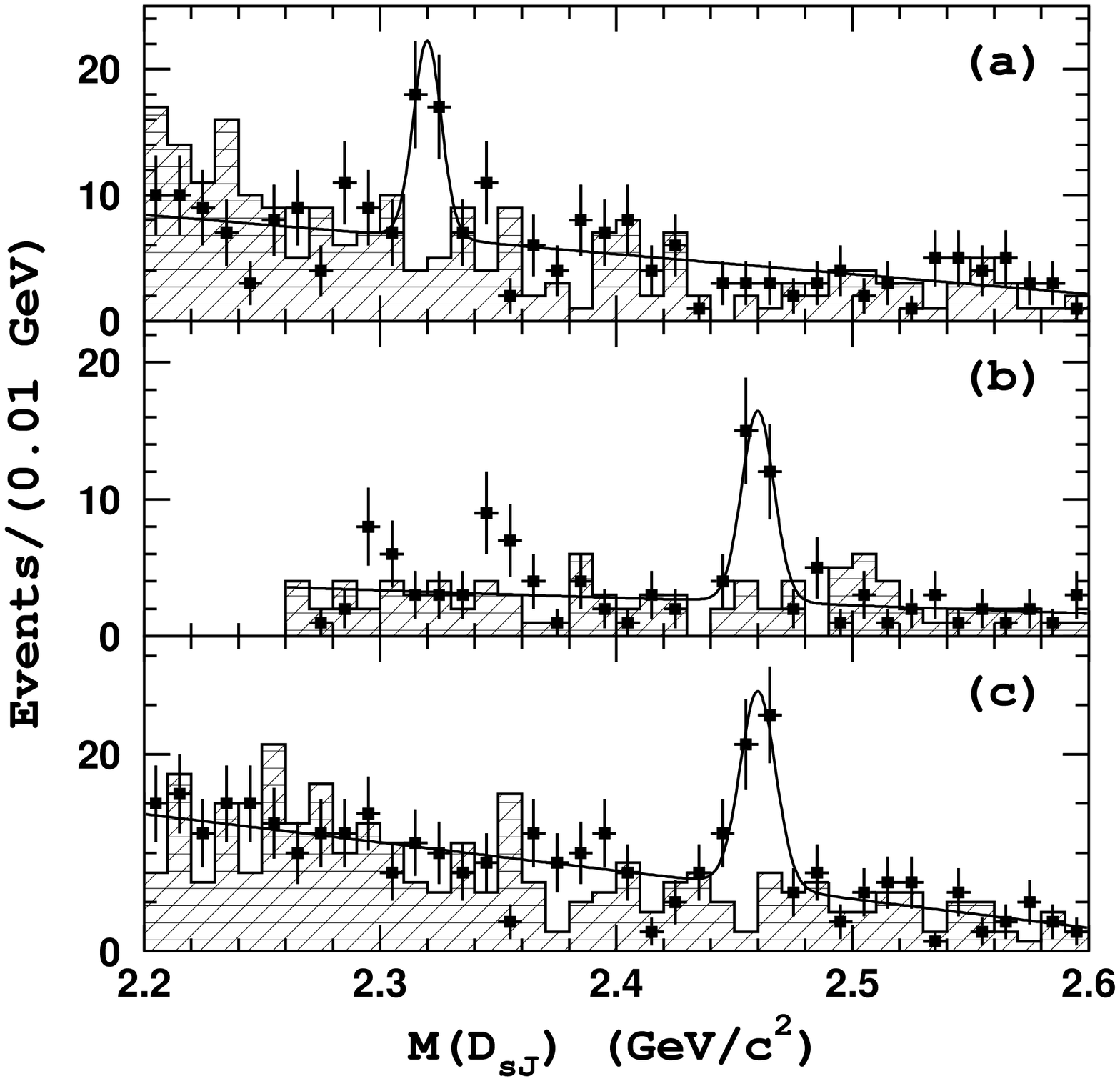}
\includegraphics[width=5.5cm,height=5.5cm]{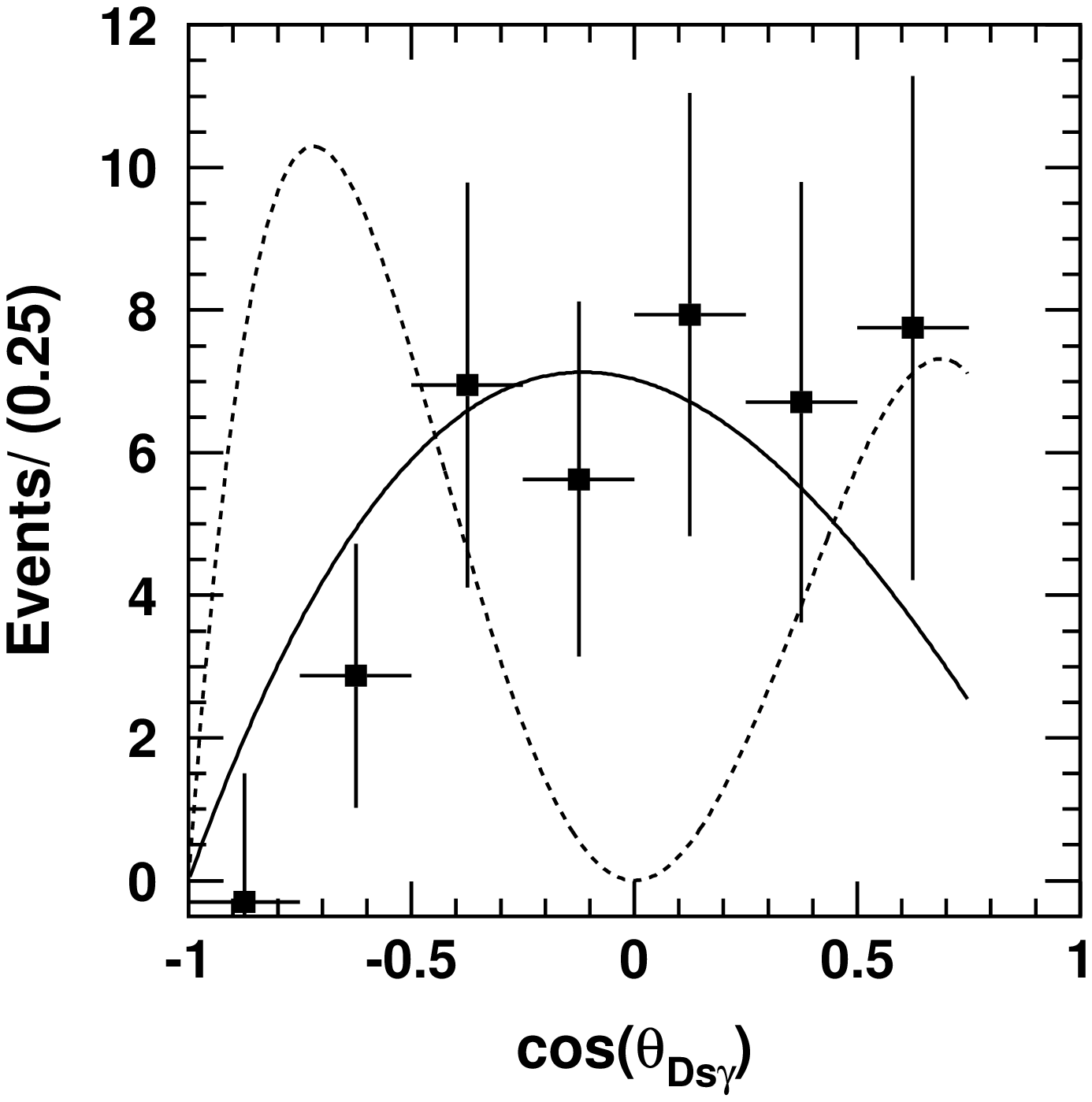}}
\caption{Left: $M(D_{sJ})$ distributions for $D_{sJ}$ final states: 
$D_s \pi^0$ (top), $D_s^{*} \pi^0$ (middle), 
$D_s \gamma$ (bottom). Hatched histograms 
show the $\Delta E$ sidebands. 
Right: the $D_{s1}(2460)\to D_s \gamma$ helicity distribution. 
Data points are compared with MC predictions 
for $J=1$ (solid line) and $J=2$ (dashed) assignments.} 
\label{fig:dsj2}
\end{figure}

To clarify the nature of these states, Belle searched for them in exclusive 
$B\to \bar{D} D_{sJ}$ decays, where $D_{sJ}$ 
denotes any excited charmed-strange meson~\cite{dsj-in-b}. 
These reactions proceed via $\bar{b} \to \bar{c} W^+ \to \bar{c} c \bar{s}$, 
which is the dominant $c\bar{s}$ production mechanism 
in $B$ decays; here $D_{s}^{* *}$ with 
$j_q=\frac{1}{2}$ are expected to be 
more readily produced than $j_q=\frac{3}{2}$ states. 
Thus, one expected to observe the $D_{s0}^{*}(2317)$ and $D_{s1}(2460)$ 
in $B \to \bar{D} D_{sJ}$, if they were the missing $c\bar{s}$ doublet.   
The $D_{sJ}$ final states studied 
were $D_s^{(*)} \pi^0$, $D_s^{(*)} \gamma$ and 
$D_s^{(*)} \pi^+ \pi^-$. Figure~\ref{fig:dsj2} shows distributions of $D_{sJ}$ 
invariant mass for $B$ candidates satisfying $\Delta E$ and $M_{bc}$ 
signal region requirements, and for the channels with significant signals 
found: $D_{s0}^*(2317) \to D_s \pi^0$, $D_{s1}(2460)\to D_s^{*}\pi^0$ and 
$D_{s1}(2460) \to D_s \gamma$. The $D_{s1}(2460)$ helicity angle distribution 
for the $D_{s1}(2460)\to D_s\gamma$ mode 
(Fig.~\ref{fig:dsj2}) showed that the data 
were consistent with the $J=1$ hypothesis. 
Study of the $D_{sJ}$ production 
rates in $B \to \bar{D} D_{sJ}$ decays seems to support the
interpretation of the $D_{s0}^*(2317)$ and $D_{s1}(2460)$ as 
the orbitally excited $c\bar{s}$ $j_q=\frac{1}{2}$ doublet. 
Although some of the models managed to reproduce the
low masses of these states~\cite{matsuki}, our understanding 
of $c\bar{s}$ spectroscopy still seems to be incomplete. 

\begin{figure}
\centerline{
  \includegraphics[width=0.245\textwidth,height=4.2cm]{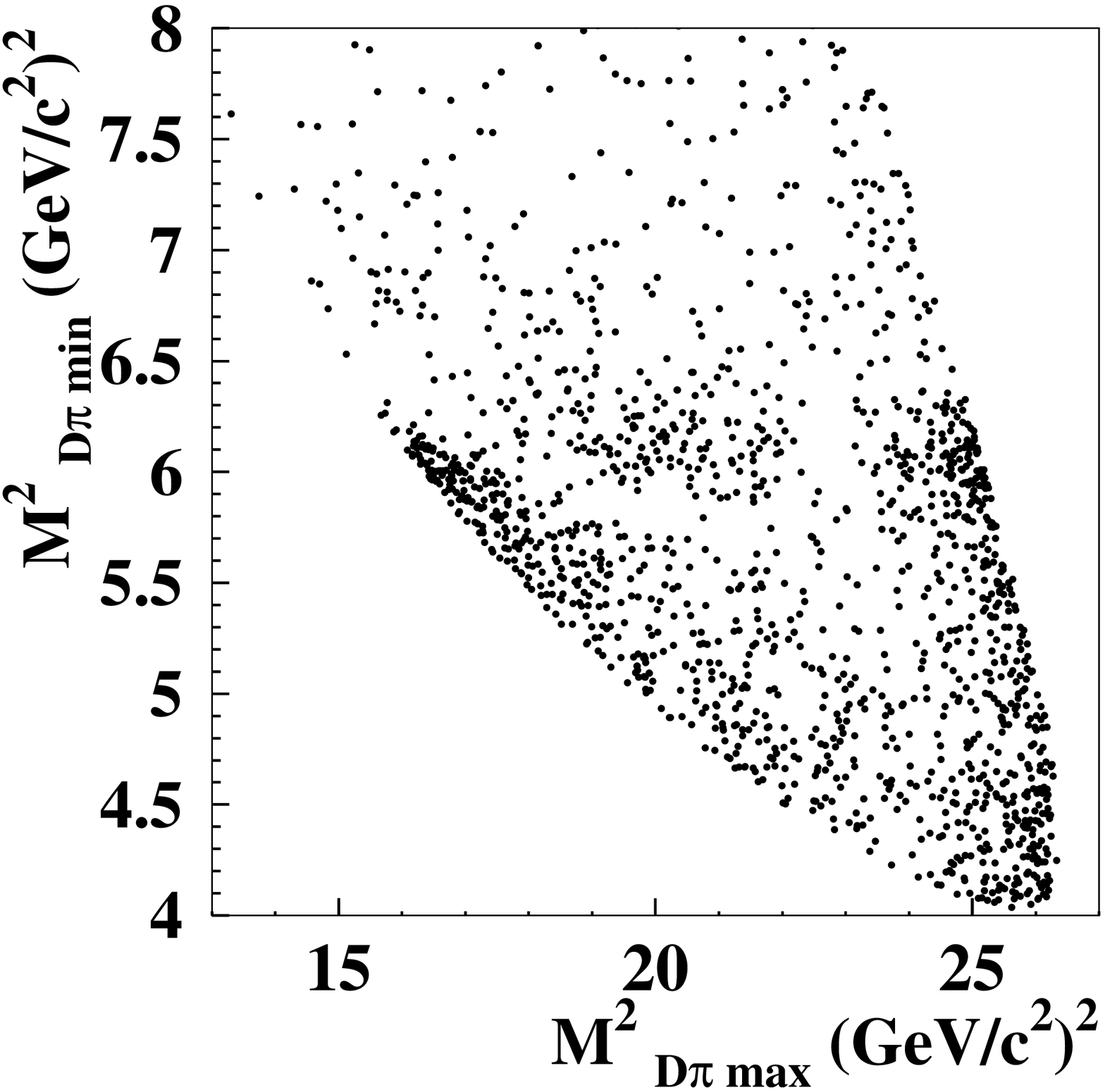}
  \includegraphics[width=0.245\textwidth,height=4.2cm]{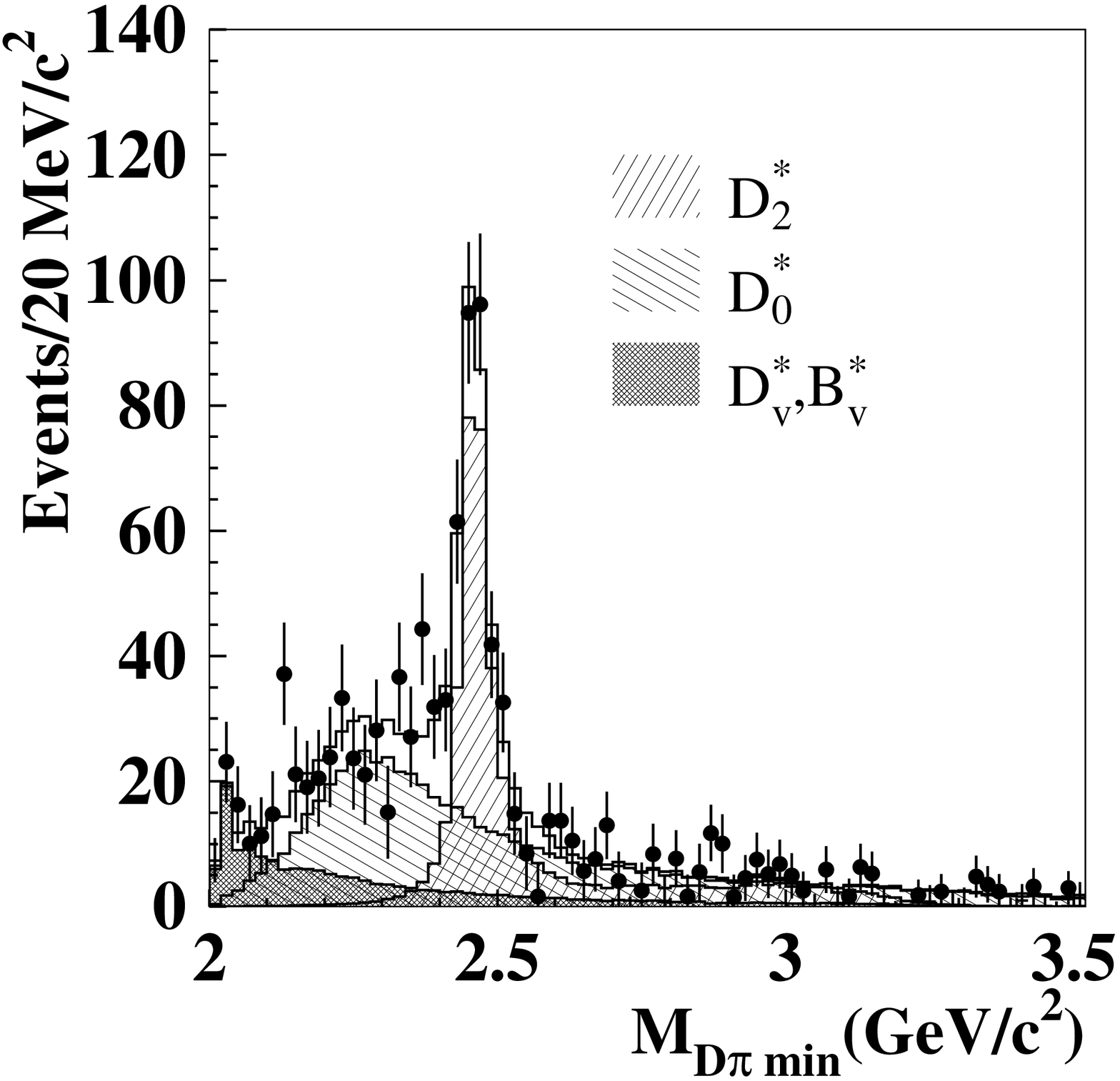}
  \includegraphics[width=0.245\textwidth,height=4.2cm]{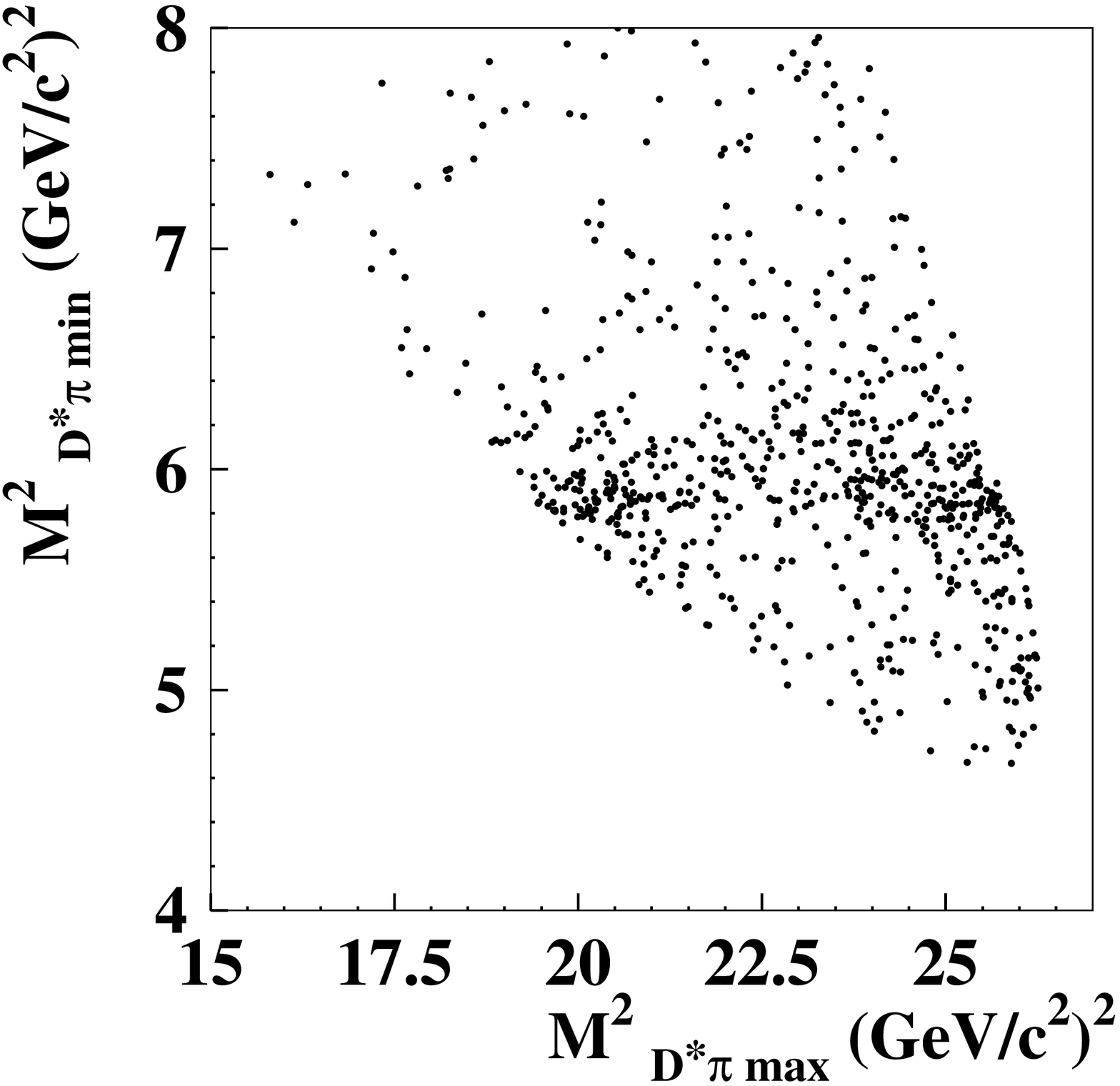}
  \includegraphics[width=0.245\textwidth,height=4.2cm]{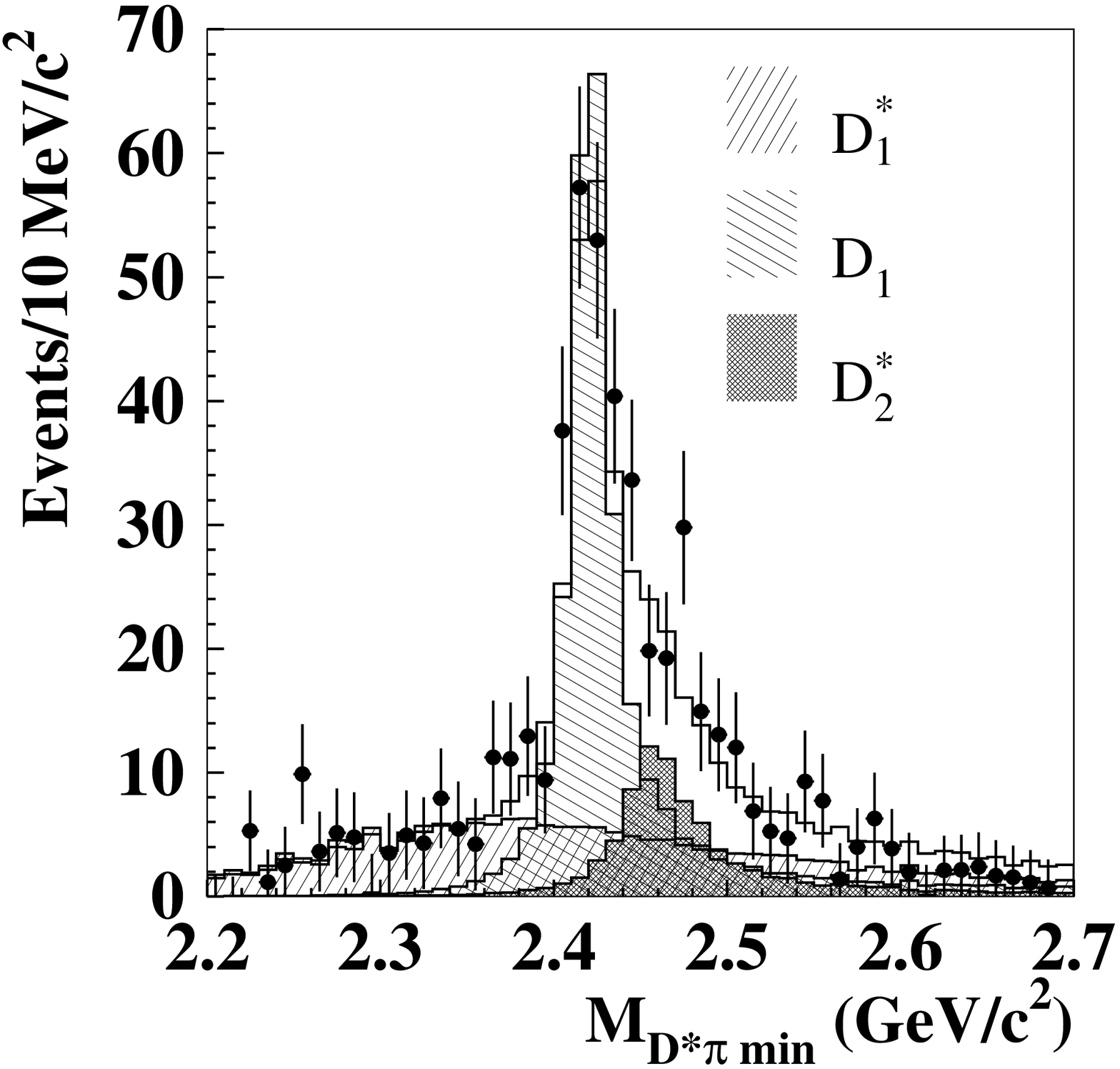}}
  \caption{Dalitz distributions for $B^- \to D^{+} \pi^- \pi^-$ 
(first from the left) and $B^- \to D^{* +} \pi^- \pi^-$ (third) signal region
  candidates. The corresponding 
  $M(D^+\pi^-)_{min}$ and $M(D^{*+}\pi^-)_{min}$ projections, 
with background subtracted, are shown as the 
second and fourth plots, respectively. 
Hatched histograms show the fitted  resonance contributions. 
The open histogram is the coherent sum of all contributions.}  
\label{fig:dpp_dalitz}
\end{figure}

On the other hand, the 
corresponding $j_q=\frac{1}{2}$ doublet in the $c \bar{u}$ spectrum, 
discovered by Belle about the same time as the narrow $D_{s}^{**}$ states, has 
properties that perfectly match potential model predictions. 
The $D^{**}$ mesons, expected to decay dominantly into
 $D^{(*)} \pi$ final states, were studied at Belle in a 
full Dalitz plot analysis of 
$B^+ \to D^{(*)-} \pi^+\pi^+$ decays~\cite{dss-Bp}. 
To distinguish between the two identical final-state pions, 
$D^{(*)-} \pi^+$ combinations 
having minimal and maximal mass values were used as the Dalitz plot variables. 
The $M^2(D^{(*)} \pi)_{min}$ vs. $M^2(D^{(*)} \pi)_{max}$ plots for
$B$ candidates within the $\Delta E$-$M_{bc}$ signal region, 
are shown in Fig.~\ref{fig:dpp_dalitz}. 
Non-uniformly distributed events 
indicate intermediate resonances emerging in the $M^2(D \pi)_{min}$ spectrum. 
The fitted resonance contributions to the $M(D^{(*)-}\pi^+)_{min}$ 
projection are shown in Fig.~\ref{fig:dpp_dalitz}. 
The $D \pi$ system was found to be composed of a tensor $D_2^{*0}$ 
and broad scalar state $D_0^{*0}$, while the $D^* \pi$ system
consists of a narrow axial $D_{1}$, a tensor 
$D^{*}_{2}$, as well as a broad axial $D^{'}_{1}$. 
The two broad states, observed for the first 
time, were consistent 
with the $j_q=\frac{1}{2}$ $P$-wave $c\bar{u}$ doublet. The measured 
parameters of the $D^{**0}$ states 
are summarized in Table~\ref{tab:dss_parameters2}; differences 
between the $D^{**}$ and $D_s^{**}$ properties are striking. 
Belle also performed a similar analysis 
for the $D^{**+}$'s produced 
in $B^0 \to \bar{D}^{(*)0} \pi^+ \pi^-$~\cite{dss-B0}.
\begin{table}
  \caption{Parameters of the $D^{**}$ mesons.} 
  \label{tab:dss_parameters2}
  \begin{center}
\begin{tabular}{ccccc} 
  \hline \hline
$J^P (j_q)$ & $D^{**}$ & Decay modes & Mass (MeV/$c^2$) & Width (MeV/$c^2$)     \\ 
  \hline
$0^+ (\frac{1}{2})$ & $D_0^*(2400)$   & $D \pi$       & $2308 \pm 17 \pm 15 \pm 28$      & $276 \pm 21 \pm 18 \pm 60$       \\
$1^+ (\frac{1}{2})$ & $D_1^{'}(2420)$ & $D^* \pi$     & $2427 \pm 26 \pm 20 \pm 15$      & $384^{+107}_{-75} \pm 24 \pm 70$ \\
$1^+ (\frac{3}{2})$ & $D_1(2420)$     & $D^* \pi$     & $2421.4 \pm 1.5 \pm 0.4 \pm 0.8$ & $23.7 \pm 2.7 \pm 0.2 \pm 4.0$ \\
$2^+ (\frac{3}{2})$ & $D_2^*(2460)$   & $D^{(*)} \pi$ & $2461.6 \pm 2.1 \pm 0.5 \pm 3.3$ & $45.6 \pm 4.4 \pm 6.5 \pm 1.6$   \\
  \hline\hline
\end{tabular}
  \end{center}
\end{table}

Studies performed by Belle allowed one 
to investigate important implications of HQS breaking. 
Theory predicts that the two $1^+$ mesons, with $j_q=\frac{1}{2}$ and 
$j_q=\frac{3}{2}$, decay into $D^* \pi$ in an $S$ and $D$ wave, respectively. 
Due to the finite $c$-quark mass, the observed (physical) $1^+$ states can be 
a mixture of such pure states and, thus, 
the resulting $D^{'}_{1}$ and $D_{1}$ amplitudes 
are superpositions of $S$- and $D$-wave amplitudes. 
The corresponding mixing angle was measured to be nonzero~\cite{dss-Bp}. 
Similarly, mixing between the two $c \bar{s}$ axial states can be expected. 
An angular analysis performed 
for the $D_{s1}(2536)^{+}\to D^{*+}K^0_S$ mode showed that, 
contrary to the HQS prediction of a pure $D$-wave decay, 
the $S$-wave decay dominates~\cite{ds2536-angular}. 

\begin{figure}
\centerline{
\includegraphics[width=4.6cm,height=3.8cm]{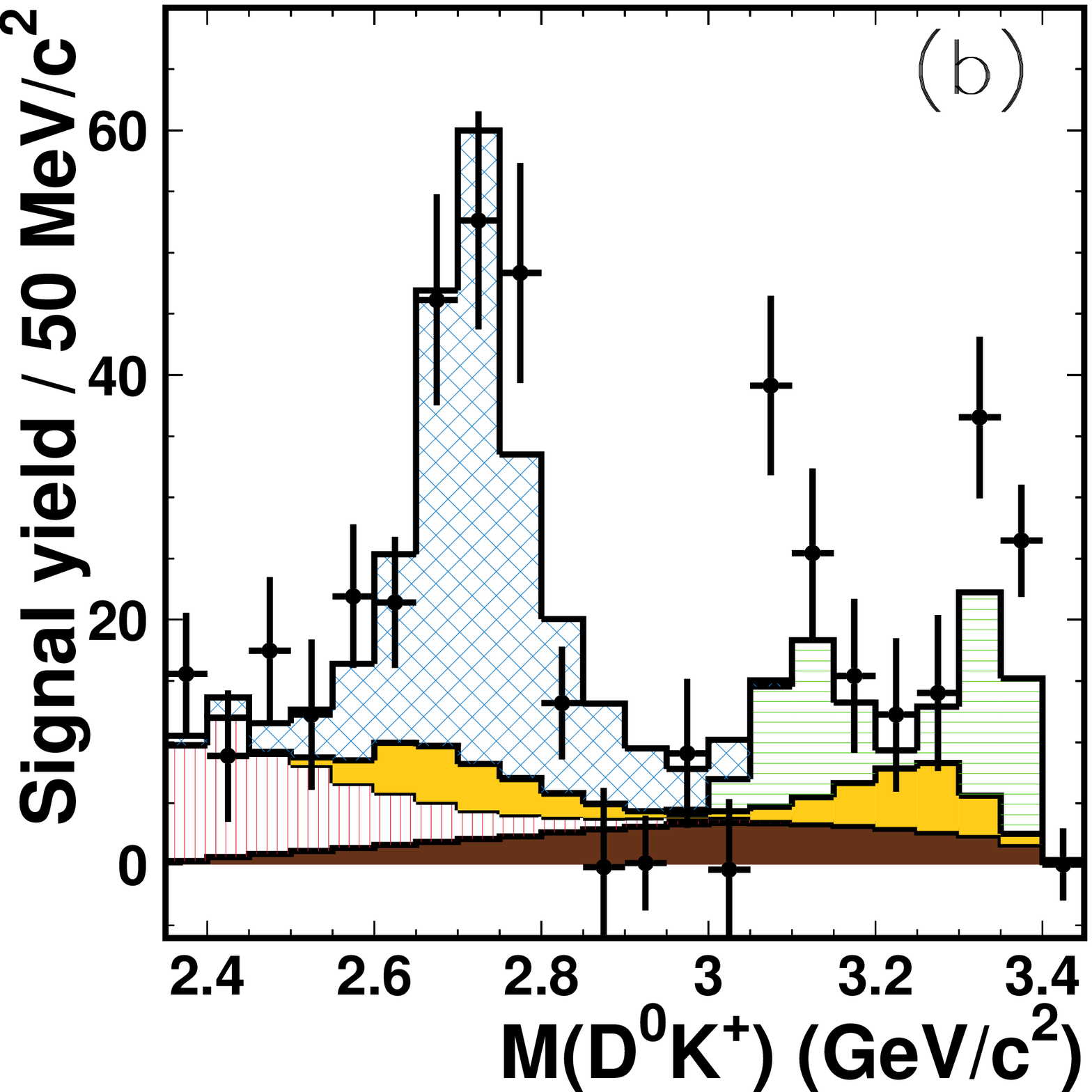}
\includegraphics[width=4.6cm,height=3.8cm]{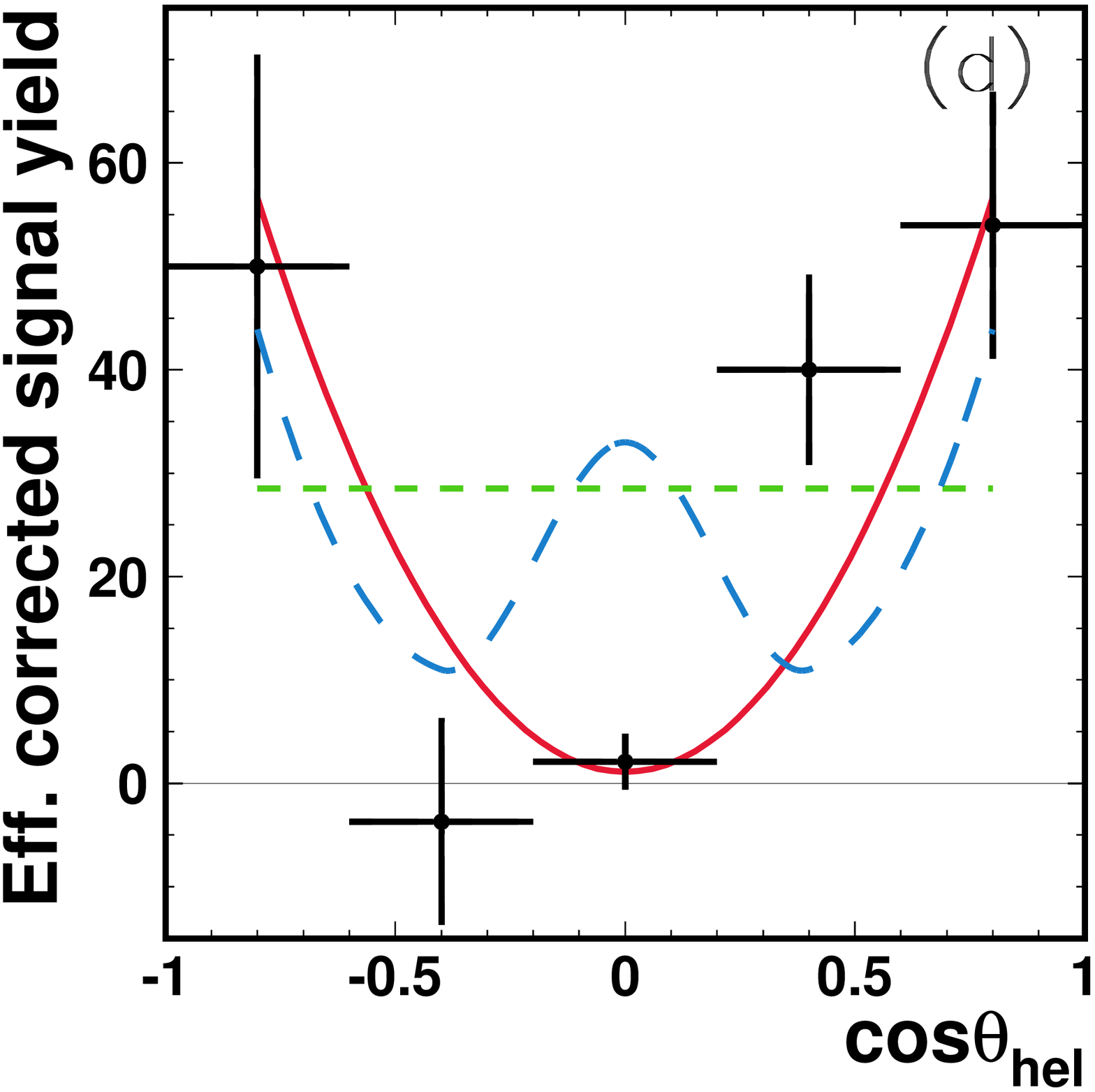}}
\caption{Left: Background-subtracted $M(D^0K^+)$ 
distribution for $B^+ \to \bar{D}^0 D^0 K^+$ with a
contribution from $D_{s1}^{*}(2700)^{+}$ (blue histogram), 
reflections from $\psi(3770)$ (green)
and $\psi(4160)$ (yellow) decaying to $\bar{D}^0 D^0$, 
non-resonant contributions (brown and red). 
Right: $D_{s1}^{*}(2700)$ helicity distribution 
compared to predictions for $J=0$ (green), 1 (red) 
and, 2 (blue) spin assignments.}
\label{fig:dsj2700}
\end{figure}

Potential models also predict 
multiplets of higher orbital and radial excitations 
of charmed mesons. The first example of such a $c \bar{s}$ meson, 
the $D_{s1}^{*}(2700)^{+}$, was observed 
in the $D^0K^+$ final state 
produced in doubly-charmed $B^+ \to \bar{D}^0 D^0 K^+$ 
decays~\cite{jolanta_ds2710}. Its mass was measured 
to be $2708 \pm 9 ^{+11}_{-10}~\text{MeV}/\text{c}^2$, while  
its width is $108 \pm 23^{+36}_{-31}~\text{MeV}/\text{c}^2$. 
The  $D_{s1}^{*}(2700)^{+}$'s spin-parity of $1^-$ was established 
from a study of its helicity angle. 
The $M(D^0K^+)$ spectrum together with the measured 
intermediate resonance contributions, 
as well as the $D_{s1}^{*}(2700)$ helicity distribution 
are shown in Fig.~\ref{fig:dsj2700}. 
Observation of the $D_{s1}^{*}(2700) \to D^*K$ decay with 
a rate comparable to that for $D K$, 
suggests that the $D_{s1}^{*}(2700)$ is 
a $D_s^*$ radial excitation~\cite{dsj2700-id}. 

\subsubsection{Charmed baryons}

Charmed baryons provide a laboratory for the 
study of the dynamics of a light diquark in the environment 
of a heavy quark and allow one to test 
many theoretical predictions~\cite{baryons-models}. 
For the charmed baryons with $cud$, $cdd$ or $cuu$ quark content, 
the only states known before the start of the $B$-Factories 
were the $\Lambda_c^+$ and $\Sigma_c(2455)^{0,+,++}$ 
ground states with $J^P=\frac{1}{2}^+$, 
the $\frac{3}{2}^+$ spin excitation $\Sigma_c(2520)$, 
as well as four $\Lambda_c$ excitations  
observed by CLEO in the $\Lambda_c \pi \pi$ final state. 
Two states, the $\Lambda_c(2595)$ 
and the $\Lambda_c(2625)$, 
were identified as orbitally excited states, while the interpretation of 
$\Lambda_c(2765)$ and $\Lambda_c(2880)$ remained unknown. 
Except for the $\Lambda_c$, quantum 
numbers of charmed baryons were not measured but, instead, 
either assigned based on model predictions or unknown. 
Since the predicted spectra are rich and dense, $J^P$ assignment 
for a given state is difficult and requires an experimental determination. 

\begin{figure}[t]
\centering
\begin{picture}(550,125)
\put(-7, 0){\includegraphics[width=5.8cm,height=0.36\textwidth]{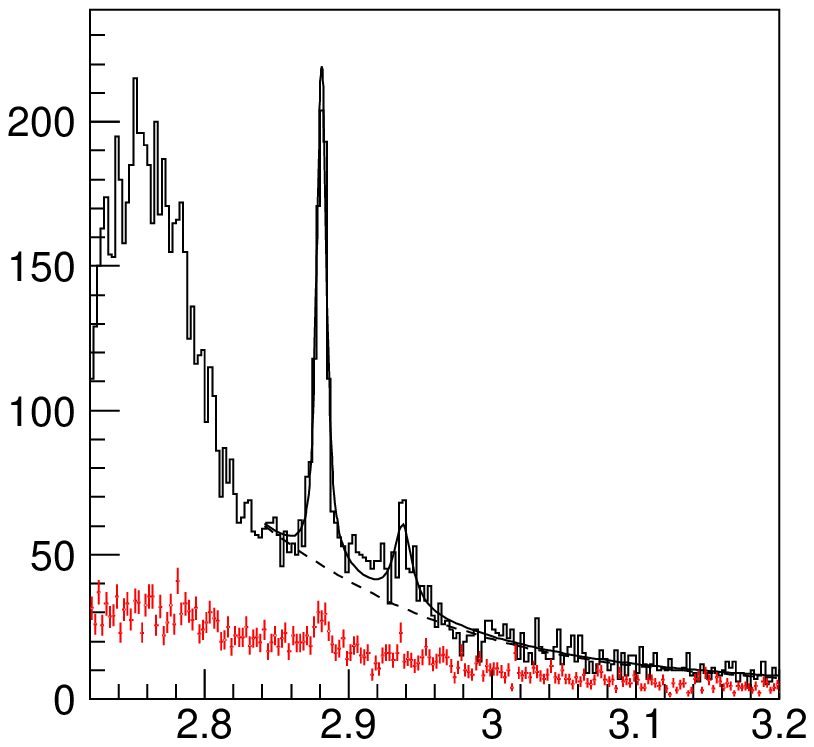}}
\put(144,15){\includegraphics[width=4.4cm,height=0.26\textwidth]{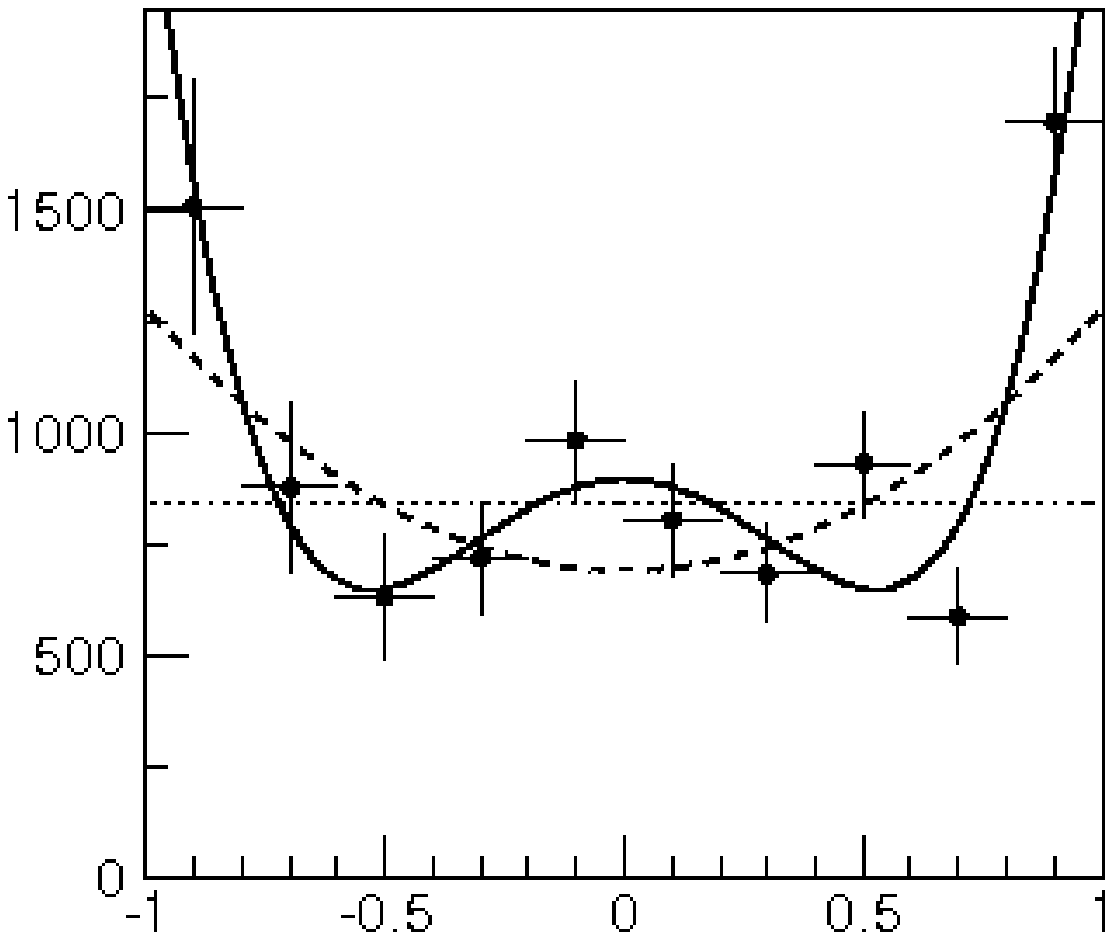}}
\put(265,-5){\includegraphics[width=5.8cm,height=0.37\textwidth]{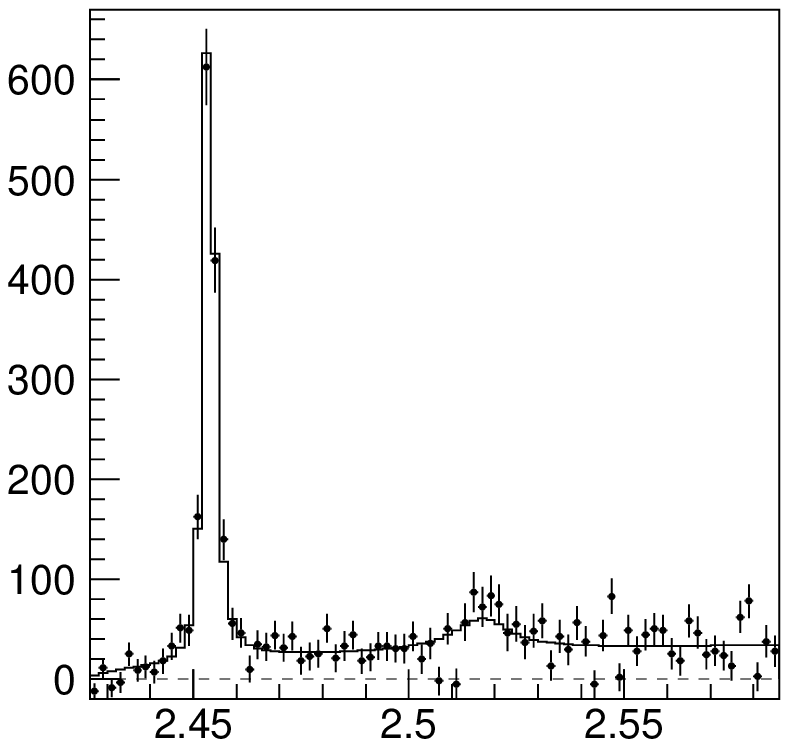}}
\put(0,30){\rotatebox{90}{$\text{Events} \ / \  2\text{MeV}/\text{c}^2$}} 
\put(25,5){$M(\Lambda_c^+\pi^+\pi^-),~\text{GeV}/\text{c}^2$} 
\put(136,45){\rotatebox{90}{$\text{Events} \ / \ 0.2$}}
\put(200,5){$\cos{\theta}$} 
\put(272,30){\rotatebox{90}{$\text{Events} \ / \  2\text{MeV}/\text{c}^2$}} 
\put(300,5){$M(\Lambda_c^+\pi^{\pm}),~\text{GeV}/\text{c}^2$} 
\end{picture}
\caption{Left: $M(\Lambda_c^+ \pi^+ \pi^-)$ distribution with $\Lambda_c^+ \pi^{\pm}$ 
within $\Sigma_c(2455)^{0,++}$ signal (black) and sideband (red) regions. 
Middle: Helicity distribution of $\Lambda_c(2880)\to \Sigma_c(2455) \pi$ 
with fit results for the $J=\frac{1}{2}$ (dotted), $\frac{3}{2}$ (dashed) and 
$\frac{5}{2}$ (solid) hypotheses. 
Right: $\Lambda_c(2880)$ yield as a function of $M(\Lambda_c^+ \pi^{\pm})$.}
\label{fig:lambdac2880}
\end{figure}

The first such measurement, 
performed for the $\Lambda_c(2880)$, is an excellent example 
of a comprehensive study of baryon properties~\cite{lc2880}. 
Figure~\ref{fig:lambdac2880} 
shows the $\Lambda_c^+ \pi^+ \pi^-$ invariant mass, 
with the $\Lambda_c^+ \to pK^-\pi^+$ mode
reconstructed. 
In addition to the $\Lambda_c(2880)$ signal, there are also peaks 
associated with the $\Lambda_c(2765)$, as well as 
the $\Lambda_c(2940)$ found by BaBar in the $D^0 p$ final state.~\cite{lc2940}
The parameters of the narrow baryons, obtained from a fit 
to the $M(\Lambda_c \pi^+ \pi^-)$ distribution
are: $M_{\Lambda_c(2880)}=2881.2 \pm 0.2 \pm 0.4~\text{MeV}/\text{c}^2$, 
$\Gamma_{\Lambda_c(2880)}=5.8\pm 0.7 \pm 1.1~\text{MeV}/\text{c}^2$, 
$M_{\Lambda_c(2940)}=2938.0 \pm 1.3^{+2.0}_{-1.4}~\text{MeV}/\text{c}^2$, 
$\Gamma_{\Lambda_c(2940)}=13^{+8 \ +27}_{-5 \ -7}~\text{MeV}/\text{c}^2$. 
The measured $\Lambda_c(2880)$ 
helicity distribution (see Fig.~\ref{fig:lambdac2880}) 
is consistent with the spin $\frac{5}{2}$ hypothesis. 
The quark model predicts the lowest 
$\frac{5}{2}^-$ and $\frac{5}{2}^+$ $\Lambda_c$ spin excitations at 
about $2900~\text{MeV}/\text{c}^2$, in agreement with 
the $\Lambda_c(2880)$ mass. 
Distribution of the $\Lambda_c(2880)$ yield as a 
function of the $M(\Lambda_c^+ \pi^{\pm})$, 
shown in Fig.~\ref{fig:lambdac2880}, 
indicates contributions from the $\Sigma_c(2455)$ 
and $\Sigma_c(2520)$. The measured $\Lambda_c(2880)$ partial width ratio,  
$\frac{\Gamma(\Sigma_c(2520)\pi)}{\Gamma(\Sigma_c(2455)\pi)}=
0.22 \pm 0.06 \pm 0.02$, is consistent with the prediction for 
the $\frac{5}{2}^+$ state.~\cite{lc2880-th} 

\begin{figure}
\centering
\begin{picture}(550,115)
\put(-5,-10){\includegraphics[width=1.07\textwidth, height=0.37\textwidth]{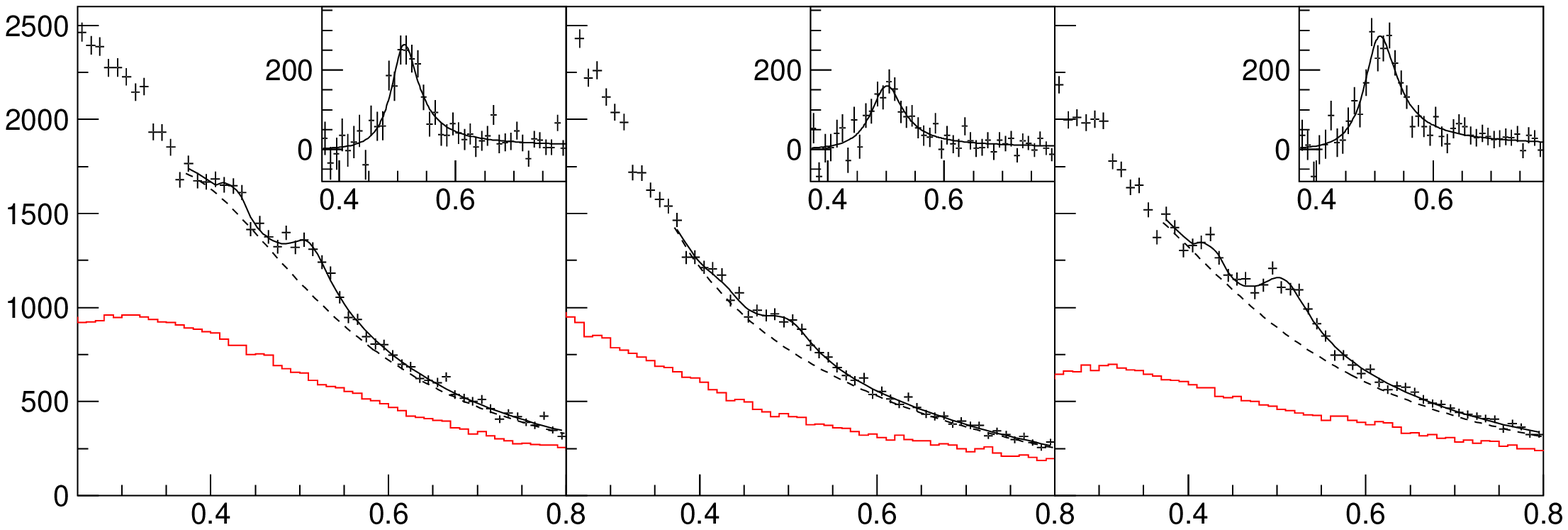}}
\put(0,45){\rotatebox{90}{$N / 10~\text{MeV}/\text{c}^2$}} 
\put(110,60){$\Lambda_c^+\pi^-$} 
\put(230,60){$\Lambda_c^+\pi^0$} 
\put(350,60){$\Lambda_c^+\pi^+$} 
\put(255,0){$M(\Lambda_c^+\pi)-M(\Lambda_c^+),~\text{GeV}/\text{c}^2$} 
\end{picture}
\caption{$M(\Lambda_c \pi) - M(\Lambda_c)$ distributions for the $\Lambda_c^+$ signal window (points) and scaled sidebands (red histogram). 
Insets show background 
subtracted distributions for the $\Sigma_c(2800)$. 
The peaks at $0.43~\text{GeV}/\text{c}^2$ 
are cross-feeds from $\Lambda_c(2880) \to \Sigma_c(2455) \pi$ 
where the pion from the $\Sigma_c(2455)$ decay is missing.}
\label{fig:lambdapi}
\end{figure}
Belle also studied 
excited charmed baryons decaying to $\Lambda_c \pi$ final states. 
Figure~\ref{fig:lambdapi} shows distributions of the 
$\Delta M(\Lambda_c \pi) \equiv M(\Lambda_c\pi)-M(\Lambda_c)$ mass differences
for the $\Lambda_c^+ \pi^-$, 
$\Lambda_c^+ \pi^0$, and $\Lambda_c^+ \pi^+$ combinations~\cite{lc-isotriplet}. Peaks near $0.51~\text{GeV}/\text{c}^2$ were attributed to 
new baryons forming an isotriplet denoted 
as $\Sigma_c(2800)^{0,+,++}$. 
The measured $\Sigma_c(2800)$ mass splittings relative to 
the $\Lambda_c$ and $\Sigma_c(2800)$ widths are summarized 
in Table~\ref{tab:baryons_parameters}.
These new states could be members 
of the $\Sigma_{c2}$ triplet with $J^P=\frac{3}{2}^-$ with
total angular momentum of the light diquark equal to two, 
which are expected to have 
$\Delta M(\Lambda_c \pi) \approx 0.5~\text{GeV}/\text{c}^2$ 
and a width of $15~\text{MeV}/\text{c}^2$. 
Mixing of the $\Sigma_{c2}$ with other states predicted to lie
nearby, could be a reason for the wider observed state. 
\begin{table}[t]
  \caption{Parameters of charmed baryons 
discovered by Belle. The $\Sigma_c(2800)$ masses  
  were measured with respect 
to the $\Lambda_c$ mass of $2286.46 \pm 0.14~\text{MeV}/\text{c}^2$.} 
  \label{tab:baryons_parameters}
  \begin{center}
\begin{tabular}{ccccc} 
  \hline \hline
 Name & Quark content & Decay mode & Mass (MeV/$c^2$) & Width (MeV/$c^2$)     \\ 
  \hline
 $\Sigma_c(2800)^0$   & $cdd$ & $\Lambda_c^+ \pi^-$ & $515.4^{+3.2+2.1}_{-3.1-6.0} \ + M_{\Lambda_c}$   & $61^{+18+22}_{-13-13}$ \\
 $\Sigma_c(2800)^+$   & $cud$ & $\Lambda_c^+ \pi^0$ & $505.4^{+5.8+12.4}_{-4.6-12.0} \ + M_{\Lambda_c}$ & $62^{+37+52}_{-23-38}$ \\
 $\Sigma_c(2800)^{++}$& $cuu$ & $\Lambda_c^+ \pi^+$ & $514.5^{+3.4+2.8}_{-3.1-4.9} \ + M_{\Lambda_c}$   & $75^{+18+12}_{-13-11}$ \\
  \hline
 $\Xi_c(2980)^+$ & $csu$ & $\Lambda_c^+ K^- \pi^+$   & $2978.5 \pm 2.1 \pm 2.0$ & $43.5 \pm 7.5 \pm 7.0$ \\
 $\Xi_c(2980)^0$ & $csd$ & $\Lambda_c^+ K^0_S \pi^+$ & $2977.1 \pm 8.8 \pm 3.5$ & $ 43.5$ (fixed) \\
 $\Xi_c(3077)^+$ & $csu$ & $\Lambda_c^+ K^- \pi^+$   & $3076.7 \pm 0.9 \pm 0.5$ & $6.2 \pm 1.2 \pm 0.8$ \\
 $\Xi_c(3077)^0$ & $csd$ & $\Lambda_c^+ K^0_S \pi^+$ & $3082.8 \pm 1.8 \pm 1.5$ & $5.2 \pm 3.1 \pm 1.8$ \\
  \hline\hline
\end{tabular}
  \end{center}
\end{table}

\begin{figure}
\centerline{
\includegraphics[width=5.8cm,height=3.7cm]{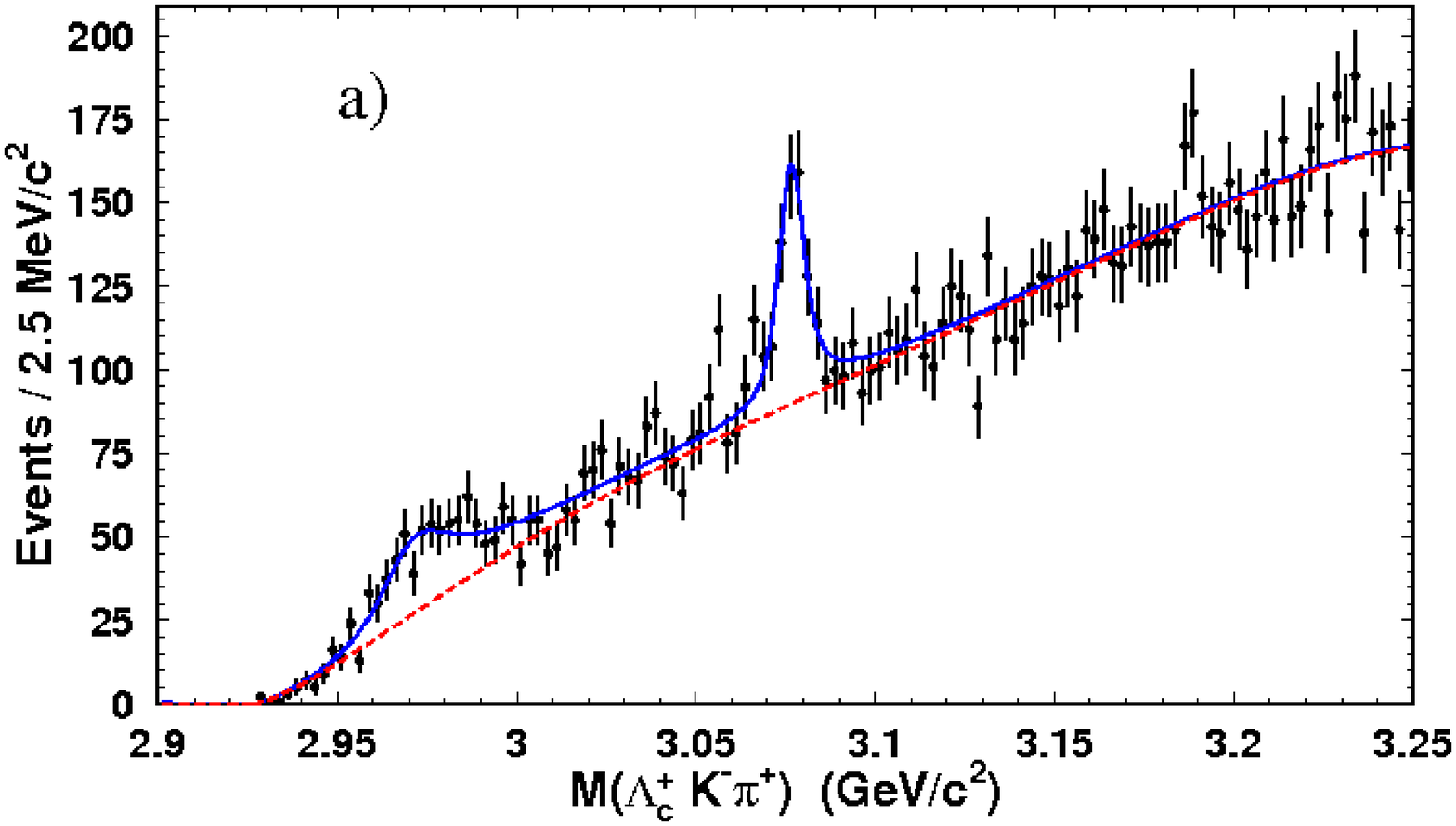}
\includegraphics[width=5.8cm,height=3.7cm]{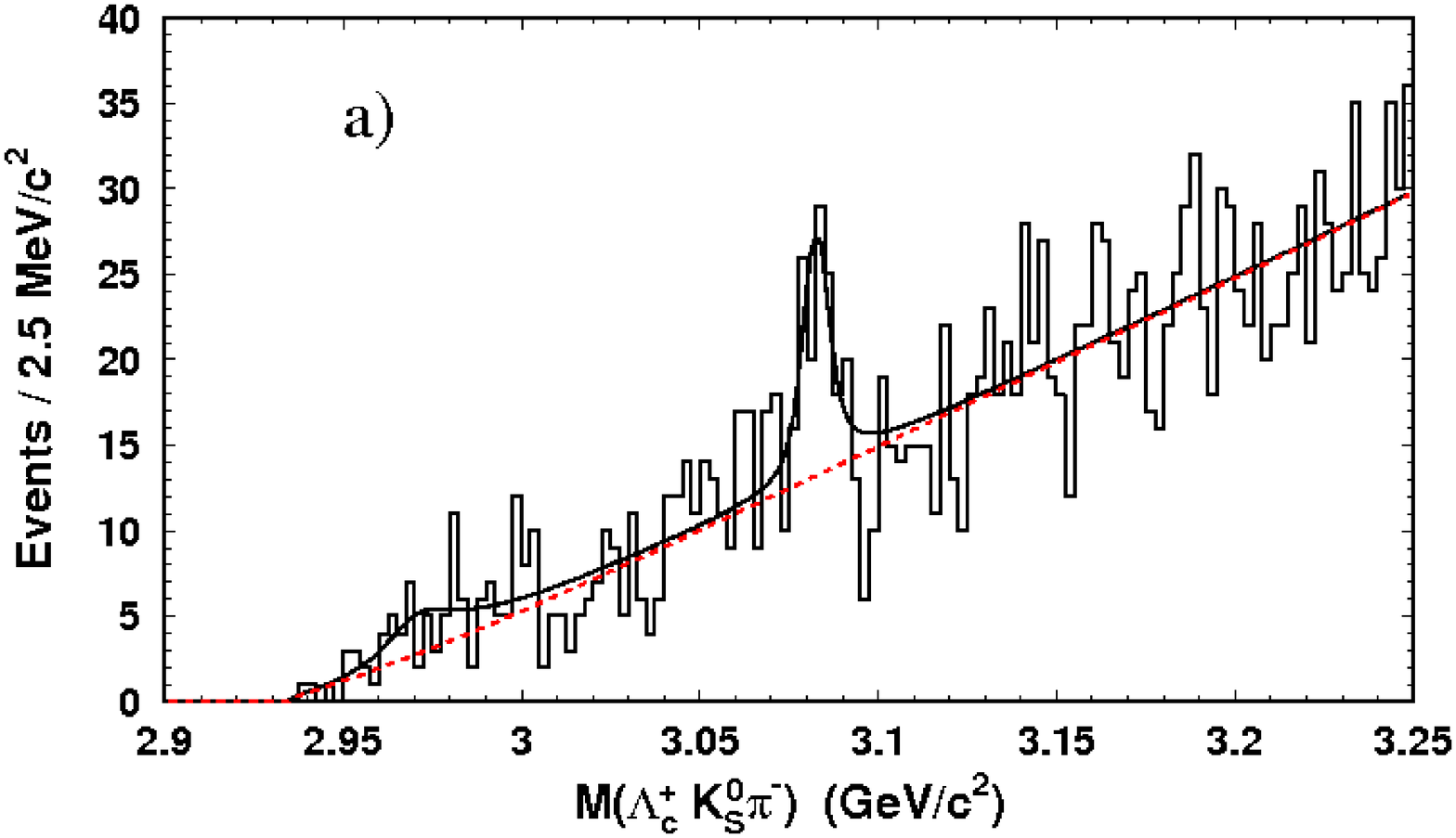}}
\caption{Distributions of $M(\Lambda_c^+ K^- \pi^+)$ (left) and 
$M(\Lambda_c^+ K^0_s \pi^-)$ (right) with the fit curves overlaid. }
\label{fig:xic}
\end{figure}

For charmed-strange baryons formed from 
$csd$ or $csu$ quarks, in addition to the ground states  
$\Xi_c^{(') 0, +}$ and the $\frac{3}{2}^+$ spin excitation 
$\Xi_c(2645)^{0, +}$, there were  
also two candidates for $P$-wave excitations, 
the $\Xi_c(2790)$ and $\Xi_c(2815)$, observed  
in the $\Xi_c^{'}\pi$ and $\Xi_c(2645) \pi$ final states,
respectively. Belle studied $\Xi_c$ states 
decaying into $\Lambda_c K \pi$~\cite{xic_lckpi}, which requires the
$c$ and $s$ quarks in the initial states to be 
carried away by different daughter particles. 
Two peaks in the $M(\Lambda_c^+ K^- \pi^+)$ spectrum 
shown in Fig.~\ref{fig:xic}, were 
attributed to the new excited baryons denoted
 $\Xi_{cx}(2980)^+$ and $\Xi_{cx}(3077)^+$.  
Their neutral isospin partners were found in $\Lambda_c^+ K^0_s \pi^-$.   
The $\Xi_{cx}(2980)$ and $\Xi_{cx}(3077)$ parameters, obtained from 
 fit to the $M(\Lambda_c K \pi)$ distribution,  
are summarized in Table~\ref{tab:baryons_parameters}. 
The spin parity assignments for these baryons remain unknown.  

\def \gg {{\gamma\gamma}}
\def \ee {e^+e^-}
\def \ggX {\gg\to X}
\def \eeX {\ee\to\ee X}
\def \Ggg {\Gamma_\gg}
\def \simgt {\stackrel{\displaystyle >}{\raisebox{-0.5ex}{$\sim$}}}
\def \simlt {\stackrel{\displaystyle <}{\raisebox{-0.5ex}{$\sim$}}}
\def \GeV {{\rm GeV}}
\def \MeV {{\rm MeV}}
\def \keV {{\rm keV}}
\def \eV {{\rm eV}}

\def \ee {e^+e^-}
\def \qq {q\bar{q}}
\def \pipi {\pi^+\pi^-}
\def \mumu {\mu^+\mu^-}
\def \KK {K^+K^-}
\def \MM {M^+M^-}
\def \ggqq {\gg\rightarrow\qq}
\def \ggpipi {\gg\rightarrow \pipi}
\def \ggKK {\gg\rightarrow \KK}
\def \B {\cal B}
\newcommand{\Gcenter}[2]{
  \dimen0=\ht\strutbox%
  \advance\dimen0\dp\strutbox%
  \multiply\dimen0 by#1%
  \divide\dimen0 by2%
  \advance\dimen0 by-.5\normalbaselineskip
  \raisebox{-\dimen0}[0pt][0pt]{#2}}

\section{Two-photon physics}
\label{chap_2photon}
An $e^+e^-$ collider is also a $\gamma\gamma$ collider.
Through measurements of two-photon collision processes, we can  
study hadron spectroscopy. Two-photon physics
at Belle includes searches for new resonances, 
tests of perturbative QCD, and measurements of photon\---meson couplings
and form factors. 
In this section, we report our investigations of scalar resonances and 
QCD tests in meson-pair production processes from two-photon collisions 
in the energy range between 1~GeV and 3~GeV, respectively. New
resonances produced in two-photon processes are discussed in Sect.~9.

\subsection{Hadron physics and QCD}
\begin{figure}[b]
\centering
\includegraphics[width=4.7cm,clip]{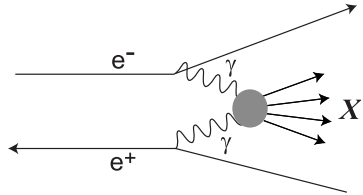}
\caption{A two-photon collision diagram for the process $e^+e^- \to e^+e^-X$.}
\label{fig:ggX}
\end{figure}

 The Feynman diagram for the two-photon process, $\gamma \gamma \to X$ 
at an $e^+e^-$ collider  is shown in Fig.~\ref{fig:ggX}, 
where the reaction is regarded as a collision of two photons, each of
which is emitted from one of the initial $e^+e^-$ beams,
i.e. $e^+e^- \to 
e^+e^-\gamma\gamma \to e^+e^- X$.  The CM energy of
the two-photon collision system covers a wide range, and
hence $\gamma \gamma$ reactions can be measured 
over a continuous and broad energy range.
Usually, two-photon measurements are performed by
exclusively reconstructing the final-state particle system $X$ 
in order to determine the collision energy 
of the two photons ($W=M_X$) for each event.

 Meson resonance formation processes are explored
in measurements in the low energy region ($W \simlt 3$~GeV). 
Since two or more overlapping resonances are often produced 
 we extract each component from a partial-wave analysis, 
which takes into account interference.    
It is known that some 
light-quark scalar mesons, such as the $f_0(980)$ and $a_0(980)$, cannot
easily be explained in a $q\bar{q}$ constituent model.
The two-photon decay width $\Gamma_{\gamma\gamma}$ of these light
quark mesons, which is measured by two-photon processes,
is the most important parameter that provides information
on the internal structure of such mesons.

In the higher energy region ($W \simgt 3$~GeV), we study the 
properties of charmonia and search for new hadronic states
with even charge-conjugation $C$. Since the contributions from resonances 
are relatively small in this region, we can test QCD 
by measuring the differential cross section of
meson-pair production processes, $\gamma \gamma \to M M'$, which
is calculated theoretically in a model with
quark-pair production $\gamma \gamma \to q \bar{q}$ followed by
quark hadronization. The hadronization part is described by several
different models based on perturbative and non-perturbative QCD.  
The Belle data sample has been used to perform such QCD tests 
with by far the highest statistics to date.

\subsection{Principles of a two-photon process measurement 
at an $e^+e^-$ collider}
In a two-photon process at an $e^+e^-$ collider, 
photons emitted from the beam particles are always virtual,
and the four-momentum squared ($q^2$, which is the same as
the invariant mass squared of the photon) is always negative.  
The virtuality of the photon $Q^2(=-q^2)$ is well approximated 
as $Q^2 = 4 E_b E' \sin^2 \frac{\theta}{2}$, where $E_b$ is the
CM beam energy, and $E'$ and $\theta$ are the recoil energy 
and the scattering angle of the beam particle, respectively, 
when $\theta$ is not very close to zero. 
However, the emission angle of the photon has a strong peak near 
$\theta \sim 0$ ($Q^2 \sim 0$). When the $\theta$ angles of both 
photons are small and the recoil $e^-$ and $e^+$ are not 
detected, the reaction is regarded to a good approximation as a collision
of two real photons (we call this case a "zero-tag event").

In a zero-tag event, the transverse momentum component ($p_t$) of the 
final-state system $X$ tends to be balanced, that is, close to 
zero. Requiring $p_t$ balance and a much smaller detected energy 
compared with that for the $e^+e^-$ beams, we can easily 
separate the two-photon 
signal process from background $e^+e^-$ annihilation processes. 
However, if $W$ is greater than about $0.5 \sqrt{s}$ for the
$e^+e^-$ beams, measurement of two-photon processes becomes difficult 
due to the large background from annihilation processes and/or the
small statistics of the signal. In the $B$-factory energy
range, measurements up to $W \simlt 4.5$~GeV are feasible.

 The measured cross section $\sigma(e^+e^- \to e^+e^- X)$ can be
translated into a two-photon collision cross section
using the relation :
\[
\sigma(e^+e^- \to e^+e^- X) = \int \sigma(\gamma \gamma \to X;W)
\frac{dL_{\gamma\gamma}}{dW} dW,
\]
where $dL_{\gamma\gamma}/dW$ is the two-photon luminosity function
calculated in QED as a probability density distribution for the CM 
energy of the two-photon systems, which are emitted from the incident $e^+e^-$.
The two-photon cross section
 depends very weakly (logarithmically) on the $e^+e^-$ beam energy. 

\subsection{Single meson formation process}
If only one meson is produced in a collision of two real photons, 
the quantum numbers of the meson, $C$, and spin-parity ($J^{PC}$) are
restricted to be $({\rm even})^{\pm +}$ or $({\rm odd}, J \neq 1)^{++}$.
The production of $J=1$ mesons is forbidden. Thus, two-photon
production is complementary to $e^+e^-$ annihilation processes
where only $1^{--}$ mesons are produced directly.
 
In these processes, the production cross section of a meson 
is proportional to its two-photon decay 
width $\Gamma_{\gamma\gamma}$ via the relation:
\[
\sigma(W) = 8\pi (2J+1)\frac{\Gamma_{\gamma\gamma}(R)\Gamma_R{\cal B}
(R \to {\rm final\ state})}{(W^2 - M_R^2)^2 + M_R^2 \Gamma_R^2},
\]
where $M_R$ and $\Gamma_R$ are the mass and total width of the meson,
and ${\cal B}$ is the branching fraction.

In the zero-tag mode, we measure only the final-state particles from
the decay of a produced meson.
This significantly reduces backgrounds compared to the case
of $\gamma\gamma$ inclusive meson production.
The two-photon decay width of neutral mesons is a direct 
and sensitive probe of their internal structure, as mentioned above. 
In addition,
detailed analyses of final states are useful to study the
branching fractions and decay structures.

\subsection{Production of light-quark mesons}
\label{sec-3}
Meson production through two-photon processes had been
studied in the past at PEP, PETRA, TRISTAN, and LEP
(see, e.g., the compilation in Ref.~\cite{past_exp}).
However, the more than three orders of magnitude larger
statistics available at a $B$-factory compared to past 
experiments have qualitatively improved the analyses, 
allowing detailed studies of resonances that were impossible 
in the past.

\begin{figure}
 \centering
\includegraphics[width=7.0cm,clip]{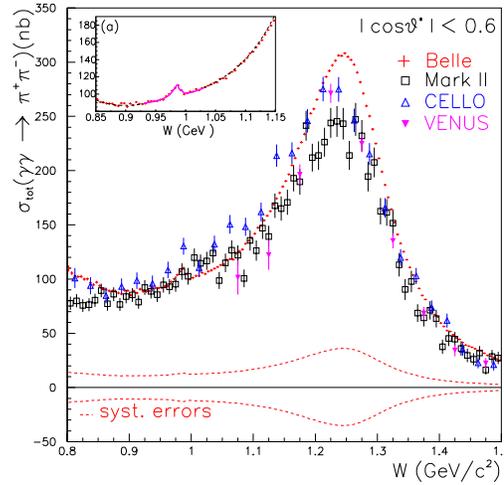}
 \caption{Integrated cross section ($|\cos \theta^*|<0.6$)
of $\gamma \gamma \to \pi^+ \pi^-$. The inset shows an
enlarged view of the Belle data near the $f_0(980)$ peak. A fit
with a resonance parameterization is superimposed.}
\label{fig:pipn1}
\end{figure}

Figure~\ref{fig:pipn1} shows an example of the large Belle 
two-photon data statistics; here we give the integrated cross section
($|\cos \theta^*|<0.6$) for $\gamma \gamma \to \pi^+ \pi^-$ as a function
of $W$, where $\theta^*$ is the angle of the produced 
particle relative to one of the incident photons in the CM system 
of the two photons and $W$ is the total CM energy~\cite{mori1, mori2}.
This analysis used an early Belle data sample with
an integrated luminosity of only 85~fb$^{-1}$, 
($\sim 9\%$ of the full data). The Belle data have
negligibly small error bars and
a structure due to the $f_0(980)$ is clearly visible
near $W \simeq 1~\GeV$, as shown in the inset.

\subsection{Measurements of pseudoscalar-meson-pair
production at Belle}
\label{sub:3-1}
Belle has performed a study of $\gamma \gamma \to P_1 P_2$, where $P_1P_2$
are $\pi^+ \pi^-$~\cite{mori1, mori2, nkzw},
$K^+ K^-$, and $K^0_S K^0_S$~\cite{nkzw, kabe, wtchen},
$\pi^0 \pi^0$~\cite{ pi0pi0, pi0pi02},
$\eta \pi^0\ $~\cite{etapi0}, and $\eta \eta$~\cite{etaeta}.
The angular coverage for charged-particle-pair production is restricted to
the range $|\cos \theta^*| < 0.6$ due to the limitations of 
the charged track triggers.
On the other hand, for $\pi^0 \pi^0$, $\pi^0 \eta$, and $\eta \eta$, we
can extend the angular range to
$|\cos \theta^*| < 0.8$ or even to $|\cos \theta^*| < 1.0$ 
(full angular coverage) owing to the wider coverage of the calorimeter
trigger for multi-photon final-state events. It should be noted that
a wider angular coverage plays an essential role in separating partial waves.

A study of resonance production in two-photon collisions 
gives several resonance ($R$) parameters: its mass, its total width, and
$\Gamma_{\gamma \gamma} {\cal B}(R \rightarrow P_1 P_2)$.
The latter is difficult to obtain otherwise.

\subsection{Differential cross sections and partial wave amplitudes}
Partial waves with even angular momenta 
contribute to the cross section of $\gamma \gamma \to P_1 P_2$.
Up to G waves may be considered 
at low energy ($W \simlt 2~\GeV$)~\cite{pw}.
The differential cross section can then be written as
\begin{equation}
\frac{d \sigma}{d \Omega} (\gamma \gamma \to P_1 P_2)
 = \left| S \: Y^0_0 + D_0 \: Y^0_2  + G_0 \: Y^0_4 \right|^2 
+ \left| D_2 \: Y^2_2  + G_2 \: Y^2_4 \right|^2 \; ,
\label{eqn:diff}
\end{equation}
where $S$ is the $J=0$ partial wave, $D_0$ and $G_0$ ($D_2$ and $G_2$) 
are the helicity-zero (-two) components of partial waves for $J=2$ 
and 4, respectively, and $Y^m_J$s are spherical harmonics; 
partial waves determine the energy ($W$) dependence, while
spherical harmonics govern the angular dependence.
Because spherical harmonics are not independent of each other, the
partial waves cannot be determined by only fitting the differential cross
section.

If we write
\begin{equation}
\frac{d \sigma}{4 \pi d |\cos \theta^*|} (\gamma \gamma \to P_1 P_2)
 = \hat{S}^2 \: |Y^0_0|^2  + \hat{D}_0^2 \: |Y^0_2|^2
+ \hat{D}_2^2  \: |Y^2_2|^2 \, 
+ \hat{G}_0^2  \: |Y^0_4|^2  \, 
+ \hat{G}_2^2  \: |Y^2_4|^2  \, ,
\label{eqn:diff2}
\end{equation}
we can determine the ``hat
amplitudes'' $\hat{S}^2$, $\hat{D}_0^2$,  $\hat{D}_2^2$,  $\hat{G}_0^2$, 
and $\hat{G}_2^2$ by fitting differential cross sections in each $W$-bin,
because the $|Y^m_J|^2$s are independent of each other.
Spectra of hat amplitudes can give useful information on partial waves even
though they contain terms arising from the interference of partial waves
($S$, $D_0$, $D_2$, $G_0$, and $G_2$)~\cite{pi0pi0}.

In order to obtain information on possible resonances, we have to 
parameterize the partial waves and then fit differential cross sections 
according to Eq.~(\ref{eqn:diff}).
Such analyses allow measurement of the two-photon widths of some mesons, 
including the $f_0(980)$ and $a_0(980)$.

The existence of the low-lying scalar nonet ($f_0(500)$ (or $\sigma$), 
$K^*(800)$ (or $\kappa$), $f_0(980)$, and $a_0(980)$) 
is a long-standing puzzle, yet these scalar mesons
are thought to play the role of a ``Higgs boson in QCD'', 
by spontaneously breaking the chiral symmetry of the QCD vacuum~\cite{amsler}.

The measured two-photon widths of $f_0(980)$ and $a_0(980)$ are small
(although the results have large systematic errors)
compared to those of the $f_2(1270)$ and $a_2(1320)$,  
as listed in Table~\ref{tab:gggb}. This pattern of widths
supports a picture in which the low-lying scalar mesons
are made of color-triplet diquark pair~\cite{amsler}.

A more satisfactory way to derive information on partial waves is to do 
partial wave analyses utilizing hadron data of the past and 
fully taking into account theoretical constraints~\cite{penn};
we eagerly await such analyses using the high-statistics data from Belle.

\begin{table}
\begin{center}
\caption{Two-photon width ($\times \B)$}
\label{tab:gggb}
\begin{tabular}{clc} \hline \hline
Meson & $\Gamma_{\gamma \gamma} (\times \B$) (eV) & Ref. \\\hline
$f_0(980)$ & $\Gamma_{\gamma \gamma} = 286 \pm 17^{+211}_{-70} 
$ & \cite{pi0pi0} \\
$a_0(980)$ & $\Gamma_{\gamma \gamma} {\cal B} (\eta \pi^0) = 128 ^{+3+502}_{-2-40}$ & \cite{etapi0} \\
$f_2(1270)$ & $\Gamma_{\gamma \gamma} = 3030 \pm 350
$ & \cite{pdg2012} \\
$a_2(1320)$ & $\Gamma_{\gamma \gamma} = 1000 \pm 60
$ & \cite{pdg2012} \\
\hline \hline
\end{tabular}
\end{center}
\end{table}

\subsection{QCD in the higher energy region}

In the higher energy region $(W\simgt 3~\GeV)$ where 
resonance contributions are small, QCD can be studied by measuring
exclusive two-body hadron production.
It is believed that QCD gives reliable predictions at sufficiently
high energy but the applicable energy is not known.
Belle can measure two-photon processes up to $W$ of 4.5~GeV.
S.J.~Brodsky and G.R.~Farrar predicted 
\begin{eqnarray}
\frac{d\sigma}{dt}=s^{2-n_c}f(\theta^*)
\end{eqnarray}
for the hadron-pair production in a two-photon process
at sufficiently high energy, 
using the Mandelstam variables $s(=W^2)$ and $t$~\cite{scaling}. 
$n_c$ is the total number of elementary particles 
involved in the initial and final states, eight for 
baryon-pair production ($\therefore$ $\sigma\!\sim\! W^{-10}$)
and six for meson-pair production ($\therefore$ $\sigma\!\sim\! W^{-6}$).
S.J.~Brodsky and G.P.~Lepage (BL) also calculated 
the differential cross section for meson-pair production~\cite{bl}.
Their calculation was based on perturbative QCD where 
the perturbatively calculable $\gg\to q\bar{q}$ part is
convoluted with the quark distribution amplitude. They obtained
\begin{eqnarray}
\frac{d\sigma}{d|\!\cos\theta^*\!|}
&=& 
16\pi\alpha^2\frac{|F_M(s)|^2}{s}
\Big\{\frac{(e_1\!-\!e_2)^4}{\sin^4\theta^*} \nonumber\\
&& +\frac{2(e_1e_2)(e_1-e_2)^2}{\sin^2\theta^*}g(\theta^*)\nonumber\\
&& +2(e_1e_2)^2g^2(\theta^*)\Big\},
\label{eq:BL1}
\end{eqnarray}
where $F_M(s)$ 
is the electromagnetic form factor for a meson $M$,
$e_i$ is the charge of a constituent quark,
and $g$ is a function that depends on the quark distribution
function.
For charged meson-pair processes this calculation predicts
$d\sigma/d\cos\theta^*\sim\sin^{-4}\theta^*$, 
and 
$d\sigma(\pipi)/d\sigma(\KK)=(f_K/f_\pi)^4$.
The first term in Eq.~(\ref{eq:BL1}), which is dominant
for charged meson pair processes, 
does not depend on $g$ because the
dependence on the quark distribution function is absorbed into $F_M$.
This prediction was improved by taking into account
the effect of the $s$ quark and modifying distribution functions~\cite{bc}.
Predictions for neutral meson-pair processes are not straightforward, 
since the terms that include $g$ are dominant.

On the other hand, a non-perturbative calculation in the 
handbag model~\cite{dkv} (DKV)  
factorizes the non-perturbative hadronization part 
and gives the differential cross section
\begin{equation}
\frac{d\sigma}{d|\cos\theta^*|}=
\frac{8\pi\alpha^2}{s}\frac{1}{\sin^4\theta^*}|R_{M\overline{M}}(s)|^2.
\end{equation}
Although this model cannot predict absolute values
for the cross sections, it gives a relation between annihilation
form factors $R_{M\bar M}(s)$ in different processes.

The Belle experiment has measured cross sections for
$\pipi$~\cite{nkzw},
$\pi^0\pi^0$~\cite{pi0pi02},
$\eta\pi^0$~\cite{etapi0},
$\eta\eta$~\cite{etaeta},
$\KK$~\cite{nkzw},
$K^0_SK^0_S$~\cite{wtchen}, and
$p\bar{p}$~\cite{kuo} production in two-photon production.
Before the Belle experiment no data were available to test these models 
due to limited statistics and poor particle identification capabilities.

The angular distribution measurements are summarized in Table~\ref{tab:sin4}.
The $W^{-n}$ dependence of the cross section and ratios of cross sections
are listed in Table~\ref{tab:val_n}.
The measured angular dependences agree with $\sin^{-4}\theta^{*}$
except for the $\eta\eta$ process.
We obtained larger $n$ values than the BL prediction
of six, and in the neutral meson-pair process the value is close to
the BC prediction of ten, which may be due to a significant higher 
order contribution in this energy region~\cite{bc}.

The ratios of cross sections asymptotically approach
a constant as energy increases, but no model 
can systematically reproduce all the measured values.

For baryon-pair processes, the measured $n$ value is larger than the 
perturbative QCD prediction of ten, but decreases as $W$ increases~\cite{kuo}.
The angular distribution above 2.5 GeV agrees qualitatively with the
perturbative QCD prediction but has a steeper rise.

\begin{table}
 \caption{%
   Comparison between measured angular distribution and perturbative
   QCD prediction of $\sin^{-4}\theta^*$}
 \label{tab:sin4}
\begin{center}
\begin{tabular}{lcccc} \hline \hline
Mode & $\sin^{-4}\theta^*$ & $W$ (GeV) & $|\cos\theta^*|$ & Ref.\\ \hline
$\pi^+\pi^-$ & Well matched & 3.0\---4.1 & $<0.6$ & \cite{nkzw}\\
$K^+K^-$ & Well matched & 3.0\---4.1 & $<0.6$ & \cite{nkzw}\\
$K^0_SK^0_S$ & Matched & 2.4\---3.3 & $<0.6$ & \cite{wtchen}\\ [.07in]
\Gcenter{2}{$\pi^0\pi^0$} & 
     Better agreement with $\sin^{-4}\theta^*\!+b\cos\theta^*$
     & \Gcenter{2}{2.4 - 4.1} & \Gcenter{2}{$< 0.8$} &
        \Gcenter{2}{\cite{pi0pi0}}\\
& Approaches $\sin^{-4}\theta^*$ above 3.1 GeV & & \\ [.07in]
$\eta\pi^0$ & Good agreement above 2.7 GeV & 3.1\---4.1 & $< 0.8$ & \cite{etapi0} \\ [.07in]
\Gcenter{2}{$\eta\eta$} & Poor agreement & \Gcenter{2}{2.4 - 3.3} &
                             \Gcenter{2}{$< 0.9$} 
& \Gcenter{2}{ \cite{etaeta}}\\
& Close to $\sin^{-6}\theta^*$ above 3.0 GeV & & \\\hline\hline
\end{tabular}\\
\end{center}
\end{table}

\begin{table}
 \caption{
   Energy dependence of the measured cross section.
 ($n$ value in $\sigma_0\propto W^{-n}$)
   and ratios of $\sigma_0$ between different processes.
   $\sigma_0$ is the cross section integrated over
   the sensitive angular region.
   An SU(3) octet (mixture of octet and singlet with mixing angle
   $-18^\circ$) is assumed for the $\eta$ meson.
   $R_f$ is the ratio of decay constants squared, $f^2_\eta/f^2_{\pi^0}$.
 }
 \label{tab:val_n}
 \begin{center}
  \begin{tabular}{lcccccc} \hline \hline
   Process & $n$ or $\sigma_0$ ratio & $W$(GeV) & $|\cos\theta^*|$ &
   BL~\cite{bl} & BC~\cite{bc} & DKV~\cite{dkv} \\\hline
   $\pi^+\pi^-$ & $7.9 \pm 0.4 \pm 1.5$ & 3.0\---4.1 & $<0.6$ & 6 & 6 & \\
   $K^+K^-$  & $7.3 \pm 0.3 \pm 1.5$ & 3.0\---4.1 & $<0.6$ & 6 & 6 & \\
   $K^0_SK^0_S$  & $10.5 \pm 0.6 \pm 0.5$ & 2.4\---4.0 &
          $<0.6$ & 6 & 10 & \\
   $\pi^0\pi^0$ & $8.0 \pm 0.5 \pm 0.4$ & 3.1\---4.1 & $<0.8$
          & 6 & 10 & \\
   $\eta\pi^0$ & $10.5 \pm 1.2 \pm 0.5$ & 3.1\---4.1 & $<0.8$ & 6 & 10 & \\
   $\eta\eta$ & $7.8 \pm 0.6 \pm 0.4$ & 2.4 -- 3.3 & $<0.8$ & 6 & 10 & \\
   $p\bar{p}$ & $12.4^{+2.4}_{-2.3}$ & 3.2\---4.0 & $<0.6$ & 10 & & \\
   \hline
   $K^{+}K^{-}/\pi^+\pi^-$ & $0.89\pm 0.04\pm 0.15$ & 3.0\---4.1 & $<
          0.6$ & 2.3 & 1.06 &  \\
   $K^0_SK^0_S/K^{+}K^{-}$ & $\sim 0.13$ to $\sim 0.01$ & 2.4\---4.0 & $< 0.6$
          & & 0.005 & 0.08 \\
   $\pi^0\pi^0/\pi^+\pi^-$ & $0.32\pm 0.03\pm 0.06$ & 3.1\---4.1 & $<
          0.6$ & & 0.04\---0.07 & 0.5  \\
   $\eta\pi^0/\pi^0\pi^0$ & $0.48\pm 0.05\pm 0.04$ & 3.1\---4.0 & $< 0.8$
          & $0.24R_f$($0.46R_f)$ & &  \\
   $\eta\eta/\pi^0\pi^0$ & $0.37\pm 0.02\pm 0.03$ & 2.4\---3.3 & $< 0.8$
          & $0.36R_f^2$($0.62R_f^2)$ & &  \\
   \hline\hline
  \end{tabular}\\
 \end{center}
\end{table}

\subsection{Summary and outlook}
Two-photon processes can be a background when
studying $CP$ violation, the main theme at a $B$-factory, as well 
as for other physics topics.
However, a detailed study of two-photon data can contribute much 
to the understanding of hadron physics in its own right as described above.
The topics that can be addressed are divided into four categories: 
the search for and study of new or exotic particles, the production 
and decay structure of charmonia, the nature of light-quark resonances, 
and tests of perturbative QCD.
The overwhelming statistics available at a $B$-factory has opened
a new era in two-photon physics.

Our study so far has mostly been limited to collisions of two real photons;
a vast unstudied region remains open for future investigation, 
in which one or both of the photons are virtual, i.e. the study of 
single and double tagged two-photon physics.


\section{Summary}

The Belle experiment at KEKB is described in 
Sect.~\ref{chap_detector}. Belle accomplished its main mission,
which was the verification of Kobayashi and Maskawa's bold proposal
that a single irreducible complex phase can explain all 
matter\---antimatter asymmetries ($CP$-violating phenomena).

As discussed in detail in Sect.~\ref{chap_angles}, Belle's observation 
of large time-dependent $CP$ asymmetries in modes such as $B\to J/\psi K_S$
(together with similar results from BaBar) in 2001 demonstrated
that the KM proposal was correct and laid the foundation
for their 2008 Nobel Prize in Physics. In addition, the results
provided a theoretically clean measurement of one of the interior
angles of the unitarity triangle, $\phi_1$ (or $\beta$). 
After the accumulation of the one ab$^{-1}$ data set, the measurements
of $CP$ asymmetries involving $\phi_1$ became precision results
and important calibrations for new physics studies.

To check the consistency of the SM of particle physics,
it is also necessary to measure the other two interior angles of
the unitarity triangle, $\phi_2$ (or $\alpha$) and $\phi_3$ (or $\gamma$). 
Although theoretical plans for the determination of these angles
were proposed at the start of the $B$-factories, the final and most
precise results were obtained by new methods that were not
originally anticipated; e.g. for $\phi_3$, the best sensitivity was
obtained from Dalitz analysis of $B\to D K$, $D\to K_S \pi^+ \pi^-$ decays.

The development of the methods for determination of the length of
the sides of the unitarity triangle also followed a somewhat
unexpected path that was determined by the convergence of high
statistics $B$-factory data and theoretical insight. 
The results and methods used
for $|V_{cb}|$ and $|V_{ub}|$ determination are described in 
Sect.~\ref{section_slb}.

The results for the sides and interior angles of the unitarity
triangle are consistent. However, reasonably large new physics
contributions, of order 10\% the size of the SM amplitude,
are still allowed. In parallel with the work on fixing the weak
interaction parameters of the unitarity triangle, Belle also completed 
a decade of studies and publications on rare decays, as described in
Sect.~\ref{chap_rare}. 

In rare decays for which the SM
amplitude contribution is highly suppressed, the effects 
of NP could be clear and dramatic. Belle established the
existence of a number of highly suppressed processes including
$b\to d\gamma$ and $b\to s \ell^+\ell^-$.
In addition, as the data sample has increased, there have been 
a number of intriguing hints of NP in various channels, e.g. 
exclusive hadronic $b\to s$ $CP$-violating modes, $B\to \tau \nu$, 
and $B\to K^*\ell^+\ell^-$, but so far there is no compelling
evidence of NP at the current level of sensitivity in Belle. 
Exploration of NP will require the luminosity of SuperKEKB 
and Belle II.

A $B$-factory is also a high energy tau-charm factory and has the 
largest samples of $\tau$ leptons and reconstructed charm. The 
results on $\tau$ lepton physics are described in Sect.~\ref{chap_tau}.
Searches for lepton-flavor-violating (LFV) decays and $CP$ violation
in the $\tau$ sector have reached an interesting sensitivity at Belle
but so far no NP signals have been found. The foundation for
Belle II explorations of this sector has been established. The results 
on charm are discussed in Sect.~\ref{chap_charm}. The highlights
include two classes of unexpected and unanticipated results:
the discovery of $D$\---$\bar{D}$ mixing and the existence of a large
number of new charmonium-like resonances (Sect.~\ref{chap_resonances}). 
The latter was completely unexpected by the theoretical community 
and was guided by Belle data.

Belle is also the world's leading two-photon facility. The results
in this domain of physics are discussed in Sect.~\ref{chap_2photon}. 
Finally, KEKB's capabilities to operate in a range of center of mass
energies allowed Belle to record a number of {\it unique} large data sets
at the $\Upsilon(1S)$, $\Upsilon(2S)$, and $\Upsilon(5S)$ resonances.
The $\Upsilon(5S)$ data were used, as expected, to study some
properties and decays of $B_s$ mesons (Sect.~\ref{chap_bs}). 
However, theorists did not anticipate
that these data could be used to discover a series of peculiar
bottomonium-like resonances or find the missing bottomonia states
such as the $\eta_{b}(2S)$, $h_b(1P)$, and $h_b(2P)$. These discoveries
in hadron spectroscopy are described in Sect.~\ref{chap_resonances}.

In addition to establishing the KM model, measuring weak 
interaction parameters, and observing suppressed SM processes,
the analysis of Belle data was marked by a series of unexpected
discoveries driven by data. At the next stage in Belle II at
SuperKEKB, the focus will shift to NP exploration. 
However, it is likely that the large increase in luminosity 
will also lead to unanticipated results and discoveries.

\section*{Acknowledgements}
We thank the KEKB group for excellent operation of the
accelerator; the KEK cryogenics group for efficient solenoid
operations; and the KEK computer group, the NII, and
PNNL/EMSL for valuable computing and SINET4 network support.
We acknowledge support from MEXT, JSPS, and Nagoya's TLPRC (Japan);
ARC and DIISR (Australia); NSFC (China); MSMT (Czechia);
DST (India); INFN (Italy); MEST, NRF, GSDC of KISTI, and WCU (Korea);
MNiSW (Poland); MES and RFAAE (Russia); ARRS (Slovenia);
SNSF (Switzerland); NSC and MOE (Taiwan); and DOE and NSF (USA).


\begin{thebibliography}{99}
\bibitem{footnote:phi1phi2phi3}
Another naming convention $\beta(=\phi_1)$, $\alpha(=\phi_2)$ and 
$\gamma(=\phi_3)$ is also used in the literatures.
\bibitem{Ref:taunuavg} The semileptonic tag result, $\BF_{\rm SL}$, is
rescaled by using the updated branching fraction of a background mode
so that a consistent set of input parameters are used in the combined
fit.
\bibitem{sm_expect}
The exclusive $K^{(*)} l^+ l^-$ was modeled according to Refs.~\cite{ali1,ali2}, 
while non-resonant $X_s l^+l^-$ with $X_s$ mass above 1.0~GeV$/c^2$ was based 
on a model described by Refs.~\cite{ali1,kruger} and the Fermi motion 
model~\cite{fermi}, following by JETSET~\cite{jetset} to hadronize the system 
with a strange quark and a spectator quark.
\bibitem{remi_full} 
This result is obtained by Belle using 121.4~fb$^{-1}$ of data and 
the method described in Ref.~\cite{Dspi}.
\bibitem{charge-conjugates} 
Charge-conjugate modes are implicitly included.
\bibitem{pw} We denote individual partial
waves by roman letters and parameterized waves by italic.
\end{thebibliography}
\end{document}